\PassOptionsToPackage{dvipsnames}{xcolor}
\let\arxivKernelLabel\label
\documentclass[10pt,showpacs,letterpaper,aps,nofootinbib,prx, twocolumn,superscriptaddress]{revtex4-2}

 \usepackage{tikz}
 \usetikzlibrary{patterns,arrows,calc,decorations.pathmorphing,positioning}

\usepackage{amsmath,amssymb,amsfonts,amsthm,gensymb,bm,physics,mathtools,graphicx,array,float,csquotes,comment,bbold,syntonly,textcomp,centernot}
\usepackage[title]{appendix}
\usepackage[format=plain,labelfont=it,textfont=it]{caption}
\usepackage{chngcntr}
\usepackage{color}
	\definecolor{mygray}{RGB}{222, 222, 222}
\usepackage{enumitem}
	\setlist{nosep}
\usepackage[none]{hyphenat}

\usepackage[indent,parfill]{parskip}%
\usepackage[nottoc,numbib]{tocbibind}
	\counterwithout{figure}{section}
\usepackage{xparse}%
\usepackage{dsfont}
\IfFormatAtLeastTF{2023-06-01}{\let\label\arxivKernelLabel}{}
\usepackage[colorlinks=false,hidelinks,hypertexnames=false]{hyperref}
\usepackage[capitalize]{cleveref}
\usepackage{soul}
\usepackage{tabstackengine}
\usepackage{tikzit}
\usepackage{changepage}
\setstackEOL{\cr}

\usepackage{thmtools}
\usepackage{thm-restate}
\makeatletter
\@ifundefined{newcounteralias}{}{%
  \renewcommand\thmt@autorefsetup{%
    \@xa\def\csname\thmt@envname autorefname\@xa\endcsname\@xa{\thmt@thmname}}%
}
\newcommand{\mainonlyfootnote}[1]{%
  \ifthmt@thisistheone
    \footnote{#1}%
  \fi
}

\newcommand{\mainonly}[1]{%
  \ifthmt@thisistheone
    #1%
  \fi
}

\newcommand{\restateonly}[1]{%
  \ifthmt@thisistheone
  \else
    #1%
  \fi
}
\makeatother

\newtheorem{theorem}{Theorem}
\newtheorem{corollary}[theorem]{Corollary}
\newtheorem{lemma}[theorem]{Lemma}
\newtheorem{proposition}[theorem]{Proposition}
\newtheorem{definition}[theorem]{Definition}
\newtheorem{remark}[theorem]{Remark}

\newtheorem{example}[theorem]{Example}

\newcommand{\beq}{\begin{equation}}
\newcommand{\eeq}{\end{equation}}
\newcommand{\bea}{\begin{align}}
\newcommand{\eea}{\end{align}}

\newcommand{\iffupdown}{\mathrel{\rotatebox[origin=c]{90}{$\iff$}}}

	\DeclarePairedDelimiterX{\Set}[1]{\{}{\}}{%
		
		#1
	}
	\makeatletter
		\let\oldSet\Set
		\def\Set{\@ifstar{\oldSet}{\oldSet*}}
	\makeatother

	\DeclarePairedDelimiterX{\Family}[1]{(}{)}{%
		
		#1
	}
	\makeatletter
\let\oldFamily\Family
\def\Family{\@ifstar{\oldFamily}{\oldFamily*}}
\makeatother

	\newsavebox{\numbox}%
	\newsavebox{\slashbox}%
	\newsavebox{\denbox}%
	\newlength{\slashlength}%
	\newlength{\faktorscale}%
	\DeclareDocumentCommand{\newfaktor}{m O{0.35} m O{-0.35}}{%
		\savebox{\numbox}{\ensuremath{#1}}%
		\savebox{\slashbox}{\ensuremath{\diagup}}%
		\savebox{\denbox}{\ensuremath{#3}}%
		\setlength{\faktorscale}{0.5\ht\numbox+0.5\ht\denbox}%
		\setlength{\slashlength}{2pt+0.8\faktorscale+#2\faktorscale-#4\faktorscale}%
		\raisebox{#2\ht\slashbox}{\usebox{\numbox}}%
		\mkern-2mu%
		\rotatebox{-30}{\rule[#4\ht\denbox]{0.4pt}{\slashlength}}%
		\mkern9mu%
		\hspace{-0.44\slashlength}%
		\raisebox{#4\ht\denbox}{\usebox{\denbox}}%
	}
	\DeclareDocumentCommand{\linefaktor}{m O{0.08} m O{-0.08}}{%
		\savebox{\numbox}{\ensuremath{#1}}%
		\savebox{\slashbox}{\ensuremath{\diagup}}%
		\savebox{\denbox}{\ensuremath{#3}}%
		\setlength{\faktorscale}{0.5\ht\numbox+0.5\ht\denbox}%
		\setlength{\slashlength}{0.2\faktorscale+0.8\baselineskip}%
		\raisebox{#2\ht\slashbox}{\usebox{\numbox}}%
		\mkern-1mu%
		\raisebox{-0.8pt}{%
			\rotatebox{-30}{\rule[#4\ht\denbox]{0.4pt}{\slashlength}}%
		}%
		\mkern-1mu%
		\hspace{-0.25\slashlength}%
		\raisebox{#4\ht\denbox}{\usebox{\denbox}}%
	}

\definecolor{js}{rgb}{0.578125,0.23828125,0.9609375}

\newcommand{\blk}{\color{black}}

\usepackage[normalem]{ulem}
\usepackage{cancel}

\def\End{{\rm End}}
\def\Hom{{\rm Hom}}

\def\dim{{\rm dim}}
\def\GL{{\rm GL}}
\def\spann{{\rm span}}
\def\CPTP{{\rm CPTP}}

\def\I{{\mathbf 1}}

\def\spann{{\rm span}}

\def\Im{{\rm Im}}
\def\Herm{{\rm Herm}}

\def\Sys{{\mathbf{Sys}}}

\def\Cor{{\rm Cor}}

\def\Semi{{\rm Semi}}

\newcommand{\SU}{\mathrm{SU}}
\newcommand{\SO}{\mathrm{SO}}

\newcommand{\ca}{\mathcal A}

\newcommand{\cc}{\mathcal C}

\newcommand{\ch}{\mathcal H}

\newcommand{\cm}{\mathcal M}

\newcommand{\cP}{\mathcal P}

\newcommand{\calr}{\mathcal R}
\newcommand{\cs}{\mathcal S}
\newcommand{\ct}{\mathcal T}
\newcommand{\cu}{\mathcal U}

  \newcommand{\Fl}{\mathfrak{L}}

\def\Nl{{\mathchoice
{\setbox0=\hbox{$\displaystyle\rm N$}\hbox{\hbox to0pt
{\kern0.4\wd0\vrule height0.9\ht0\hss}\box0}}
{\setbox0=\hbox{$\textstyle\rm N$}\hbox{\hbox to0pt
{\kern0.4\wd0\vrule height0.9\ht0\hss}\box0}}
{\setbox0=\hbox{$\scriptstyle\rm N$}\hbox{\hbox to0pt
{\kern0.4\wd0\vrule height0.9\ht0\hss}\box0}}
{\setbox0=\hbox{$\scriptscriptstyle\rm N$}\hbox{\hbox to0pt
{\kern0.4\wd0\vrule height0.9\ht0\hss}\box0}}}}
\def\Zl{{\mathchoice
{\setbox0=\hbox{$\displaystyle\rm Z$}\hbox{\hbox to0pt
{\kern0.4\wd0\vrule height0.9\ht0\hss}\box0}}
{\setbox0=\hbox{$\textstyle\rm Z$}\hbox{\hbox to0pt
{\kern0.4\wd0\vrule height0.9\ht0\hss}\box0}}
{\setbox0=\hbox{$\scriptstyle\rm Z$}\hbox{\hbox to0pt
{\kern0.4\wd0\vrule height0.9\ht0\hss}\box0}}
{\setbox0=\hbox{$\scriptscriptstyle\rm Z$}\hbox{\hbox to0pt
{\kern0.4\wd0\vrule height0.9\ht0\hss}\box0}}}}
\def\Ql{{\mathchoice
{\setbox0=\hbox{$\displaystyle\rm Q$}\hbox{\hbox to0pt
{\kern0.4\wd0\vrule height0.9\ht0\hss}\box0}}
{\setbox0=\hbox{$\textstyle\rm Q$}\hbox{\hbox to0pt
{\kern0.4\wd0\vrule height0.9\ht0\hss}\box0}}
{\setbox0=\hbox{$\scriptstyle\rm Q$}\hbox{\hbox to0pt
{\kern0.4\wd0\vrule height0.9\ht0\hss}\box0}}
{\setbox0=\hbox{$\scriptscriptstyle\rm Q$}\hbox{\hbox to0pt
{\kern0.4\wd0\vrule height0.9\ht0\hss}\box0}}}}
\def\Rl{{\mathchoice
{\setbox0=\hbox{$\displaystyle\rm R$}\hbox{\hbox to0pt
{\kern0.4\wd0\vrule height0.9\ht0\hss}\box0}}
{\setbox0=\hbox{$\textstyle\rm R$}\hbox{\hbox to0pt
{\kern0.4\wd0\vrule height0.9\ht0\hss}\box0}}
{\setbox0=\hbox{$\scriptstyle\rm R$}\hbox{\hbox to0pt
{\kern0.4\wd0\vrule height0.9\ht0\hss}\box0}}
{\setbox0=\hbox{$\scriptscriptstyle\rm R$}\hbox{\hbox to0pt
{\kern0.4\wd0\vrule height0.9\ht0\hss}\box0}}}}
\def\Cl{{\mathchoice
{\setbox0=\hbox{$\displaystyle\rm C$}\hbox{\hbox to0pt
{\kern0.4\wd0\vrule height0.9\ht0\hss}\box0}}
{\setbox0=\hbox{$\textstyle\rm C$}\hbox{\hbox to0pt
{\kern0.4\wd0\vrule height0.9\ht0\hss}\box0}}
{\setbox0=\hbox{$\scriptstyle\rm C$}\hbox{\hbox to0pt
{\kern0.4\wd0\vrule height0.9\ht0\hss}\box0}}
{\setbox0=\hbox{$\scriptscriptstyle\rm C$}\hbox{\hbox to0pt
{\kern0.4\wd0\vrule height0.9\ht0\hss}\box0}}}}
\def\Hl{{\mathchoice
{\setbox0=\hbox{$\displaystyle\rm H$}\hbox{\hbox to0pt
{\kern0.4\wd0\vrule height0.9\ht0\hss}\box0}}
{\setbox0=\hbox{$\textstyle\rm H$}\hbox{\hbox to0pt
{\kern0.4\wd0\vrule height0.9\ht0\hss}\box0}}
{\setbox0=\hbox{$\scriptstyle\rm H$}\hbox{\hbox to0pt
{\kern0.4\wd0\vrule height0.9\ht0\hss}\box0}}
{\setbox0=\hbox{$\scriptscriptstyle\rm H$}\hbox{\hbox to0pt
{\kern0.4\wd0\vrule height0.9\ht0\hss}\box0}}}}
\def\Ol{{\mathchoice
{\setbox0=\hbox{$\displaystyle\rm O$}\hbox{\hbox to0pt
{\kern0.4\wd0\vrule height0.9\ht0\hss}\box0}}
{\setbox0=\hbox{$\textstyle\rm O$}\hbox{\hbox to0pt
{\kern0.4\wd0\vrule height0.9\ht0\hss}\box0}}
{\setbox0=\hbox{$\scriptstyle\rm O$}\hbox{\hbox to0pt
{\kern0.4\wd0\vrule height0.9\ht0\hss}\box0}}
{\setbox0=\hbox{$\scriptscriptstyle\rm O$}\hbox{\hbox to0pt
{\kern0.4\wd0\vrule height0.9\ht0\hss}\box0}}}}

\def\Fl{{\mathchoice
{\setbox0=\hbox{$\displaystyle\rm F$}\hbox{\hbox to0pt
{\kern0.4\wd0\vrule height0.9\ht0\hss}\box0}}
{\setbox0=\hbox{$\textstyle\rm F$}\hbox{\hbox to0pt
{\kern0.4\wd0\vrule height0.9\ht0\hss}\box0}}
{\setbox0=\hbox{$\scriptstyle\rm F$}\hbox{\hbox to0pt
{\kern0.4\wd0\vrule height0.9\ht0\hss}\box0}}
{\setbox0=\hbox{$\scriptscriptstyle\rm F$}\hbox{\hbox to0pt
{\kern0.4\wd0\vrule height0.9\ht0\hss}\box0}}}}

\usetikzlibrary{shapes.multipart}
\colorlet{green}{Green}

\tikzstyle{xor}=[circuit ee IEC, bulb,fill=white,rotate=45]
\tikzstyle{merge}=[circuit ee IEC, bulb,fill=white]
\tikzstyle{dmerge}=[circuit ee IEC, bulb,fill=black!20]
\tikzset{->-/.style={decoration={
  markings,
  mark=at position .5 with {\arrow{>}}},postaction={decorate}}}
\tikzset{-<-/.style={decoration={
  markings,
  mark=at position .5 with {\arrow{<}}},postaction={decorate}}}

\tikzstyle{bwSpider}=[
       rectangle split,
       rectangle split parts=2,
       rectangle split part fill={black,white},
 minimum size=3.6 mm, inner sep=-2mm, draw=black,scale=0.5,rounded corners=0.8 mm
       ]
 \tikzstyle{wbSpider}=[
       rectangle split,
       rectangle split parts=2,
       rectangle split part fill={white,black},
 minimum size=3.6 mm, inner sep=-2mm, draw=black,scale=0.5,rounded corners=0.8 mm
       ]

\tikzstyle{epiCopoint}=[regular polygon,regular polygon sides=3,draw,scale=0.75,inner sep=-0.5pt,minimum width=5mm,fill=white,regular polygon rotate=0,line width=1pt]
\tikzstyle{epiPoint}=[regular polygon,regular polygon sides=3,draw,scale=0.75,inner sep=-0.5pt,minimum width=5mm,fill=white,regular polygon rotate=180,line width=1pt]
\tikzstyle{epiPointWide}=[regular polygon,regular polygon sides=3,draw,scale=0.75,inner sep=-0.5pt,minimum width=8mm,fill=white,regular polygon rotate=180,line width=1pt]
\tikzstyle{epiBox}=[fill=white,draw, line width = 1pt,inner sep=0.6mm,font=\footnotesize,minimum height=3mm,minimum width=3mm]
\tikzstyle{epiBoxWide}=[fill=white,draw, line width = 1pt,inner sep=0.6mm,font=\footnotesize,minimum height=3mm,minimum width=5mm]
\tikzstyle{epiBoxVeryWide}=[fill=white,draw, line width = 1pt,inner sep=0.6mm,font=\footnotesize,minimum height=3mm,minimum width=7mm]

\tikzstyle{qWire}=[line width = 1pt, color=black]
\tikzstyle{cWire}=[color=gray,line width = .75pt]
\tikzstyle{CqWire}=[color=gray,line width = .75pt,->-]
\tikzstyle{CcWire}=[color=gray,line width = .75pt,->-]
\tikzstyle{RqWire}=[line width = 1pt, color=black,-<-]
\tikzstyle{RcWire}=[color=gray,line width = .75pt,-<-]
\tikzstyle{env}=[copoint,regular polygon rotate=0,minimum width=0.2cm, fill=black]

\tikzstyle{probs}=[shape=semicircle,fill=white,draw=black,shape border rotate=180,minimum width=1.2cm]

\tikzstyle{every picture}=[baseline=-0.25em,scale=0.5]
\tikzstyle{dotpic}=[] 
\tikzstyle{diredges}=[every to/.style={diredge}]
\tikzstyle{math matrix}=[matrix of math nodes,left delimiter=(,right delimiter=),inner sep=2pt,column sep=1em,row sep=0.5em,nodes={inner sep=0pt},text height=1.5ex, text depth=0.25ex]

\tikzstyle{inline text}=[text height=1.5ex, text depth=0.25ex,yshift=0.5mm]
\tikzstyle{label}=[font=\footnotesize,text height=1.5ex, text depth=0.25ex,yshift=0.5mm]
\tikzstyle{left label}=[label,anchor=east,xshift=1.5mm]
\tikzstyle{right label}=[label,anchor=west,xshift=-1mm]
\tikzstyle{up label}=[label,anchor=south,yshift=-1mm]

\tikzstyle{braceedge}=[decorate,decoration={brace,amplitude=2mm,raise=-1mm}]
\tikzstyle{small braceedge}=[decorate,decoration={brace,amplitude=1mm,raise=-1mm}]

\tikzstyle{doubled}=[line width=1.6pt] 
\tikzstyle{boldedge}=[doubled,shorten <=-0.17mm,shorten >=-0.17mm]
\tikzstyle{boldedgegray}=[doubled,gray,shorten <=-0.17mm,shorten >=-0.17mm]
\tikzstyle{singleedgegray}=[gray]

\tikzstyle{semidoubled}=[line width=1.4pt] 
\tikzstyle{semiboldedgegray}=[semidoubled,gray,shorten <=-0.17mm,shorten >=-0.17mm]

\tikzstyle{boxedge}=[semiboldedgegray]

\tikzstyle{boldedgedashed}=[very thick,dashed,shorten <=-0.17mm,shorten >=-0.17mm]
\tikzstyle{vboldedgedashed}=[doubled,dashed,shorten <=-0.17mm,shorten >=-0.17mm]
\tikzstyle{left hook arrow}=[left hook-latex]
\tikzstyle{right hook arrow}=[right hook-latex]
\tikzstyle{sembracket}=[line width=0.5pt,shorten <=-0.07mm,shorten >=-0.07mm]

\tikzstyle{causal edge}=[->,thick,gray]
\tikzstyle{causal nondir}=[thick,gray]
\tikzstyle{timeline}=[thick,gray, dashed]

\tikzstyle{cedge}=[<->,thick,gray!70!white]

\tikzstyle{empty diagram}=[draw=gray!40!white,dashed,shape=rectangle,minimum width=1cm,minimum height=1cm]
\tikzstyle{empty diagram small}=[draw=gray!50!white,dashed,shape=rectangle,minimum width=0.6cm,minimum height=0.5cm]

\tikzstyle{dot}=[inner sep=0mm,minimum width=2mm,minimum height=2mm,draw,shape=circle]
\tikzstyle{bigdot}=[inner sep=0mm,minimum width=5mm,minimum height=5mm,draw,shape=circle]
\tikzstyle{leak}=[white dot, shape=regular polygon, minimum size=3.3 mm, regular polygon sides=3, outer sep=-0.2mm, regular polygon rotate=270]
\tikzstyle{proj}=[regular polygon,regular polygon sides=4,draw,scale=0.75,inner sep=-0.5pt,minimum width=6mm,fill=white]
\tikzstyle{projOut}=[regular polygon,regular polygon sides=3,draw,scale=0.75,inner sep=-0.5pt,minimum width=7.5mm,fill=white,regular polygon rotate=180]
\tikzstyle{projIn}=[regular polygon,regular polygon sides=3,draw,scale=0.75,inner sep=-0.5pt,minimum width=7.5mm,fill=white]
\tikzstyle{Vleak}=[white dot, shape=regular polygon, minimum size=3.3 mm, regular polygon sides=3, outer sep=-0.2mm, regular polygon rotate=90]
\tikzstyle{dleak}=[white dot, line width=1.6pt, shape=regular polygon, minimum size=3.3 mm, regular polygon sides=3, outer sep=-0.2mm, regular polygon rotate=270]

\tikzstyle{Wsquare}=[white dot, shape=regular polygon, rounded corners=0.8 mm, minimum size=3.3 mm, regular polygon sides=3, outer sep=-0.2mm]
\tikzstyle{Wsquareadj}=[white dot, shape=regular polygon, rounded corners=0.8 mm, minimum size=3.3 mm, regular polygon sides=3, outer sep=-0.2mm, regular polygon rotate=180]
\tikzstyle{ddot}=[inner sep=0mm, doubled, minimum width=2.5mm,minimum height=2.5mm,draw,shape=circle]

\tikzstyle{clear dot}=[dot,fill=none,text depth=-0.2mm,draw=gray, line width = .75pt]
\tikzstyle{tall clear dot}=[dot,fill=none,text depth=-0.2mm,draw=gray, line width = .75pt,shape=ellipse, minimum height=5mm]
\tikzstyle{wide clear dot}=[dot,fill=none,text depth=-0.2mm,draw=gray, line width = .75pt, shape=ellipse, minimum width = 5mm]
\tikzstyle{very wide clear dot}=[dot,fill=none,text depth=-0.2mm,draw=gray, line width = .75pt, shape=ellipse, minimum width = 7mm ]

\tikzstyle{black dot}=[dot,fill=black]
\tikzstyle{white dot}=[dot,fill=white,,text depth=-0.2mm]
\tikzstyle{white Wsquare}=[Wsquare,fill=gray,,text depth=-0.2mm]
\tikzstyle{white Wsquareadj}=[Wsquareadj,fill=white,,text depth=-0.2mm]
\tikzstyle{green dot}=[white dot] 
\tikzstyle{gray dot}=[dot,fill=gray!40!white,,text depth=-0.2mm]
\tikzstyle{red dot}=[gray dot] 

\tikzstyle{black ddot}=[ddot,fill=black]
\tikzstyle{white ddot}=[ddot,fill=white]
\tikzstyle{gray ddot}=[ddot,fill=gray!40!white]

\tikzstyle{gray edge}=[gray!60!white]

\tikzstyle{small dot}=[inner sep=0.2mm,minimum width=0pt,minimum height=0pt,draw,shape=circle]

\tikzstyle{small black dot}=[small dot,fill=black]
\tikzstyle{small white dot}=[small dot,fill=white]
\tikzstyle{small gray dot}=[small dot,fill=gray,draw=gray]

\tikzstyle{causal dot}=[inner sep=0.4mm,minimum width=0pt,minimum height=0pt,draw=white,shape=circle,fill=gray!40!white]

\tikzstyle{phase dimensions}=[minimum size=5mm,font=\footnotesize,rectangle,rounded corners=2.5mm,inner sep=0.2mm,outer sep=-2mm]
\tikzstyle{dphase dimensions}=[minimum size=5mm,font=\footnotesize,rectangle,rounded corners=2.5mm,inner sep=0.2mm,outer sep=-2mm]

\tikzstyle{white phase dot}=[dot,fill=white,phase dimensions]
\tikzstyle{white phase ddot}=[ddot,fill=white,dphase dimensions]

\tikzstyle{white rect ddot}=[draw=black,fill=white,doubled,minimum size=5mm,font=\footnotesize,rectangle,rounded corners=2.5mm,inner sep=0.2mm]
\tikzstyle{gray rect ddot}=[draw=black,fill=gray!40!white,doubled,minimum size=6mm,font=\footnotesize,rectangle,rounded corners=3mm]

\tikzstyle{gray phase dot}=[dot,fill=gray!40!white,phase dimensions]
\tikzstyle{gray phase ddot}=[ddot,fill=gray!40!white,dphase dimensions]
\tikzstyle{grey phase dot}=[gray phase dot]
\tikzstyle{grey phase ddot}=[gray phase ddot]

\tikzstyle{small phase dimensions}=[minimum size=4mm,font=\tiny,rectangle,rounded corners=2mm,inner sep=0.2mm,outer sep=-2mm]
\tikzstyle{small dphase dimensions}=[minimum size=4mm,font=\tiny,rectangle,rounded corners=2mm,inner sep=0.2mm,outer sep=-2mm]

\tikzstyle{small gray phase dot}=[dot,fill=gray!40!white,small phase dimensions]
\tikzstyle{small gray phase ddot}=[ddot,fill=gray!40!white,small dphase dimensions]

\tikzstyle{small map}=[draw,shape=rectangle,minimum height=4mm,minimum width=4mm,fill=white]

\tikzstyle{cnot}=[fill=white,shape=circle,inner sep=-1.4pt]

\tikzstyle{asym hadamard}=[fill=white,draw,shape=NEbox,inner sep=0.6mm,font=\footnotesize,minimum height=4mm]
\tikzstyle{asym hadamard conj}=[fill=white,draw,shape=NWbox,inner sep=0.6mm,font=\footnotesize,minimum height=4mm]
\tikzstyle{asym hadamard dag}=[fill=white,draw,shape=SEbox,inner sep=0.6mm,font=\footnotesize,minimum height=4mm]

\tikzstyle{hadamard}=[fill=white,draw,inner sep=0.6mm,font=\footnotesize,minimum height=4mm,minimum width=4mm]
\tikzstyle{small hadamard}=[fill=white,draw,inner sep=0.6mm,minimum height=1.5mm,minimum width=1.5mm]
\tikzstyle{small hadamard rotate}=[small hadamard,rotate=45]
\tikzstyle{dhadamard}=[hadamard,doubled]
\tikzstyle{small dhadamard}=[small hadamard,doubled]
\tikzstyle{small dhadamard rotate}=[small hadamard rotate,doubled]
\tikzstyle{antipode}=[white dot,inner sep=0.3mm,font=\footnotesize]

\tikzstyle{scalar}=[diamond,draw,inner sep=0.5pt,font=\small]
\tikzstyle{dscalar}=[diamond,doubled, draw,inner sep=0.5pt,font=\small]

\tikzstyle{small box}=[rectangle,inline text,fill=white,draw,minimum height=5mm,yshift=-0.5mm,minimum width=5mm,font=\small]
\tikzstyle{small gray box}=[small box,fill=gray!30]
\tikzstyle{medium box}=[rectangle,inline text,fill=white,draw,minimum height=5mm,yshift=-0.5mm,minimum width=10mm,font=\small]
\tikzstyle{square box}=[small box] 
\tikzstyle{medium gray box}=[small box,fill=gray!30]
\tikzstyle{semilarge box}=[rectangle,inline text,fill=white,draw,minimum height=5mm,yshift=-0.5mm,minimum width=12.5mm,font=\small]
\tikzstyle{large box}=[rectangle,inline text,fill=white,draw,minimum height=5mm,yshift=-0.5mm,minimum width=15mm,font=\small]
\tikzstyle{large gray box}=[small box,fill=gray!30]

\tikzstyle{Bayes box}=[rectangle,fill=black,draw, minimum height=3mm, minimum width=3mm]

\tikzstyle{gray square point}=[small box,fill=gray!50]

\tikzstyle{dphase box white}=[dhadamard]
\tikzstyle{dphase box gray}=[dhadamard,fill=gray!50!white]
\tikzstyle{phase box white}=[hadamard]
\tikzstyle{phase box gray}=[hadamard,fill=gray!50!white]

\tikzstyle{point}=[regular polygon,regular polygon sides=3,draw,scale=0.75,inner sep=-0.5pt,minimum width=9mm,fill=white,regular polygon rotate=180]
\tikzstyle{infpoint}=[regular polygon,regular polygon sides=3,draw,scale=0.75,inner sep=-0.5pt,minimum width=9mm,fill=white,regular polygon rotate=90]
\tikzstyle{point nosep}=[regular polygon,regular polygon sides=3,draw,scale=0.75,inner sep=-2pt,minimum width=9mm,fill=white,regular polygon rotate=180]
\tikzstyle{infcopoint}=[regular polygon,regular polygon sides=3,draw,scale=0.75,inner sep=-0.5pt,minimum width=9mm,fill=white,regular polygon rotate=270]
\tikzstyle{copoint}=[regular polygon,regular polygon sides=3,draw,scale=0.75,inner sep=-0.5pt,minimum width=9mm,fill=white]
\tikzstyle{dpoint}=[point,doubled]
\tikzstyle{dcopoint}=[copoint,doubled]

\tikzstyle{pointgrow}=[shape=cornerpoint,kpoint common,scale=0.75,inner sep=3pt]
\tikzstyle{pointgrow dag}=[shape=cornercopoint,kpoint common,scale=0.75,inner sep=3pt]

\tikzstyle{wide copoint}=[fill=white,draw,shape=isosceles triangle,shape border rotate=90,isosceles triangle stretches=true,inner sep=0pt,minimum width=1.5cm,minimum height=6.12mm]
\tikzstyle{wide point}=[fill=white,draw,shape=isosceles triangle,shape border rotate=-90,isosceles triangle stretches=true,inner sep=0pt,minimum width=1.5cm,minimum height=6.12mm,yshift=-0.0mm]
\tikzstyle{wide point plus}=[fill=white,draw,shape=isosceles triangle,shape border rotate=-90,isosceles triangle stretches=true,inner sep=0pt,minimum width=1.74cm,minimum height=7mm,yshift=-0.0mm]

\tikzstyle{wide dpoint}=[fill=white,doubled,draw,shape=isosceles triangle,shape border rotate=-90,isosceles triangle stretches=true,inner sep=0pt,minimum width=1.5cm,minimum height=6.12mm,yshift=-0.0mm]

\tikzstyle{tinypoint}=[regular polygon,regular polygon sides=3,draw,scale=0.55,inner sep=-0.15pt,minimum width=6mm,fill=white,regular polygon rotate=180]

\tikzstyle{white point}=[point]
\tikzstyle{white dpoint}=[dpoint]
\tikzstyle{green point}=[white point] 
\tikzstyle{white copoint}=[copoint]
\tikzstyle{gray point}=[point,fill=gray!40!white]
\tikzstyle{gray dpoint}=[gray point,doubled]
\tikzstyle{red point}=[gray point] 
\tikzstyle{gray copoint}=[copoint,fill=gray!40!white]
\tikzstyle{gray dcopoint}=[gray copoint,doubled]

\tikzstyle{white point guide}=[regular polygon,regular polygon sides=3,font=\scriptsize,draw,scale=0.65,inner sep=-0.5pt,minimum width=9mm,fill=white,regular polygon rotate=180]

\tikzstyle{black point}=[point,fill=black,font=\color{white}]
\tikzstyle{black copoint}=[copoint,fill=black,font=\color{white}]

\tikzstyle{tiny gray point}=[tinypoint,fill=gray!40!white]

\tikzstyle{diredge}=[->]
\tikzstyle{ddiredge}=[<->]
\tikzstyle{rdiredge}=[<-]
\tikzstyle{thickdiredge}=[->, very thick]
\tikzstyle{pointer edge}=[->,very thick,gray]
\tikzstyle{pointer edge part}=[very thick,gray]
\tikzstyle{dashed edge}=[dashed]
\tikzstyle{thick dashed edge}=[very thick,dashed]
\tikzstyle{thick gray dashed edge}=[thick dashed edge,gray!40]
\tikzstyle{thick map edge}=[very thick,|->]

\makeatletter
\newcommand{\boxshape}[3]{%
\pgfdeclareshape{#1}{
\inheritsavedanchors[from=rectangle] 
\inheritanchorborder[from=rectangle]
\inheritanchor[from=rectangle]{center}
\inheritanchor[from=rectangle]{north}
\inheritanchor[from=rectangle]{south}
\inheritanchor[from=rectangle]{west}
\inheritanchor[from=rectangle]{east}
\backgroundpath{
\southwest \pgf@xa=\pgf@x \pgf@ya=\pgf@y
\northeast \pgf@xb=\pgf@x \pgf@yb=\pgf@y

\@tempdima=#2
\@tempdimb=#3

\pgfpathmoveto{\pgfpoint{\pgf@xa - 5pt + \@tempdima}{\pgf@ya}}
\pgfpathlineto{\pgfpoint{\pgf@xa - 5pt - \@tempdima}{\pgf@yb}}
\pgfpathlineto{\pgfpoint{\pgf@xb + 5pt + \@tempdimb}{\pgf@yb}}
\pgfpathlineto{\pgfpoint{\pgf@xb + 5pt - \@tempdimb}{\pgf@ya}}
\pgfpathlineto{\pgfpoint{\pgf@xa - 5pt + \@tempdima}{\pgf@ya}}
\pgfpathclose
}
}}

\boxshape{NEbox}{0pt}{3pt}
\boxshape{SEbox}{0pt}{-3pt}
\boxshape{NWbox}{5pt}{0pt}
\boxshape{SWbox}{-5pt}{0pt}
\boxshape{EBox}{-3pt}{3pt}
\boxshape{WBox}{3pt}{-3pt}
\makeatother

\tikzstyle{cloud}=[shape=cloud,draw,minimum width=1.5cm,minimum height=1.5cm]

\tikzstyle{map}=[draw,shape=NEbox,inner sep=1pt,minimum height=4mm,fill=white]
\tikzstyle{dashedmap}=[draw,dashed,shape=NEbox,inner sep=2pt,minimum height=6mm,fill=white]
\tikzstyle{mapdag}=[draw,shape=SEbox,inner sep=1pt,minimum height=4mm,fill=white]
\tikzstyle{mapadj}=[draw,shape=SEbox,inner sep=2pt,minimum height=6mm,fill=white]
\tikzstyle{maptrans}=[draw,shape=SWbox,inner sep=2pt,minimum height=6mm,fill=white]
\tikzstyle{mapconj}=[draw,shape=NWbox,inner sep=2pt,minimum height=6mm,fill=white]

\tikzstyle{medium map}=[draw,shape=NEbox,inner sep=2pt,minimum height=6mm,fill=white,minimum width=7mm]
\tikzstyle{medium map dag}=[draw,shape=SEbox,inner sep=2pt,minimum height=6mm,fill=white,minimum width=7mm]
\tikzstyle{medium map adj}=[draw,shape=SEbox,inner sep=2pt,minimum height=6mm,fill=white,minimum width=7mm]
\tikzstyle{medium map trans}=[draw,shape=SWbox,inner sep=2pt,minimum height=6mm,fill=white,minimum width=7mm]
\tikzstyle{medium map conj}=[draw,shape=NWbox,inner sep=2pt,minimum height=6mm,fill=white,minimum width=7mm]
\tikzstyle{semilarge map}=[draw,shape=NEbox,inner sep=2pt,minimum height=6mm,fill=white,minimum width=9.5mm]
\tikzstyle{semilarge map trans}=[draw,shape=SWbox,inner sep=2pt,minimum height=6mm,fill=white,minimum width=9.5mm]
\tikzstyle{semilarge map adj}=[draw,shape=SEbox,inner sep=2pt,minimum height=6mm,fill=white,minimum width=9.5mm]
\tikzstyle{semilarge map dag}=[draw,shape=SEbox,inner sep=2pt,minimum height=6mm,fill=white,minimum width=9.5mm]
\tikzstyle{semilarge map conj}=[draw,shape=NWbox,inner sep=2pt,minimum height=6mm,fill=white,minimum width=9.5mm]
\tikzstyle{large map}=[draw,shape=NEbox,inner sep=2pt,minimum height=6mm,fill=white,minimum width=12mm]
\tikzstyle{large map conj}=[draw,shape=NWbox,inner sep=2pt,minimum height=6mm,fill=white,minimum width=12mm]
\tikzstyle{very large map}=[draw,shape=NEbox,inner sep=2pt,minimum height=6mm,fill=white,minimum width=17mm]

\tikzstyle{medium dmap}=[draw,doubled,shape=NEbox,inner sep=2pt,minimum height=6mm,fill=white,minimum width=7mm]
\tikzstyle{medium dmap dag}=[draw,doubled,shape=SEbox,inner sep=2pt,minimum height=6mm,fill=white,minimum width=7mm]
\tikzstyle{medium dmap adj}=[draw,doubled,shape=SEbox,inner sep=2pt,minimum height=6mm,fill=white,minimum width=7mm]
\tikzstyle{medium dmap trans}=[draw,doubled,shape=SWbox,inner sep=2pt,minimum height=6mm,fill=white,minimum width=7mm]
\tikzstyle{medium dmap conj}=[draw,doubled,shape=NWbox,inner sep=2pt,minimum height=6mm,fill=white,minimum width=7mm]
\tikzstyle{semilarge dmap}=[draw,doubled,shape=NEbox,inner sep=2pt,minimum height=6mm,fill=white,minimum width=9.5mm]
\tikzstyle{semilarge dmap trans}=[draw,doubled,shape=SWbox,inner sep=2pt,minimum height=6mm,fill=white,minimum width=9.5mm]
\tikzstyle{semilarge dmap adj}=[draw,doubled,shape=SEbox,inner sep=2pt,minimum height=6mm,fill=white,minimum width=9.5mm]
\tikzstyle{semilarge dmap dag}=[draw,doubled,shape=SEbox,inner sep=2pt,minimum height=6mm,fill=white,minimum width=9.5mm]
\tikzstyle{semilarge dmap conj}=[draw,doubled,shape=NWbox,inner sep=2pt,minimum height=6mm,fill=white,minimum width=9.5mm]
\tikzstyle{large dmap}=[draw,doubled,shape=NEbox,inner sep=2pt,minimum height=6mm,fill=white,minimum width=12mm]
\tikzstyle{large dmap conj}=[draw,doubled,shape=NWbox,inner sep=2pt,minimum height=6mm,fill=white,minimum width=12mm]
\tikzstyle{large dmap trans}=[draw,doubled,shape=SWbox,inner sep=2pt,minimum height=6mm,fill=white,minimum width=12mm]
\tikzstyle{large dmap adj}=[draw,doubled,shape=SEbox,inner sep=2pt,minimum height=6mm,fill=white,minimum width=12mm]
\tikzstyle{large dmap dag}=[draw,doubled,shape=SEbox,inner sep=2pt,minimum height=6mm,fill=white,minimum width=12mm]
\tikzstyle{very large dmap}=[draw,doubled,shape=NEbox,inner sep=2pt,minimum height=6mm,fill=white,minimum width=19.5mm]

\tikzstyle{muxbox}=[draw,shape=rectangle,minimum height=3mm,minimum width=3mm,fill=white]
\tikzstyle{dmuxbox}=[muxbox,doubled]

\tikzstyle{box}=[draw,shape=rectangle,inner sep=2pt,minimum height=6mm,minimum width=6mm,fill=white]
\tikzstyle{dbox}=[draw,doubled,shape=rectangle,inner sep=2pt,minimum height=6mm,minimum width=6mm,fill=white]
\tikzstyle{dmap}=[draw,doubled,shape=NEbox,inner sep=2pt,minimum height=6mm,fill=white]
\tikzstyle{dmapdag}=[draw,doubled,shape=SEbox,inner sep=2pt,minimum height=6mm,fill=white]
\tikzstyle{dmapadj}=[draw,doubled,shape=SEbox,inner sep=2pt,minimum height=6mm,fill=white]
\tikzstyle{dmaptrans}=[draw,doubled,shape=SWbox,inner sep=2pt,minimum height=6mm,fill=white]
\tikzstyle{dmapconj}=[draw,doubled,shape=NWbox,inner sep=2pt,minimum height=6mm,fill=white]

\tikzstyle{ddmap}=[draw,doubled,dashed,shape=NEbox,inner sep=2pt,minimum height=6mm,fill=white]
\tikzstyle{ddmapdag}=[draw,doubled,dashed,shape=SEbox,inner sep=2pt,minimum height=6mm,fill=white]
\tikzstyle{ddmapadj}=[draw,doubled,dashed,shape=SEbox,inner sep=2pt,minimum height=6mm,fill=white]
\tikzstyle{ddmaptrans}=[draw,doubled,dashed,shape=SWbox,inner sep=2pt,minimum height=6mm,fill=white]
\tikzstyle{ddmapconj}=[draw,doubled,dashed,shape=NWbox,inner sep=2pt,minimum height=6mm,fill=white]

\boxshape{sNEbox}{0pt}{3pt}
\boxshape{sSEbox}{0pt}{-3pt}
\boxshape{sNWbox}{3pt}{0pt}
\boxshape{sSWbox}{-3pt}{0pt}
\tikzstyle{smap}=[draw,shape=sNEbox,fill=white]
\tikzstyle{smapdag}=[draw,shape=sSEbox,fill=white]
\tikzstyle{smapadj}=[draw,shape=sSEbox,fill=white]
\tikzstyle{smaptrans}=[draw,shape=sSWbox,fill=white]
\tikzstyle{smapconj}=[draw,shape=sNWbox,fill=white]

\tikzstyle{dsmap}=[draw,dashed,shape=sNEbox,fill=white]
\tikzstyle{dsmapdag}=[draw,dashed,shape=sSEbox,fill=white]
\tikzstyle{dsmaptrans}=[draw,dashed,shape=sSWbox,fill=white]
\tikzstyle{dsmapconj}=[draw,dashed,shape=sNWbox,fill=white]

\boxshape{mNEbox}{0pt}{10pt}
\boxshape{mSEbox}{0pt}{-10pt}
\boxshape{mNWbox}{10pt}{0pt}
\boxshape{mSWbox}{-10pt}{0pt}
\tikzstyle{mmap}=[draw,shape=mNEbox]
\tikzstyle{mmapdag}=[draw,shape=mSEbox]
\tikzstyle{mmaptrans}=[draw,shape=mSWbox]
\tikzstyle{mmapconj}=[draw,shape=mNWbox]

\tikzstyle{mmapgray}=[draw,fill=gray!40!white,shape=mNEbox]
\tikzstyle{smapgray}=[draw,fill=gray!40!white,shape=sNEbox]

\makeatletter

\pgfdeclareshape{cornerpoint}{
\inheritsavedanchors[from=rectangle] 
\inheritanchorborder[from=rectangle]
\inheritanchor[from=rectangle]{center}
\inheritanchor[from=rectangle]{north}
\inheritanchor[from=rectangle]{south}
\inheritanchor[from=rectangle]{west}
\inheritanchor[from=rectangle]{east}
\backgroundpath{
\southwest \pgf@xa=\pgf@x \pgf@ya=\pgf@y
\northeast \pgf@xb=\pgf@x \pgf@yb=\pgf@y

\pgfmathsetmacro{\pgf@shorten@left}{\pgfkeysvalueof{/tikz/shorten left}}
\pgfmathsetmacro{\pgf@shorten@right}{\pgfkeysvalueof{/tikz/shorten right}}

\pgfpathmoveto{\pgfpoint{0.5 * (\pgf@xa + \pgf@xb)}{\pgf@ya - 5pt}}
\pgfpathlineto{\pgfpoint{\pgf@xa - 8pt + \pgf@shorten@left}{\pgf@yb - 1.5 * \pgf@shorten@left}}
\pgfpathlineto{\pgfpoint{\pgf@xa - 8pt + \pgf@shorten@left}{\pgf@yb}}
\pgfpathlineto{\pgfpoint{\pgf@xb + 8pt - \pgf@shorten@right}{\pgf@yb}}
\pgfpathlineto{\pgfpoint{\pgf@xb + 8pt - \pgf@shorten@right}{\pgf@yb - 1.5 * \pgf@shorten@right}}
\pgfpathclose
}
}

\pgfdeclareshape{cornercopoint}{
\inheritsavedanchors[from=rectangle] 
\inheritanchorborder[from=rectangle]
\inheritanchor[from=rectangle]{center}
\inheritanchor[from=rectangle]{north}
\inheritanchor[from=rectangle]{south}
\inheritanchor[from=rectangle]{west}
\inheritanchor[from=rectangle]{east}
\backgroundpath{
\southwest \pgf@xa=\pgf@x \pgf@ya=\pgf@y
\northeast \pgf@xb=\pgf@x \pgf@yb=\pgf@y

\pgfmathsetmacro{\pgf@shorten@left}{\pgfkeysvalueof{/tikz/shorten left}}
\pgfmathsetmacro{\pgf@shorten@right}{\pgfkeysvalueof{/tikz/shorten right}}

\pgfpathmoveto{\pgfpoint{0.5 * (\pgf@xa + \pgf@xb)}{\pgf@yb + 5pt}}
\pgfpathlineto{\pgfpoint{\pgf@xa - 8pt + \pgf@shorten@left}{\pgf@ya + 1.5 * \pgf@shorten@left}}
\pgfpathlineto{\pgfpoint{\pgf@xa - 8pt + \pgf@shorten@left}{\pgf@ya}}
\pgfpathlineto{\pgfpoint{\pgf@xb + 8pt - \pgf@shorten@right}{\pgf@ya}}
\pgfpathlineto{\pgfpoint{\pgf@xb + 8pt - \pgf@shorten@right}{\pgf@ya + 1.5 * \pgf@shorten@right}}
\pgfpathclose
}
}

\makeatother

\pgfkeyssetvalue{/tikz/shorten left}{0pt}
\pgfkeyssetvalue{/tikz/shorten right}{0pt}

\tikzstyle{kpoint common}=[draw,fill=white,inner sep=1pt,minimum height=4mm]
\tikzstyle{kpoint sc}=[shape=cornerpoint,kpoint common]
\tikzstyle{kpoint adjoint sc}=[shape=cornercopoint,kpoint common]
\tikzstyle{kpoint}=[shape=cornerpoint,shorten left=5pt,kpoint common]
\tikzstyle{kpoint adjoint}=[shape=cornercopoint,shorten left=5pt,kpoint common]
\tikzstyle{kpoint conjugate}=[shape=cornerpoint,shorten right=5pt,kpoint common]
\tikzstyle{kpoint transpose}=[shape=cornercopoint,shorten right=5pt,kpoint common]
\tikzstyle{kpoint symm}=[shape=cornerpoint,shorten left=5pt,shorten right=5pt,kpoint common]

\tikzstyle{wide kpoint sc}=[shape=cornerpoint,kpoint common, minimum width=1 cm]
\tikzstyle{wide kpointdag sc}=[shape=cornercopoint,kpoint common, minimum width=1 cm]

\tikzstyle{black kpoint}=[shape=cornerpoint,shorten left=5pt,kpoint common,fill=black,font=\color{white}]

\tikzstyle{black kpoint sm}=[shape=cornerpoint,shorten left=5pt,kpoint common,fill=black,font=\color{white},scale=0.75]

\tikzstyle{black kpoint adjoint}=[shape=cornercopoint,shorten left=5pt,kpoint common,fill=black,font=\color{white}]
\tikzstyle{black kpointadj}=[shape=cornercopoint,shorten left=5pt,kpoint common,fill=black,font=\color{white}]

\tikzstyle{black kpointadj sm}=[shape=cornercopoint,shorten left=5pt,kpoint common,fill=black,font=\color{white},scale=0.75]

\tikzstyle{black dkpoint}=[shape=cornerpoint,shorten left=5pt,kpoint common,fill=black, doubled,font=\color{white}]
\tikzstyle{black dkpoint adjoint}=[shape=cornercopoint,shorten left=5pt,kpoint common,fill=black, doubled,font=\color{white}]
\tikzstyle{black dkpointadj}=[shape=cornercopoint,shorten left=5pt,kpoint common,fill=black, doubled,font=\color{white}]

\tikzstyle{black dkpoint sm}=[shape=cornerpoint,shorten left=5pt,kpoint common,fill=black, doubled,font=\color{white},scale=0.75]
\tikzstyle{black dkpointadj sm}=[shape=cornercopoint,shorten left=5pt,kpoint common,fill=black, doubled,font=\color{white},scale=0.75]

\tikzstyle{kpointdag}=[kpoint adjoint]
\tikzstyle{kpointadj}=[kpoint adjoint]
\tikzstyle{kpointconj}=[kpoint conjugate]
\tikzstyle{kpointtrans}=[kpoint transpose]

\tikzstyle{big kpoint}=[kpoint, minimum width=1.2 cm, minimum height=8mm, inner sep=4pt, text depth=3mm]

\tikzstyle{wide kpoint}=[kpoint, minimum width=1 cm, inner sep=2pt]
\tikzstyle{wide kpointdag}=[kpointdag, minimum width=1 cm, inner sep=2pt]
\tikzstyle{wide kpointconj}=[kpointconj, minimum width=1 cm, inner sep=2pt]
\tikzstyle{wide kpointtrans}=[kpointtrans, minimum width=1 cm, inner sep=2pt]

\tikzstyle{wider kpoint}=[kpoint, minimum width=1.25 cm, inner sep=2pt]
\tikzstyle{wider kpointdag}=[kpointdag, minimum width=1.25 cm, inner sep=2pt]
\tikzstyle{wider kpointconj}=[kpointconj, minimum width=1.25 cm, inner sep=2pt]
\tikzstyle{wider kpointtrans}=[kpointtrans, minimum width=1.25 cm, inner sep=2pt]

\tikzstyle{gray kpoint}=[kpoint,fill=gray!50!white]
\tikzstyle{gray kpointdag}=[kpointdag,fill=gray!50!white]
\tikzstyle{gray kpointadj}=[kpointadj,fill=gray!50!white]
\tikzstyle{gray kpointconj}=[kpointconj,fill=gray!50!white]
\tikzstyle{gray kpointtrans}=[kpointtrans,fill=gray!50!white]

\tikzstyle{gray dkpoint}=[kpoint,fill=gray!50!white,doubled]
\tikzstyle{gray dkpointdag}=[kpointdag,fill=gray!50!white,doubled]
\tikzstyle{gray dkpointadj}=[kpointadj,fill=gray!50!white,doubled]
\tikzstyle{gray dkpointconj}=[kpointconj,fill=gray!50!white,doubled]
\tikzstyle{gray dkpointtrans}=[kpointtrans,fill=gray!50!white,doubled]

\tikzstyle{white label}=[draw,fill=white,rectangle,inner sep=0.7 mm]
\tikzstyle{gray label}=[draw,fill=gray!50!white,rectangle,inner sep=0.7 mm]
\tikzstyle{black label}=[draw,fill=black,rectangle,inner sep=0.7 mm]

\tikzstyle{dkpoint}=[kpoint,doubled]
\tikzstyle{wide dkpoint}=[wide kpoint,doubled]
\tikzstyle{dkpointdag}=[kpoint adjoint,doubled]
\tikzstyle{wide dkpointdag}=[wide kpointdag,doubled]
\tikzstyle{dkcopoint}=[kpoint adjoint,doubled]
\tikzstyle{dkpointadj}=[kpoint adjoint,doubled]
\tikzstyle{dkpointconj}=[kpoint conjugate,doubled]
\tikzstyle{dkpointtrans}=[kpoint transpose,doubled]

\tikzstyle{kscalar}=[kpoint common, shape=EBox, inner xsep=-1pt, inner ysep=3pt,font=\small]
\tikzstyle{kscalarconj}=[kpoint common, shape=WBox, inner xsep=-1pt, inner ysep=3pt,font=\small]

\tikzstyle{spekpoint}=[kpoint sc,minimum height=5mm,inner sep=3pt]
\tikzstyle{spekcopoint}=[kpoint adjoint sc,minimum height=5mm,inner sep=3pt]

\tikzstyle{dspekpoint}=[spekpoint,doubled]
\tikzstyle{dspekcopoint}=[spekcopoint,doubled]

 \tikzstyle{upground}=[circuit ee IEC,thick,ground,rotate=90,scale=2.5]
 \tikzstyle{downground}=[circuit ee IEC,thick,ground,rotate=-90,scale=2.5]
 \tikzstyle{infupground}=[circuit ee IEC,thick,ground,rotate=90,scale=2.5]
 \tikzstyle{infdownground}=[circuit ee IEC,thick,ground,rotate=270,scale=2.5]
 \tikzstyle{bigground}=[regular polygon,regular polygon sides=3,draw=gray,scale=0.50,inner sep=-0.5pt,minimum width=10mm,fill=gray]

\tikzstyle{arrs}=[-latex,font=\small,auto]
\tikzstyle{arrow plain}=[arrs]
\tikzstyle{arrow dashed}=[dashed,arrs]
\tikzstyle{arrow bold}=[very thick,arrs]
\tikzstyle{arrow hide}=[draw=white!0,-]
\tikzstyle{arrow reverse}=[latex-]
\tikzstyle{cdnode}=[]

\tikzstyle{tilde}=[draw=blue]
\tikzstyle{tildelabel}=[text=blue]
\tikzstyle{down arrow node} = [
    regular polygon,
    regular polygon sides=3,
    draw=BurntOrange,
    fill=BurntOrange,
    minimum size=2.2mm,
    inner sep=0pt,
    shape border rotate=180
]
\tikzstyle{aWire}=[line width = 0.9pt, color=BurntOrange]
\tikzstyle{black box} = [draw=black, fill=black,rectangle,minimum size=2mm,inner sep=0pt]
\tikzstyle{gWire}=[line width = .9pt, color=green]
\tikzstyle{sWire}=[line width = 1.1pt, color=violet]
\tikzstyle{qWire}=[line width = 0.5pt, color=violet]

\tikzstyle{nsWire}=[line width = .8pt, color=violet]
\tikzstyle{ndWire}=[line width = 1.1pt, color=violet, double, double distance = 1pt]
\tikzstyle{muWire}=[line width = .8pt, color=black]
\tikzstyle{pWire}=[line width = 1pt, color=blue]
\tikzstyle{gcWire}=[line width = 1.1pt, color=gray]

\usetikzlibrary{circuits.ee.IEC}

\usepackage{setspace}
\usepackage{cached-diagrams}

\allowdisplaybreaks

\renewcommand{\selectlanguage}[1]{}

\makeatletter

\let\OriginalLSubsubsection\l@subsubsection

\newcommand{\HideSubsubsectionsFromTOC}{%
  \let\l@subsubsection\@gobbletwo
}

\newcommand{\RestoreSubsubsectionsToTOC}{%
  \let\l@subsubsection\OriginalLSubsubsection
}

\newcommand{\HideSublevelsFromTOC}{%
  \let\l@subsection\@gobbletwo
  \let\l@subsubsection\@gobbletwo
}

\makeatother
\AtBeginDocument{%
  \setlength{\abovedisplayskip}{5pt plus 1pt minus 1pt}
  \setlength{\belowdisplayskip}{5pt plus 1pt minus 1pt}
  \setlength{\abovedisplayshortskip}{3pt plus 1pt minus 1pt}
  \setlength{\belowdisplayshortskip}{3pt plus 1pt minus 1pt}
  \setlength{\jot}{2pt}
}
\makeatletter
\renewcommand{\frontmatter@abstractwidth}{450pt}
\makeatother
\begin{document}

\title{Twirled and swirled worlds: a compositional approach}
 \author{Sina Soltani}
  \affiliation{International Centre for Theory of Quantum Technologies, University of Gda\'nsk, 80-309 Gda\'nsk, Poland}

  \author{Thomas D. Galley}
  \affiliation{Institute for Quantum Optics and Quantum Information, Austrian Academy of Sciences, Boltzmanngasse 3, A-1090 Vienna, Austria}
  \affiliation{Vienna Center for Quantum Science and Technology (VCQ),  Faculty of Physics, University of Vienna, Vienna, Austria}
  \affiliation{Université Paris-Saclay, Laboratoire des Méthodes Formelles, 91190 Gif-sur-Yvette, France}

  \author{Marco Erba}
  \affiliation{International Centre for Theory of Quantum Technologies, University of Gda\'nsk, 80-309 Gda\'nsk, Poland}

 \author{John H.~Selby}
 \affiliation{International Centre for Theory of Quantum Technologies, University of Gda\'nsk, 80-309 Gda\'nsk, Poland}

\begin{abstract}

\vspace{0.4cm}

In recent work two constructions have been defined for generalised probabilistic theories known as \emph{twirled} and \emph{swirled} worlds. Specifically, these can be seen as theories with symmetry-imposed superselection rules. In this paper, we generalise these constructions to a fully compositional setting for finite-dimensional GPTs and compact symmetry groups, where process spaces may involve arbitrary collections of input and output systems, and states, effects, and transformations are treated uniformly as processes. We distinguish two independent features of a theory symmetry: whether it is specified through actions on individual systems or directly through actions on the process spaces of the theory, and whether it is physical or nonphysical. System actions can be used to induce actions on the process spaces, while one may also specify the actions on each process space directly, without deriving them from system actions. We define an action on the theory as a whole by requiring the collection of actions on its process spaces to respect the compositional structure of the theory. In this way, system actions can give rise to an action on the theory through their induced process-space actions, while more general theory actions may be defined from process-space actions that do not arise from system actions. Averaging such a theory action over the symmetry group then defines the corresponding invariant subtheory. We call this averaging \emph{twirling} when the theory action is physical and \emph{swirling} when it is nonphysical, so these labels refer to the physicality of the underlying theory action rather than to the averaged map itself. 

A central question in this compositional setting is whether tomography survives the restriction to invariant processes. Unlike state and effect tomography, process tomography is not automatically guaranteed by the invariant-subtheory construction. We then identify conditions under which these compositional invariant subtheories satisfy process tomography, namely through dynamically faithful states and internal classical reference frames. We also introduce a framework for unifying twirled subtheories associated with different physical system actions, and separately sketch a proposal for unifying invariant subtheories associated with different process-space actions, without assuming their physicality. Within this framework, we also formulate incoherent simulations in our compositional GPT setting and make their extension to arbitrary compact groups explicit.

For invariant subtheories of locally tomographic GPTs, we obtain substantially stronger results. We show that for this class of GPTs, every theory symmetry can be represented by linear actions on the individual systems, although the representing transformations need not be physical, and that these system actions are physical if and only if the corresponding actions on the process spaces are physical. We also prove that process tomography is always satisfied for the corresponding invariant subtheories. While previous work establishes that local tomography fails in twirled subtheories without determining precisely which process spaces exhibit this failure, we show that local tomography fails for an invariant multipartite process space if and only if there exists a bipartition whose two sides share a non-trivial irreducible representation of the group under consideration. We also quantify the extent of this failure by the number of invariant degrees of freedom that are inaccessible to local tomography. Our analysis applies to invariant subtheories arising from both physical and nonphysical symmetry actions. We further study the operational consequences of this failure, characterising when process tomography can be achieved without ancillary systems, when entanglement is necessary for such ancilla-free tomography, when ancillary systems are necessary, and which irreducible representations an ancillary system must carry. We relate these conditions to the irreducible representations shared by the systems. Finally, we illustrate the framework by showing that bilocal classical theory arises as a twirled subtheory of classical theory and using our unification scheme to unify it with classical theory viewed as the trivial twirled theory. Our general representation-theoretic results also recover the bilocal state, effect, and process tomography of BCT from its $\mathbb Z_2$ representation structure. We then use real quantum theory as an example of $\mathbb Z_2$-swirling, recovering its known description and tomography properties from our general results.

\end{abstract}
\maketitle

\onecolumngrid

{\small

\vspace{.22cm}
\begin{center}
\emph{This work is dedicated to the memory of Sadra Soltani (2002–2026), who contributed \\ to the coloring of the diagrams in this paper and whose dedication to freedom remained \\ a lasting source of inspiration throughout the preparation of this manuscript.}
\end{center}
}

\clearpage
\twocolumngrid

\begin{spacing}{0.0}
\tableofcontents
\end{spacing}

\newpage

 \section{Introduction}
 \label{Sec:intro}

Symmetries play a central role in modern physics, from particle physics to quantum metrology. Particularly within quantum information theory, a key tool in the study of symmetries is the notion of \emph{twirling} with respect to the representation of some symmetry group providing a way to characterise group invariant subspaces. Twirling also arises naturally in the study of superselection rules and quantum reference frames~\cite{Bartlett_2007}. In quantum theory, the absence of a reference frame associated with a compact symmetry group $G$ can be described by averaging over the corresponding group action, yielding a $G$-invariant description of states and operations~\cite{Bartlett_2007,palmer2014changing,krumm2021quantum,hoehn2023quantum,castro2025interpreting}. The resulting restriction is operationally equivalent to a superselection rule associated with $G$~\cite{wick1952intrinsic,aharonov1967charge,doplicher1969fields,kitaev2004superselection,giulini2016superselection}.

In recent work, Ref.~\cite{centeno2024twirled}, twirling was extended to the framework of generalised probabilistic theories (GPTs). GPTs provide an operational framework in which physical theories are described in terms of their states, effects, transformations, and the rules governing their composition, with probabilities assigned to the outcomes of experiments~\cite{hardy2001quantum,barrett2007information,chiribella2010probabilistic,plavala2023general}. Since classical theory, quantum theory, and a broad range of post-quantum theories can all be formulated within this framework, GPTs place these theories within a common landscape in which their structural similarities and differences can be systematically compared. Extending twirling to GPTs therefore allows one to study the consequences of imposing a symmetry restriction across this broader landscape, and in particular to distinguish features that arise from the twirling construction itself from those that depend specifically on the structure of quantum theory. This extension of twirling to the GPT framework led to two key conceptually interesting results: that local tomography generally fails in twirled subtheories; and that such a failure does not imply any notion of ontological holism. This situates twirled worlds within the broader foundational study of local tomography, which has been used as a structural principle for comparing and reconstructing physical theories~\cite{hardy2001quantum,masanes2011derivation,hardy2013reconstructing,barnum2014local}, and whose failure arises in a variety of settings, including real and quaternionic quantum theories~\cite{Hardy2011Limited,barnum2020composites}, bilocal classical theory~\cite{d2020classicality,soltani2026decoupling}, and a toy theory in which the individual system types are described classically while their composites can exhibit nonclassical features~\cite{chiribella2024bell}. Particularly instructive is fermionic quantum theory~\cite{darianoFermionic2014,Dariano2014feynmanproblem}, where the failure of local tomography shows that the principle is relevant not only to abstract or deliberately constructed foil theories, but also to physically motivated theories describing fermionic systems.

Beyond these results on local tomography, Ref.~\cite{ying2025whether} introduced the related notion of \emph{swirling}, in which the group representation does not correspond to physically realisable transformations of the system. In particular, real quantum theory can be viewed as the subtheory of complex quantum theory obtained by imposing invariance under the collective action of a $\mathbb{Z}_2$ time-reversal symmetry~\cite{haake1991quantum,strasberg2022quantum}. Since the nontrivial element of this symmetry is represented antiunitarily rather than by a physical unitary transformation, this provides a natural example of a swirled quantum world. This construction was motivated in part by the possibility of experimentally distinguishing real quantum theory from quantum theory under assumptions on causal structure~\cite{Renou_2021}, and provided additional symmetry-restricted foil theories in which to investigate causal compatibility gaps and incoherent simulations. A familiar example of the distinction between physical and nonphysical symmetry transformations arises already in quantum theory, where Wigner's theorem states that symmetries of the pure-state space preserving transition probabilities are represented by either unitary or antiunitary operators~\cite{wigner2012group,wigner1960normal,bargmann1964note}, while antiunitary symmetries need not correspond to physically realisable transformations of the system~\cite{Chiribella2021Symmetries}.

A crucial limitation of Ref.~\cite{centeno2024twirled}, however, is that although it developed twirled GPTs and considered covariant transformations, it did not formulate the construction as a fully compositional theory of arbitrary multipartite processes. Ref.~\cite{ying2025whether} subsequently introduced swirled theories, treating arbitrary quantum operations and circuits and providing a GPT extension for locally tomographic parent theories. Here, we overcome these remaining limitations by developing a fully compositional framework for twirling and swirling in arbitrary finite-dimensional GPTs, without assuming local tomography. Our compositional generalisation is formulated using the language of process theories and category theory, which provides a natural framework for treating states, effects, and transformations uniformly as processes and for describing their sequential and parallel composition~\cite{abramsky2004categorical,selinger2007dagger,Coecke2006IntroducingCategories,DArianoChiribellaPerinotti2017QuantumTheory,CoeckeKissinger2017PicturingQuantumProcesses,FongSpivak2019SevenSketches,SelbyScandoloCoecke2021ReconstructingQuantumTheory,Gogioso2017FantasticQuantumTheories,selby2025generalised}. In so doing, we distinguish whether a theory symmetry arises from actions on individual systems or is specified directly through actions on its process spaces, and independently whether it is physical or nonphysical. In either case, not every choice of such actions need extend to an action on the theory as a whole; when they do, the resulting theory action must be compatible with the compositional structure of the theory. We use twirling and swirling for averaging physical and nonphysical theory actions, respectively. This perspective naturally leads to interpreting twirling and swirling as constructions of invariant subtheories of a GPT. Somewhat surprisingly, while Ref.~\cite{centeno2024twirled} showed that invariant states and effects suffice for tomography within prepare–measure scenarios, this property does not automatically extend to the fully compositional framework. In particular, process tomography is not automatically guaranteed within a twirled/swirled world. We identify two natural conditions, however, that ensure process tomography, namely the existence of invariant states that are dynamically faithful with respect to the invariant subtheory, or when the group action can be ``internalised'' with a classical reference frame. Exploring a more radical possibility is left to future work, namely, that the possibility of such a failure of process tomography tells us that we should be incorporating ``testers'', i.e., 1-combs mapping processes to scalars, as part of the defining data of a GPT. Allowing arbitrary twirled/swirled testers would then salvage process tomography but at the expense of having a substantially more complicated definition of a theory.

For invariant subtheories of locally tomographic GPTs, our main representation-theoretic results provide a substantially more general characterisation of the failure of local tomography and its operational consequences for process tomography, while also quantifying the extent of this failure. Previous results establish the failure of local tomography for bipartite states of identical systems, and hence systems carrying the same group representation, but do not determine when such a failure occurs more generally. Our approach instead applies to arbitrary multipartite settings, allowing the individual input and output systems to carry different representations of the group, and determines whether local tomography holds or fails for the corresponding process space of interest. We further characterise the entanglement and ancillary resources required for process tomography, showing in particular that whenever complete process tomography is possible without ancillary systems in a non-locally-tomographic invariant process space, entanglement is nevertheless necessary in general.

In the remainder of this section, we introduce the compositional GPT formalism that will be used throughout the paper. Section II reviews the necessary group-theoretic background, introducing a diagrammatic notation that interacts naturally with the GPT framework. In Section III, we define twirled subtheories arising from physical system actions. In Section IV, we generalise the invariant-subtheory construction to arbitrary theory actions, without assuming their physicality. Averaging over a physical theory action is called \emph{twirling}, whereas averaging over a nonphysical theory action is called \emph{swirling}. In this sense, twirling induced by physical system actions is a special case of twirling. We then study the status of process tomography in the resulting invariant subtheories. Section V focuses on these invariant subtheories for tomographically local GPTs, where we develop our representation-theoretic analysis of the failure of local tomography and its implications for process tomography. Finally, in Section VI, we provide a detailed analysis of how Bilocal Classical Theory (BCT)~\cite{d2020classicality,soltani2026decoupling} arises as the twirled version of classical theory, and then analyse real quantum theory (RQT) as an example of a swirled quantum theory within the framework we developed.

 \subsection{Compositional generalised probabilistic theories}

In this section we take a (slightly) nonstandard approach to the framework of generalised probabilistic theories (GPTs) based on forthcoming work \cite{subtheories}. In particular, for the purposes of this paper it will be important to work compositionally, and to treat all kinds of processes (states, effects, transformations) on an equal footing. These desiderata lead us to a framework which is most naturally expressed in the language of enriched category theory. We, however, will keep the definitions ``unpacked'' for pedagogical clarity.

\label{Sec:GPT-intro}
\subsubsection{Process theories}
A \emph{process theory} is a collection of systems $A,B,C,\ldots$ and a collection of processes, where a process with inputs and outputs is depicted as boxes with wires. For example,
   \begin{equation}
         \input{Diagrams/12_proc.tikz}
    \end{equation}
    is a process with inputs $A, B$ and $B$, and outputs $B$ and $C$. This collection of processes is required to be closed under wiring. For example, the circuit
\begin{equation}
       \scalebox{0.9}{$  \input{Diagrams/12_proc1.tikz}$}\;,
    \end{equation}
    corresponds to a process with inputs $A$ and $B$, and outputs $B$ and $C$. Note that a process may have no input or no output (or even neither). Furthermore, when composing processes, system types must match and the wiring must remain acyclic. In a process theory, two diagrams are considered equal if they share the same connectivity. For example, for the above diagram one can have
    \begin{equation}
        \scalebox{0.9}{$ \input{Diagrams/12_proc1.tikz}= \input{Diagrams/12_proc2.tikz}$}\;.
    \end{equation}
    In this paper, we intend to impose additional structure on process theories, since our goal is to study the framework of generalised probabilistic theories through the lens of process theories.

    \subsubsection{Process spaces and process maps}

    As a first step, we define \emph{process spaces} as follows:

\begin{definition}[Process spaces]\label{def:process-spaces}
A process space is a triple $\left(\bar{P}, P, U\right)$, where $U$ is a real finite-dimensional vector space. $\bar{P} \subset U$ is the set of all normalized processes, which is taken to be convex, with the affine span not containing the origin, i.e., $\mathbf{0} \notin \mathrm{AffSpan}[\bar{P}]$. $P \subset U$ is the set of all subnormalized processes, which is convex, contains the origin $\mathbf{0}$, and includes $\bar{P}$ as a subset.
\end{definition}
When clear from context, we sometimes use $P$ to denote the process space $\left(\bar{P}, P, U \right)$.

Important special cases of process spaces are state spaces, effect spaces, and the scalar space. A \emph{state space} is a process space $(\bar{\Omega},\Omega,V)$, where $V$ is a finite-dimensional real vector space, $\bar{\Omega}\subset V$ is the convex set of normalised states with $\mathbf{0}\notin\operatorname{AffSpan}(\bar{\Omega})$, and $\Omega:=\{p\omega\mid \omega\in\bar{\Omega},\,p\in[0,1]\}$ is the set of subnormalised states. In particular, for every $s\in\Omega$, one can write $s=p\omega$ for some $\omega\in\bar{\Omega}$ and define $s^\perp:=(1-p)\omega$, so that $s+s^\perp=\omega\in\bar{\Omega}$. An \emph{effect space} is a process space $(\{u\},E,W)$, where $W$ is a finite-dimensional real vector space, $u\in W$ is nonzero and is the unique deterministic effect, and for every $e\in E$ there exists a complementary effect $e^\perp\in E$ such that $e+e^\perp=u$; this is a special case of a process space with normalised set given by the singleton $\{u\}$. Finally, the \emph{scalar space} is the process space $(\{1\},[0,1],\mathbb{R})$.

\begin{definition}[Subprocess spaces]
A subprocess space $\left(\bar{Q}, Q, V\right)$ of a process space $\left(\bar{P}, P, U\right)$ is itself a process space, where $\bar{Q} \subseteq \bar{P}$, $Q \subseteq P$, and $V \subseteq U$.
\end{definition}

\begin{definition}[Process maps]\label{def:processMap}
 A process map $\eta : (\bar{P}, P, U) \to (\bar{Q}, Q, V)$ is a linear map $\eta : \mathrm{LinSpan}(P) \to \mathrm{LinSpan}(Q)$ such that $\eta(P) \subseteq Q $.

\end{definition}
A map is said to be \textit{faithful} if and only if it is injective, and is said to be \textit{full} if and only if it is surjective, i.e., $\eta(P) = Q$. It is \textit{invertible} if and only if it has a linear inverse that is itself a process map, i.e., $\eta^{-1}(Q) \subseteq P$.

\begin{definition}[Multi process maps]
    A multi-process map
    $$\tau: (\bar{P}_1, P_1, U_1)\times\cdots \times (\bar{P}_n, P_n, U_n)\to (\bar{Q}, Q, V) $$
    is a multi-linear map
    $$\tau: \mathrm{LinSpan}(P_1)\times\cdots\times \mathrm{LinSpan}(P_n)\to\mathrm{LinSpan}(Q)$$
    such that $\tau (P_1\times\cdots \times P_n)\subseteq Q$.
\end{definition}

\subsubsection{Generalised probabilistic theories}
A \emph{generalised probabilistic theory (GPT)} $\mathcal{P}$ is defined by a collection of systems, denoted as $\textbf{Sys}[\mathcal{P}]$, which represents the physical systems described by the GPT $\mathcal{P}$, with each individual system denoted by an uppercase Latin letter, $A \in \textbf{Sys}[\mathcal{P}]$, and the trivial system denoted by $I$. In this framework, a process can be viewed as an operation mapping a list of input systems—that is, a composite system $(A_1, \ldots, A_n)$—to a list of output systems $(B_1, \ldots, B_m)$, where each system belongs to the set of systems, i.e., $A_i, B_j \in \textbf{Sys}[\mathcal{P}]$.

Each pair of input and output systems defines a process space for a GPT $\mathcal{P}$, which we denote as
\begin{equation}
\label{eq:process-space-GPT}
    \left( \bar{P}^{B_1, \ldots, B_m}_{A_1, \ldots, A_n}, P^{B_1, \ldots, B_m}_{A_1, \ldots, A_n}, U^{B_1, \ldots, B_m}_{A_1, \ldots, A_n} \right).
\end{equation}
Note that the order in which systems in a list appear matters, as we require that their corresponding process spaces be distinct. For example, for two arbitrary systems $A_1, A_2 \in \textbf{Sys}[\mathcal{P}]$, we consider $(A_1, A_2)$ to be different from $(A_2, A_1)$. The only exception is the trivial system, where for any system $A \in \textbf{Sys}[\mathcal{P}]$, one has $(A, I) = A = (I, A)$.

A GPT $\mathcal{P}$ should specify every set of process spaces associated with a list of inputs $(A_1, \ldots, A_n)$ and a list of outputs $(B_1, \ldots, B_m)$. We denote the process space $P_I^{A_1, \ldots, A_n}$ associated with the trivial input by $P^{A_1, \ldots, A_n}$, and we require it to be a state space. Similarly, we denote the process space $P^I_{B_1, \ldots, B_m}$ associated with the trivial output by $P_{B_1, \ldots, B_m}$, and we require it to be an effect space. When both the input and output are trivial, we denote the corresponding process space as $P^I_I$ and require it to be the scalar space $\left(\{1\}, [0,1],\mathbb{R}\right)$.

We depict a system $A \in \textbf{Sys}[\mathcal{P}]$ as a thick violet wire
\begin{equation}
    \input{Diagrams/6_GPT3.tikz}\;,
\end{equation}
and a composite system, denoted as a list $(A_1, \dots, A_n)$, is depicted as
\begin{equation}
     \input{Diagrams/6_GPT4.tikz}.
\end{equation}
A process $f \in P^B_A$ is depicted as
\begin{equation}
\input{Diagrams/2_theorysym14f.tikz}\;,
\end{equation}
and a process with composite input and composite output, $f \in P^{B_1, \ldots, B_m}_{A_1, \ldots, A_n}$, is depicted as
\begin{equation}
\input{Diagrams/6_GPT.tikz}\;.
\end{equation}
For a state and an effect, we leave a blank space to represent the trivial system, and use a triangle for both states and effects. For example
\begin{equation}
    \input{Diagrams/6_GPT1.tikz} \in P_{A_1, A_2} \;, \;\input{Diagrams/6_GPT2.tikz} \in P^{B}.
\end{equation}
\textit{Sequential composition} of two processes $f \in P^B_A$ and $h \in P^C_B$ in a GPT $\mathcal{P}$, where the output of one matches the input of the other, is given by the bi-process map $\circ : P^B_A \times P^C_B \to P^C_A$, and is depicted as
\begin{equation}
    \circ: \input{Diagrams/2_theorysym14f.tikz},\input{Diagrams/6_GPT5.tikz}\to \input{Diagrams/6_GPT6.tikz}\;.
\end{equation}
Sequential composition is associative, that is $(f\circ h)\circ k= f\circ (h\circ k)$ for every process $f$, $h$, and $k$ that can be composed sequentially in a GPT $\mathcal{P}$.

An \textit{identity} process for a system $A \in \textbf{Sys}[\mathcal{P}]$, denoted by $\mathcal{I}_A$, is a process with the same system as input and output, i.e., $\mathcal{I}_A \in P^A_A$, and is depicted as
\begin{equation}
    \input{Diagrams/6_GPT9.tikz}\;,
\end{equation}
which, for every $f\in P^B_A$ and every $h\in P_B^A$, satisfies
\begin{equation}
    \input{Diagrams/6_GPT10.tikz}=\input{Diagrams/2_theorysym14f.tikz}\;\;,\;\;\input{Diagrams/6_GPT11.tikz}=\input{Diagrams/6_GPT12.tikz}\;.
\end{equation}
We depict an identity process as a wire without a dashed grey box. We demand that in a GPT $\mathcal{P}$, for every system $A \in \textbf{Sys}[\mathcal{P}]$, the identity process $\mathcal{I}_A$ exists, i.e., $\mathcal{I}_A \in P^A_A$.

\textit{Parallel composition} of two processes $f \in P^B_A$ and $h \in P^D_C$ in a GPT $\mathcal{P}$ is given by the bi-process map $\boxtimes : P^B_A \times P^D_C \to P^{B,D}_{A,C}$, and is depicted as%
\begin{equation}
\boxtimes:\input{Diagrams/2_theorysym14f.tikz},\input{Diagrams/6_GPT7.tikz}\to \input{Diagrams/6_GPT8.tikz}\;.
\end{equation}
 The parallel composition map in quantum theory and classical theory—and, as we will explain later, in locally tomographic theories—is given by the usual tensor product rule, denoted as $\otimes$. However, in general, one can have theories where the parallel composition rule is not given by the tensor product~\cite{erba2024compositionrulequantumsystems,d2020classicality}. Therefore, we use the symbol $\boxtimes$ to account for this more general possibility.
Parallel composition is also associative, i.e., one has $(f\boxtimes h)\boxtimes k=f\boxtimes (h\boxtimes k)$, for every process $f$, $h$, and $k$ in a GPT $\mathcal{P}$.

A \textit{swap} process for a pair of systems $A, B \in \textbf{Sys}[\mathcal{P}]$, denoted as $\sigma_{A,B}$, is a process that exchanges the systems, and is depicted as
\begin{equation}
     \input{Diagrams/6_GPT13.tikz}\;,
\end{equation}
which is a self-inverse process, that is
\begin{equation}
\input{Diagrams/6_GPT14.tikz}=\input{Diagrams/6_GPT15.tikz}\;.
\end{equation}
For every pair of systems, we demand that the swap process exists; that is, in a GPT $\mathcal{P}$, for every $A, B \in \textbf{Sys}[\mathcal{P}]$, one has $\sigma_{A,B} \in P^{B,A}_{A,B}$.

Sequential and parallel composition must satisfy bifunctoriality; that is, in a GPT $\mathcal{P}$, for every $f \in P^B_A$, $h \in P^C_B$, $k \in P^E_D$, and $l \in P^F_E$, one has $(f \circ h) \boxtimes (k \circ l) = (f \boxtimes k) \circ (h \boxtimes l)$, which can be depicted as
\begin{equation}
    \label{eq:bifunctoriality}\scalebox{0.9}{$\input{Diagrams/6_GPT16.tikz}=\input{Diagrams/6_GPT17.tikz}$}\;.
\end{equation}
A GPT $\mathcal{P}$ is closed under composition; that is, for every arbitrary composition of processes, the resulting process is itself a valid process.

 A GPT $\mathcal{P}$ is closed under composition if and only if it satisfies the closure condition
\begin{equation}
\label{eq:composition-closure1}
\scalebox{0.9}{$\input{Diagrams/4_twirl18.tikz}=\input{Diagrams/6_GPT18.tikz}$}\;,
\end{equation}
that is, for every $f \in P_A^{D,C}$ and every $h \in P_{C,B}^E$, one has $k \in P_{A,B}^{D,E}$.
 If we set $D = B = I$, we preserve closure under sequential composition, and if we set $C = I$, we preserve closure under parallel composition. Note that, when $A = B = E = I$, this requirement reduces to the well-known property of conditional states being valid states (also known as closure under steering), namely that composing a bipartite state with an effect on one of the systems results in a valid state. Similarly, when $A = D = E = I$, we obtain the property of conditional effects being valid effects.

We note that the requirement that $P_I^I$ is the real scalar process space implies that any closed circuit is associated to a probability, thus defining a probabilistic theory.

Finally, we require that a GPT $\mathcal{P}$ be tomographically complete such that distinct processes are operationally distinguishable.
\begin{definition}[Tomographic completeness]\label{def:tomographic-completness}
A theory is tomographically complete if and only if any process can be fully characterized by the probabilities it generates; that is, two processes $f \in P^B_A$ and $h \in P^B_A$ are equal if and only if they yield the same probabilities, i.e.,
\begin{equation}
\label{eq:process-tomography}
\scalebox{0.9}{$\input{Diagrams/2_theorysym14f.tikz}=\input{Diagrams/6_GPT7h.tikz}\iff\input{Diagrams/6_GPT19.tikz}=\input{Diagrams/6_GPT20.tikz}$}\;,
\end{equation}
for every $E \in \textbf{Sys}[\mathcal{P}]$, every $s \in P^{A,E}$, and every $e \in P_{B,E}$. Process tomography also implies that if two processes $f,h \in P^B_A$ are distinct, $f \neq h$, then there exists a state $s \in P^{A,E}$ and an effect $e \in P_{B,E}$ for a system $E \in \textbf{Sys}[\mathcal{P}]$, such that one obtains different statistics.
\end{definition}

 To summarize what we define as a fully compositional GPT so far, we have the following definition for a GPT $\mathcal{P}$:

\begin{definition}[Compositional GPTs]
\label{def:GPT}
A fully compositional GPT $\mathcal{P}$ is given by:
\begin{enumerate}[label=\roman*), itemsep=0.6em, topsep=0.6em]
    \item A set of systems denoted as $\textbf{Sys}[\mathcal{P}]$ which includes a trivial system $I$.

    \item For each pair of lists $(A_1, \ldots, A_n) \in \textbf{Sys}[\mathcal{P}]^{\times n}$ and $(B_1, \ldots, B_m)\in \textbf{Sys}[\mathcal{P}]^{\times m}$ representing input and output systems, a process space $P^{B_1, \ldots, B_m}_{A_1, \ldots, A_n}$  satisfying:
    \begin{enumerate}[label=\alph*), itemsep=0.5em, topsep=0.3em]
        \item $\mathcal{I}_A \in P^A_A$ for all $A \in \textbf{Sys}[\mathcal{P}]$.
        \item $\sigma_{A,B} \in P^{B,A}_{A,B}$ for all $A, B \in \textbf{Sys}[\mathcal{P}]$.
        \item The process space $P^I_I$ (with both trivial input and output) is the scalar process space $\left(\{1\}, [0,1], \mathds{R}\right)$.
        \item The process space with trivial input is a state space, and the process space with trivial output is an effect space.
    \end{enumerate}

    \item Parallel and sequential compositions between two process spaces are defined as associative, normalization-preserving bi-process maps.

    \item Parallel and sequential compositions obey bifunctoriality.

    \item The set of processes is tomographic, i.e., for processes $f \in P^{B}_A$, there exist systems $E\in\textbf{Sys}[\mathcal{P}]$, states $s\in P^{A,E}$, and effects $e\in P_{B,E}$ that one can use to perform process tomography.
\end{enumerate}
\end{definition}

Note that what we define as sequential and parallel composition, together with the requirement that they obey bifunctoriality, implies closure under composition.

\begin{definition}[Subtheories and GPT-subtheories] \label{def:subGPT}
  A subtheory $\mathcal{Q}$ of a GPT $\mathcal{P}$  is given by:
  \begin{enumerate}[label=\roman*), itemsep=0.6em, topsep=0.6em]
    \item A set of systems, denoted by $\textbf{Sys}[\mathcal{Q}]$, which forms a subset of the systems of the original GPT, i.e., $\textbf{Sys}[\mathcal{Q}] \subseteq \textbf{Sys}[\mathcal{P}]$, and includes the trivial system $I$.

    \item For each pair of lists $(A_1, \ldots, A_n) \in \textbf{Sys}[\mathcal{Q}]^{\times n}$ and $(B_1, \ldots, B_m) \in \textbf{Sys}[\mathcal{Q}]^{\times m}$, representing input and output systems, a subprocess space of the corresponding process space of the original GPT, $Q^{B_1, \ldots, B_m}_{A_1, \ldots, A_n} \subseteq P^{B_1, \ldots, B_m}_{A_1, \ldots, A_n}$, satisfying:

    \begin{enumerate}[label=\alph*), itemsep=0.5em, topsep=0.3em]
        \item $\mathcal{I}_A \in Q^A_A$ for all $A \in \textbf{Sys}[\mathcal{Q}]$.
        \item $\sigma_{A,B} \in Q^{B,A}_{A,B}$ for all $A, B \in \textbf{Sys}[\mathcal{Q}]$.
        \item The process space $Q^I_I$ (with both trivial input and output) is a subprocess space of the scalar process space $\left(\{1\}, [0,1], \mathds{R}\right)$.
        \item The subprocess space with trivial input is a state space, and the subprocess space with trivial output is an effect space.
    \end{enumerate}

    \item Parallel and sequential compositions between two subprocess spaces of $\mathcal{Q}$ are given by the parallel and sequential compositions of the original GPT $\mathcal{P}$, which implies bifunctoriality in $\mathcal{Q}$ as well.

\item Subprocess spaces are closed under composition; that is, the composition of any processes from the subprocess spaces is again a process in the corresponding subprocess space.

\end{enumerate}
Moreover, a subtheory $\mathcal{Q}$ is a GPT-subtheory, if the set of subprocesses is tomographic, that is, for every process of the subtheory $f \in Q^{B}_A$, there exist systems $E\in\textbf{Sys}[\mathcal{Q}]$, states $s\in Q^{A,E}$, and effects $e\in Q_{B,E}$ that one can use to perform process tomography.

\end{definition}

\subsubsection{Maps between different GPTs}
\label{subsubsec:maps}

In order to talk about theory-level symmetries and group actions, it is important to introduce maps between GPTs.

\begin{definition}[Maps between two GPTs]\label{def:map-between-gpts}
A map $\input{Diagrams/11_maps.tikz}$ from a GPT $\mathcal{P}$ to another GPT $\mathcal{Q}$ assigns each system $A \in \textbf{Sys}[\mathcal{P}]$ to a system $\Phi(A) \in \textbf{Sys}[\mathcal{Q}]$, with $\Phi(I) = I$ for the trivial system, such that, for every process space $P^B_A$, it defines a process map (\cref{def:processMap}) to the process space $Q^{\Phi(B)}_{\Phi(A)}$, where for an arbitrary process $f \in P^B_A$, we depict the action of $\Phi$ as
\begin{equation}
	\Phi::\input{Diagrams/2_theorysym14f.tikz}\mapsto \input{Diagrams/11_maps1.tikz}\;.
\end{equation}
\end{definition}

A map is said to be \emph{diagram preserving} if applying it to an entire circuit is equivalent to applying it to each process individually and then composing the resulting processes. For example,
\begin{equation}\label{eq:connectivity-preservation}
   \scalebox{0.95}{$ \input{Diagrams/11_maps2.tikz}= \input{Diagrams/11_maps3.tikz}$}\;.
\end{equation}
Diagram-preserving maps also preserve the identities and swaps, that is
\begin{equation}
\label{eq:swap-identity-preservation}
  \input{Diagrams/11_maps10.tikz}= \input{Diagrams/11_maps11.tikz} \quad,\quad \input{Diagrams/11_maps12.tikz}= \input{Diagrams/11_maps13.tikz}\quad .
\end{equation}
A map is said to be \emph{scalar-preserving}  if for every scalar \(p \in [0,1]\), one has \(\Phi(p) = p\), depicted as
\begin{equation}
    \input{Diagrams/11_maps6.tikz}= \input{Diagrams/11_maps7.tikz}\;.
\end{equation}

A map \(\Phi\) is said to be \emph{linear} on the processes of \(\mathcal{P}\) if for every \(r_x \in \Rl\), every system \(A,B \in \textbf{Sys}[\mathcal{P}]\), and every process \(f_x \in P^B_A\) such that \(\sum_x r_x f_x \in P^B_A\), one has
\begin{equation}
\Phi{\left(\sum_{x}r_{x}f_{x}\right)}=\sum_{x} r_{x}\Phi{\left(f_{x}\right)}.
\end{equation}

If the domain and codomain of a map $\Phi$ coincide with the same GPT, i.e., $\Phi:\mathcal{P}\to\mathcal{P}$, then $\Phi$ is said to act \emph{trivially on objects} if for every system $A\in \mathbf{Sys}[\mathcal{P}]$ one has $\Phi(A)=A$, and for every composite system $S=(A_1,\ldots,A_n)$ one has $\Phi(S)=S$.

\section{Symmetries, group actions, and  generalised probabilistic theories}
\label{Sec:Symmetries}

In this section we introduce tools for studying symmetries and group actions for generalised probabilistic theories. We start by providing a convenient diagrammatic representation of groups and then leverage this to describe group actions on systems within a GPT, processes within a GPT, and on a GPT as a whole.

\subsection{Diagrammatic representation of groups}
\label{Sec:SystemSymmetries}

In this paper, we denote a group $G$ diagrammatically as
\begin{equation}
\input{Diagrams/1_sys1.tikz}\;,
\end{equation}
where for a group element $g\in G$, we have
\begin{equation}
\input{Diagrams/1_sys2.tikz}\;,
\end{equation}
and for the identity element, we have
\begin{equation}
\input{Diagrams/1_sys3.tikz}\;.
\end{equation}
Group multiplication is depicted as
\begin{equation}
\input{Diagrams/1_sys4.tikz}\;,
\end{equation}
such that for every $g \in G$ and $g' \in G$, one has
\begin{equation}
\scalebox{0.9}{$\input{Diagrams/1_sys7.tikz}=\input{Diagrams/1_sys8.tikz}$}\;.
\end{equation}
The inverse operation is depicted as
\begin{equation}
\input{Diagrams/1_sys6.tikz}\;,
\end{equation}
such that for every $g\in G$, it gives
\begin{equation}
\scalebox{0.9}{$\input{Diagrams/1_sys9.tikz}=\input{Diagrams/1_sys10.tikz}$}\;,
\end{equation}
 where we can show
\begin{equation}
\label{eq:inverse-swap}
    \scalebox{0.9}{$\input{Diagrams/1_sys57.tikz}=     \input{Diagrams/1_sys58.tikz}$}\;.
\end{equation}
We introduce copying as
\begin{equation}
\input{Diagrams/1_sys5.tikz}\;,
\end{equation}
such that for every $g\in G$, one has
\begin{equation}
\scalebox{0.9}{$\input{Diagrams/1_sys11.tikz}=\input{Diagrams/1_sys12.tikz}$}\;.
\end{equation}
We can generalize copying to more than two branches, that is
\begin{equation}
\input{Diagrams/1_sys13.tikz}\;,
\end{equation}
such that for every $g\in G$, it gives
\begin{equation}
\scalebox{0.9}{$\input{Diagrams/1_sys14.tikz}=\input{Diagrams/1_sys15.tikz}$}\;.
\end{equation}
We can now verify that
\begin{equation}
\label{eq:bialgebra}
\scalebox{0.9}{$\input{Diagrams/1_sys61.tikz}=\input{Diagrams/1_sys62.tikz}$}\;.
\end{equation}
Finally, we denote discarding in group $G$ as:
\begin{equation}
\input{Diagrams/1_sys29.tikz}\;,
\end{equation}
such that for every $g\in G$, one has
\begin{equation}
    \input{Diagrams/1_sys30.tikz}=    \input{Diagrams/1_sys31.tikz}\;,
\end{equation}
Applying discarding after copying on one of the branches gives
\begin{equation}
  \scalebox{0.9}{$  \input{Diagrams/1_sys43.tikz}=    \input{Diagrams/1_sys45.tikz}=\input{Diagrams/1_sys44.tikz}$}\;,
\end{equation}
Thus, one has
\begin{equation}
\label{eq:copy-discard}
  \scalebox{0.9}{$\input{Diagrams/1_sys46.tikz} =\input{Diagrams/1_sys29.tikz}$}\;.
\end{equation}
\begin{remark}
   Formally, at this stage, we can think about this all as a convenient diagrammatic representation of a group object living in the category $\mathbf{Set}$.
\end{remark}
Having this diagrammatic notation of a group allows us to elegantly depict group-theoretic equations diagrammatically. For example, we can depict group associativity as
\begin{equation}
\label{eq:associativity}
    \scalebox{0.9}{$\input{Diagrams/1_sys32.tikz}=   \input{Diagrams/1_sys33.tikz}$}\;,
\end{equation}
and the identity element being the unit as
\begin{equation}
   \label{eq:identity-unit}\scalebox{0.9}{$\input{Diagrams/1_sys37.tikz}=\input{Diagrams/1_sys39.tikz}=\input{Diagrams/1_sys38.tikz}$}\;.
\end{equation}
Also, since multiplying any group element by its inverse yields the identity element, one has
\begin{equation}
\label{copy-inverse}
\scalebox{0.9}{$\input{Diagrams/1_sys34.tikz}=\input{Diagrams/1_sys36.tikz}=\input{Diagrams/1_sys35.tikz}$}\;.
\end{equation}

 In this paper, however, we are not just interested in group elements and group operations on them, but also in probability measures over the group, $\mathcal{D}[G]$, and group operations acting on them. We, in particular, restrict our attention to compact groups.
 \begin{remark}\label{rem:hopf}
     Formally, we can see all of this as defining a Hopf algebra in the category of Markov kernels,  $\mathbf{Stoch}$, via the Dirac embedding from $\delta:\mathbf{Meas}\to\mathbf{Stoch}::f\mapsto \delta_f$. The fact that the group object now lives in $\mathbf{Meas}$ rather than $\mathbf{Set}$ means that we must restrict to measurable groups, which is implied by our restriction to compact groups. With slight abuse of notation, we will use the same diagrams to represent the image of $\delta$ as the domain, as we will not require anything in the domain for the remainder of the paper.
 \end{remark}

 We denote the Haar probability measure as $\mu\in \mathcal{D}[G]$, and define
\begin{equation}
\label{eq:haar-measure}
 \input{Diagrams/1_sys56.tikz}:=\int_{G} d\mu (g)\input{Diagrams/1_sys2.tikz}\;.
 \end{equation}
As it is a probability measure it is normalised, i.e.,
\begin{equation}\label{eq:normalised-haar-measure}
\int_{ G} d\mu (g)= 1 ,
\end{equation}
which means that
\begin{equation}\label{eq:haar-is-normalised}
\input{Diagrams/1_sys53.tikz}=    \input{Diagrams/1_sys31.tikz}\;.
\end{equation}

Moreover, as we have restricted our attention to compact groups, the Haar measure is invariant under both left and right multiplication, i.e.,
\begin{equation}
\label{eq:haar-invariance}
\scalebox{0.9}{$\input{Diagrams/1_sys22.tikz}=\input{Diagrams/1_sys24.tikz}=\input{Diagrams/1_sys23.tikz}$}\;,
\end{equation}
and the group inverse, i.e.,
\begin{equation}
\label{eq:haar-invariance-inverse}
\scalebox{0.9}{$\input{Diagrams/1_sys54.tikz}=\input{Diagrams/1_sys55.tikz}$}\;.
\end{equation}

 In this paper, we are not just interested in abstract group theory, but in the action of these groups on physical systems, processes, and on whole physical theories. We address this in the following subsections.

\subsection{Group actions on systems}
\label{subsec:symmetries-of-systems}

One way to define the action of the group $G$ on a system $A$ would be to say that for each element of the group $g\in G$ we should have some element of the real linear span of its physical transformations, $\pi_g\in\spann(P_A^A)$, and that these transformations are chosen to preserve the group structure and topology. In this paper, however, we find it more convenient to group all of these transformations $\{\pi_g\}_{g\in G}$ together into a single map $\pi$ which has a group system $G$ as an extra (not necessarily physical) input which can be thought of as a control for which transformation is applied to the system $A$. 

\begin{definition}[System actions ($S$-actions)]\label{def:system-symmetry}
Let $A$ be a system in a GPT $\mathcal{P}$ and $G$ be some group. An $S$-action of $G$ on $A$, $\pi:G\to\spann(P_A^A)$, is depicted as
\begin{equation}\label{eq:system-symmetry}
\input{Diagrams/1_sys.tikz}\;,
\end{equation}
and for a given group element $g$, we depict $\pi_g$ as
\begin{equation}
\label{eq:sys-rep}
\input{Diagrams/1_sys25.tikz}:=\input{Diagrams/1_sys52.tikz}\;.
\end{equation}
Composition involving elements of $\spann(P_A^A)$ is understood through the bilinear extensions of the GPT composition maps. To ensure that this is indeed a valid group action, we demand that the action of the identity group element $e\in G$ is the identity transformation:
\begin{equation}\label{eq:ident}
\input{Diagrams/1_sys26.tikz}=\input{Diagrams/1_sys27.tikz}\;,
\end{equation}
and that group multiplication is represented by sequential composition:
\begin{equation}
\label{eq:rep-sequential}
\scalebox{0.8}{$\input{Diagrams/1_sys16.tikz}=\input{Diagrams/1_sys17.tikz}$}\;.
\end{equation}
Moreover, we demand that $\pi$ is a continuous  function implying that its action on probability measures $\pi:\mathcal{D}[G]\to\spann(P_A^A)$ is well defined by integration, that is, for any $\nu\in\mathcal{D}[G]$ we have that $\pi_\nu:=\int_Gd\nu(g) \pi_g$.
\end{definition}

\begin{definition}[Physical and nonphysical $S$-actions]
An $S$-action is \emph{physical} if $\pi_g\in P_A^A$ for every $g\in G$, and is \emph{nonphysical} otherwise.
\end{definition}
\begin{remark}
Beyond local tomography (see~\cref{def:local_tomography}), distinct elements of $\spann(P_A^A)$ may act identically on states of $A$ but differently on joint states of $AE$. Once $\pi_g\in\spann(P_A^A)$ is specified, whether physical or nonphysical, its ancillary extensions are fixed by $s\mapsto(\pi_g\boxtimes\mathcal I_E)\circ s$ for every $E$ and $s\in\spann(P^{A,E})$\footnote{An alternative, and potentially more general, approach to defining nonphysical actions for a tomographically nonlocal GPT is to define them as ancillary-system-indexed families of linear maps without assuming that these are generated by elements of $\spann(P_A^A)$ -- we explore this in follow up work.}. 
\end{remark}
 \blk

\blk
\begin{remark} \label{rem:integration}

Let \(G\) be a compact Hausdorff group, equipped with its Borel \(\sigma\)-algebra \(\mathcal B(G)\). We equip the finite-dimensional real vector space \(\spann(P_A^A)\) with its norm topology (which is independent of the choice of norm) and \(P_A^A \subseteq \spann(P_A^A)\) with the corresponding subspace topology. Every continuous map \(G \to \spann(P_A^A)\) is then Borel measurable.

Since $\pi : G \to \spann(P_A^A)$ is continuous and \(G\) is compact, \(\pi(G)\) is compact and hence bounded. Therefore \(\pi\), regarded as a \(\spann(P_A^A)\)-valued map, is integrable with respect to every
\(\nu \in \mathcal D[G]\), and
\[
    \pi_\nu := \int_G d\nu(g)\,\pi_g
\]
is well defined. Moreover, \(\pi_\nu\) lies in the closed convex hull of \(\pi(G)\). Since \(\spann(P_A^A)\) is finite dimensional, the convex hull of the compact set \(\pi(G)\) is compact, hence closed. Thus $\pi_\nu\in \spann(P_A^A)$. If the $S$-action is physical, then
\[
    \pi_\nu \in \operatorname{conv}(\pi(G)) \subseteq P_A^A,
\]
where the last inclusion follows from the convexity of \(P_A^A\). A detailed treatment of integration in finite-dimensional compositional GPTs, including the construction used here, will be given in forthcoming work~\cite{SalzgerSelbySoltaniIntegration}.

\end{remark}

We can then define the automorphism group for a given system:
\begin{definition}(Automorphism group of a system).
Let $A$ be a system in a GPT $\mathcal{P}$. The automorphism group
$\big(\spann(P_A^A)\big)^\times$ is the group of invertible elements
of the real algebra $\spann(P_A^A)$.
\end{definition}

With this definition in hand we can then prove the following.
\begin{proposition}
Given an arbitrary $S$-action $\pi$, it is necessarily the case that $\pi_g\in\big(\spann(P_A^A)\big)^\times$.
\end{proposition}
\proof
Using our basic definitions one has that
\begin{equation}
\label{eq:identity-discard}
\scalebox{0.9}{$\input{Diagrams/1_sys47.tikz}\stackrel{\eqref{eq:rep-sequential}}{=}\input{Diagrams/1_sys48.tikz}\stackrel{\eqref{copy-inverse}}{=}\input{Diagrams/1_sys49.tikz}\stackrel{\eqref{eq:ident}}{=}\input{Diagrams/1_sys50.tikz}$}\;,
\end{equation}
 and similarly, one has
\begin{equation}
 \label{eq:identity-discard1}
\scalebox{0.8}{$\input{Diagrams/1_sys51.tikz}=\input{Diagrams/1_sys50.tikz}$}\;.
\end{equation}
Then, for every $g\in G$ \cref{eq:identity-discard,eq:identity-discard1} imply that one has
\begin{equation}
\label{eq:invertibility-of-representation}
   \scalebox{0.8}{$ \input{Diagrams/1_sys40.tikz}=  \input{Diagrams/1_sys42.tikz}=\input{Diagrams/1_sys41.tikz}$}\;.
\end{equation}
Therefore, $\pi_g$ is necessarily invertible for every system $A$ and every $g \in G$ with the inverse $\left(\pi_g\right)^{-1}$ given by
\begin{equation}
    \input{Diagrams/1_sys28.tikz}\;,
\end{equation}
that is, by $\pi_{g^{-1}}$.
\endproof

This implies that every $S$-action defines a group homomorphism $\pi:G\to\big(\spann(P_A^A)\big)^\times$. If the $S$-action is physical,
then each $\pi_g$ is moreover a reversible physical process.

Note, however, that when we consider the action of $\pi$ on general probability measures, it is no longer necessarily the case that these will be mapped to invertible elements of $\spann(P_A^A)$.

\begin{example}[Symmetries of quantum systems]
For a quantum system $A$ associated to $\Cl^a$, the process space $P_A^A$ is given by completely positive trace-nonincreasing maps in $U_A^A=\Hom(\Herm(\Cl^a),\Herm(\Cl^a))$, with $\bar P_A^A$ given by the $\CPTP$ maps. In this case, $\spann(P_A^A)=U_A^A$. General $S$-actions therefore take values in the invertible elements of $\spann(P_A^A)$, which need not be physical quantum transformations. Physical $S$-actions, on the other hand, take values in $P_A^A$, and are therefore given by unitary transformations $\cu_g:X\mapsto U_g X U_g^\dagger$.
\end{example}

Two further related concepts are important for this work, collective actions, and intertwiners.

\begin{definition}(Collective actions). A collective action on a composite system $S=(A_1,A_2,\cdots , A_n)$ associated with the group $G$ is given by $\pi_{S}:G\to\spann(P^{A_1\cdots A_n}_{A_1\cdots A_n})$, and is depicted as
\begin{equation}
\label{eq:collective-representation}
\scalebox{0.9}{$\input{Diagrams/1_sys19.tikz}$}\;,
\end{equation}
where the $\pi^{(i)}$ are valid $S$-actions for the systems $A_i$. We denote this symbolically as $\pi_S= \pi^{(1)}\boxtimes\cdots\boxtimes\pi^{(n)}$. It is straightforward to verify that this is indeed a valid group action. If all the $\pi^{(i)}$ are physical $S$-actions, then $\pi_S$ is physical.
\end{definition}

\begin{definition}\label{def:intertwiners}(Intertwiners). An intertwiner between a pair of $S$-actions on systems, $(\pi_A:G\to\spann(P_A^A),\pi_B:G\to\spann(P_B^B))$, is a process $f\in P_A^B$ such that:
\begin{equation}
\input{Diagrams/J4_twirl61.tikz}=\input{Diagrams/J4_twirl62.tikz}\;.
\end{equation}
Special cases of intertwiners characterise invariant states and effects by letting $A$ or $B$ be trivial systems (with the trivial representation) respectively.
\end{definition}

Group actions on systems can, in fact, be used to induce group actions on process spaces, and hence on theories as a whole, as we will see in the subsequent sections.

\subsection{Group action on process spaces}

In the previous subsection we saw how groups could act on the systems within a theory. In this section we instead turn to group actions specified directly on the process spaces of a theory. To do so, it is convenient to denote the space of process maps from $(\bar{P},P,U)$ to $(\bar{Q},Q,V)$ as $\mathsf{Proc}(P,Q)$, whose span can be identified with the finite dimensional vector space of linear maps $\mathcal{L}(\text{LinSpan}(P),\text{LinSpan}(Q))$.

\begin{definition}[Process actions ($P$-actions)]\label{def:group_action_process_space}
A $P$-action $\eta$ of a group $G$ on a process space $(\bar{P},P,U)$ is given by a map $\eta: G\to \mathsf{Proc}(P,P)::g\mapsto \eta_g$  where the identity element is represented by the identity process map, $\eta_e=\mathds{1}$ and group multiplication is preserved, $\eta_{g}\circ \eta_{g'}=\eta_{g\cdot g'}$. Moreover, we demand that $\eta$ is a continuous function, such that its action on probability measures $\eta:\mathcal{D}[G]\to \mathsf{Proc}(P,P)$ is well defined by integration, that is, for any $\nu\in\mathcal D[G]$ we have that $\eta_\nu := \int_Gd\nu(g) \eta_g$.
\end{definition}

Note that this definition automatically implies that the $\eta_g$ are in fact process space isomorphisms, with inverse given by $\eta_{g^{-1}}$. Moreover, this is in fact a special kind of group representation on $\mathcal{L}(\text{LinSpan}(P),\text{LinSpan}(P))$. It is special in the sense that not every group representation on this space consists of process space isomorphisms, i.e., they will not necessarily map processes to processes.

\begin{definition}[Physical and nonphysical $P$-actions]\label{def:physical-process-action}
Let $\eta$ be a $P$-action on a process space $P_A^B$. It is \emph{physical} if, for every $g\in G$, its action can be realised by a fixed physical circuit with a single open slot into which any process $f:A\to B$ can be inserted, where the circuit may depend on $g$ and on the process type $A\to B$, but not on the particular process $f$.

Diagrammatically, if we depict $\eta_g(f)$ as
\begin{equation}
\input{Diagrams/17_physicality.tikz}\;,
\end{equation}
then physicality of $\eta$ implies that for every $g\in G$, there exist a system $E_g\in\mathsf{Sys}[\mathcal P]$, $h_g\in P_A^{A,E_g}$, and $l_g\in P_{B,E_g}^{B}$ such that, for every $f\in P_A^B$, one has\footnote{For notational brevity, one could equivalently depict $\eta_g$, $h_g$, and $l_g$ as boxes $\eta$, $h$, and $l$ controlled by the same group element $g$. We will use this controlled notation later in the paper.}
\begin{equation}
\input{Diagrams/17_physicality.tikz}
\quad=\quad
\input{Diagrams/17_physicality2.tikz}\;.
\end{equation}
The $P$-action is \emph{nonphysical} otherwise.
\end{definition}
Physicality is stronger than merely mapping physical processes to physical processes, which every $P$-action already does by definition.
\begin{remark}

 Since $\mathrm{LinSpan}(P)$ is a finite-dimensional real vector space, so is $\mathcal{L}(\mathrm{LinSpan}(P),\mathrm{LinSpan}(P))$. Hence, the same argument as in~\cref{rem:integration} applies here, with $\spann(P_A^A)$ replaced by $\mathcal{L}(\mathrm{LinSpan}(P),\mathrm{LinSpan}(P))$ and $P_A^A$ replaced by $\mathsf{Proc}(P,P)$. In particular, for every $\nu\in\mathcal{D}[G]$, the integral $\eta_\nu:=\int_G d\nu(g)\,\eta_g$, understood componentwise in any basis, is well defined and independent of the choice of basis. Moreover, since $\mathsf{Proc}(P,P)$ is convex, one has $\eta_\nu\in\mathsf{Proc}(P,P)$.
\end{remark}

\begin{definition}[Intertwiners]
Given a pair of process spaces $P$ and $Q$ each with a group action $\eta^P$ and $\eta^Q$ respectively, we can define intertwiners as process maps $\varphi\in \mathsf{Proc}(P,Q)$ such that $\eta^Q_g\circ \varphi = \varphi\circ \eta^P_g$ for all $g\in G$.
\end{definition}

 The group actions on systems from the previous section can be leveraged to define group actions on process spaces $P_{A_1,...,A_n}^{B_1,...,B_m}$ within the theory as follows:

\begin{definition}[$P$-actions induced by $S$-actions]
\label{def:process-symmetries-induced-by-system-symmetries}
Given a choice of $S$-actions
$\boldsymbol{\pi}:=\left\{\pi_X:G\to\spann(P_X^X)\middle|X\in\mathbf{Sys}[\mathcal{P}]\right\}$,
we can define an induced action on the process space
$P_{A_1,...,A_n}^{B_1,...,B_m}$ via
\begin{equation}
\label{eq:SystemToProcess}
\scalebox{0.9}{$\input{Diagrams/j1.tikz}$}\;.
\end{equation}
Whenever a collection of $S$-actions is used in this way, we take its action on composite systems to be collective in the sense defined above, and its action on the trivial system to be trivial. Whenever the resulting maps preserve the physical process space, they define a $P$-action. If the $S$-actions are physical, the induced $P$-action is physical. An action on states is the special case with no inputs and on effects as the special case with no outputs.
\end{definition}

A general $P$-action, however, need not be induced by $S$-actions.

\subsection{Group actions on theories}
\label{Sec:TheorySymmetries}

A family of $P$-actions specified directly on the process spaces defines an action on the theory as a whole when it is compatible with the compositional structure of the theory. Equivalently, this amounts to providing a suitable diagram-preserving map $\Xi_g:\mathcal{P}\to\mathcal{P}$ for each $g\in G$ such that these preserve the group structure and topology. Like with group actions on systems, however, we find it more convenient to instead take $G$ as a control system which we can use to select which diagram-preserving map $\Xi_g$ is applied.

\begin{definition}[Theory actions ($T$-actions)]\label{def:theorySym}
A $T$-action of a group $G$ on $\mathcal{P}$ is a $G$-controlled diagram-preserving map $\Xi$. This is defined by its actions on the process spaces making up $\mathcal{P}$. That is, for every $P_A^B$ we have a $P$-action (see \cref{def:group_action_process_space})
depicted on a process $f \in P^B_A$ as 
\begin{equation}
\input{Diagrams/2_theorysym.tikz}\;.
\end{equation}
Given that this is a process space action we can define it both on elements of the group $g\in G$ as well as probability distributions on the group $\nu\in \mathcal D[G]$ by integration.
We denote these as
\begin{equation}
\input{Diagrams/2_theorysym1.tikz}\quad \text{and} \quad\input{Diagrams/new_2_theorysym1.tikz}
\end{equation}
respectively.

For each group element $g \in G$, we obtain a  diagram preserving map  $\Xi_g:\mathcal{P}\to\mathcal{P}$  that is identity on systems.
The diagram-preserving property of the $\Xi_g$ can be captured in terms of $\Xi$, for example, by
\begin{equation}
\label{eq:diagram-preservation}
\scalebox{0.9}{$\input{Diagrams/2_theorysym4.tikz}=\input{Diagrams/2_theorysym5.tikz}$}\;.
\end{equation}
This recovers the diagram-preservation of the $\Xi_g$ via, for example,
\begin{align}
\label{eq:diagram-preservation1}
\scalebox{0.9}{$\input{Diagrams/2_theorysym2.tikz}$}&=\scalebox{0.9}{$\input{Diagrams/2_theorysym3.tikz}$}\nonumber\\
&=\scalebox{0.9}{$\input{Diagrams/2_theorysym6.tikz}$}\;.
\end{align}
When we consider probability distributions $\nu\in\mathcal D[G]$, we can still define $\Xi_\nu$ by its action on the process spaces $P_A^B$ by integration, but
it is no longer (necessarily) the case that $\Xi_\nu$ will be diagram preserving.

Since $\Xi$ is a group action, for the identity element $e\in G$, we have that $\Xi_e$ is the identity, such that
\begin{equation}
\label{eq:identity-theory-sym}
\input{Diagrams/2_theorysym13.tikz}=\input{Diagrams/2_theorysym14f.tikz}\;.
\end{equation}
and group multiplication is represented by sequential application of the diagram preserving maps:
\begin{equation}
\label{eq:seq-map}
\input{Diagrams/2_theorysym11.tikz}=\input{Diagrams/2_theorysym12.tikz}\;.
\end{equation}
\end{definition}\blk
Note that, equivalently, one can define a process action $\eta^B_A$ for every process space $P_A^B$ of the theory of interest, such that (i) composition is an intertwiner, that is, in a GPT $\mathcal{P}$ with sequential composition $\mathsf{seq}_{A,B,C}:P_A^B\times P_B^C\to P_A^C$ one has
\begin{equation}
    \mathsf{seq}_{A,B,C}\circ\left( \eta_A^B (g) \times \eta_B^C (g)\right)=\eta _A^C (g) \circ   \mathsf{seq}_{A,B,C} \quad, \forall g\in G,
\end{equation}
(ii) for parallel composition $\mathsf{par}_{A,C}^{B,D}:P_A^B\times P_C^D\to P_{A,C}^{B,D}$, one has
\begin{equation}
    \mathsf{par}_{A,C}^{B,D}\circ\left( \eta_A^B (g) \times \eta_C^D (g) \right)=\eta _{A,C}^{B,D} (g)\circ  \mathsf{par}_{A,C}^{B,D} \quad, \forall g\in G,
\end{equation}
and (iii) it preserves the identity and the swap for all process spaces.
 To see that this is indeed equivalent, one direction is obtained by relabelling the action of $\Xi$ on each process space $P_A^B$ as $\eta_A^B$, while the diagram-preserving property of $\Xi$ ensures that the resulting process actions satisfy conditions (i) to (iii). Conversely, if each process space $P_A^B$ is equipped with a process action $\eta_A^B$ of the same group $G$ satisfying conditions (i) to (iii), then these actions together define a group action $\Xi$ on $\mathcal{P}$. Thus, compatibility with the compositional structure is the additional condition required for a family of $P$-actions $\eta_A^B$ on the process spaces to define a $T$-action.

We can also define the group of reversible diagram preserving maps for a given theory:
\begin{definition}(Automorphism group of a theory). For a GPT $\mathcal{P}$, the group of symmetries $\mathsf{Sym}[\mathcal{P}]$ is defined as the set of all invertible diagram preserving maps of $\mathcal{P}$ into itself that are linear, and act trivially on objects.
\end{definition}
\begin{proposition}\label{prop:automophismsPreserveScalars}
    Automorphisms, $\Xi\in \mathsf{Sym}[\mathcal{P}]$, of a theory are necessarily scalar preserving.
\end{proposition}
\proof
Since for each process space of $\mathcal{P}$, $\Xi$ defines a process map, consider the scalar process space $P_I^I = (\{1\}, [0,1], \mathds{R})$. Then $\Xi : P_I^I \to P_I^I$ is given by an invertible linear map $f : \mathbb{R} \to \mathbb{R}$. By linearity, there exists $r_f \in \mathbb{R}$ such that $f(x) = r_f x$. Invertibility implies $r_f \neq 0$. Moreover, since $f([0,1]) = [0,1]$, it follows that $r_f = 1$, hence $f(x) = x$, which implies that $\Xi$ preserves scalars.
\endproof

Analogously to group action on systems, we can show that:
\begin{proposition}\label{prop:xi-is-symmetry}
    Given an arbitrary group action $\Xi$ on theory $\mathcal{P}$ it is necessarily the case that $\Xi_g\in \mathsf{Sym}[\mathcal{P}]$.
\end{proposition}
\proof
\cref{eq:seq-map,eq:identity-theory-sym} imply that $\Xi_g$ is an isomorphism, since one has
\begin{equation}
\scalebox{0.9}{$\input{Diagrams/2_theorysym15.tikz}=\input{Diagrams/2_theorysym17.tikz}= \input{Diagrams/2_theorysym16.tikz}$}\;.
\end{equation}
\endproof
An immediate corollary of these two propositions is that the group action is necessarily scalar preserving.
\begin{definition}[Physical and nonphysical $T$-actions]\label{def:physical-theory-action}
A $T$-action $\Xi$ is \emph{physical} if its restriction to every process space is a physical $P$-action in the sense of~\cref{def:physical-process-action}, and is \emph{nonphysical} otherwise.
\end{definition}

Physicality of a $T$-action is therefore stronger than the requirement that each $\Xi_g$ preserve the process spaces of the theory, which is already part of the definition of a $T$-action. It additionally requires that, on every process space, the corresponding $P$-action itself be physically implementable in the sense of~\cref{def:physical-process-action}. In particular, every $T$-action induced by physical $S$-actions is physical.

 \begin{definition}[$T$-actions induced by $S$-actions]\label{def:theory-sym-induced-by-sys}
Given a group $G$ and a choice of $S$-actions
$\boldsymbol{\pi}:=\left\{\pi_X:G\to\spann(P_X^X)\middle|X\in\mathbf{Sys}[\mathcal{P}]\right\}$,
suppose that the induced actions of~\cref{def:process-symmetries-induced-by-system-symmetries} define a $P$-action on every process space. These $P$-actions are compatible with composition and hence define a $T$-action $\Pi$:
\begin{equation}
    \scalebox{0.9}{$\input{Diagrams/john4.tikz}\quad:=\quad\input{Diagrams/john3.tikz}$}\;.
\end{equation}
\end{definition}

It is straightforward to verify that the associated maps $\Pi_g$ for each $g\in G$ are indeed diagram preserving, for example, using~\cref{eq:sys-rep}, one has
\begin{align}
\scalebox{0.70}{\input{Diagrams/2_theorysym18.tikz}}&= \scalebox{0.70}{\input{Diagrams/2_theorysym19.tikz}}\nonumber\\[10pt]
\stackrel{\eqref{eq:invertibility-of-representation}}{=}\scalebox{0.70}{\input{Diagrams/2_theorysym20.tikz}}
&=\scalebox{0.70}{\input{Diagrams/2_theorysym21.tikz}}\;.
\end{align}

\begin{remark}[Hierarchy of actions]\label{rem:symmetry-terminology}
We distinguish three levels of group actions. $S$-actions act on systems, $P$-actions act on individual process spaces, and $T$-actions act on the theory as a whole.

Given a collection of $S$-actions, we take the action on composite systems to be collective and the action on the trivial system to be trivial. If, moreover, the induced action on a process $f\in P_A^B$ is defined by
\begin{equation}
\eta_g(f)
=
\pi_B^g\circ f\circ\pi_A^{g^{-1}},
\end{equation}
with the corresponding collective actions used whenever $A$ or $B$ is composite, then, whenever these induced maps send valid processes to valid processes and hence define $P$-actions, they are automatically compatible with the compositional structure of the theory. Consequently, the induced $P$-actions on all process spaces together define a $T$-action.

A general $P$-action, however, need not be induced by $S$-actions. More generally, a family of $P$-actions on the process spaces defines a $T$-action precisely when it is compatible with the compositional structure of the theory.

Physicality is a separate distinction at each level. $S$-, $P$-, and $T$-actions may each be physical or nonphysical in the senses defined above. Physical $S$-actions necessarily induce physical $P$-actions, and hence a physical $T$-action. Nonphysical $S$-actions, on the other hand, may induce either physical or nonphysical $P$-actions and, correspondingly, physical or nonphysical $T$-actions.

\end{remark}

Importantly, there exist $T$-actions which are not induced by physical $S$-actions, an important example of which is global complex conjugation in quantum theory. In locally tomographic quantum theory, this $T$-action admits $S$-actions given by transposition on each subsystem, which are nonphysical since transposition is not itself a physical quantum process.

Given a $T$-action on a theory, it is of particular interest to study the invariant processes of that theory, which can be used to define an invariant subtheory as follows:

    \begin{definition}[Invariant subtheories]\label{def:invariant-subtheories}
    An invariant subtheory of a GPT $\mathcal{P}$ with respect to group $G$ and the action $\Xi: G\to \text{Sym}[\mathcal{P}]$ is given by the subtheory that for each $g\in G$, every process remains invariant under the action of $\Xi_g:\mathcal{P}\to \mathcal{P}$, i.e.,
    \begin{equation}
        \label{eq:invariant-subtheories}
        \input{Diagrams/4_twirl102.tikz}=\input{Diagrams/2_theorysym14f.tikz}\;,\;\forall g \in G.
    \end{equation}

    \end{definition}

Note that, as we will show in subsequent sections, composing invariant processes yields another invariant process, so that the invariant subtheory is indeed closed under composition, and satisfies all requirements posited for a subtheory in~\cref{def:subGPT}. In general it is not clear that this will be a GPT-subtheory, as we discuss further in~\cref{Sec:Twirled}.

In the following sections, we construct an invariant subtheory given a $T$-action on a theory. To this end, we first construct invariant subtheories with respect to $T$-actions induced by physical $S$-actions. We then show how this construction of invariant subtheories can be further generalised to arbitrary $T$-actions.

 \section{System-induced twirled worlds}

In this section we develop the notion of system-induced twirled worlds~\cite{centeno2024twirled} in the fully compositional setting considered here. Given a choice of physical $S$-actions, we consider the $T$-action $\Pi$ induced by them and denote the corresponding averaging map by $\mathcal{T}$, which we call \emph{$S$-twirling}, or equivalently \emph{system-induced twirling}. The key subtlety is that the invariant-subtheory construction does not by itself guarantee process tomography (\cref{def:tomographic-completness}). We provide two distinct conditions which suffice to ensure process tomography. Moreover, we show how all of these system-induced twirled worlds for a given group $G$ acting on a given theory $\mathcal{P}$ can be unified within one common theory $\mathsf{Twirl}_G[\mathcal{P}]$.

\subsection{System-induced twirling on processes}\label{Sec:Twirled}

Before we get to defining $S$-twirled worlds, let us define $S$-twirling on processes.

\begin{definition}\label{def:twirl-on-process-space}($S$-twirling on processes). In a GPT $\mathcal{P}$, for a process $f \in P^{B_1\cdots B_m}_{A_1\cdots A_n}$, given a chosen physical $S$-action on each system, $\boldsymbol{\pi}$, the corresponding \textit{$S$-twirling} $\mathcal{T}$ on $f$ is defined as: 
\begin{align}\label{eq:twirling-def}
\scalebox{0.9}{$\input{Diagrams/4_twirl1.tikz}$}&:=\scalebox{0.9}{$\input{Diagrams/4_twirl3.tikz}$}\nonumber\\%
&=\scalebox{0.9}{$ \input{Diagrams/john2.tikz}$}\;,
\end{align}
where $\mu$ is the Haar probability measure for the compact group $G$, and $\Pi$ is (following~\cref{def:theory-sym-induced-by-sys}) the $T$-action induced by the physical $S$-actions $\boldsymbol{\pi}$. Since $\Pi$ is physical, this averaging is an $S$-twirling in the terminology introduced above.
\end{definition}

Next, let us prove some important properties of $S$-twirling and $S$-twirled processes. We begin with the following useful lemma:
\begin{lemma}
Applying $\Pi$ after $S$-twirling leaves all processes invariant:
     \begin{equation}
\label{eq:haar-measure-invariance}
  \scalebox{0.9}{$  \input{Diagrams/4_twirl86.tikz}= \input{Diagrams/4_twirl87.tikz}$}\;.
\end{equation}
 \end{lemma}
 \proof
 Using the group structure and the invariance of the Haar measure, one can show
\begin{equation}
\scalebox{0.9}{$\input{Diagrams/4_twirl6.tikz}\stackrel{\eqref{eq:seq-map}}{=}\input{Diagrams/4_twirl8.tikz}
\stackrel{\eqref{eq:haar-invariance}}{=}\input{Diagrams/4_twirl7.tikz}$}\;,
\end{equation}
or equivalently
\begin{equation}
\label{eq:haar-measure-invariance1}
\scalebox{0.9}{$\input{Diagrams/4_twirl22.tikz}=\input{Diagrams/4_twirl23.tikz}$}\;.
\end{equation}
Recalling that we can write twirling as
\begin{equation}\label{eq:twirl_rep}
\input{Diagrams/4_twirl.tikz}=\input{Diagrams/4_twirl7.tikz}\;,
\end{equation}
we can rewrite~\cref{eq:haar-measure-invariance1} as
\begin{equation}
  \scalebox{0.9}{$ \input{Diagrams/4_twirl86.tikz}= \input{Diagrams/4_twirl87.tikz}$}\;,
\end{equation}
which completes the proof.
\endproof

Using this lemma it is straightforward to show that $S$-twirling is idempotent:
\begin{proposition}\label{lem:idempotent}
The $S$-twirling map is idempotent, that is for every pair of systems $A,B\in\textbf{Sys}[\mathcal{P}]$, and every process $f \in P^B_A$, one has $\mathcal{T}\big(\mathcal{T}(f)\big) = \mathcal{T}(f)$.
\end{proposition}
\begin{proof}
This can be shown using the properties of the Haar measure, as for every pair of systems $A,B\in\textbf{Sys}[\mathcal{P}]$, for every $f\in P^B_A$, one has
\begin{equation}
\scalebox{0.9}{$\input{Diagrams/4_twirl63.tikz}=    \input{Diagrams/4_twirl64.tikz}
\stackrel{\eqref{eq:haar-measure-invariance}}{=}\input{Diagrams/4_twirl65.tikz}
=\input{Diagrams/4_twirl66.tikz}$}\;.
\end{equation}
\end{proof}

Next we can show that invariance under $S$-twirling is equivalent to being invariant under the group action which, later, will be essential to connecting $S$-twirled worlds and invariant subtheories.
\begin{proposition}
\label{prop:group-invariance}
For every pair of systems $A,B \in \textbf{Sys}[\mathcal{P}]$, a process $f \in P^B_A$ is invariant under $S$-twirling if and only if it is invariant under $\Pi$ for all group elements $g \in G$, i.e.,

\begin{equation}
\input{Diagrams/4_twirl.tikz}=\input{Diagrams/2_theorysym14f.tikz} \iff \input{Diagrams/4_twirl53.tikz}=\input{Diagrams/2_theorysym14f.tikz}\;,\;\forall g \in G.
\end{equation}
\end{proposition}
\begin{proof}
To see that invariance under $S$-twirling implies invariance under $\Pi$ for any arbitrary group element, consider applying $\Pi$ for a group element $g \in G$ on a process $f$ that is invariant under $S$-twirling, i.e.,
\begin{align}
\scalebox{0.7}{$\input{Diagrams/4_twirl53.tikz}$} =& \scalebox{0.7}{$\input{Diagrams/4_twirl92.tikz}$}
\stackrel{\eqref{eq:haar-measure-invariance}}{=}&\scalebox{0.7}{$\input{Diagrams/4_twirl93.tikz}$}\;,
\end{align}
For the other direction, first note that, for any $g\in G$, the action of $\Pi$ on a $\Pi$-invariant process $f$ is given by
\begin{equation}
\label{eq:invariance-under-group}
\input{Diagrams/4_twirl53.tikz}=\input{Diagrams/2_theorysym14f.tikz}\Rightarrow \input{Diagrams/4_twirl4.tikz}= \input{Diagrams/4_twirl94.tikz}\;,
\end{equation}
which in turn implies invariance under $S$-twirling as one has
\begin{equation}
\input{Diagrams/4_twirl.tikz}=\input{Diagrams/4_twirl52.tikz}=\input{Diagrams/4_twirl95.tikz}=\input{Diagrams/2_theorysym14f.tikz}\;.
\end{equation}
\end{proof}

Now, let us unpack what invariance under $\Pi$ means back at the level of the $S$-actions $\boldsymbol{\pi}$. 
\begin{proposition}
    $\Pi$-invariant processes are $\boldsymbol{\pi}$ intertwiners. In particular, this means that states and effects are invariant under the corresponding $S$-actions.
\end{proposition} 
\proof
For every $\Pi$-invariant process $f\in \mathcal{T}(P^B_A)$, and any $g\in G$, one has
\begin{equation}
  \scalebox{0.9}{$  \input{Diagrams/2_theorysym14f.tikz} = \input{Diagrams/4_twirl53.tikz} = \input{Diagrams/4_twirl60.tikz}=\input{Diagrams/4_twirl108.tikz}$}\;.
\end{equation}
Composing $\pi_g$ sequentially with both the left-hand side and the right-hand side gives
\begin{equation}
  \scalebox{0.9}{$  \input{Diagrams/4_twirl110.tikz}= \input{Diagrams/4_twirl109.tikz}\;,\quad \forall g\in G $},
\end{equation}
where using~\cref{eq:invertibility-of-representation} implies that $\Pi$-invariant processes are intertwiners (\cref{def:intertwiners}), i.e.,
\begin{equation}
\scalebox{0.9}{$\input{Diagrams/4_twirl61.tikz}=\input{Diagrams/4_twirl62.tikz}\;,\qquad \forall g \in G $}.
\end{equation}
In the special case of states, for every $\Pi$-invariant state $\omega\in P^A$, one has
\begin{equation}
\input{Diagrams/4_twirl57.tikz}=\input{Diagrams/4_twirl58.tikz}=\input{Diagrams/4_twirl59.tikz}\;,\quad \forall g \in G,
\end{equation}
so that one has $\pi^{A}_{g} \circ \omega = \omega$ for every $g \in G$. With the same logic, for every $\Pi$-invariant effect $e\in P_A$, one has $e\circ\pi^{A}_{g} = e$ for any $g \in G$.
\endproof

Finally, we obtain the following:
\begin{corollary}\label{cor:twirlingIsScalarPreserving}
    $S$-twirling is scalar-preserving, i.e., for every $p\in [0,1]$, one has $\mathcal{T}(p)=p$.
\end{corollary}
\begin{proof}
Since $\Pi$ is scalar-preserving (see~\cref{prop:automophismsPreserveScalars} and \cref{prop:xi-is-symmetry}), that is,
\begin{equation}
    \label{eq:pi-is-scalar-preserving}\input{Diagrams/4_twirl118.tikz}=\input{Diagrams/4_twirl119.tikz}\;,
\end{equation}
it follows that
\begin{equation}
\input{Diagrams/4_twirl114.tikz}=\input{Diagrams/4_twirl115.tikz}\stackrel{\eqref{eq:pi-is-scalar-preserving}}{=}\input{Diagrams/4_twirl116.tikz}\stackrel{\eqref{eq:haar-is-normalised}}{=}\input{Diagrams/4_twirl117.tikz}\;.
\end{equation}
\end{proof}

It is important to note that $S$-twirling, unlike each $\Pi_g$, is not a diagram preserving map, as we have
\begin{align}
&\scalebox{0.7}{$\input{Diagrams/4_twirl141.tikz}$}=\scalebox{0.7}{$\input{Diagrams/4_twirl142.tikz}$}=\scalebox{0.7}{$\input{Diagrams/4_twirl143.tikz}$}\nonumber\\
&\neq \scalebox{0.7}{$\input{Diagrams/4_twirl144.tikz}$}
=\scalebox{0.7}{$\input{Diagrams/4_twirl145.tikz}$}\;.
\end{align}
Nonetheless, in the following subsection we will see that it can be used to define a valid subtheory.

\subsection{System-induced twirled subtheories}

We now turn to the theory as a whole, showing that the $S$-twirled processes defined in the previous subsection can be combined in a coherent way. We give a detailed example in \cref{sec:BCT} showing how the theory known as Bilocal Classical Theory can be constructed via $S$-twirling of classical theory. As mentioned above, the key challenge to address here is the fact that $S$-twirled subtheories are not guaranteed to satisfy process tomography (\cref{def:tomographic-completness}), and hence, whilst they are necessarily subtheories of the original theory, they might not be GPT subtheories (\cref{def:subGPT}).

First we will provide a definition of an $S$-twirled subtheory, and then we will show that it does indeed define a valid subtheory of our initial theory.

\begin{definition}\label{def:invariant-subspace}
($S$-twirled subtheories)
   An \textit{$S$-twirled subtheory} $\mathcal{T}(\mathcal{P})$ has the same systems as $\mathcal{P}$ and, for every pair of systems $A,B\in\textbf{Sys}[\mathcal{P}]$, its process space $\mathcal{T}(P^B_A)$ is the subprocess space of $P^B_A$ consisting of processes that are invariant under $S$-twirling. That is, for every pair of systems $A,B\in\textbf{Sys}[\mathcal{P}]$, and every process $f \in \mathcal{T}(P_A^B)$, one has $\mathcal{T}(f) = f$, depicted as
\begin{equation}
\input{Diagrams/4_twirl.tikz}=\input{Diagrams/2_theorysym14f.tikz}\;.
\end{equation}
\end{definition}

\cref{lem:idempotent} implies that applying the $S$-twirling map to an arbitrary process yields a $\mathcal{T}$-invariant process. That is, for every pair of systems $A, B \in \textbf{Sys}[\mathcal{P}]$, applying the $S$-twirling map to any process $f \in P^B_A$ results in an invariant process $\mathcal{T}(f)$, since one has $\mathcal{T}\big(\mathcal{T}(f)\big) = \mathcal{T}(f)$.

Furthermore, the $S$-twirling map is surjective: All $\mathcal{T}$-invariant processes of an $S$-twirled subtheory are also in the original theory, where the $S$-twirling map acts as the identity on them; hence, the $S$-twirling map is surjective, and therefore full. Thus, one can obtain an $S$-twirled subtheory by applying the $S$-twirling map to the process spaces of an arbitrary GPT.

\begin{proposition}[
  restate=twirledSubspaceBeingSubtheory,
  name={}
]
\label{prop:twirledSubspaceBeingSubtheory}
An $S$-twirled subtheory $\mathcal{T}(\mathcal{P})$, obtained by applying the $S$-twirling map to a GPT $\mathcal{P}$ as defined in~\cref{def:invariant-subspace}, is indeed a subtheory of $\mathcal{P}$.
\end{proposition}
\begin{proof}
See~\cref{proof:twirled_is_subtheory}.
\end{proof}
As a consequence of~\cref{prop:group-invariance}, this subtheory is equivalent to the $\Pi$-invariant subtheory (see \cref{def:invariant-subtheories}). Hence, in this sense, $S$-twirling can be viewed as a specific map from an arbitrary theory to an invariant subtheory under $\Pi$. Furthermore, since we have shown that an $S$-twirled subtheory is a valid subtheory of the original GPT, the corresponding equivalent $\Pi$-invariant subtheory is indeed a valid subtheory as well, that is, it satisfies all the conditions posited for a subtheory in~\cref{def:subGPT}. It is, however, not necessarily the case that this will be a GPT-subtheory as it may not necessarily satisfy process tomography.

\subsection{Tomography in system-induced twirled worlds}
We investigate tomography in the $S$-twirled subtheory. In particular we show that in any $S$-twirled world we can do tomography of states and effects, but that this does not immediately generalise to general processes. Throughout the sections concerning tomography, we use a tilde to denote invariant processes whenever this helps to clarify the discussion.

\begin{proposition}\label{prop:state-tomography}
  An $S$-twirled subtheory admits state/effect tomography.
\end{proposition}
\begin{proof}
  Tomographic completeness of the original GPT (\cref{def:tomographic-completness}) implies that for any two distinct invariant states $\tilde{\omega}_1\neq\tilde{\omega}_2\in\mathcal{T}(P^A)$ in an $S$-twirled subtheory, one can find $a\in P_{A,E}$ and $s\in P^E$ from the original GPT such that
\begin{equation}
    \label{eq:state-tomography2}\scalebox{0.9}{$\input{Diagrams/4_twirl30.tikz}\neq \input{Diagrams/4_twirl31.tikz}\Rightarrow \input{Diagrams/4_twirl32.tikz}\neq\input{Diagrams/4_twirl33.tikz}$}\;.
\end{equation}
Therefore, one can find an effect $e$ from the original GPT such that
\begin{equation}
\label{eq:state-tomography}
   \scalebox{0.9}{$ \input{Diagrams/4_twirl30.tikz}\neq \input{Diagrams/4_twirl31.tikz}\Rightarrow \input{Diagrams/4_twirl34.tikz}\neq\input{Diagrams/4_twirl35.tikz}$}\;.
\end{equation}
Scalar preservation of $\mathcal{T}$ (\cref{cor:twirlingIsScalarPreserving}) implies that for an invariant state $\omega_1$, one has
\begin{align}
\scalebox{0.8}{$\input{Diagrams/4_twirl97.tikz}=\input{Diagrams/4_twirl36.tikz}=\input{Diagrams/4_twirl37.tikz}$}
=\scalebox{0.8}{$\input{Diagrams/4_twirl38.tikz}$}\nonumber\\
\stackrel{\eqref{eq:haar-measure-invariance}}{=}\scalebox{0.8}{$\input{Diagrams/4_twirl98.tikz}$}
=\scalebox{0.8}{$\input{Diagrams/4_twirl39.tikz}$}\;.
\end{align}
Hence, from~\eqref{eq:state-tomography}, one has
\begin{equation}
\label{eq:state-tomography-twirl}
\scalebox{0.8}{$\input{Diagrams/4_twirl42.tikz}\neq \input{Diagrams/4_twirl41.tikz}$}\;.
\end{equation}
Therefore, there exists an effect in an $S$-twirled subtheory that can distinguish any pair of invariant distinct states, which implies the completeness of state tomography in an $S$-twirled subtheory. The same reasoning also applies to effect tomography: for every pair of invariant distinct effects in an $S$-twirled subtheory, one can use the effect tomography of the original theory, and then, using the $S$-twirling map, show that in an $S$-twirled subtheory there exists an invariant state that allows effect tomography. Hence, an $S$-twirled subtheory satisfies effect tomography as well.
\end{proof}

It is natural to ask whether we can make a similar argument for process-tomography. Indeed a similar argument works, but we find the following:
\begin{proposition}[
  restate=twirledCombTomography,
  name={}
]
\label{prop:twirledcombtomography}
$S$-twirled combs, i.e.,
\begin{equation}
\scalebox{0.7}{$\input{Diagrams/4_twirl112.tikz}$}\;,
\end{equation}
suffice for $S$-twirled process tomography. That is, for any pair of distinct $S$-twirled processes $\tilde{f}\neq\tilde h \in \mathcal{T}(P_A^B)$ we have that there exists a state, $s\in P^{A,E}$, and effect, $e\in P_{B,E}$, in the original theory such that:
\begin{equation}
\label{eq:process-tomography-in-twiled-worlds}
\scalebox{0.70}{$
\input{Diagrams/4_twirl91.tikz} \neq
\input{Diagrams/4_twirl111.tikz}\quad .
$}
\end{equation}
\end{proposition}

\proof
See \cref{proof:twirledcombtomography}.
\endproof
It is not clear, however, that this always means that we can do process tomography within the $S$-twirled subtheory. Specifically, it is not clear that all $S$-twirled combs are realisable using states and effects within the $S$-twirled subtheory, that is, there may not exist a state $s'\in P^{A,E'}$ and effect $e'\in P_{B,E'}$ such that
\begin{equation}
\scalebox{0.7}{$\input{Diagrams/4_twirl112.tikz}$}
 = \scalebox{0.7}{$\input{Diagrams/4_twirl113.tikz}$}\quad.
\end{equation}
Hence, using only invariant effects and states is not necessarily sufficient to ensure the completeness of process tomography in an $S$-twirled subtheory, and hence, we do not have a guarantee that the $S$-twirled subtheory is a GPT-subtheory.

In the following two subsections we show, however, that under certain conditions we can indeed show that $S$-twirled worlds are GPT-subtheories. A direction for future work is to determine whether these conditions can be weakened or removed, or whether one can construct an explicit example of a well-defined GPT whose $S$-twirled subtheory fails to satisfy process tomography.

\begin{remark}
  An alternative approach which we considered would be to elevate 1-combs to being part of the defining data of a GPT, and then to define $S$-twirled subtheories as having all $S$-twirled 1-combs in their ``1-comb spaces''. This all seems like it should work out as one would hope, but doing it in full detail is beyond the scope of this paper. In particular, adding in 1-combs as additional data leads to tricky issues about composition, see, for example, \cite{JulesBlog}.
\end{remark}

\subsubsection{Dynamically faithful states}

Process tomography in general physical theories has been studied systematically in~\cite{chiribella2021process}, where processes are characterised through their action on states, including states involving ancillary systems. This is directly related to the notion of process tomography considered here. A particularly useful tool for process tomography is a dynamically faithful state, namely a bipartite state such that distinct transformations acting on one subsystem give rise to distinct bipartite output states. Many well-known GPTs admit dynamically faithful states~\cite{chiribella2010probabilistic}. A dynamically faithful state therefore provides an embedding of transformations into bipartite states, abstracting the injective direction of the Choi-Jamio{\l}kowski correspondence.
\begin{definition}[Dynamically faithful states]\label{def:dynamically-faithful-state}
    A dynamically faithful state $\sigma\in P^{A,E}$ is a state that for every pair of processes $f\in P^B_A$ and $f'\in P^B_A$ one has the following implication:
    \begin{equation}   \scalebox{0.9}{$\input{Diagrams/4_twirl103f.tikz}=\input{Diagrams/4_twirl104.tikz} \Rightarrow  \input{Diagrams/4_twirl105.tikz}=\input{Diagrams/4_twirl106.tikz}$}\;.
    \end{equation}
\end{definition}
If such a state exists, then for two distinct processes $f\in P_A^B$ and $h\in P _A^B$ one has
\begin{equation}
   \scalebox{0.9}{$ \input{Diagrams/4_twirl103f.tikz}\neq  \input{Diagrams/4_twirl107.tikz}$}\;,
\end{equation}
such that the mapping defined by
\begin{equation}
\scalebox{0.9}{$ P_A^B\to P^{B,E}::\input{Diagrams/4_twirl105.tikz} \mapsto \input{Diagrams/4_twirl103f.tikz}$}\;,
\end{equation}
is injective. We now show that if an $S$-twirled subtheory admits a dynamically faithful state, then it is tomographically complete:
\begin{proposition}\label{prop:twirl-process-tomography-faithful}
If, in an $S$-twirled subtheory, for every system $A$ there exists an ancillary system $E$ such that the subtheory admits a dynamically faithful state on system $AE$, then the $S$-twirled subtheory is tomographically complete.
\end{proposition}

\begin{proof}
    Consider two arbitrary distinct invariant processes $\tilde f \neq \tilde h \in \mathcal{T}(P^B_A)$. By assumption, the $S$-twirled subtheory admits a dynamically faithful state $\tilde \sigma \in \mathcal{T}(P^{A,E})$, and hence one has
    \begin{equation}
\scalebox{0.9}{$\input{Diagrams/4_twirl103.tikz}\neq\input{Diagrams/4_twirl122.tikz}$}\;.
    \end{equation}
   Since an $S$-twirled subtheory satisfies state tomography (\cref{prop:state-tomography}), there exists an invariant effect $\tilde e \in \mathcal{T}(P_{B,E})$ such that
    \begin{equation}
\scalebox{0.9}{$\input{Diagrams/4_twirl120.tikz}\neq\input{Diagrams/4_twirl121.tikz}$}\;.
    \end{equation}
    This shows that any pair of distinct processes in the $S$-twirled subtheory can be distinguished by the statistics they produce, using dynamically faithful states of the subtheory and invariant effects. Hence, the $S$-twirled subtheory is tomographically complete.
\end{proof}
We emphasise that a dynamically faithful state of the $S$-twirled subtheory need only distinguish processes within that subtheory. This is weaker than requiring an invariant dynamically faithful state of the original theory, by which we mean a state that is dynamically faithful in the original theory and is also invariant under $S$-twirling. Every invariant dynamically faithful state of the original theory is therefore dynamically faithful for the $S$-twirled subtheory, but the converse need not hold. This distinction is reflected in the following immediate corollary:
\begin{corollary}\label{cor:twirled-invariant-dynamically-faithful-states}
    If, in the original theory of an $S$-twirled subtheory, for every system $A$ there exists an ancillary system $E$ such that the theory admits an invariant dynamically faithful state on system $AE$, then the $S$-twirled subtheory is tomographically complete.
\end{corollary}

\begin{example}\label{ex:choi-classcial}
    In classical theory, states are probability distributions. For a classical system $A$ associated to $\Rl^a$, let $\{e_i\}_{i=0}^{a-1}$ be the standard basis of $\Rl^a$. Then the uniformly correlated state $\frac{1}{a}\sum_{i=0}^{a-1} e_i \otimes e_i \in \Rl^a \otimes \Rl^a$ of system $AA$ is an invariant dynamically faithful state of system $A$ under any collective action $\pi_A\otimes \pi_A$.
\end{example}
\begin{example}\label{ex:choi-quantum}
In quantum theory, for a quantum system $A$ associated to $\Cl^a$, an invariant dynamically faithful state is given by the maximally entangled state on $AA^*$. Here $A^*$ denotes the dual system of $A$, which transforms under the conjugate representation. Hence, the collective action is $\pi_A\otimes \pi_{A^*}$. With respect to an orthonormal basis ${\vert i\rangle}_{i=0}^{a-1}$ of $\Cl^a$ and the corresponding dual basis of ${\Cl^a}^*$, this state is $\frac{1}{\sqrt{a}}\sum_{i=0}^{a-1}\vert i \rangle_A \otimes \langle i\vert_{A^*} \in \Cl^a \otimes {\Cl^a}^*$. Equivalently, using the isomorphism $(\Cl^a)^* \cong \overline{\Cl^a}$, after identifying $A^*$ with the conjugate Hilbert space, it can be written as $\frac{1}{\sqrt{a}}\sum_{i=0}^{a-1}\vert i \rangle_A\vert i \rangle_{A^*}$. See Ref.~\cite[App.~A]{selby2022time} for more details.
\end{example}

We note that an even weaker requirement than the existence of an invariant dynamically faithful state ensures that process tomography is satisfied in an $S$-twirled subtheory, as shown in the following proposition.

\begin{proposition}\label{prop:dynamically-faithful-vector}
In an $S$-twirled subtheory, process tomography is possible if, for every system $A$, there exists an ancillary system $E$ such that the span of the invariant state space on $AE$, i.e., $\spann \left(\mathcal{T}(P^{A,E})\right)$, contains a dynamically faithful vector for the $S$-twirled subtheory.
\end{proposition}

\begin{proof}
    Let $\bar{\sigma}\in \spann \left(\mathcal{T}(P^{A,E}) \right)$ be a dynamically faithful vector for the $S$-twirled subtheory on $AE$. Then, given any two distinct invariant processes $\tilde f\neq \tilde h\in\mathcal{T}(P^B_A)$, one has
\begin{equation}
\scalebox{0.9}{$\input{Diagrams/4_twirl179.tikz}=\input{Diagrams/4_twirl177.tikz}\neq\input{Diagrams/4_twirl178.tikz}=\input{Diagrams/4_twirl180.tikz}$}\;,
\end{equation}
so that one has the implication
\begin{align}
\scalebox{0.78}{$\input{Diagrams/4_twirl181.tikz}$}\neq\scalebox{0.78}{$\input{Diagrams/4_twirl182.tikz}$}
\Rightarrow\scalebox{0.78}{$\input{Diagrams/4_twirl183.tikz}$}\neq\scalebox{0.78}{$\input{Diagrams/4_twirl184.tikz}$}\;.
\end{align}
Now, let $\{ \omega_i \}_i$ be a basis for $\spann (P^{A,E})$, where $\omega_i \in P^{A,E}$. Then, the above non-equality implies that there is at least one $\omega_i$ such that
\begin{align}
\scalebox{0.8}{$\input{Diagrams/4_twirl185.tikz}$}\neq \scalebox{0.8}{$\input{Diagrams/4_twirl186.tikz}$}
\Rightarrow  \scalebox{0.8}{$\input{Diagrams/4_twirl188.tikz}$}\neq \scalebox{0.9}{$\input{Diagrams/4_twirl189.tikz}$}\;.
\end{align}
Therefore, there exists an invariant state $\tilde{\omega}_i :=\mathcal{T}(\omega_i)$ such that, when acted on by $\tilde f\boxtimes\mathcal{I}_E$ and $\tilde h\boxtimes\mathcal{I}_E$, it results in two distinct invariant states. Since any $S$-twirled subtheory admits state tomography (see~\cref{prop:state-tomography}), one can distinguish these two distinct invariant states by invariant effects within the $S$-twirled subtheory, so that eventually one can distinguish between $\tilde f$ and $\tilde h$. Hence, process tomography holds in the $S$-twirled subtheory as well.
\end{proof}

Unlike in the previous case where we found there was a single state that sufficed for process tomography, here we find that we need a basis of states.

We emphasise that the dynamically faithful vector in the proposition need only distinguish processes within the $S$-twirled subtheory. This is weaker than requiring an invariant dynamically faithful vector of the original theory. Any such vector of the original theory lies in $\spann\!\left(\mathcal{T}(P^{A,E})\right)$ by invariance and linearity of $\mathcal{T}$, yielding the following immediate corollary:
\begin{corollary}\label{cor:dynamically-faithful-vector}
If, in the original GPT of an $S$-twirled subtheory, for every system $A$ there exists an ancillary system $E$ such that the GPT admits an invariant dynamically faithful vector in the span of the state space on $AE$, i.e., in $\spann \left( P^{A,E} \right)$, then the $S$-twirled subtheory is tomographically complete, and hence is a valid GPT-subtheory.
\end{corollary}
Later on, we show that in locally tomographic GPTs, such an invariant dynamically faithful vector always exists in the span of the invariant state space for any finite-dimensional representation of any compact group on any system. Hence, these theories satisfy process tomography.

\blk

\subsubsection{Internal classical reference frames}

So far in the paper we have treated the group systems as abstract mathematical systems. However, commonly these are viewed as physical systems in their own right -- in particular, as physical reference-frame systems encoding the group element. This motivates the question as to whether such reference frames should be treated as classical or quantum systems. Here, we show that if they are treated as classical systems, then we obtain process tomography in the $S$-twirled world.

To begin, let us define the classical GPT system corresponding to the compact group $G$.
\begin{definition}[Classical system $\Delta_G$]
Given a compact group $G$, the classical system $\Delta_G$ has normalized state space $\bar P^{\Delta_G}$ given by the probability measures defined over $G$, state space $P^{\Delta_G}$ given by the subprobability measures over $G$, and effect space $P_{\Delta_G}$ given by the measurable functions $G\to[0,1]$. We depict the classical system $\Delta_G$ as a grey wire labelled by $G$,
\begin{equation}
\input{Diagrams/15_internalRef2.tikz}\;.
\end{equation}
\end{definition}

When clear from context, we use $G$, both in notation and in diagrams, to denote the classical system $\Delta_G$ associated with the group $G$.

We, moreover, will assume that in this GPT we have universal classical control:
\begin{definition}[Universal classical control]
We say that we have universal classical control if for any continuous function  $f:G\to P_A^B$ there exists a process $\hat{f}\in P_{G,A}^B$ such that
\begin{equation}
   \scalebox{0.9}{$ \input{Diagrams/15_internalRef4.tikz}= \input{Diagrams/15_internalRef5.tikz}\;,\quad \forall g\in G $}.
\end{equation}
\end{definition}
With a slight abuse of notation, we denote $\delta_g$ simply by $g$.  The mixed process spaces $P^B_{G,A}$ involving the generally
infinite-dimensional classical system $G$, together with the compositions
used here, are constructed and shown to be well defined in forthcoming
work~\cite{SalzgerSelbySoltaniIntegration}.

Note that, since for every physical $S$-action $\pi$ on system $A$, one has $\pi: G \to P^A_A$, for GPTs with internal classical system $G$ and with universal classical control, there exists $\hat{\pi} \in P_{G,A}^A$ such that 
\begin{equation}
    \label{eq:pihat-twirl}\scalebox{0.9}{$\input{Diagrams/15_internalRef7.tikz}=\input{Diagrams/15_internalRef6.tikz}\; ,\quad\forall g\in G$ }.
\end{equation}
The discussion of measurability and the integration of $S$-actions over probability measures on $G$ in~\cref{rem:integration} applies to $\hat{\pi}$ as well.

This is exactly what we mean by an internal classical reference frame. We have replaced the abstract group system (i.e., the green wire) with a physical classical system (the grey wire) but otherwise it behaves more or less the same. The exception, however, is that as we are treating it as a physical system it should also be acted on by the group. Specifically, we can define an $S$-action on the classical system $G$ as 
\begin{definition}\label{def:action-on-classical-system}
In a GPT $\mathcal{P}$ with internal classical system $G$ and with universal classical control, the $S$-action of $G$ on the classical system $G$ is defined as 
\begin{equation}
  \label{eq:action-on-classical-system}\input{Diagrams/15_internalRef.tikz}:=  \input{Diagrams/15_internalRef3.tikz}\;,
\end{equation}
where
\begin{equation}
      \scalebox{0.9}{$\input{Diagrams/15_internalRef104.tikz}:=\input{Diagrams/15_internalRef11.tikz}\;,\quad \forall g,g'\in G$}.
\end{equation}
This will naturally lead us to define $\hat{\pi}$ (\cref{eq:pihat-twirl}) on a classical system $G$ as
\begin{equation}
   \input{Diagrams/15_internalRef105.tikz}:= \input{Diagrams/15_internalRef106.tikz} \;,
\end{equation}
where
\begin{equation}
 \scalebox{0.9}{$\input{Diagrams/15_internalRef10.tikz}:=\input{Diagrams/15_internalRef11.tikz}\;,\quad \forall g,g'\in G$}.
\end{equation}
\end{definition}
It is quite straightforward to show that, similar to~\cref{eq:rep-sequential}, one has
\begin{equation}
\label{eq:rep-sequential-controlled}
   \scalebox{0.9}{$ \input{Diagrams/15_internalRef41.tikz}= \input{Diagrams/15_internalRef42.tikz}$}\;.
\end{equation}

$S$-twirling on an arbitrary process $f$ is then given by
\begin{align}
    \label{eq:twirl-controlled}\scalebox{0.8}{$\input{Diagrams/4_twirl1.tikz}=\input{Diagrams/15_internalRef70.tikz}$}\;,
\end{align}
where
\begin{equation}\label{eq:pi-prime}
\scalebox{0.9}{$\input{Diagrams/15_internalRef71.tikz}:=\input{Diagrams/15_internalRef72.tikz}$}\;,
\end{equation}
and
\begin{equation}
\label{eq:haar-controlled}
\input{Diagrams/15_internalRef14.tikz}:=\int_{G} d\mu (g) \input{Diagrams/15_internalRef15.tikz}\;.
\end{equation}
The inverse and copy are also defined as
\begin{equation}
\scalebox{0.9}{$\input{Diagrams/15_internalRef16.tikz}:=\input{Diagrams/15_internalRef17.tikz}$}\;,
\end{equation}
and
\begin{equation}
 \scalebox{0.9}{$\input{Diagrams/15_internalRef18.tikz}:=\input{Diagrams/15_internalRef19.tikz}$}\;,
\end{equation}
 where one can verify
\begin{equation}
\label{eq:haar-invariance-twirl-controlled}
 \scalebox{0.9}{$\input{Diagrams/15_internalRef80.tikz}=\input{Diagrams/16_internalRefSwirl11.tikz}$}\;.
\end{equation}
It is straightforward to verify that a classical system $G$ in $\Delta_G$ satisfies all the properties of the group $G$ listed in~\cref{Sec:SystemSymmetries}. Without loss of generality, in the following, whenever we invoke a corresponding property of the group $G$ for systems in $\Delta_G$, we refer to the associated equation in~\cref{Sec:SystemSymmetries}.
\begin{proposition}[
  restate=piHatInvariance,
  name={}
]
\label{prop:pihatInvariance}
Applying $\hat{\Pi}$ to an arbitrary process $f\in P^B_A$ results in a $\hat{\Pi}$  invariant process, in $P^B_{G,A}$, i.e.,
\begin{equation}
    \label{eq:pihat-invariance}\scalebox{0.8}{$\input{Diagrams/15_internalRef81.tikz}=\input{Diagrams/15_internalRef82.tikz}$}\;.
\end{equation}
\end{proposition}
\proof
See \cref{app:pihat-invariance}.
\endproof
Note that, similarly to $\Pi$, one can show that a process is $\hat{\Pi}$-invariant if and only if it is invariant under $S$-twirling (with a proof analogous to~\cref{prop:group-invariance}, except that one twirls with respect to $\hat{\Pi}$ as in~\cref{eq:twirl-controlled} rather than $\Pi$); thus, an immediate corollary of the above proposition is the following:

\begin{corollary}\label{cor:invariance-under-twirl-pihat}
    Applying $\hat{\Pi}$ to any arbitrary process in $P_A^B$ yields a process, in $P^B_{G,A}$, that is invariant under both $S$-twirling and $\Pi$.
\end{corollary}
\proof
Since applying $\hat{\Pi}$ to any arbitrary process yields a $\hat{\Pi}$-invariant process (\cref{prop:pihatInvariance}), it follows that the resulting process is also invariant under $S$-twirling, and hence under $\Pi$ as well.
\endproof

\begin{theorem}[
  restate=twirlInternalRef,
  name={}
]
\label{thm:twirlInternalRef}
In a GPT $\mathcal{P}$ with internal classical system $G$ and with universal classical control, any $S$-twirled world with respect to $G$ will satisfy process tomography.
\end{theorem}
\proof
See \cref{app:twirlInternalRef}.
\endproof

\begin{example}[Reference frame quantization  via internal classical reference frames  in quantum theory]
Given a finite group $G$, then the quantum system $\Cl^{|G|}$ transforming under the regular representation of $G$, given by $g\ket h \mapsto \ket{gh}$ with $\{\ket{g}\}_{g \in G}$, is an ideal quantum reference frame for $G$. Given some quantum system $\ch_S \cong \Cl^d$ acted on by a representation $U_S(g)$ of $G$, reference frame quantization is given by  the maps  $\rho_S\mapsto \frac{1}{|G|}\sum_{g\in G}\ketbra{g}{g}\otimes U_S(g)\rho_S U_S(g)^\dagger$~\cite[Sec.~IV.B]{Bartlett_2007}. In accordance with our construction above, this realises an internal classical reference frame within quantum theory. Although the reference frame is represented by the quantum system $\Cl^{|G|}$, the group element $g$ is encoded only in the perfectly distinguishable states $\{\ketbra{g}{g}\}_{g\in G}$, so the reference frame carries classical information about the group element and moreover is only ever classically correlated with the system $\ch_S$, hence it is effectively classical. In this example, \cref{prop:pihatInvariance} can be seen directly by acting with an arbitrary $h\in G$ on the reference frame and the system. This maps the reference-frame-quantized state to $\frac{1}{|G|}\sum_{g\in G}\ketbra{hg}{hg}\otimes U_S(hg)\rho_S U_S(hg)^\dagger$, which, after the relabelling $g'=hg$, is equal to the original state. Thus the reference-frame-quantized state is invariant under the corresponding action, providing the quantum-theoretic instance of \cref{prop:pihatInvariance}. The same relabelling argument extends from states to arbitrary processes.
\end{example}

\subsection{Unification of system-induced twirled worlds}

Given some GPT $\mathcal{P}$ and some group $G$ we can construct a variety of different $S$-twirled worlds by choosing different physical $S$-actions. In this subsection we show how all of these can be combined into a unified theory which contains them all. Due to the challenges with process tomography discussed in the previous subsection, it isn't clear, however, that this unified theory will in fact be a GPT.

To be more precise, we want to define a way, given any compact group $G$, and any GPT $\mathcal{P}$, to construct a new theory $\mathsf{Twirl}_G[\mathcal{P}]$ such that for any $S$-twirling map $\mathcal{T}$ we have that $\mathcal{T}(\mathcal{P})\subseteq \mathsf{Twirl}_G[\mathcal{P}]$. Moreover, we want to do this in a way that the different $S$-twirled subtheories can non-trivially interact, i.e., we are not content simply to take the product of all of the different $S$-twirled subtheories.
\begin{definition}\label{def:twirlUnified}
    Given a compact group $G$ and a GPT $\mathcal{P}$ we define the theory $\mathsf{Twirl}_G[\mathcal{P}]$ as follows.

  Systems in $\mathsf{Twirl}_G[\mathcal{P}]$ are arbitrary pairs $(A,\pi)$ where $A\in\mathbf{Sys}[\mathcal{P}]$ and $\pi$ is a physical $S$-action on $A$. Here, $\pi$ and $\chi$ distinguish input and output physical $S$-actions, respectively.

    Processes in $\mathsf{Twirl}_G[\mathcal{P}]_{(A_1,\pi_1),...,(A_n,\pi_n)}^{(B_1,\chi_1),...,(B_m,\chi_m)}$ are processes in $\mathcal{P}$ which are invariant under the appropriate process-space action induced by these $S$-actions, that is, are processes \blk
    $f\in P_{A_1,...,A_n}^{B_1,...,B_m}$ such that
    \begin{equation}
        \label{eq:appropriate-induced-process action}\scalebox{0.9}{$\input{Diagrams/john6.tikz}\quad =\quad\input{Diagrams/john5.tikz}$}\;.
    \end{equation}
    Composition of processes in $\mathsf{Twirl}_G[\mathcal{P}]$ is given by composition of the underlying processes in $\mathcal{P}$.
\end{definition}

First, let us verify that this is indeed a valid theory:
\begin{proposition}[
  restate=twirlGTheory,
  name={}
]
\label{prop:TwirlGTheory}
   $\mathsf{Twirl}_G[\mathcal{P}]$ satisfies all of the conditions to be a GPT, except that it might not satisfy process tomography.
\end{proposition}
\proof
See \cref{proof:TwirlGTheory}.
\endproof
Next, let us verify that it does indeed contain arbitrary $S$-twirled subtheories as subtheories:
\begin{theorem} Any $S$-twirled subtheory $\mathcal{T}(\mathcal{P})$ is a subtheory of $\mathsf{Twirl}_G[\mathcal{P}]$. Note that this includes $\mathcal{P}$ itself as a special case of trivial twirling.
\end{theorem}
\proof
Defining an $S$-twirled subtheory $\mathcal{T}(\mathcal{P})$ amounts to a choice of physical $S$-action for each irreducible system $\boldsymbol{\pi}:=\{\pi_S|S\in\mathbf{Sys}[\mathcal{P}]\}$. If we consider the subtheory of $\mathsf{Twirl}_G[\mathcal{P}]$ generated by the systems $(S,\pi_S)$, then it is immediate from the definitions that this is isomorphic to $\mathcal{T}(\mathcal{P})$ where the isomorphism is given by a relabeling of systems $S\cong (S,\pi_S)$.
\endproof

Let us note that if we have two different $S$-twirled subtheories $\mathcal{T}_1(\mathcal{P})$ and $\mathcal{T}_2(\mathcal{P})$, then these can both be embedded as subtheories of $\mathsf{Twirl}_G[\mathcal{P}]$. This allows for nontrivial interactions between these two subtheories. For example, suppose we have a system $(A_1,\pi_1)$ coming from the first subtheory, and a system $(A_2,\pi_2)$ coming from the second, then within $\mathsf{Twirl}_G[\mathcal{P}]$ we can transform one into the other using arbitrary intertwiners between $\pi_1$ and $\pi_2$ in $P_{A_1}^{A_2}$.

Our framework should also make it possible to give an alternative operational formulation of dual systems within the setting constructed here. From an operational point of view, sometimes it can be more natural in finite-dimensional GPTs to interpret the dual copy of a system not as a distinct kind of physical system, but as the same underlying system equipped with a different physical $S$-action. In quantum theory, for example, a qubit and its dual are both realised by the same two-level system. The essential distinction is that one copy carries the unitary action $U$, while the other carries the conjugate unitary action $\overline{U}$. This is precisely what gives the canonical Choi-Jamio{\l}kowski vector on $AA^*$ its usual form and ensures its invariance under the collective action. The choice of action on each system therefore depends on the operational scenario that one aims to describe. For example, if one wants to preserve the canonical maximally entangled state of two qubits, one copy must transform under $U$ and the other under $\bar{U}$. This suggests a reformulation in which one removes dual systems as primitive objects and instead minimally extends each system $A$ so that it admits physical $S$-actions, denoted by $(A,\pi)$ and $(A,\pi^*)$, where $( \pi^* )^*=\pi$. The unification construction $\mathsf{Twirl}_G[\mathcal{P}]$ already provides the appropriate setting for this idea, since its systems are pairs $(A,\pi)$, with $\pi$ a physical $S$-action, and its processes are specified by the systems together with the actions they carry. In this formulation, the role usually played by $A^*$ is taken by $(A,\pi^*)$, while the distinction between $AA$ and $AA^*$ is preserved as the distinction between $(A,\pi)\otimes(A,\pi)$ and $(A,\pi)\otimes(A,\pi^*)$. This reformulation therefore makes clearer that these two copies correspond, operationally, to the same physical system of the starting GPT, distinguished only by the symmetry action they carry depending on the operational scenario.

Finally, we note that this provides a unification of all $S$-twirled subtheories of $\mathcal{P}$ for a fixed compact group $G$. A natural question to ask is whether we can consider a broader unification which allows for also varying the group. A simple option would be to have a unification of $S$-twirled subtheories with respect to $G_1$ and $G_2$ by considering $\mathsf{Twirl}_{G_1\times G_2}[\mathcal{P}]$. In particular cases, however, this will be bigger than what we want, for example, if $G_1\subseteq G_2$ then really we obtain everything just by considering $\mathsf{Twirl}_{G_2}[\mathcal{P}]$. We leave a comprehensive study of this for future work.

Another interesting direction for future work is to investigate the categorical structure of $\mathsf{Twirl}_G$. In particular, it is quite plausible that this is defining a monad on the category of `GPTs without process tomography', or more abstractly, on the category of convex-enriched coloured PROPs.

\subsection{Incoherent simulations of system-induced twirled worlds}

Ref.~\cite{ying2025whether} defines a notion of an \emph{incoherent simulation} of an $S$-twirled world. In our language, this can be thought of as a map $\$_\mathcal{T}:\mathcal{P}\to \mathcal{T}(\mathcal{P})$ which is linear and \emph{connectivity preserving}. That is, it satisfies \cref{eq:connectivity-preservation} of the diagram preservation condition, but not (necessarily) \cref{eq:swap-identity-preservation}. The key purpose of this in Ref.~\cite{ying2025whether} was to show that any correlations which could be produced in any causal scenario within $\mathcal{P}$ could also be produced within $\mathcal{T}(\mathcal{P})$, and, hence, the two theories cannot be distinguished using ideas coming from the study of causal networks. Here we formulate their notion in our compositional GPT setting and make its extension to arbitrary compact groups explicit.

\begin{definition}[Incoherent simulation $\$_\mathcal{T}$]
Given some choice of physical $S$-actions $\boldsymbol{\pi}$ and associated $S$-twirling map $\mathcal{T}$, we can define an incoherent simulation $\$_\mathcal{T}:\mathcal{P}\to \mathcal{T}(\mathcal{P})$ by its action on arbitrary processes:

\begin{equation}
    \label{eq:incohorent-simulation}\scalebox{0.8}{$\input{Diagrams/4_twirl151.tikz}=\input{Diagrams/4_twirl152.tikz}$}\;,
\end{equation}
where the $S$-action on the classical wires is given by \cref{def:action-on-classical-system}. 
\end{definition}

One can then straightforwardly verify the following:
\begin{proposition}[
  restate=incoherentSimulation,
  name={}
]
\label{prop:IncoherenSimulation}
    Incoherent simulations are linear and connectivity preserving maps, $\$_\mathcal{T}:\mathcal{P}\to \mathcal{T}(\mathcal{P})$
\end{proposition}
\proof
See \cref{app:IncorherenSimulation}.
\endproof
Note, however, that in general incoherent simulations are not diagram preserving as they do not preserve identities or swaps
\begin{equation}
    \scalebox{0.8}{$\input{Diagrams/4_twirl171.tikz}=\input{Diagrams/4_twirl172.tikz}\stackrel{\eqref{eq:rep-sequential-controlled}}{=}\input{Diagrams/4_twirl173.tikz}$}\;.
\end{equation}
 To see why the RHS is not the identity, suppose we feed the $g_1$ state as input and take the $g_2$ effect as output. Then, for finite groups (using the spider-like notation for a finite group $G$ introduced in~\cref{def:spiders}), the above equation gives $\hat{\pi}_{g_2\cdot g_1 ^{-1} }$ on system $A$, whereas if the RHS were the identity, it would have to give $\delta_{g_1 g_2}\cdot \mathcal{I}_A $. The swap is likewise not preserved, as we have
\begin{equation}
    \scalebox{0.8}{$\input{Diagrams/4_twirl174.tikz}=\input{Diagrams/4_twirl175.tikz}=\input{Diagrams/4_twirl176.tikz}$}\;.
\end{equation}
 Note, however, that (as is the case for all connectivity preserving maps) this is still an idempotent map which can be verified by direct computation.
 
As an immediate corollary we therefore have a generalisation of Theorem~1 of Ref.~\cite{ying2025whether} to the case of compact groups:
\begin{theorem}
   For all causal structures and $S$-twirling maps, there is no gap between the correlations realisable by $\mathcal{T}(\mathcal{P})$ and $\mathcal{P}$.
\end{theorem}
\proof
Follows immediately from \cref{prop:IncoherenSimulation} and the proof of Theorem 1 in Ref.~\cite{ying2025whether}.
\endproof
Our framework should also make it possible to formulate coherent simulations of the $S$-twirled worlds introduced in~\cite{ying2025whether}. This, however, would require extending the classical construction of internal references to quantum systems by formulating group actions for internal quantum references. The reference frame would remain internal in the same sense, namely, as a physical system inside the GPT rather than an external background structure. The difference is that the group information would no longer be carried only as classical data, but could instead be encoded in quantum states of a reference system, allowing coherent reference frame degrees of freedom and quantum transformations between descriptions. Developing this notion of group elements as internal physical systems for quantum systems should then make it straightforward to extend coherent simulation of $S$-twirled worlds from finite groups to compact groups. We leave this for future work.

\section{Invariant subtheories from theory actions}\label{Sec:Swirled}

In this section we extend the invariant-subtheory construction to arbitrary $T$-actions. We first define averaging over an arbitrary $T$-action and establish the general properties of the resulting invariant subtheories, without assuming that the action is physical. 

As in the physical system-induced case, the key subtlety is that process tomography is not automatically guaranteed for these invariant subtheories (\cref{def:tomographic-completness}). We find that, again, the two conditions discussed in the case of system-induced twirling suffice to ensure process tomography. For general $T$-actions, however, it does not seem that there is a nontrivial way to unify all of the resulting invariant subtheories within some super-theory in the standard categorical framework. Nevertheless, we provide a preliminary sketch of a possible unification in which composition is subject to additional constraints, leaving its full formalisation to future work.

\subsection{Averaging on processes}

Before we get to defining the corresponding invariant subtheories, let us define averaging on processes.

\begin{definition}
(Averaging on processes). In a GPT $\mathcal{P}$, for a process
$f \in P^{B_1\cdots B_m}_{A_1\cdots A_n}$, and a given $T$-action $\Xi$, averaging $\mathcal{A}$ on $f$ is defined as:
\begin{equation}
\scalebox{0.8}{$\input{Diagrams/5_swirl1n.tikz}:=\input{Diagrams/5_swirl2.tikz}$}\;,
\end{equation}
where $\mu$ is the Haar probability measure for the compact group $G$, and $\Xi$ is the $T$-action on the theory $\mathcal{P}$ (\cref{def:theorySym}). We call $\mathcal{A}$ the averaging map associated with $\Xi$.
\end{definition}

Next let us prove some useful properties of averaging and invariant processes.

\begin{lemma} Applying $\Xi$ after averaging leaves all processes invariant:
\begin{equation}
\label{eq:haar-measure-invariance-swirl}
\scalebox{0.8}{$\input{Diagrams/5_swirl3.tikz}=\input{Diagrams/5_swirl4.tikz}$}\;,
\end{equation}
or equivalently
\begin{equation}
    \label{eq:haar-measure-invariance-swirl1}
\scalebox{0.8}{$\input{Diagrams/5_swirl16n.tikz}=   \input{Diagrams/5_swirl17n.tikz}$}\;.
\end{equation}
\end{lemma} 

\proof
Since the restriction of $\Xi$ to each process space is a $P$-action,~\cref{eq:haar-measure-invariance1} also holds for averaging, i.e., invariance of the Haar measure implies
\begin{equation}
\scalebox{0.8}{$\input{Diagrams/5_swirl3.tikz}=\input{Diagrams/5_swirl4.tikz}$}\;,
\end{equation}
\endproof

The averaging map is idempotent:
\begin{proposition}
\label{lem:idempotent-swirl}
For every pair of systems $A,B\in\textbf{Sys}[\mathcal{P}]$, and every process $f \in P^B_A$, one has $\mathcal{A}\big(\mathcal{A}(f)\big) = \mathcal{A}(f)$.
\end{proposition}
\begin{proof}
The proof is the same as in~\cref{lem:idempotent}, except that instead of~\cref{eq:haar-measure-invariance}, one must use the analogous equation for averaging, namely~\cref{eq:haar-measure-invariance-swirl1}.
\end{proof}
In particular, this means that a process is invariant under averaging if and only if it is an averaged process.

Next we can show that invariance under averaging is equivalent to being invariant under the $T$-action, which, later, will be essential to connecting invariant processes and invariant subtheories.
\begin{proposition}
\label{prop:group-invariance1}
   For every pair of systems $A,B \in\textbf{Sys}[\mathcal{P}]$, a process $f \in P^B_A$ is invariant under averaging if and only if it is invariant under $\Xi$ for all group elements $g \in G$, i.e.,
   \begin{equation}
 \scalebox{0.9}{$\input{Diagrams/5_swirl23n.tikz}=\input{Diagrams/2_theorysym14f.tikz} \iff \input{Diagrams/5_swirl24.tikz}=\input{Diagrams/2_theorysym14f.tikz}\;,\;\forall g \in G$}.
\end{equation}
\end{proposition}
\begin{proof}
    The proof is similar to~\cref{prop:group-invariance}, just that one needs to replace $\mathcal{T}$ with $\mathcal{A}$, and $\Pi$ with $\Xi$.
\end{proof}

\begin{corollary}\label{cor:swirlingIsScalarPreserving}
    Averaging is scalar-preserving, i.e., for every $p\in [0,1]$, one has $\mathcal{A}(p)=p$.
\end{corollary}
\begin{proof} The proof is similar to~\cref{cor:twirlingIsScalarPreserving}, that is, since $\Xi$ is scalar-preserving (\cref{prop:automophismsPreserveScalars}), one can use it to show that $\mathcal{A}$ is scalar-preserving by replacing $\mathcal{T}$ with $\mathcal{A}$ and $\Pi$ with $\Xi$ in~\cref{cor:twirlingIsScalarPreserving}.
\end{proof}

Note that averaging is not, in general, a diagram-preserving map, as one has
\begin{align}
\label{eq:diagram-preservation2}
&\scalebox{0.7}{$\input{Diagrams/5_swirl25n.tikz}$}=\scalebox{0.7}{$\input{Diagrams/2_theorysym8.tikz}$}=\scalebox{0.7}{$\input{Diagrams/2_theorysym9.tikz}$}\nonumber\\
&\neq \scalebox{0.7}{$\input{Diagrams/2_theorysym10.tikz}$}=\scalebox{0.7}{$\input{Diagrams/5_swirl26n.tikz}$}\;.
\end{align}
Nevertheless, it can still be used to define a valid subtheory as we see in the following subsection.

\subsection{Invariant subtheories from averaging}

We now turn to the theory as a whole and show how the averaged processes defined in the previous subsection give rise to invariant subtheories. As mentioned above, the key challenge is that these invariant subtheories are not guaranteed to satisfy process tomography (\cref{def:tomographic-completness}), and hence, whilst they are necessarily subtheories of the original theory, they are not necessarily GPT-subtheories (\cref{def:subGPT}).

First we provide the corresponding invariant-subtheory construction and then show that it indeed defines a valid subtheory of the original theory.

\begin{definition}\label{def:swirled-subspace}
(Invariant subtheories from averaging). Given a $T$-action $\Xi$ and its corresponding averaging map $\mathcal{A}$, the invariant subtheory $\mathcal{A}(\mathcal{P})$ has the same systems as $\mathcal{P}$ and, for every pair of systems $A,B\in\textbf{Sys}[\mathcal{P}]$, its process space $\mathcal{A}(P^B_A)$ is the subprocess space of $P^B_A$ consisting of processes that are invariant under averaging. That is, for every pair of systems $A,B\in\textbf{Sys}[\mathcal{P}]$, and every process $f \in \mathcal{A}(P^B_A)$, one has $\mathcal{A}(f)=f$, depicted as
\begin{equation}
\input{Diagrams/5_swirl0n.tikz}=\input{Diagrams/2_theorysym14f.tikz}\;.
\end{equation}
\end{definition}
As a consequence of \cref{lem:idempotent-swirl}, together with the fact that averaging defines a surjective map from the original GPT to the corresponding invariant subtheory, starting from an arbitrary GPT one can obtain this invariant subtheory by applying the averaging map.

\begin{proposition}[
  restate=swirledSubspaceBeingSubtheory,
  name={}
]
\label{prop:swirledSubspaceBeingSubtheory}
The invariant subtheory $\mathcal{A}(\mathcal{P})$, obtained by applying the averaging map to a GPT $\mathcal{P}$ as defined in~\cref{def:swirled-subspace}, is indeed a subtheory of $\mathcal{P}$.
\end{proposition}
\begin{proof}
    See~\cref{proof:swirled_is_subtheory}.
\end{proof}
As a consequence of~\cref{prop:group-invariance1}, this subtheory is equivalent to the $\Xi$-invariant subtheory (see~\cref{def:invariant-subtheories}). Hence, averaging defines a map from an arbitrary theory to a $\Xi$-invariant subtheory. Since $\mathcal{A}(\mathcal{P})$ is a valid subtheory, the corresponding $\Xi$-invariant subtheory is also a valid subtheory. However, it is not necessarily a GPT-subtheory, as it may fail to satisfy process tomography.

\subsection{Tomography in invariant subtheories}

We investigate tomography in the invariant subtheory obtained from averaging. As in the $S$-twirling case, we show that state and effect tomography are always possible, but that this does not immediately generalise to arbitrary processes. 

\begin{proposition} \label{prop:state-tomography-swirl}
   an invariant subtheory obtained from averaging admits state/effect tomography.
\end{proposition}
\begin{proof}
   The proof is similar to that of~\cref{prop:state-tomography}, replacing the $S$-twirling map $\mathcal{T}$ with the averaging map $\mathcal{A}$, and using~\cref{eq:haar-measure-invariance-swirl1} instead of~\cref{eq:haar-measure-invariance}.

    This implies that for any pair of distinct invariant states $\tilde \omega_1\neq\tilde \omega_2\in \mathcal{A}(P^A)$, there exists an invariant effect $\mathcal{A}(e)$ in the invariant subtheory (with $e$ from the original theory) such that
\begin{equation}
\scalebox{0.9}{$\input{Diagrams/4_twirl30.tikz} \neq \input{Diagrams/4_twirl31.tikz} \Rightarrow \input{Diagrams/5_swirl6n.tikz} \neq \input{Diagrams/5_swirl7n.tikz}$}\;,
\end{equation}
and by the same logic, for any pair of distinct invariant effects $\tilde e_1\neq\tilde e_2\in \mathcal{A}(P_A)$, there exists a state $\mathcal{A}(\omega)$ in the invariant subtheory (with $\omega$ from the original theory) that distinguishes them.
\end{proof} 

As in the $S$-twirling case, lifting this argument to the level of general processes tells us that averaged combs suffice for process tomography.

\begin{proposition}\label{prop:swirledcombtomography}
    Averaged combs, i.e.,
  \begin{equation}
\scalebox{0.7}{$\input{Diagrams/5_swirl27.tikz}$}\;,
\end{equation}
    suffice for process tomography of the invariant subtheory. That is, for any pair of distinct invariant processes $\tilde f\neq \tilde h \in \mathcal{A}(P_A^B)$, there exist a state $s\in P^{A,E}$ and an effect $e\in P_{B,E}$ in the original theory such that
\begin{equation}
\label{eq:process-tomography-in-swiled-worlds}
\scalebox{0.7}{$
\input{Diagrams/5_swirl28.tikz} \neq
\input{Diagrams/5_swirl29.tikz}\quad .
$}
\end{equation}
\end{proposition}
\proof
The proof is analogous to that of the $S$-twirling case in~\cref{prop:twirledcombtomography}: using process tomography of the original theory, scalar preservation of averaging (\cref{cor:swirlingIsScalarPreserving}), and~\cref{eq:haar-measure-invariance-swirl1}, one can follow the same argument as in~\cref{prop:twirledcombtomography}, replacing $\mathcal{T}$ with $\mathcal{A}$ and $\Pi$ with $\Xi$.
\endproof

As in the $S$-twirling case, this implies that we might not be able to perform process tomography within the invariant subtheory, since these averaged combs may not be realisable as a composite of an invariant state and an invariant effect. The same two conditions as in the $S$-twirling case suffice to ensure process tomography within the invariant subtheory. We discuss these in the following two subsections.

\subsubsection{Dynamically faithful states}

Once again, we find that the existence of dynamically faithful states in an invariant subtheory suffices for process tomography:

\begin{proposition}\label{prop:swirl-process-tomography-faithful}
If, in an invariant subtheory obtained from averaging, for every system $A$ there exists an ancillary system $E$ such that the subtheory admits a dynamically faithful state on system $AE$, then it is tomographically complete.
\end{proposition}

\begin{proof}
   The proof is similar to that of~\cref{prop:twirl-process-tomography-faithful}, where using state tomography of the invariant subtheory (\cref{prop:state-tomography-swirl}), one can show that a dynamically faithful state allows process tomography.
\end{proof} 
As in the $S$-twirling case (\cref{cor:twirled-invariant-dynamically-faithful-states}), the proposition above only requires dynamic faithful states within the invariant subtheory. Any invariant dynamically faithful state of the original GPT also satisfies this, yielding the following corollary.
\begin{corollary}
If, in the original GPT of an invariant subtheory obtained from averaging, for every system $A$ there exists an ancillary system $E$ such that the GPT contains an invariant dynamically faithful state on system $AE$, then the invariant subtheory is tomographically complete.
\end{corollary}

Since, in Classical theory, as we show later in~\cref{sec:CT}, every $T$-action is induced by physical $S$-actions, the uniformly correlated state in~\cref{ex:choi-classcial} is an invariant dynamically faithful state contained in any invariant subtheory obtained from averaging classical theory.

Also, as we show later in~\cref{Sec:LocalTomog}, for locally tomographic theories including quantum theory, every $T$-action admits a system-level representation, although the representing transformations need not be physical. When $A^*$ carries the corresponding conjugate representation, this shows that the maximally entangled state in~\cref{ex:choi-quantum} is also an invariant dynamically faithful state contained in the corresponding invariant subtheory obtained from quantum theory.

Similar to the $S$-twirling case, an even weaker requirement than the existence of an invariant dynamically faithful state ensures that process tomography is satisfied in an invariant subtheory obtained from averaging, as shown in the following proposition.

\begin{proposition}\label{prop:dynamically-faithful-vector-swirl}
In an invariant subtheory obtained from averaging, process tomography is possible if, for every system $A$, there exists an ancillary system $E$ such that the span of the invariant state space on $AE$, i.e., $\spann \left(\mathcal{A}(P^{A,E})\right)$, contains a dynamically faithful vector for the averaged subtheory.
\end{proposition}
\begin{proof}
Let $\bar{\sigma}\in \spann\left(\mathcal{A}(P^{A,E})\right)$ be a dynamically faithful vector on $AE$ in the invariant subtheory. Then, for any two distinct invariant processes $\tilde f\neq \tilde h\in \mathcal{A}(P^B_A)$, one has
\begin{equation}
\scalebox{0.9}{$\input{Diagrams/5_swirl43n.tikz}
=
\input{Diagrams/4_twirl177.tikz}
\neq
\input{Diagrams/4_twirl178.tikz}
=
\input{Diagrams/5_swirl44n.tikz}$}\;.
\end{equation}
The left- and right-hand sides can be used to define the linear maps $\phi_{\tilde f},\phi_{\tilde h}:\spann(P^{A,E})\to \spann(P^{B,E})$, diagrammatically depicted as
\begin{equation}
\phi_{\tilde f}(\cdot):=\scalebox{0.9}{$\input{Diagrams/5_swirl46n.tikz}$}\quad\textit{and}\quad
\phi_{\tilde h}(\cdot):=\scalebox{0.9}{$\input{Diagrams/5_swirl45n.tikz}$}\;.
\end{equation}
Since $\phi_{\tilde f}(\bar{\sigma})\neq\phi_{\tilde h}(\bar{\sigma})$, the maps $\phi_{\tilde f}$ and $\phi_{\tilde h}$ are distinct. Now, let $\{\omega_i\}_i$ be a basis for $\spann\left(P^{A,E}\right)$, with $\omega_i\in P^{A,E}$. Then there exists at least one $\omega_i$ such that the two maps yield distinct invariant states, namely $\phi_{\tilde f}(\omega_i)\neq\phi_{\tilde h}(\omega_i)$, as depicted by
\begin{equation}
\scalebox{0.9}{$\input{Diagrams/5_swirl48n.tikz}
\neq
\input{Diagrams/5_swirl47n.tikz}$}\;.
\end{equation}
Since the invariant subtheory admits state tomography (see~\cref{prop:state-tomography-swirl}), these two distinct invariant states can be distinguished by invariant effects within the invariant subtheory. Consequently, the processes $\tilde f$ and $\tilde h$ can be distinguished, which implies process tomography.
\end{proof} 
This immediately implies the following:
\begin{corollary}\label{cor:dynamically-faithful-vector-swirl}
If, in the original GPT of an invariant subtheory obtained from averaging, for every system $A$ there exists an ancillary system $E$ such that the GPT admits an invariant dynamically faithful vector in the span of the state space on $AE$, i.e., in $\spann \left(P^{A,E}\right)$, then the invariant subtheory is tomographically complete, and hence is a valid GPT-subtheory.
\end{corollary}
Based on the above result, we will show in \cref{sec:FailureLT} that any invariant subtheory obtained from averaging a locally tomographic GPT satisfies process tomography.

\subsubsection{Internal classical reference frames}

The notion of an internal classical reference frame needs a little refinement over the $S$-twirling case. In particular, we require that the $T$-action $\Xi$ extends consistently to the process spaces involving the measurable classical system $G$.\footnote{In forthcoming work~\cite{SalzgerSelbySoltaniIntegration}, we show how process spaces involving measurable classical systems can be defined even when the classical systems are infinite dimensional. A process in $P_{A,G}^{B,G}$ is specified by a GPT process $K_x(S)\in P_A^B$ for each classical input $x\in G$ and measurable classical output event $S\subseteq G$. This makes the extension of the action natural: the group action simply changes the classical input and output event being considered, while the $T$-action transforms the GPT process associated with that classical data. The result is therefore again a process-valued kernel.}

\begin{definition}\label{def:CTisTL_Representation}
For a GPT with internal classical system $G$, universal classical control, and a $T$-action $\Xi$ of the group $G$, we define the action of $\Xi$ on the process space $P^{B,G}_{A,G}$ as follows:
\begin{equation}\label{eq:CTisTL_Representation}
\scalebox{0.8}{$\input{Diagrams/16_internalRefSwirl2.tikz}:=\input{Diagrams/16_internalRefSwirl3.tikz}$}\;,
\end{equation} 
where the black dots represent the action of the group $G$ on classical systems (see~\cref{def:action-on-classical-system}).
\end{definition}
With this extension of $\Xi$ in place, we can now define the corresponding controlled action $\hat{\Xi}$.
\begin{definition}
For a GPT $\mathcal{P}$ with internal classical system $G$ and with universal classical control, equipped with a $T$-action $\Xi$ of a group $G$, one can define a map $\hat{\Xi}$, such that, for every process space {\small $P^B_A$}, it induces a $G$-controlled process map {\small $\hat{\Xi}: P^B_A \to P^B_{G,A}$}, where one has
  \begin{equation}\label{eq:xihat}
       \scalebox{0.9}{$\input{Diagrams/16_internalRefSwirl.tikz}:=\input{Diagrams/2_theorysym1.tikz}\;,\quad \forall g\in G $}.
  \end{equation}
  \end{definition}

Note that for any $g\in G$ we have that
 \begin{align}
     \input{Diagrams/16_internalRefSwirl21.tikz}
\stackrel{\eqref{eq:CTisTL_Representation}}{=}\scalebox{0.8}{$\input{Diagrams/16_internalRefSwirl23.tikz}$}\nonumber\\\stackrel{\eqref{eq:action-on-classical-system}}{=}\scalebox{0.8}{$\input{Diagrams/16_internalRefSwirl24.tikz}$}\;.
 \end{align}
Thus, for any $g\in G$, we can define $\hat{\Xi}^B_A$ as
 \begin{equation}
\scalebox{0.8}{$\input{Diagrams/16_internalRefSwirl25.tikz}:=\input{Diagrams/16_internalRefSwirl24.tikz} $}\;,
 \end{equation}
which implies that the action of $\hat{\Xi}$ (as defined in~\cref{eq:xihat}) on the process space $P^{B,G}_{A,G}$ is given by
 \begin{equation}\label{eq:xihat-action}
\input{Diagrams/16_internalRefSwirl29.tikz}=\scalebox{0.8}{$\input{Diagrams/16_internalRefSwirl30.tikz}$} \;,
 \end{equation}
  It is straightforward to show that
 \begin{equation}
     \label{eq:seq-compoistion-xihat}\scalebox{0.8}{$\input{Diagrams/16_internalRefSwirl36.tikz}=\input{Diagrams/16_internalRefSwirl32.tikz}$}\;,
 \end{equation}
where the LHS should be interpreted as two successive applications of $\hat{\Xi}_A^B$ to the pair of systems $(A,B)$ (note that we do not denote this simply as a ``nesting'' of the shaded regions to make it clear that the second application does not touch the $G$ system created by the first application).

Averaging can then be written as
 \begin{equation}
\scalebox{0.8}{$ \label{eq:swirl-controlled}
\input{Diagrams/5_swirl1n.tikz}=\input{Diagrams/16_internalRefSwirl4.tikz}$}\;.
\end{equation}
Also, using~\cref{eq:CTisTL_Representation}, one has
\begin{equation}\label{eq:haar-invariance-swirl-controlled}
 \scalebox{0.9}{$\input{Diagrams/16_internalRefSwirl10.tikz}=\input{Diagrams/16_internalRefSwirl11.tikz}$}\;.
      \end{equation}

\begin{proposition}[
  restate=xiHatInvariance,
  name={}
]
\label{prop:xihatInvariance}
Applying $\hat{\Xi}$ to an arbitrary process $f\in P^B_A$ results in a $\hat{\Xi}$ invariant process, in $P^B_{G,A}$, i.e.,
\begin{equation}
\scalebox{0.8}{$\input{Diagrams/16_internalRefSwirl26.tikz}=\input{Diagrams/16_internalRefSwirl27.tikz}$}\;.
\end{equation}
\end{proposition}
\begin{proof}
 See \cref{app:xihatInvariance}
\end{proof}
As in the $S$-twirling case (\cref{cor:invariance-under-twirl-pihat}), the following corollary follows immediately:

\begin{corollary}\label{cor:invariance-under-swirl-xihat}
Applying $\hat{\Xi}$ to any arbitrary process in $P_A^B$ yields a process, in $P^B_{G,A}$, that is invariant under both averaging and $\Xi$.
\end{corollary} 

\begin{theorem}[
  restate=swirlInternalRef,
  name={}
]
\label{thm:swirlInternalRef}
In a GPT $\mathcal{P}$ with internal classical system $G$ and with universal classical control, any swirled theory whose $T$-action on process spaces involving $G$ is given by~\cref{def:CTisTL_Representation} satisfies process tomography.
\end{theorem}
\proof
See~\cref{app:swirlInternalRef}.
\endproof

\subsection{Towards unification of invariant subtheories}

Given some GPT $\mathcal{P}$ and some group $G$ we can construct a variety of different invariant subtheories associated with different $T$-actions. In this subsection we offer some preliminary ideas on how to unify these into a single theory. This theory, however, has some unusual compositional properties, which appear to take it out of the standard categorical framework, but perhaps can be formalised with ideas coming from Refs.~\cite{selby2025generalised,wilson2021composable}. This, however, is beyond the scope of this current work. 

To be more precise, what we aim to achieve by such a unification is, given any compact group $G$, and any GPT $\mathcal{P}$, to construct a new theory\footnote{Which may not be a theory in any standard framework due to the aforementioned compositionality issues.} $\mathsf{Inv}_G[\mathcal{P}]$ such that for any averaging map $\mathcal{A}$, $\mathcal{A}(\mathcal{P})\subseteq \mathsf{Inv}_G[\mathcal{P}]$. Moreover, we want to do this in a way such that the different invariant subtheories can (potentially) non-trivially interact, i.e., we do not want to simply take the product of all of the different invariant subtheories.

\begin{definition}[sketch]\label{def:swirlUnified}
Given a compact group $G$ and a GPT $\mathcal{P}$ we define the theory $\mathsf{Inv}_G[\mathcal{P}]$ via a specification of its process spaces and composition rules.

Process spaces in $\mathsf{Inv}_G[\mathcal{P}]$ are labelled by arbitrary pairs $[P_{A_1,...,A_n}^{B_1,...,B_m}]^\eta:=(P_{A_1,...,A_n}^{B_1,...,B_m},\eta)$ where $P_{A_1,...,A_n}^{B_1,...,B_m}$ is a process space in $\mathcal{P}$ and $\eta$ is a process space action thereon. The processes in $[P_{A_1,...,A_n}^{B_1,...,B_m}]^\eta$ are the subset of processes in $P_{A_1,...,A_n}^{B_1,...,B_m}$ which are invariant under $\eta$.

Composition of processes is typically defined as a multiprocess map from the input process spaces to the output process space, with the caveat that composition is defined only on valid diagrams, i.e., those in which the diagram is acyclic and in which system types match. Here we must add in an extra constraint to ensure that composition is consistent with the group actions on the process spaces, i.e., composition of process spaces is allowed if and only if it is a process space action intertwiner for the relevant process space actions.

For example, within $\mathcal{P}$ we can compose process spaces $P_A^B$ and $P_B^C$ using sequential composition $\mathsf{seq}_{A,B,C}:P_A^B\times P_B^C\to P_A^C$. We say that this is a valid composition rule within $\mathsf{Inv}_G[\mathcal{P}]$ for process spaces $[P_A^B]^{\eta_1}$ and $[P_B^C]^{\eta_2}$ to process space $[P_A^C]^{\eta_3}$ if and only if
\begin{equation}
\label{eq:sequential-composition-unification-swirl}\mathsf{seq}_{A,B,C}\circ\eta_1(g)\times\eta_2(g) = \eta_3(g)\circ \mathsf{seq}_{A,B,C}
\end{equation}
for all $g\in G$.
\end{definition} 

This definition ensures that, when restricting to process-space actions induced by physical $S$-actions, $\mathsf{Inv}_G[\mathcal{P}]$ reduces to $\mathsf{Twirl}_G[\mathcal{P}]$. To see this, consider $(P_{A_1,\ldots,A_n}^{B_1,\ldots,B_m},\eta)$, where $\eta$ is generated by the collective action $\pi^{(1)}\boxtimes\cdots\boxtimes\pi^{(n)}$ on the input and $\pi'^{(1)}\boxtimes\cdots\boxtimes\pi'^{(m)}$ on the output, with the $\pi^{(i)}$ and $\pi'^{(j)}$ physical $S$-actions. Then one can relabel $(P_{A_1,\ldots,A_n}^{B_1,\ldots,B_m},\eta)\cong (P_{(A_1,\pi_1),\ldots,(A_n,\pi_n)}^{(B_1,\pi'_1),\ldots,(B_m,\pi'_m)})$, and verify that composing process spaces using the composition rules of the underlying theory $\mathcal{P}$ (with matching input and output labels for sequential composition) satisfies~\cref{eq:sequential-composition-unification-swirl}.

However, in general, it seems implausible that this constraint on composition can be boiled down to local data on the system types, and, hence, unlikely that we can reduce this kind of theory to a symmetric monoidal category. We leave further formalisation of this theory to future work. However, initial ideas on how to go about such a formalisation can be taken from Refs.~\cite{selby2025generalised,wilson2021composable,barrett2019computational}.

We can, however, prove the following key theorem:
\begin{theorem}\label{thrm:swirl-subtheory-of-swirlg}
Any invariant subtheory $\mathcal A(\mathcal P)$ obtained by averaging a $T$-action is a subtheory of $\mathsf{Inv}_G[\mathcal{P}]$. Note that this includes $\mathcal{P}$ itself as the special case of the trivial $T$-action.
\end{theorem}
\proof
For every $T$-action $\Xi$, one can relabel each process space as $P^{B_1,\cdots,B_m}_{A_1,\cdots,A_n}\cong(P^{B_1,\cdots,B_m}_{A_1,\cdots,A_n},\eta^{B_1,\cdots,B_m}_{A_1,\cdots,A_n})$ such that the restriction of $\Xi$ to $P^{B_1,\cdots,B_m}_{A_1,\cdots,A_n}$ is given by the $P$-action $\eta^{B_1,\cdots,B_m}_{A_1,\cdots,A_n}$. The subtheory generated by $\Xi$-invariant processes is then equal to a subtheory of $\mathsf{Inv}_G[\mathcal{P}]$ generated by invariant process spaces of the form $(P^{B_1,\cdots,B_m}_{A_1,\cdots,A_n},\eta^{B_1,\cdots,B_m}_{A_1,\cdots,A_n})$, where diagram-preservation of $\Xi$ ensures that it satisfies the composition rule posited in~\cref{eq:sequential-composition-unification-swirl}.
\endproof

\subsection{Twirling and swirling}

In \cref{Sec:Twirled}, we introduced $S$-twirling in the more restrictive setting where the relevant $T$-action is induced by physical $S$-actions. The general averaging construction developed in this section allows us to distinguish whether physicality is assessed at the process or system level, leading respectively to the notions of $P$-twirling and $P$-swirling, and $S$-twirling and $S$-swirling.

\begin{figure}[H]
\centering
\resizebox{0.80\columnwidth}{!}{%
\begin{tikzpicture}[
    reset cm,
    x=1cm, y=1cm,
    every node/.style={},
    every path/.style={},
    box/.style={
        draw,
        thick,
        rounded corners=5pt,
        minimum width=2.6cm,
        minimum height=1.0cm,
        align=center,
        font=\normalsize
    },
    arrow/.style={
        ->,
        >=stealth,
        thick
    }
]

\node[box] (Stwirl) at (0,1.8) {$S$-twirling};
\node[box] (Ptwirl) at (6.4,1.8) {$P$-twirling};

\node[box] (Sswirl) at (0,0) {$S$-swirling};
\node[box] (Pswirl) at (6.4,0) {$P$-swirling};

\draw[arrow]
(Stwirl.east) -- node[above,font=\small] {always} (Ptwirl.west);

\draw[arrow]
(Sswirl.east) -- node[above,font=\small] {possible} (Pswirl.west);

\draw[arrow]
(Sswirl.north east) --
node[pos=0.40,above=4pt,font=\small] {possible}
(Ptwirl.south west);

\end{tikzpicture}%
}

\vspace{2pt}

{\scriptsize
In locally tomographic theories:
$S$-twirling $\leftrightarrow$ $P$-twirling,
and $S$-swirling $\leftrightarrow$ $P$-swirling.
}

\caption{Hierarchy of $S$- and $P$-twirling and swirling.}
\label{fig:SP-twirl-swirl}
\end{figure}
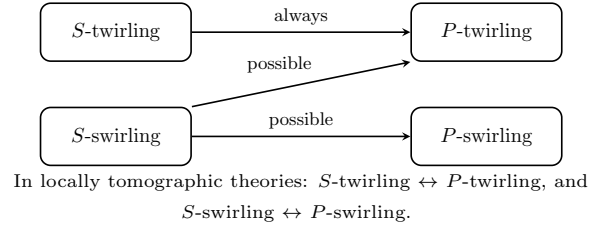

\begin{definition}[$P$-twirling and $P$-swirling]\label{def:twirling-swirling}
Let $\Xi$ be a $T$-action defined by a family of $P$-actions, and let $\mathcal{A}$ be the corresponding averaging map. If the $P$-actions defining $\Xi$ are physical, equivalently if $\Xi$ is a physical $T$-action in the sense of~\cref{def:physical-theory-action}, then we call $\mathcal{A}$ a \emph{$P$-twirling}, and the corresponding invariant subtheory a \emph{$P$-twirled subtheory}. If $\Xi$ is nonphysical, equivalently if at least one of the $P$-actions defining it is nonphysical, then we call $\mathcal{A}$ a \emph{$P$-swirling}, and the corresponding invariant subtheory a \emph{$P$-swirled subtheory}. Without loss of generality, we denote the $P$-twirling of a process $f$ by $\mathcal{T}(f)$ and the $P$-swirling of $f$ by $\mathcal{S}(f)$.
\end{definition}
The previous definition classifies averaging in terms of the physicality of the $P$-actions defining the $T$-action. When the same $T$-action is induced by $S$-actions, one can likewise classify it at the system level. i.e.:
\begin{definition}[$S$-twirling and $S$-swirling]\label{def:S-twirling-swirling}
Let $\Pi$ be a $T$-action induced by a choice of $S$-actions $\boldsymbol{\pi}$, and let $\mathcal{A}$ be the corresponding averaging map. If the $S$-actions $\boldsymbol{\pi}$ are physical, then we call $\mathcal{A}$ an \emph{$S$-twirling}, and the corresponding invariant subtheory an \emph{$S$-twirled subtheory}. If at least one of the $S$-actions in $\boldsymbol{\pi}$ is nonphysical, then we call $\mathcal{A}$ an \emph{$S$-swirling}, and the corresponding invariant subtheory an \emph{$S$-swirled subtheory}.
\end{definition}
The $S$- and $P$-classifications need not coincide in general. Their relation is summarised in the following remark:
\begin{remark}[Hierarchy of $S$- and $P$-twirling and swirling]
Since physical $S$-actions induce physical $P$-actions, every $S$-twirling is also a $P$-twirling. An $S$-swirling, however, may induce either physical or nonphysical $P$-actions, and hence may correspond to either a $P$-twirling or a $P$-swirling. Conversely, a $P$-twirling or $P$-swirling need not generally admit an $S$-action representation; whenever a $P$-swirling does, it is necessarily an $S$-swirling. This hierarchy is summarised in~\cref{fig:SP-twirl-swirl}. For locally tomographic theories, as we show in~\cref{Sec:LocalTomog}, the correspondence becomes one-to-one: $S$-twirling corresponds to $P$-twirling, and $S$-swirling to $P$-swirling.
\end{remark}

Throughout the remainder of the paper, we use $\mathcal{A}$ and $\mathcal{A}(\mathcal{P})$ as umbrella notation for averaging and the corresponding invariant subtheory whenever a statement applies independently of these distinctions. We use $\mathcal{T}$ and $\mathcal{S}$ for twirling and swirling, respectively, and specify $P$- or $S$-twirling and swirling whenever the distinction between the process- and system-level notions is relevant.

\section{Local tomography}\label{Sec:LocalTomog}

One of the original motivations for studying Twirled Worlds~\cite{centeno2024twirled} was to understand the notion of local tomography, in particular to see how twirling a tomographically local theory could result in one in which tomographic locality failed. Here we will consider tomographic locality in the fully compositional setting of this paper.

In the following we assume, without loss of generality, that $U^B_A = \spann( P^B_A)$ for every pair of systems $A,B$ in $\cP$.

\begin{definition}\label{def:local_tomography}(Local Tomography). A GPT $\mathcal{P}$ satisfies local tomography if and only if, for every pair of processes $f\in P_{A_1...A_n}^{B_1...B_m}$ and $h\in P_{A_1...A_n}^{B_1...B_m}$, one has
\begin{align}
 \scalebox{0.9}{$\input{Diagrams/7_LT2.tikz}$} &=  \scalebox{0.9}{$\input{Diagrams/7_LT3.tikz}$} \nonumber \\
& \iffupdown \nonumber \\
 \scalebox{0.9}{$\input{Diagrams/7_LT.tikz}$} &=  \scalebox{0.9}{$\input{Diagrams/7_LT1.tikz}$},\\  &\forall s_1,...,s_n,e_1,...,e_m.
\end{align}
\end{definition}
In other words, when local tomography holds, using local states and local effects suffices to implement process tomography; that is, the statistics obtained from local states and local effects fully determine a process.

 A particular process theory that is relevant to our discussion of locally tomographic GPTs is $\mathbf{FVect_{\mathbb{R}}}$, defined as follows
\begin{definition}\label{def:fvectr}
$\mathbf{FVect_{\mathbb{R}}}$ is a process theory when systems are labelled by finite-dimensional real vector spaces $V$, where the composition of two systems $V$ and $W$ is given by their tensor product $V \otimes W$. Processes are defined as real linear maps from the input state space to the output state space. Sequential composition of two processes corresponds to the usual composition of linear maps, while parallel composition corresponds to their tensor product. Processes with no input correspond to vectors in $V$ (with $V$ denoting the output system), and processes with no output correspond to covectors, i.e., elements of $V^*$ (with $V$ denoting the input system). Processes with neither input nor output correspond to scalars, represented by real numbers.
\end{definition}

Tomographically local GPTs are particularly simple to study as they have an elegant representation in terms of $\mathbf{FVect}_\mathds{R}$:
\begin{theorem}[
restate=FVectRep,
name={}
]\label{thm:FVectRep}
    A GPT, $\mathcal{P}$, is tomographically local if and only if it is isomorphic to a spanning subtheory \mainonlyfootnote{ I.e., one in which the hom-sets of the subtheory span the full space of linear maps. More precisely, for every pair of objects $V,W$ in the subtheory, the hom-set $\Hom_{\mathcal{P}}(V,W)$, consisting of the allowed processes from $V$ to $W$, satisfies $\spann\!\left(\Hom_{\mathcal{P}}(V,W)\right)=\Hom_{\mathbf{FVect}_{\mathds{R}}}(V,W)$.} of $\mathbf{FVect}_\mathds{R}$, the symmetric monoidal category of (finite dimensional) real vector spaces and linear maps.
\end{theorem}
\proof See App.~\ref{app:RepofTLGPT}.
\endproof

In what follows in this section we will, for notational simplicity, identify the tomographically local GPT, $\mathcal{P}$ with the isomorphic spanning subtheory of $\mathbf{FVect}_\mathds{R}$. As an immediate consequence of~\cref{thm:FVectRep}, the process spaces of a locally tomographic GPT admit the following canonical vector-space identifications:

\begin{corollary}\label{cor:LT_proc_isom}
    In locally tomographic theories $U_{A_1,...,A_n}^{B_1,...,B_m}\cong \Hom(U^{A_1} \otimes ... U^{A_n}, U^{B_1} \otimes ... \otimes U^{B_m})$, in particular this implies $U^{AB} \cong U^A \otimes U^B$, $U^A \cong (U_A)^*$ and $U_A^A \cong \End(U^A)$.
\end{corollary}

\subsection{Structure theorem for symmetries of locally tomographic theories}
\label{Sec:StructureTheorem}

In tomographically local theories, theory symmetries take a particularly simple form following from Refs.~\cite{schmid2024structure,wagner2025structure}. In the following, for notational simplicity, we identify $\mathcal{P}$ with its representation as a subtheory of $\mathbf{FVect_\mathbb{R}}$ given by~\cref{thm:FVectRep}. The results would equally hold if one explicitly retained the inclusion $\mathcal{P}\hookrightarrow\mathbf{FVect_\mathbb{R}}$ as a diagram-preserving map; see~\cite[Sec.~4.6]{schmid2024structure}.

\begin{theorem}
\label{thm:structure-theorem}
Any map $\input{Diagrams/3_ST2.tikz}$ that is linear, scalar-preserving, and diagram-preserving, acting on a process $f \in P^{B_1 \cdots B_m}_{A_1 \cdots A_n}$, where $\mathcal{P}$ is locally tomographic, and that sends a system $X \in \textbf{Sys}[\mathcal{P}]$ to the system $V_X \in \textbf{Sys}[\mathbf{FVect_{\mathbb{R}}}]$, can be represented as
\begin{equation}
\label{eq:structure-theorem}
\scalebox{0.8}{$\input{Diagrams/3_ST3.tikz} \;=\; \input{Diagrams/3_ST4.tikz}$}\;,
\end{equation}
where, for each system $X \in \textbf{Sys}[\mathcal{P}]$, $\chi_X:U^X\to V_X$ is an invertible linear transformation in $\mathbf{FVect}_{\mathbb{R}}$.
\end{theorem}
\proof
See~\cite[Thm.~4.1]{schmid2024structure} or~\cite[Thm.~III.1]{wagner2025structure}.
\endproof

We can now apply~\cref{thm:structure-theorem} to each symmetry $\Xi_g:\mathcal{P}\to\mathcal{P}$ of a $T$-action $\Xi$. Since $\Xi_g$ acts trivially on systems, for every system $X$ and every $g\in G$ we obtain an invertible linear transformation $\chi_X^g:U^X\to U^X$, depicted as
\begin{equation}
\input{Diagrams/3_ST7.tikz}\;.
\end{equation}
By~\cref{cor:LT_proc_isom}, $\spann(P_X^X)=U_X^X\cong\End(U^X)$, so these transformations belong to the required process span. As $g$ varies, they define an $S$-action $\chi_X:G\to\spann(P_X^X)$, whose image is identified with a subgroup of $\GL(U^X)$. Indeed, its action on states is the restriction of $\Xi$, so the group law and continuity follow from those of $\Xi$ and the fact that states span $U^X$. The action of $\Xi_g$ on an arbitrary process can therefore be written as
\begin{equation}
\label{eq:structure-theorem-swirling1}
\scalebox{0.8}{$\input{Diagrams/3_ST5.tikz} = \input{Diagrams/3_ST6.tikz}$}\;.
\end{equation}
Using the diagrammatic notation
\begin{equation}
\label{eq:chi}
\scalebox{0.9}{$\input{Diagrams/3_ST9.tikz} := \input{Diagrams/3_ST7.tikz}$}\;,
\end{equation}
we can equivalently depict the $T$-action as
\begin{equation}
\label{eq:structure-theorem-swirling}
\scalebox{0.8}{$\input{Diagrams/3_ST.tikz} = \input{Diagrams/3_ST1.tikz}$}\;.
\end{equation}
Thus, every $T$-action on a locally tomographic GPT admits a representation by $S$-actions $\boldsymbol{\chi}$. The structure theorem imposes no requirement that the transformations $\chi_X^g$ be physical processes, so these $S$-actions need not be physical.
\blk
\subsection{Twirling and swirling in locally tomographic theories}
\label{subsec:swirled-tomographically-local-worlds}

In \cref{def:twirling-swirling,def:S-twirling-swirling}, we distinguished $P$-twirling and $P$-swirling from $S$-twirling and $S$-swirling. The structure theorem above shows that, in locally tomographic theories, every $T$-action admits a representation by $S$-actions. We now show that physicality at the $P$- and $S$-action levels coincides, so that the two classifications become one-to-one.

Using~\cref{eq:structure-theorem-swirling}, the averaging map associated with any $T$-action on a locally tomographic theory can therefore be written as
\begin{equation}\label{eq:structure-theorem-swirling2}
\scalebox{0.8}{$\input{Diagrams/5_swirl1n.tikz} = \input{Diagrams/3_st10.tikz}$}\;.
\end{equation}
This has the same form as $S$-twirling, except that the $S$-actions $\boldsymbol{\chi}$ need not be physical. It therefore remains to determine how the physicality of these $S$-actions relates to that of the corresponding $P$-actions, which is established in the following proposition
\begin{proposition}[$S$-$P$ correspondence in locally tomographic theories]
\label{prop:SP-correspondence-LT}
Let $\mathcal{P}$ be a locally tomographic GPT and let $\Xi$ be a $T$-action of a compact group $G$. Then the $P$-actions defining $\Xi$ are physical if and only if the corresponding $S$-actions are physical. Consequently, the $S$- and $P$-classifications coincide: an averaging map is an $S$-twirling if and only if it is a $P$-twirling, and it is an $S$-swirling if and only if it is a $P$-swirling.
\end{proposition}
\begin{proof}

If the $S$-actions are physical, then by~\cref{def:process-symmetries-induced-by-system-symmetries} the induced $P$-action on every process space is physical, and hence the corresponding $T$-action is physical.

Conversely, suppose that the $P$-actions defining $\Xi$ are physical. By the structure theorem, $\Xi$ is completely determined by a collection of $S$-actions $\boldsymbol{\chi}$. It therefore suffices to show that, for every system $A$ and every $g\in G$, the corresponding transformation $\chi_A^g$ is physical.

Fix a system $A$ and $g\in G$. By the structure theorem, for every state $\omega\in P^A$, one has $\Xi_g(\omega)=\chi_A^g\circ\omega$. On the other hand, the physicality condition of~\cref{def:physical-process-action}, applied to the restriction of $\Xi_g$ to the state space, implies that this action is realised by a fixed physical one-slot circuit. For a state input, such a circuit is equivalently given by a physical transformation $t_A^g\in P_A^A$, independent of $\omega$, such that $\Xi_g(\omega)=t_A^g\circ\omega$ for every $\omega\in P^A$. Hence, $\chi_A^g\circ\omega=t_A^g\circ\omega$ for every $\omega\in P^A$. Since the states span $U^A$, it follows that $\chi_A^g=t_A^g$, and therefore $\chi_A^g$ is physical. Since this holds for every $A$ and every $g\in G$, the $S$-actions $\boldsymbol{\chi}$ are physical.

Thus, the $P$-actions defining $\Xi$ are physical if and only if the corresponding $S$-actions are physical, and the stated equivalence between the $S$- and $P$-classifications follows.
\end{proof}
Thus, the possibility that an $S$-swirling corresponds to a $P$-twirling, which can occur in general theories, does not occur in locally tomographic theories. \blk

Since $\mathcal{P}\subseteq\mathbf{FVect}_{\mathbb{R}}$ and every twirled or swirled subtheory is a subtheory of $\mathcal{P}$, both admit representations in terms of linear maps on real vector spaces. This does not imply that these invariant subtheories remain locally tomographic: as we show in~\cref{sec:FailureLT}, local tomography can fail after either twirling or swirling. In such cases, the representation in $\mathbf{FVect}_{\mathbb{R}}$ remains valid, but the states and effects of the invariant subtheory no longer necessarily span the vector spaces associated with its systems.

\subsection{Representation-theoretic tools for locally tomographic GPTs}

In locally tomographic theories, the $S$-actions $\pi$ and $\chi$ introduced above correspond to real group representations $\Gamma:G\to\GL(U^A)$, and can therefore be studied using the tools of group representation theory. Since these representations act on real vector spaces, $U^A\cong\Rl^n$, some standard results for complex representations require modification, most notably Schur's lemma. We avoid these technicalities in the main text whenever possible and refer to~\cref{app:group-rep-theory} when they are required. For notational convenience, throughout this section we use $\pi$ for physical $S$-actions and $\chi$ for potentially nonphysical $S$-actions.

We let \(V\) and \(W\) denote finite-dimensional real vector spaces equipped with representations
\[
\Gamma_V:G\to\mathrm{GL}(V),
\qquad
\Gamma_W:G\to\mathrm{GL}(W).
\]
When applying these results to a system \(A\) of a locally tomographic GPT, we identify its ambient vector space $U^A$ with the corresponding vector space $V_A$ in $\mathbf{FVect}_{\mathbb{R}}$. We use $\Gamma$ for the corresponding representation on this underlying real vector space, reserving $\pi$ and $\chi$ for the $S$-actions in the GPT itself. In the representation-theoretic analysis that follows, we use $V_A$, retaining $U^A$ only when referring explicitly to the original GPT notation. By~\cref{prop:SP-correspondence-LT}, in the $S$-twirling, equivalently $P$-twirling, case we may take $\Gamma_V$ to be the representation $\Gamma_\pi$ underlying the physical $S$-action $\pi_A$. In the $S$-swirling, equivalently $P$-swirling, case, $\Gamma_V$ is the representation $\Gamma_\chi$ underlying the nonphysical $S$-action $\chi_A$ obtained from the structure theorem, where for the family $\boldsymbol{\chi}$ there must exist some system $A$ and some $g\in G$ such that $\chi_A^g$ is nonphysical. Unless explicitly stated, the representation-theoretic results below make no assumption on physicality and therefore apply to both physical and nonphysical actions.

Note that~\cref{thm:FVectRep} implies that the unified system-induced twirled theory $\mathsf{Twirl}_G[\mathcal{P}]$ of a locally tomographic GPT $\mathcal{P}$ is isomorphic to a spanning subtheory of $\mathsf{Rep}_G[\mathbf{FVect_\mathbb{R}}]$, with $\Gamma_V=\Gamma_\pi$. More generally, the structure theorem implies that invariant subtheories obtained from arbitrary $T$-actions on locally tomographic GPTs, including $P$-swirled, equivalently $S$-swirled, subtheories, can be represented within $\mathsf{Rep}_G[\mathbf{FVect_\mathbb{R}}]$, with $\Gamma_V=\Gamma_\chi$ in the swirling case.

\begin{definition}\label{def:repFvect}
Given a compact group $G$, and the process theory $\mathbf{FVect_\mathbb{R}}$, we define $\mathsf{Rep}_G[\mathbf{FVect_\mathbb{R}}]$ as a process theory such that:
\begin{itemize}
\item Systems in $\mathsf{Rep}_G[\mathbf{FVect_\mathbb{R}}]$ are arbitrary pairs $(V,\Gamma)$ where $V\in\Sys[\mathbf{FVect}_{\mathbb R}]$, and $\Gamma$ is a finite-dimensional real representation $\Gamma:G\to\mathrm{GL}(V)$.

\vspace{0.4em}

\item Processes in $\mathsf{Rep}_G[\mathbf{FVect_\mathbb{R}}]^{(W_1,\Gamma_{W_1}),\ldots,(W_m,\Gamma_{W_m})}_{(V_1,\Gamma_{V_1}),\ldots,(V_n,\Gamma_{V_n})}$ are processes in $\mathbf{FVect_\mathbb{R}}$ that are $G$-invariant under the appropriate induced process action, that is, real linear transformations $L$ such that
\begin{equation}
\scalebox{0.9}{$\input{Diagrams/7_LT64.tikz}$}\;.
\end{equation}
We denote the space of $G$-invariant real linear processes as $\mathsf{Hom}_G(\bigotimes_{i=1}^n V_i,\bigotimes_{j=1}^m W_j)$. In the special case of trivial inputs, we denote the space of $G$-invariant linear processes as $\left(\bigotimes_{j=1}^m W_j\right)^G$.

\vspace{0.4em}

\item Composition of processes in $\mathsf{Rep}_G[\mathbf{FVect_\mathbb{R}}]$ is given by composition of the underlying processes in $\mathbf{FVect_\mathbb{R}}$.
\end{itemize}
\end{definition}

\begin{remark}
Categorically, $\mathsf{Rep}_G[\mathbf{FVect_\mathbb{R}}]$ is the symmetric monoidal category $\operatorname{Rep}^{\mathrm{fd}}_{\mathbb{R}}(G)$ of finite-dimensional real $G$-representations and $G$-equivariant linear maps, i.e., the category of $G$-objects in $\mathbf{FVect_\mathbb{R}}$~\cite{Riehl2016}.
\end{remark}
Note that~\cref{thm:FVectRep} implies that the unified system-induced twirled theory $\mathsf{Twirl}_G[\mathcal{P}]$ of a locally tomographic GPT $\mathcal{P}$ is isomorphic to a spanning subtheory of $\mathsf{Rep}_G[\mathbf{FVect_\mathbb{R}}]$, with $\Gamma_V=\Gamma_\pi$. More generally, the structure theorem implies that invariant subtheories obtained from arbitrary $T$-actions on locally tomographic GPTs, including swirled subtheories, can be represented within $\mathsf{Rep}_G[\mathbf{FVect_\mathbb{R}}]$, with $\Gamma_V=\Gamma_\chi$ in the swirling case. 

Every finite-dimensional real representation of a compact group admits a $G$-invariant inner product, or equivalently can be represented by orthogonal matrices, as stated in the following lemma:

\begin{lemma}
\label{lem:orthogonality}
Every real (complex) linear representation of a compact topological group $G$, including finite groups, is orthogonal (unitary) with respect to some choice of inner product.
\end{lemma}
\begin{proof}
    The proof is given in~\cite[Thm.~2, Ch.~2]{vinberg2010linear}, and also in~\cite[Lem.~6, App.~E]{centeno2024twirled}, where it is presented in a more familiar language within the framework of GPTs.
\end{proof}

Given the lemma above, let us fix, for every finite-dimensional real representation,  a $G$-invariant inner product denoted as $\langle -,-\rangle_V$. Let $\Phi_V \colon V \longrightarrow V^*$ where $\Phi_V(v)=\langle v,-\rangle_V$, be the corresponding \(G\)-equivariant Riesz isomorphism. The above lemma implies that for such an orthogonal representation, one has $\Gamma_{V}(g)\Gamma _V^{T}(g)=\I= \Gamma^T_{V}(g)\Gamma _V(g)$, for every $g\in G$.

For such a representation, let $\{e_i\}_i$ be an orthonormal basis of $V$, and let $\{e^i\}_i$ denote the corresponding dual basis of $V^*$. In what follows, when referring to $G$-representations we write $\Gamma_V\cong\Gamma_W$, while $V\cong W$ denotes an isomorphism of the underlying vector spaces.\footnote{The distinction is that $V\cong W$ only requires the existence of an invertible linear map between the vector spaces, whereas $\Gamma_V\cong\Gamma_W$ requires such a map to also intertwine the $G$-actions. Thus, isomorphic representations necessarily have isomorphic underlying vector spaces, but isomorphic vector spaces need not carry isomorphic representations.}. Then we have:

\begin{corollary}
For real linear representations of a compact group $G$,
\begin{enumerate}
    \item $V\cong V^*$.
    \item $\operatorname{Hom}(V,W)\cong V\otimes W$.
\end{enumerate}
More precisely, the isomorphisms above can be chosen to be $G$-equivariant, and hence also yield
$\Gamma_V\cong\Gamma_V^*$ and
$\Gamma_{\operatorname{Hom}(V,W)}\cong\Gamma_V\otimes\Gamma_W$. Whenever these identifications, or their tensor-product extensions, are used below at the level of representations, we use their $G$-equivariant forms.
\end{corollary}

\begin{proof}
    \begin{enumerate}
        \item Explicitly the isomorphism is $\Phi_V: V \to V^*$, $v \to \langle v, \cdot \rangle$, which is equivariant: $\Phi_V(\Gamma_V(g) v) = \langle \Gamma_V (g) v, \cdot \rangle = \langle v, \Gamma_V(g)^{-1} \cdot\rangle = \Phi_V(v) \circ \Gamma_V(g)^{-1} = \Gamma_V^\vee (g^{-1}) \Phi_V(v) = \Gamma_V^*(g) \Phi_V(v)$, where we used the fact that the inner product is invariant.
        \item The isomorphism $\Hom(V,W) \cong V^* \otimes W$ is $\vec \ : L \mapsto \sum_i e^i \otimes L(e_i)$ and is $G$-equivariant. It can be composed with $\Phi_V^{-1}$ above to give a $G$-equivariant isomorphism $\Hom(V,W) \cong V \otimes W$. Explicitly, $(\Phi_V^{-1}\otimes I_W)\circ\vec \ : L \mapsto
(\Phi_V^{-1}\otimes I_W)\left(\sum_i e^i\otimes L(e_i)\right)
= \sum_i e_i\otimes L(e_i)$.
    \end{enumerate}
\end{proof}
Now let us define the Choi vector $\ket{\Omega}:=\sum_i e_i\otimes e_i \in V\otimes V$, and similarly define $\bra{\Omega}:=\sum_i e^i\otimes e^i \in V^*\otimes V^*$, which we depict as
\begin{equation}
   \scalebox{0.85}{$ \input{Diagrams/7_LT15.tikz}$}\;.
\end{equation}
We can then represent a $G$-invariant inner product $\langle v,w\rangle_V$ diagrammatically as
\begin{equation}
   \scalebox{0.85}{$ \input{Diagrams/7_LT16.tikz}$}\;,
\end{equation}
and the corresponding $G$-equivariant Riesz isomorphism $\Phi_V(v)=\langle v,-\rangle_V$ as
\begin{equation}
  \scalebox{0.85}{$  \input{Diagrams/7_LT33.tikz}$}\;.
\end{equation}
One can also verify that
\begin{equation}
\scalebox{0.85}{$    \input{Diagrams/7_LT4.tikz}
    =
    \input{Diagrams/7_LT5.tikz}
    =
    \input{Diagrams/7_LT6.tikz}$}\;,
\end{equation}
so that for a real linear transformation $L\in \mathsf{Hom}(V,W)$ of $\mathbf{FVect_\mathbb{R}}$, the mapping defined by
\begin{equation}\label{eq:choi-isomorphism}
\scalebox{0.9}{$\input{Diagrams/7_LT66.tikz} \mapsto \input{Diagrams/7_LT65.tikz} $}\;,
\end{equation}
is injective, which diagrammatically represents the direction $\Hom(V,W) \to V\otimes W$ of the isomorphism $\Hom(V,W)\cong V\otimes W$.
\blk
\begin{corollary}\label{cor:Inv_Choi}
    The Choi vector $\ket \Omega =  \sum_i e_i \otimes e_i \in V \otimes V$ is invariant under $\Gamma_V(g) \otimes \Gamma_V(g)$: $\ket \Omega \in (V \otimes V)^G$.
\end{corollary}
\begin{proof}
    $\Hom(V,V) \cong V \otimes V$ is implemented by $L \mapsto \sum_i e_i \otimes L(e_i)$. Since $\I$ is in $\Hom_G(V,V)$ this implies that $\vec I = \ket \Omega$ is in $(V \otimes V)^G$.
\end{proof}

\begin{remark}\label{rem:compactclosed}
For each finite-dimensional real representation $(V,\Gamma_V)$ of a compact group $G$, fix a $G$-invariant inner product. The associated Choi state $|\Omega_V\rangle$ and effect $\langle\Omega_V|$ are $G$-invariant and satisfy the snake equations. Hence, every object of $\mathsf{Rep}_G[\mathbf{FVect_\mathbb{R}}]$ is self-dual, and the category underlying this process theory is compact closed.
\end{remark} 
 For clarity, in the representation-theoretic analysis that follows, we restrict attention to a single chosen representation $\Gamma_V$ for each underlying vector space $V$, and hence to a single pair $(V,\Gamma_V)$ for each $V$.

\begin{lemma}\label{lem:twirl_map}
    Given a representation $\Gamma_V:G\to\GL(V)$ of a compact group $G$, the Reynolds operator $\mathcal{R}_V\in\End(V)$ defined as
    \begin{align}
        \mathcal{R}_V:=\int_G \Gamma_V(g)\,d\mu(g),
    \end{align}
    where $\mu$ is the normalised Haar measure, has the following properties:
    \begin{enumerate}
        \item $\mathcal{R}_V\in\End(V)^{G\times G}$.
        \item $\mathcal{R}_V$ is an idempotent mapping $V\to V^G$ (the subspace of invariant vectors) with $\Im(\mathcal{R}_V)=V^G$.
        \item $\mathcal{R}_V$ is self-adjoint relative to any $G$-invariant inner product.
    \end{enumerate}
\end{lemma}

\begin{proof}
\begin{enumerate}
    \item $\End(V)$ transforms under $G\times G$ as
    \begin{align}
        L\mapsto\Gamma_V(g_2)L\Gamma_V(g_1^{-1}).
    \end{align}
    One has
    \begin{align}
        \Gamma_V(g)\mathcal{R}_V
        &=\Gamma_V(g)\int_G\Gamma_V(h)\,d\mu(h)\\
        &=\int_G\Gamma_V(gh)\,d\mu(h)\\
        &=\int_G\Gamma_V(h')\,d\mu(g^{-1}h')
        =\mathcal{R}_V,
    \end{align}
    where we used the left invariance of the Haar measure in the last step. Similarly, by right invariance, $\mathcal{R}_V\Gamma_V(g)=\mathcal{R}_V$. Hence,
    $\Gamma_V(g_2)\mathcal{R}_V\Gamma_V(g_1^{-1})=\mathcal{R}_V$ for all $g_1,g_2\in G$, so $\mathcal{R}_V\in\End(V)^{G\times G}$.

    \item $\mathcal{R}_V:V\to V^G$ follows from $\Gamma_V(g)\mathcal{R}_V(v)=\mathcal{R}_V(v)$. Idempotency follows from
    \begin{align}
        \mathcal{R}_V\circ\mathcal{R}_V
        =\int_G\mathcal{R}_V\Gamma_V(g)\,d\mu(g)
        =\int_G\mathcal{R}_V\,d\mu(g)
        =\mathcal{R}_V,
    \end{align}
    where we used $\mathcal{R}_V\Gamma_V(g)=\mathcal{R}_V$. Moreover, given any $v\in V^G\subseteq V$, $\mathcal{R}_V(v)=v$, and therefore $\Im(\mathcal{R}_V)=V^G$.

    \item
    \begin{align}
        \langle\mathcal{R}_V(v),w\rangle
        &=\left\langle\int_G\Gamma_V(g)v\,d\mu(g),w\right\rangle\\
        &=\left\langle v,\int_G\Gamma_V(g^{-1})w\,d\mu(g)\right\rangle
        =\langle v,\mathcal{R}_V(w)\rangle,
    \end{align}
    where in the last step we used invariance of the Haar measure under inversion.
\end{enumerate}
\end{proof}
In our diagrammatic notion, $\mathcal{R}_V$ can be represented as
\begin{equation}
     \input{Diagrams/7_LT10.tikz} :=  \input{Diagrams/7_LT11.tikz}\;.
\end{equation}
Note that in the case of twirling, i.e., $\Gamma_V=\Gamma_\pi$, the Reynolds operator is equivalent to the corresponding $S$-twirling on systems. Likewise, in the case of swirling, i.e., $\Gamma_V=\Gamma_\chi$, the Reynolds operator is equivalent to the corresponding $S$-swirling on systems by~\cref{prop:SP-correspondence-LT}.

\begin{corollary}
\begin{enumerate}
    \item Let $H=\Hom(U,W)$ and let the induced representation of $G$ on $H$ be given by
    $\Gamma_H(g)(L)=\Gamma_W(g)L\Gamma_U(g^{-1})$. Then
    \begin{align}
         \mathcal{R}_{H}^G(L):=\int_G \Gamma_W(g)L\Gamma_U(g^{-1})\,d\mu(g)
    \end{align}
    is the idempotent projector onto $H^G:=\Hom_G(U,W)$, the space of $G$-equivariant maps.

    \item Let $H=\Hom(U,W)$ and let the induced representation of $G'\times G'$ on $H$ be given by
    $\Gamma_H(g_1',g_2')(L)=\Gamma_W(g_2')L\Gamma_U((g_1')^{-1})$. Then
    \begin{align}
         \mathcal{R}_{H}^{G'\times G'}(L):=
         \int_{G'\times G'}
         \Gamma_W(g_2')L\Gamma_U((g_1')^{-1})
         \,d\mu(g_1')\,d\mu(g_2')
    \end{align}
    is the idempotent projector onto
    $H^{G'\times G'}:=\Hom_{G'\times G'}(U,W)$, the space of $G'\times G'$-invariant elements of $\Hom(U,W)$.
\end{enumerate}
\end{corollary}
In diagrammatic notation, we extend the idempotent projector $\mathcal{R}_{\Hom(V,W)}^G$ to multipartite processes. That is, we depict the projector that maps processes in $\mathbf{FVect_\mathbb{R}}$ onto the $G$-invariant subspace
$\Hom_G\!\left(\bigotimes_{i=1}^n V_i,\bigotimes_{j=1}^m W_j\right)$ as
\begin{definition}
   \begin{equation}\label{eq:rcor}
\mathcal{R}^{Cor}(\cdot):=\scalebox{0.9}{$\input{Diagrams/7_LT27v.tikz}:=\input{Diagrams/7_LT29v.tikz}$}\;.
\end{equation}
\end{definition}
 Next, we show that this indeed maps to $G$-invariant processes as one has
\begin{lemma}[
restate=rcorIdempotence,
name={}
]
\label{lem:rcorIdempotence}

For any real linear transformation $L\in \mathsf{Hom} (\bigotimes_{i=1}^n V_i,\bigotimes_{j=1}^m W_j)$, one has

\begin{itemize}

\item $\mathcal{R}^{Cor}(L)\in \mathsf{Hom}^G \left(\bigotimes_{i=1}^n V_i,\bigotimes_{j=1}^m W_j\right).$

\item $\mathcal{R}^{Cor}(\cdot)$ is idempotent.

\end{itemize}

\end{lemma}

\proof
See~\cref{proof:rcorIdempotence}.
\endproof
Note that, when $\Gamma_V=\Gamma_\pi$, $\mathcal{R}^{Cor}$ coincides with twirling $\mathcal{T}$, while for locally tomographic theories with $\Gamma_V=\Gamma_\chi$, $\mathcal{R}^{Cor}$ coincides with the swirling map $\mathcal{S}$ given in~\cref{eq:structure-theorem-swirling2}.

For the idempotent projector $\mathcal{R}_{\Hom(V,W)}^{G' \times G'}(L)$, the multipartite generalisation relevant here is the idempotent projector that maps processes in $\mathbf{FVect_\mathbb{R}}$ onto $\mathsf{Hom}\left(\bigotimes_{i=1}^n (V_{i})^G , \bigotimes_{j=1}^m (W_{j})^G\right)$, defined as
\begin{definition}
\begin{equation}\label{eq:rprod}
    \mathcal{R}^{Prod}(\cdot)\scalebox{0.9}{$:=\input{Diagrams/7_LT28.tikz}:=\input{Diagrams/7_LT30v.tikz}$}\;.
\end{equation}
\end{definition}
\begin{lemma}[
restate=rprodIdempotence,
name={}
]\label{lem:rprodIdempotence}

    For any real linear transformation $L\in \mathsf{Hom} (\bigotimes_{i=1}^n V_i,\bigotimes_{j=1}^m W_j)$, one has
    \begin{itemize}
        \item $\mathcal{R}^{Prod}(L)\in \mathsf{Hom}\left(\bigotimes_{i=1}^n (V_{i})^G , \bigotimes_{j=1}^m (W_{j})^G\right)$.
        \item $\mathcal{R}^{Prod}(\cdot)$ is idempotent.
    \end{itemize}
\end{lemma}
\proof
See~\cref{proof:rprodIdempotence}
\endproof

We introduce another generalisation of the projectors later in the following sections.
\begin{lemma}[
restate=projection,
name={}
]\label{lem:projection}

    $\mathcal{R}^{Prod}\left(\mathcal{R}^{Cor}(\cdot)\right)=\mathcal{R}^{Prod}(\cdot)=\mathcal{R}^{Cor}\left(\mathcal{R}^{Prod}(\cdot)\right)$, depicted as
    \begin{equation}\label{eq:projection}
        \scalebox{0.6}{$\input{Diagrams/7_LT87r.tikz}=\input{Diagrams/7_LT88.tikz}=\input{Diagrams/7_LT89.tikz}$}
    \end{equation}
\end{lemma}
\proof
See~\cref{proof:projection}
\endproof

This implies that%
the image of a linear map under $\mathcal{R}^{Prod}(\cdot)$ will be invariant under $\mathcal{R}^{Cor}(\cdot)$ and hence, a $G$-invariant linear map as well.

\subsection{Process tomography in twirled and swirled subtheories}

Recall that, in locally tomographic theories, for each system $A$, one has $U_A\cong (U^A)^*$ and, under the identification of $U^A$ with $V_A$, $U^{B_1,\ldots,B_m}_{A_1,\ldots,A_n}\cong \mathsf{Hom}\!\left(\bigotimes_{i=1}^n V_{A_i},\bigotimes_{j=1}^m V_{B_j}\right)$ (see~\cref{cor:LT_proc_isom}). Hence, the tools developed for $\mathsf{Rep}_G[\mathbf{FVect_\mathbb{R}}]$ apply to invariant subtheories obtained from locally tomographic GPTs by either twirling, with $\Gamma_V=\Gamma_\pi$, or swirling, with $\Gamma_V=\Gamma_\chi$. We now use these tools to study process tomography in such subtheories.

\begin{lemma}\label{lem:vec_space_twirl_loc}
Given a locally tomographic GPT $\cP$, for a twirled subtheory, with $\Gamma=\Gamma_\pi$, one has
\begin{equation}
\spann(\ct(P_{A_1,\ldots,A_n}^{B_1,\ldots,B_m}))\cong\Hom_G\!\left(\bigotimes_{i=1}^n V_{A_i},\bigotimes_{j=1}^m V_{B_j}\right),
\end{equation}
while for a swirled subtheory, with $\Gamma=\Gamma_\chi$, one has
\begin{equation}
\spann(\cs(P_{A_1,\ldots,A_n}^{B_1,\ldots,B_m}))\cong\Hom_G\!\left(\bigotimes_{i=1}^n V_{A_i},\bigotimes_{j=1}^m V_{B_j}\right).
\end{equation}
In each case, $\Hom_G$ is understood with respect to the corresponding representation $\Gamma$. Equivalently, under the identification of the ambient process space with $\Hom\!\left(\bigotimes_{i=1}^n V_{A_i},\bigotimes_{j=1}^m V_{B_j}\right)$, one has $\operatorname{Im}\mathcal R^{\mathrm{Cor}}=\Hom_G\!\left(\bigotimes_{i=1}^n V_{A_i},\bigotimes_{j=1}^m V_{B_j}\right)$.
\end{lemma}

\begin{proof}
Fix either case and let $\Gamma$ denote the corresponding representation. Since $\cP$ is locally tomographic, one has $\spann(P_{A_1,\ldots,A_n}^{B_1,\ldots,B_m})=U_{A_1,\ldots,A_n}^{B_1,\ldots,B_m}\cong\Hom\!\left(\bigotimes_{i=1}^n V_{A_i},\bigotimes_{j=1}^m V_{B_j}\right)$. On this space, $\mathcal R^{\mathrm{Cor}}$ is the idempotent projector onto $\Hom_G\!\left(\bigotimes_{i=1}^n V_{A_i},\bigotimes_{j=1}^m V_{B_j}\right)$. Since the corresponding twirling or swirling map agrees with $\mathcal R^{\mathrm{Cor}}$ on physical processes, and the physical processes span the ambient process space, linearity gives the claimed equality. Taking $\Gamma=\Gamma_\pi$ gives the twirling case, while $\Gamma=\Gamma_\chi$ gives the swirling case.
\end{proof}

\begin{lemma}
 The Choi vector $\ket \Omega=\sum_i e_i\otimes e_i$ on any system $AA$ is a dynamically faithful vector.
\end{lemma}
\begin{proof}
This follows directly from the isomorphism induced by the Choi vector, as stated in~\cref{eq:choi-isomorphism}. When restricted to physical processes, one obtains the injective map
\beq
P_A^B\to U^{B,A}::\input{Diagrams/4_twirl105.tikz} \mapsto \input{Diagrams/7_LT7.tikz}\;.
\eeq
Hence, for any two distinct processes $f\neq h\in P_A^B$, the Choi vector distinguishes them and is therefore dynamically faithful.
\end{proof}
By~\cref{cor:LT_proc_isom}, $U^{A,A}\cong U^A\otimes U^A$. Hence, after fixing an inner product on $U^A$ and choosing an orthonormal basis $\{e_i\}_i$, the Choi vector
$|\Omega_A\rangle:=\sum_i e_i\otimes e_i$
belongs to
$U^A\otimes U^A\cong U^{A,A}=\spann(P^{A,A})$. Thus, every locally tomographic GPT admits a Choi vector in the span of its bipartite state space that is both invariant and dynamically faithful. By~\cref{cor:dynamically-faithful-vector-swirl}, this gives the following:

\begin{proposition}\label{prop:process-tomography-from-local-tomography}
Any invariant subtheory of a locally tomographic GPT $\mathcal{P}$ satisfies process tomography and is a valid GPT-subtheory. In particular, this applies to both twirled and swirled subtheories.
\end{proposition} 
This implies that, in any invariant subtheory $\mathcal{A}(\mathcal{P})$ obtained from a locally tomographic theory, for any two distinct invariant processes $\tilde f\neq\tilde h\in\mathcal{A}(P_A^B)$, there exist an ancilla $E\cong A$, an invariant state $\tilde{\omega}$, and an invariant effect $\tilde e$ such that one has (see~\cref{prop:dynamically-faithful-vector-swirl}) 
\begin{equation}
  \input{Diagrams/7_LT8.tikz}\neq \input{Diagrams/7_LT9.tikz}  \;.
\end{equation}

\subsection{When do invariant subtheories exhibit a failure of local tomography}\label{sec:FailureLT}

Ref.~\cite{centeno2024twirled} shows that a non-trivial system representation in a locally tomographic GPT leads to a failure of local tomography in the corresponding twirled subtheory, by considering two copies of a system carrying the same non-trivial representation. Since the proof relies only on properties of real group representations and does not use the physicality of the action, the structure-theorem representation of a general averaging map $\mathcal{A}$ in~\eqref{eq:structure-theorem-swirling2} implies that the same argument applies more generally to invariant subtheories of locally tomographic GPTs. This can therefore be stated as follows:

\begin{theorem}\label{thm:ubiquity}
Let $\mathcal{P}$ be a locally tomographic GPT and let $\mathcal{A}(\mathcal{P})$ be a $G$-invariant subtheory for a compact group $G$, with corresponding representations $\Gamma_X:G\to\GL(V_X)$. If $\Gamma_A$ is non-trivial for at least one system $A$, then $\mathcal{A}(\mathcal{P})$ fails to satisfy local tomography.
\end{theorem} 

The proof can be found in \cite[Thm.~8, App.~E]{centeno2024twirled} and applies here because it depends only on the corresponding real group representations. It considers two copies of a system carrying the same non-trivial representation and shows that there exist invariant bipartite states that cannot be distinguished locally. Hence, it establishes failure of local tomography whenever some system carries a non-trivial representation, but does not determine which individual multipartite process spaces exhibit this failure. 

In this section, we give a criterion for when local tomography fails for an invariant process space of a locally tomographic GPT, and determine when this criterion is satisfied in terms of the representations carried by the systems. As we will see, the criterion can be stated using the projectors defined in the previous section. We then study its operational consequences, including when ancillary systems are required to distinguish processes and, when they are not, when entanglement is necessary and what kinds of entangled states and effects suffice for process tomography.

\begin{proposition}\label{prop:LT_twirl_gen}
Let $\mathcal{A}(\mathcal{P})$ be an invariant subtheory of a locally tomographic GPT $\mathcal{P}$. Then the process space $\mathcal{A}(P_{A_1,\ldots,A_n}^{B_1,\ldots,B_m})$ is locally tomographic if and only if
\begin{equation}
\operatorname{Im}\mathcal{R}^{\mathrm{Cor}}=\operatorname{Im}\mathcal{R}^{\mathrm{Prod}}.
\end{equation}
Equivalently, local tomography holds if and only if $\mathcal{R}^{\mathrm{Cor}}(\cdot)=\mathcal{R}^{\mathrm{Prod}}(\cdot)$. Diagrammatically, this condition is depicted as
\begin{align}
\scalebox{0.9}{$\input{Diagrams/7_LT29.tikz}=\input{Diagrams/7_LT30.tikz}$}\;.
\end{align}
\end{proposition}

\begin{proof}
In a locally tomographic theory one has $U_{A_1,\ldots,A_n}^{B_1,\ldots,B_m}\cong\Hom\!\left(\bigotimes_{i=1}^n V_{A_i},\bigotimes_{j=1}^m V_{B_j}\right)$. By~\Cref{lem:vec_space_twirl_loc}, the span of the invariant process space is $\operatorname{Im}\mathcal R^{\mathrm{Cor}}$. On the other hand, $\operatorname{Im}\mathcal R^{\mathrm{Prod}}$ is precisely the tomographically local subspace, obtained by restricting independently to the invariant sectors of the individual input and output systems. Therefore, the invariant process space is locally tomographic if and only if $\operatorname{Im}\mathcal R^{\mathrm{Cor}}=\operatorname{Im}\mathcal R^{\mathrm{Prod}}$. Since $\mathcal R^{\mathrm{Cor}}$ and $\mathcal R^{\mathrm{Prod}}$ are orthogonal projectors with respect to the same invariant inner product, equality of their images is equivalent to $\mathcal R^{\mathrm{Cor}}(\cdot)=\mathcal R^{\mathrm{Prod}}(\cdot)$.
\end{proof} 

By~\Cref{prop:LT_twirl_gen}, a process space $\mathcal{A}(P_{A_1,\ldots,A_n}^{B_1,\ldots,B_m})$ fails to be locally tomographic if and only if $\operatorname{Im}\mathcal{R}^{\mathrm{Cor}}\neq\operatorname{Im}\mathcal{R}^{\mathrm{Prod}}$, or, equivalently, $\mathcal{R}^{Cor}(\cdot)\neq\mathcal{R}^{Prod}(\cdot)$, i.e., 
\begin{align}
        \scalebox{0.9}{$\input{Diagrams/7_LT29.tikz}\neq\input{Diagrams/7_LT30.tikz}$}
        \;.
\end{align}
Note that, in invariant subtheories of locally tomographic GPTs, $\mathcal{R}^{Prod}(\cdot)$ generalises, from bipartite state and effect spaces to arbitrary process spaces, the projection onto the tomographically local subspace introduced in Ref.~\cite{baldijao2026tomographically}. In particular, it projects $\mathcal{A}(P_{A_1,\ldots,A_n}^{B_1,\ldots,B_m})$ onto its tomographically local subspace. 

 \subsection{Process tomography when local tomography fails}
Before studying process tomography when local tomography fails, we first define another projector in $\mathsf{Rep}_G[\mathbf{FVect_\mathbb{R}}]$. We previously generalised the idempotent projector $\mathcal{R}_{\Hom(V,W)}^{G'\times G'}(L)$ to $\mathcal{R}^{Prod}(\cdot)$. An alternative multipartite generalisation is obtained by defining the idempotent projector onto $\Hom^{G\times G}\!\left(\bigotimes_{i=1}^n V_i,\bigotimes_{j=1}^m W_j\right)$, namely: \
\begin{definition}
\begin{equation}\label{eq:rsemi}
    \mathcal{R}^{Semi}(\cdot)=\scalebox{0.9}{$\input{Diagrams/7_LT31.tikz}:=\input{Diagrams/7_LT32.tikz}$}\;.
\end{equation}
\end{definition}
\begin{lemma}[
restate=rsemiIdempotence,
name={}
]\label{lem:rsemiIdempotence}

    For any real linear transformation $L\in \mathsf{Hom} (\bigotimes_{i=1}^n V_i,\bigotimes_{j=1}^m W_j)$, one has
    \begin{itemize}
        \item $\mathcal{R}^{Semi}(L)\in \mathsf{Hom}^{G\times G} \left(\bigotimes_{i=1}^n V_i,\bigotimes_{j=1}^m W_j\right).$
        \item $\mathcal{R}^{Semi}(\cdot)$ is idempotent.
    \end{itemize}
\end{lemma}
\proof
See~\cref{proof:rsemiIdempotence}
\endproof

\begin{lemma}[
restate=projectionSemi,
name={}
]\label{lem:projectionSemi}

    \begin{align}\label{eq:projectionSemi}
       \mathcal{R}^{Semi}\left(\mathcal{R}^{Cor}(\cdot)\right) =\mathcal{R}^{Semi}(\cdot)=\mathcal{R}^{Cor}\left(\mathcal{R}^{Semi}(\cdot)\right)\nonumber\\
        \mathcal{R}^{Prod}\left(\mathcal{R}^{Semi}(\cdot)\right)=\mathcal{R}^{Prod}(\cdot)=\mathcal{R}^{Semi}\left(\mathcal{R}^{Prod}(\cdot)\right)\;,
    \end{align}
    depicted as
    \begin{equation*}
\scalebox{0.6}{$\input{Diagrams/7_LT97.tikz}=\input{Diagrams/7_LT96.tikz}=\input{Diagrams/7_LT95.tikz}$}
    \end{equation*}
    \begin{equation*}
        \scalebox{0.6}{$\input{Diagrams/7_LT115.tikz}=\input{Diagrams/7_LT88.tikz}=\input{Diagrams/7_LT116.tikz}$}
    \end{equation*}
\end{lemma}
\proof
See~\cref{proof:projectionSemi}
\endproof

This implies that any linear map obtained by applying $\mathcal{R}^{Semi}(\cdot)$ is invariant under $\mathcal{R}^{Cor}(\cdot)$, and hence is also a valid $G$-linear map.

\begin{remark}[Ambient projectors and physical maps]
$\mathcal{R}^{Cor}$ coincides with the averaging map $\mathcal{A}$ and therefore always preserves physical processes. For twirling, $\mathcal{R}^{Semi}$ and $\mathcal{R}^{Prod}$ also preserve the corresponding physical process spaces, since they are obtained by averaging compositions with $S$-actions. For swirling, however, $\mathcal{R}^{Semi}$ and $\mathcal{R}^{Prod}$ are in general only linear projectors on the ambient process vector space and need not map physical processes to physical processes.
\end{remark} 

For our purposes here, particularly when accounting for the potential nonphysicality of $\mathcal{R}^{Semi}$ and $\mathcal{R}^{Prod}$ in the swirling case, the following lemma will be useful:

\begin{lemma}[
restate=InvariantSubprojection,
name={}
]\label{lem:InvariantSubprojection}
Consider a swirled process space $\cs(P^{B_1,\cdots,B_m}_{A_1,\cdots,A_n})$ obtained from a locally tomographic theory $\mathcal{P}$. Let $Q(\cdot)$ be an idempotent projector on this process space satisfying
\begin{equation}\label{eq:absorption}
Q\!\left(\mathcal{R}^{Cor}(\cdot)\right)
=
Q(\cdot)
=
\mathcal{R}^{Cor}\!\left(Q(\cdot)\right).
\end{equation}
If $Q(\cdot)\neq\mathcal{R}^{Cor}(\cdot)$, then there exist two distinct
physical invariant processes
$\tilde f_1,\tilde f_2\in \cs(P^{B_1,\cdots,B_m}_{A_1,\cdots,A_n})$
such that
\begin{equation}
Q(\tilde f_1)=Q(\tilde f_2).
\end{equation}
\end{lemma}
\proof
See~\cref{proof:InvariantSubprojection}
\endproof
The same argument applies to the twirling case and, more generally, to the corresponding subprojections of $\mathcal{R}^{Cor}$. 

Since, as we will show, entanglement can sometimes be used to achieve process tomography when local tomography fails to hold, it is useful to first define what we mean by entangled invariant states and entangled invariant effects.
\begin{definition}[Entangled invariant states and effects]
A state $\tilde{s}\in \mathcal{A}(P^{A_1,\cdots,A_n})$ is separable if and only if it can be written as
\begin{equation}
    \input{Diagrams/7_LT139.tikz}\;,
\end{equation}
where $\tilde{s}^i_l\in \mathcal{A}(P^{A_l})$ are local invariant states for every $1\leq i\leq I$, with $I\in\mathbb{N}$, and $\{p_i\}_{i=1}^{I}$ is a probability distribution. Any invariant state $\tilde{s}$ that cannot be written in this form is an entangled invariant state.

An effect $\tilde{e}\in \mathcal{A}(P_{A_1,\cdots,A_n})$ is separable if and only if it can be written as
\begin{equation}
    \input{Diagrams/7_LT140.tikz}\;,
\end{equation}
where $\tilde{e}^i_l\in \mathcal{A}(P_{A_l})$ are local invariant effects for every $1\leq i\leq I$, with $I\in\mathbb{N}$. Any invariant effect $\tilde{e}$ that cannot be written in this form is an entangled invariant effect.
\end{definition}
In non-locally tomographic theories, two distinct forms of entanglement can arise, termed \emph{tomographically local} ($TL$) and \emph{tomographically nonlocal} ($TNL$) entanglement in Ref.~\cite{baldijao2026tomographically}. $TL$ entanglement corresponds to the usual form of entanglement that can already occur in locally tomographic theories, whereas $TNL$ entanglement specifically captures nonseparable states associated with degrees of freedom that are inaccessible to local tomography. Since the invariant subtheories considered here can fail to hold local tomography, one may therefore expect $TNL$ entanglement to arise in them alongside $TL$ entanglement. We now show that this is indeed the case and, adapting the results of Ref.~\cite{baldijao2026tomographically} to our framework, characterize both forms of entanglement directly in terms of the projectors introduced above.

\begin{definition}[Tomographically local and nonlocal entanglement]
Let $\tilde{s}$ be an invariant state in a multipartite invariant state space $\mathcal{A}(P^{A_1,\ldots,A_n})$. Following Ref.~\cite{baldijao2026tomographically}, $\tilde{s}$ is said to have \emph{tomographically local entanglement} ($TL$ entanglement) if its component in the tomographically local subspace does not belong to the set of separable invariant states, i.e.,
\begin{equation}
\mathcal{R}^{Prod}(\tilde{s})
\notin \operatorname{Sep}.
\end{equation}
Here, a component that is not itself a physical state is, in particular, outside the set of separable states. It is said to have \emph{tomographically nonlocal entanglement} ($TNL$ entanglement) if it has a nonzero component outside the tomographically local subspace, i.e.,
\begin{equation}
\mathcal{R}^{Prod}(\tilde{s})\neq\tilde{s}.
\end{equation}

Analogously, an invariant effect $\tilde{e}$ has $TL$ entanglement if $\mathcal{R}^{Prod}(\tilde{e})$ does not belong to the set of separable invariant effects, and has $TNL$ entanglement if $\mathcal{R}^{Prod}(\tilde{e})\neq\tilde{e}$. A component that is not itself a physical effect is, in particular, outside the set of separable effects.

\end{definition}

A particular class of invariant states and effects that can exhibit $TNL$ entanglement is obtained as follows:
\begin{definition}\label{def:TNLstateEffect}
Let $s_i\in P^{A_i}$ and $e_j\in P_{B_j}$ be local states and effects of the original theory, with each system carrying a representation $\Gamma$. We then define:
\begin{align}\label{eq:entangled-state-effect}
   \scalebox{0.9}{$ \input{Diagrams/7_LT52.tikz}$}\;,\nonumber\\
   \scalebox{0.9}{$\input{Diagrams/7_LT53.tikz}$}\;.
\end{align}
\end{definition}

\begin{proposition}
In multipartite invariant state and effect spaces obtained from a locally tomographic theory in which local tomography fails to hold, the states and effects $\tilde s_{TNL}$ and $\tilde e_{TNL}$ defined in~\cref{def:TNLstateEffect} are invariant and do not exhibit $TL$ entanglement. They can, however, exhibit $TNL$ entanglement depending on the choice of the local states and effects. In particular, letting $s$ and $e$ denote the corresponding product state and effect constructed from those local states and effects, $\tilde s_{TNL}=\mathcal{R}^{Cor}(s)$ is $TNL$-entangled if and only if $\mathcal{R}^{Prod}(s)\neq\mathcal{R}^{Cor}(s)$, and, analogously, $\tilde e_{TNL}=\mathcal{R}^{Cor}(e)$ is $TNL$-entangled if and only if $\mathcal{R}^{Prod}(e)\neq\mathcal{R}^{Cor}(e)$.
\end{proposition}
\begin{proof}
By definition, $\tilde s_{TNL}=\mathcal{R}^{Cor}(s)$, and is therefore invariant. Moreover,
\begin{align}
\mathcal{R}^{Prod}(\tilde s_{TNL})
=\mathcal{R}^{Prod}\left(\mathcal{R}^{Cor}(s)\right)
\stackrel{\eqref{eq:projection}}{=}\mathcal{R}^{Prod}(s).
\end{align}
Hence, $\mathcal{R}^{Prod}(\tilde s_{TNL})\neq \tilde s_{TNL}$ if and only if $\mathcal{R}^{Prod}(s)\neq\mathcal{R}^{Cor}(s)$, giving the stated condition for $TNL$ entanglement. Since local tomography fails, by~\cref{prop:LT_twirl_gen} one has $\mathcal{R}^{Prod}\neq\mathcal{R}^{Cor}$, and since product states span the corresponding state vector space of the original locally tomographic theory, this inequality holds for some product state $s$. Moreover, $\mathcal{R}^{Prod}(s)$ is itself a product state, so $\mathcal{R}^{Prod}(\tilde s_{TNL})$ is separable, ruling out $TL$ entanglement. The same reasoning holds for $\tilde e_{TNL}$.

In the swirling case, although $\mathcal{R}^{Prod}$ need not preserve physicality in general, it does so for the product states and effects considered here, since it acts as the product of the corresponding single-system Reynolds maps, each of which coincides with the single-system swirling map.
\end{proof}

Having defined entangled invariant states and effects, it follows that:
\begin{proposition}[
restate=fexistence,
name={}
]\label{prop:fexistence}

   Given a non-locally-tomographic invariant process space $\mathcal{A}(P_{A_1,\ldots,A_n}^{B_1,\ldots,B_m})$ of a locally tomographic theory $\mathcal{P}$, there exist two distinct invariant processes $\tilde f_1\neq\tilde f_2\in\mathcal{A}(P_{A_1,\ldots,A_n}^{B_1,\ldots,B_m})$ that yield identical statistics for all product (separable) invariant states and effects in $\mathcal{A}(\mathcal{P})$.
\end{proposition} 
\proof
See~\cref{proof:fexistence}
\endproof

\begin{proposition}[
restate=entagledSuffice,
name={}
]\label{prop:entagledSuffice}
Given a non-locally-tomographic invariant process space $\mathcal{A}(P_{A_1,\ldots,A_n}^{B_1,\ldots,B_m})$ of a locally tomographic theory $\mathcal{P}$, process tomography can be achieved without ancillary systems if and only if $\mathcal{R}^{Cor}(\cdot)=\mathcal{R}^{Semi}(\cdot)$, i.e.,
\begin{equation}
    \scalebox{0.9}{$\input{Diagrams/7_LT29.tikz}=\input{Diagrams/7_LT32.tikz}$}
    \;.
\end{equation}
\mainonly{In the ``if'' direction, process tomography can be achieved using the invariant states and effects defined in~\cref{def:TNLstateEffect}.}

\end{proposition}

\proof
see~\cref{proof:entagledSuffice}.
\endproof

\begin{corollary}[
restate=entanglementNecessary,
name={}
]\label{cor:entanglementNecessary}
Consider a non-locally-tomographic invariant process space obtained from a locally tomographic theory. If complete process tomography can be achieved without ancillary systems, then it cannot be achieved using only separable invariant states and effects. Hence, any complete ancilla-free tomographic scheme must include entanglement in at least one input state or output effect.
\end{corollary}
\begin{proof}
By~\cref{prop:fexistence}, since local tomography fails, there exist two distinct physical invariant processes $\tilde f_1\neq\tilde f_2$ that cannot be distinguished using only separable invariant states and effects. If complete process tomography can nevertheless be achieved without ancillary systems, then there must exist an ancilla-free tester that distinguishes $\tilde f_1$ and $\tilde f_2$. Such a tester cannot consist of both a separable invariant state and a separable invariant effect. Hence, at least one of its input state or output effect must be entangled.

\end{proof}

Note, however, that $\mathcal{R}^{Cor}(\cdot)\neq\mathcal{R}^{Semi}(\cdot)$, namely, the negation of the projector equality appearing in the above proposition, does not imply that every pair of distinct invariant processes $\tilde{f}_1\neq\tilde{f}_2$ necessarily requires ancillary systems to be distinguished. A more operationally relevant question is therefore to determine when a particular pair of distinct invariant processes can be distinguished without ancillary systems. We clarify this in the following proposition, where we also show that, whenever this is possible, invariant states and effects of the form defined in~\cref{def:TNLstateEffect} suffice:
\begin{proposition}[
restate=pairwiseProcessesEntangled,
name={}
]\label{prop:pairwiseProcessesEntangled}

Two distinct invariant processes $\tilde{f}_1\neq\tilde{f}_2$ can be distinguished without ancillary systems if and only if
$\mathcal{R}^{Semi}(\tilde f_1)\neq \mathcal{R}^{Semi}(\tilde f_2)$. 
In the ``if'' direction, they can be distinguished using invariant states and effects of the form defined in~\cref{def:TNLstateEffect}. 

\end{proposition}
\proof
See~\cref{proof:pairwiseProcessesEntangled}
\endproof

However, in general, if $\mathcal{R}^{Cor}(\cdot)\neq\mathcal{R}^{Semi}(\cdot)$, then based on~\cref{prop:entagledSuffice}, there exist distinct invariant processes that can be distinguished only by using ancillary systems, that is:
\begin{proposition}[
restate=ancillaNeeded,
name={}
]\label{prop:ancillaNeeded}
  Given a non-locally-tomographic invariant process space $\mathcal{A}(P_{A_1,\ldots,A_n}^{B_1,\ldots,B_m})$ of a locally tomographic theory $\mathcal{P}$, if $\mathcal{R}^{Cor}(\cdot)\neq\mathcal{R}^{Semi}(\cdot)$, then there exist distinct invariant processes that yield identical statistics for all invariant states and invariant effects without ancillary systems, but can be distinguished using ancillary systems \mainonly{, i.e.,
  \begin{equation}
\exists\,E,\tilde{s},\tilde{e}\ \text{such that}\
\scalebox{0.9}{$
\input{Diagrams/13_ubq57.tikz}
\neq
\input{Diagrams/13_ubq56.tikz}
$}\;.
\end{equation}
}%
\restateonly{.}
\end{proposition}
\proof
See~\cref{proof:ancillaNeeded}
\endproof

The following proposition characterises precisely when a given pair of distinct invariant processes requires ancillary systems to be distinguished:
\begin{proposition}[
restate=pairwiseProcessesAncilla,
name={}
]\label{prop:pairwiseProcessesAncilla}

Two distinct invariant processes $\tilde{f}_1\neq\tilde{f}_2$ require ancillary systems to be distinguished if and only if $\mathcal{R}^{Semi}(\tilde f_1)= \mathcal{R}^{Semi}(\tilde f_2)$.
\end{proposition}
\proof
See~\cref{proof:pairwiseProcessesAncilla}
\endproof

We can summarise the results of this section as follows. In an invariant subtheory of a locally tomographic GPT, when local tomography fails to hold, one has:
\begin{itemize}
    \item Ancillary systems always suffice to ensure the completeness of process tomography.

   \item On an invariant process space, process tomography can be achieved without ancillary systems if and only if
$\mathcal{R}^{Cor}(\cdot)=\mathcal{R}^{Semi}(\cdot)$.

\item Whenever process tomography can be achieved without ancillary systems in a non-locally-tomographic invariant process space, entanglement is necessary in general: complete process tomography cannot be achieved using only separable invariant states and effects.

\item Two distinct invariant processes can be distinguished without ancillary systems if and only if their images under $\mathcal{R}^{Semi}(\cdot)$ are distinct.

    \item If, on an invariant process space, one has
    $\mathcal{R}^{Cor}(\cdot)\neq\mathcal{R}^{Semi}(\cdot)$, then there exist distinct invariant processes that can be distinguished only by using ancillary systems.

    \item Two distinct invariant processes require ancillary systems to be distinguished if and only if their images under $\mathcal{R}^{Semi}(\cdot)$ are equal.
\end{itemize}

Note that, so far in this section, our analysis of process tomography in invariant subtheories of locally tomographic theories has relied only on the properties of the projectors that follow from the group structure and group averaging, rather than on the detailed representation theory of the group. The only point at which representation-theoretic structure entered directly was in establishing that process tomography holds in such invariant subtheories (see~\cref{prop:process-tomography-from-local-tomography}), where we used the existence of an invariant Choi vector associated with a basis relative to which one can define a $G$-invariant inner product.

Here, we have restricted our attention to two particular ways of partitioning the correlations captured by $\mathcal{R}^{Cor}$, namely those associated with $\mathcal{R}^{Semi}$ and $\mathcal{R}^{Prod}$. It is conceivable to define further partitions and extend the preceding analysis to all such possibilities. We do not pursue this direction, however, since it would largely amount to a technical classification without providing substantial additional conceptual insight.

A more interesting question is under what circumstances the projection maps introduced above are distinct or equivalent on a given process space, so that the operational implications discussed above arise. In the following section, we study the representation-theoretic conditions under which these projection maps are distinct and discuss their operational consequences. This representation-theoretic analysis also sheds light on which partitions are ultimately relevant for the process theory under consideration. We will see that the key representation-theoretic condition relevant to the projector relations studied here is whether the systems in the process theory of interest share common irreducible representations.

\subsection{Representation theoretic conditions for failure of local tomography, and process tomography}\label{sec:OperationalRep}
\addtocontents{toc}{\protect\HideSubsubsectionsFromTOC}
We now turn to a representation-theoretic analysis of the conditions under which local tomography fails in invariant subtheories and the implications of this failure for process tomography. Rather than beginning immediately with the most general multipartite process space, we proceed gradually through increasingly complex examples. We first consider bipartite state spaces, where we generalise~\Cref{thm:ubiquity}, and then move to single-input/single-output processes, three-wire processes, two-input/two-output processes with four wires, and finally the general case of processes with multipartite inputs and outputs. One could instead begin with the general multipartite setting and recover each of these cases by specialisation. We adopt the present order because it makes the operational significance of the representation-theoretic conditions more transparent, and in particular clarifies how the representations carried by the individual wires determine both the failure of local tomography and the form of process tomography that remains possible. We begin with a simple representation-theoretic observation that will be used throughout the analysis:

\begin{lemma}\label{lem:trivial-irrep-system}
For every system $A$, let $\Gamma_A:G\to\GL(V_A)$ be the representation
of $G$ on $V_A$. Then $\Gamma_A$ contains the trivial representation.
Equivalently, $V_A^G\neq\{0\}$.
\end{lemma}
\proof
Choose a non-zero physical state $\omega\in P^A$, so that $u_A(\omega)>0$. Define
\begin{equation}
\bar\omega:=\int_G\Gamma_A(g)\omega\,d\mu(g).
\end{equation}
Since this is the average of $\omega$ over the group, $\bar\omega$ is $G$-invariant. Moreover, since $\Gamma_A(g)$ represents the action of the corresponding $T$-action on the state space, it maps physical states to physical states, so $u_A(\Gamma_A(g)\omega)\geq 0$ for all $g\in G$. For the identity element $e_G\in G$, one has $u_A(\Gamma_A(e_G)\omega)=u_A(\omega)>0$. By continuity,
\begin{equation}
u_A(\bar\omega)=\int_G u_A(\Gamma_A(g)\omega)\,d\mu(g)>0,
\end{equation}
so $\bar\omega\neq0$. Hence $V_A^G\neq\{0\}$. In particular, $\bar\omega$ spans a one-dimensional invariant subspace on which $G$ acts trivially, so $\Gamma_A$ contains the trivial representation.
\endproof

\subsubsection{Bi-partite systems}

\Cref{prop:LT_twirl_gen} can be used to generalise \Cref{thm:ubiquity} to arbitrary pairs of systems:
\begin{theorem}
\label{prop:failure-LT-reduicible}
In an invariant subtheory $\mathcal{A}(\mathcal{P})$ of a locally tomographic GPT, local tomography fails to hold for the states of a bipartite system $(A,B)$ if and only if the systems $A$ and $B$ share at least one non-trivial irreducible representation of the group $G$.
\end{theorem}

\begin{proof}
By~\Cref{prop:LT_twirl_gen}, $AB$ fails to be locally tomographic if and only if $\mathcal{R}^{Cor}\neq\mathcal{R}^{Prod}(\cdot)$ which for a bipartite state space $P^{A,B}$ implies that $(V_A \otimes V_B)^G \neq V_A^G \otimes V_B^G$.

By~\Cref{lem:G_inv_GtimesG} $(V_A \otimes V_B)^G \neq V_A^G \otimes V_B^G$ if and only if $V_A$ and $V_B$ share at least one non-trivial irreducible representation of $G$.

\end{proof}

\subsubsection{Single input and output processes}

We now turn our attention to failures of local tomography for invariant processes $\tilde f\in\mathcal{A}(P_A^B)$. We denote by $(V_A)_{(0)}:=(V_A)^G$ the trivial isotypic component of $V_A$, namely the subspace left invariant by every element of $G$, and similarly for $(V_B)_{(0)}$.

\begin{proposition}[
restate=singleInputOutputLT,
name={}
]\label{prop:singleInputOutputLT}
An invariant process space $\mathcal{A}(P_A^B)$ of a locally tomographic GPT is locally tomographic if and only if $\Hom_G(V_A,V_B)=\Hom_G((V_A)_{(0)},(V_B)_{(0)})$, or equivalently $V_A$ and $V_B$ do not share any non-trivial irreducible representations of $G$.
\end{proposition} 

\begin{proof}
This follows directly from~\Cref{prop:LT_twirl_gen} by setting $A_1=A$, $B_1=B$, $A_j=I$ for $2\leq j\leq n$, and $B_l=I$ for $2\leq l\leq m$. In this setting one has $\operatorname{Im}\mathcal R^{Cor}=\Hom_G(V_A,V_B)$ and $\operatorname{Im}\mathcal R^{Prod}=\Hom((V_A)^G,(V_B)^G)$, since $\mathcal R^{Prod}$ projects the input and output independently onto their $G$-invariant subspaces. Moreover, $(V_A)^G=(V_A)_{(0)}$ and $(V_B)^G=(V_B)_{(0)}$, since the $G$-invariant subspace is precisely the trivial representation sector. Since every map between trivial representation sectors is automatically $G$-equivariant, $\Hom((V_A)_{(0)},(V_B)_{(0)})=\Hom_G((V_A)_{(0)},(V_B)_{(0)})$. Hence, local tomography holds if and only if $\Hom_G(V_A,V_B)=\Hom_G((V_A)_{(0)},(V_B)_{(0)})$. By~\Cref{lem:G_inv_GtimesG}, this holds if and only if $V_A$ and $V_B$ do not share any non-trivial irreducible representation of $G$.
\end{proof}

For a single-input/single-output process space $\ca(P_A^B)$, since $\mathcal R^{Semi}=\mathcal R^{Prod}$, the condition for ancillary systems to be required for complete process tomography from~\Cref{prop:ancillaNeeded} reduces to $\mathcal R^{Cor}(\cdot)\neq\mathcal R^{Prod}(\cdot)$. By~\Cref{prop:singleInputOutputLT}, this occurs if and only if $V_A$ and $V_B$ share a non-trivial irrep of $G$. Hence, complete tomography of the process space requires an ancilla precisely when the input and output systems share a non-trivial irrep. For a particular pair of distinct invariant processes $\tilde f\neq\tilde h\in\ca(P_A^B)$, the criterion of~\Cref{prop:pairwiseProcessesEntangled} becomes especially simple: an ancilla is required precisely when $\tilde f$ and $\tilde h$ have the same restriction to the trivial sectors, meaning that they induce the same map from $(V_A)^{(0)}=V_A^G$ to $(V_B)^{(0)}=V_B^G$. Equivalently, they can be distinguished without an ancilla precisely when these restricted maps differ.

\subsubsection{Three wire processes}

Given a three-wire process (such as a tripartite state or a two-input/one-output process), we can use~\Cref{prop:LT_twirl_gen} to provide explicit conditions for when the corresponding invariant process space exhibits a failure of local tomography.

\begin{proposition}[
restate=thereeWire,
name={}
]\label{prop:thereeWire}

   Given a three-wire process space $P$ with wires $C_1C_2C_3$, the corresponding invariant process space $\mathcal{A}(P)$ fails to be locally tomographic if and only if 
    \begin{enumerate}
        \item $(V_{C_i} \otimes V_{C_j})^G \neq V_{C_i}^G \otimes V_{C_j}^G$ for some $i,j$.
        \item (or) $(V_{C_1} \otimes V_{C_2} \otimes V_{C_3})^G \neq (V_{C_i} \otimes V_{C_j})^G \otimes V_{C_k}^G$ for some $i,j,k$.
    \end{enumerate}
       equivalently
    \begin{enumerate}
        \item $V_{C_i} $ and $V_{C_j} $ share a non-trivial irreducible representation for some $i,j$.
        \item $(V_{C_i} \otimes V_{C_j})$ and $V_{C_k}$ share a non-trivial irreducible representation for some $i,j,k$.
    \end{enumerate}

\end{proposition}

\begin{proof}
    Since, based on~\Cref{prop:LT_twirl_gen}, local tomography holds if and only if $\Im \mathcal{R}^{Cor}=\Im \mathcal{R}^{Prod}$, and since the ambient real vector space of every process space with three wires $C_1C_2C_3$ is isomorphic to $V_{C_1}\otimes V_{C_2}\otimes V_{C_3}$, local tomography is equivalent to $(V_{C_1} \otimes V_{C_2} \otimes V_{C_3})^G = V_{C_1}^G \otimes V_{C_2}^G\otimes V_{C_3}^G$.

    The statement of the proposition is then equivalent to stating that $(V_{C_1} \otimes V_{C_2} \otimes V_{C_3})^G = V_{C_1}^G \otimes V_{C_2}^G\otimes V_{C_3}^G$  $\Leftrightarrow$

    \begin{enumerate}
        \item $(V_{C_i} \otimes V_{C_j})^G = V_{C_i}^G \otimes V_{C_j}^G$ for all  $i,j$.
        \item (and) $(V_{C_1} \otimes V_{C_2} \otimes V_{C_3})^G = (V_{C_i} \otimes V_{C_j})^G \otimes V_{C_k}^G$ for all $i,j,k$.
    \end{enumerate}

    Since, by~\Cref{lem:trivial-irrep-system}, $V_{C_i}^G \neq 0$, for each partition $(C_iC_j)|C_k$, let $\mathcal R^{(ij)|k}$ denote the projector onto $(V_{C_i}\otimes V_{C_j})^G\otimes V_{C_k}^G$. Since $V_{C_i}^G\neq 0$ we have the following subspace inclusions:
    \begin{align}
     \scalebox{0.9}{$  V_{C_1}^G \otimes V_{C_2}^G\otimes V_{C_3}^G \subseteq  (V_{C_i} \otimes V_{C_j})^G \otimes V_{C_k}^G \subseteq  (V_{C_1} \otimes V_{C_2} \otimes V_{C_3})^G$}
    \end{align}
    for all $i,j,k$. In terms of projectors, these inclusions are $\Im\mathcal R^{Prod}\subseteq\Im\mathcal R^{(ij)|k}\subseteq\Im\mathcal R^{Cor}$, as can also be seen directly from the absorption relations $\mathcal R^{Cor}\mathcal R^{(ij)|k}=\mathcal R^{(ij)|k}=\mathcal R^{(ij)|k}\mathcal R^{Cor}$ and $\mathcal R^{(ij)|k}\mathcal R^{Prod}=\mathcal R^{Prod}=\mathcal R^{Prod}\mathcal R^{(ij)|k}$.

    Clearly $(V_{C_1} \otimes V_{C_2} \otimes V_{C_3})^G = V_{C_1}^G \otimes V_{C_2}^G\otimes V_{C_3}^G$ implies that the equalities in 1. and 2. hold, since the subspace inclusions collapse to equalities.

    We now show that 1. and 2. imply $(V_{C_1} \otimes V_{C_2} \otimes V_{C_3})^G = V_{C_1}^G \otimes V_{C_2}^G\otimes V_{C_3}^G$.

    Taking 1. and substituting into 2. gives $(V_{C_1} \otimes V_{C_2} \otimes V_{C_3})^G = V_{C_1}^G \otimes V_{C_2}^G\otimes V_{C_3}^G$.

    We have established that $LT \Leftrightarrow c_1 \land ...\land c_n$ where $c_i$ are the conditions in 1. and 2.

    This is equivalent to $\neg LT \Leftrightarrow \neg c_1 \lor ... \lor \neg c_n$.
\end{proof}

This shows that for three wires failure of local tomography can occur due to a pairwise coupling between wires: $V_{C_1}^G \otimes V_{C_2}^G\otimes V_{C_3}^G  \subsetneq (V_{C_i} \otimes V_{C_j})^G \otimes V_{C_k}^G $ but can also occur due to a genuine triple coupling $\sum_{i<j,  \{i,j,k\}}
(V_{C_i}\otimes V_{C_j})^G\otimes V_{C_k}^G
\subsetneq (V_{C_1}\otimes V_{C_2}\otimes V_{C_3})^G = V_0$, namely at least one of the trivial irreps in $V_0$ occurs from the coupling of a non-trivial irrep $\lambda$ in $V_{C_i} \otimes V_{C_j}$ with the non-trivial irrep $\lambda \in V_{C_k}$.

\begin{example}
    For representations of $\SO(3)$, labelled by non-negative integers
$n \in \mathbb{Z}_{\geq 0}$ (with $n = 0$ the trivial representation), the tensor product $V_{n_A} \otimes V_{n_B} \otimes V_{n_C}$ contains
the trivial representation $V_0$ if and only if
\begin{equation}
    |n_A - n_B| \leq n_C \leq n_A + n_B,
\end{equation}
in which case the multiplicity is exactly one. When $n_A, n_B, n_C \geq 1$, and the triangle inequality holds then $V_0 \cong (V_{n_A} \otimes V_{n_B} \otimes V_{n_C})^G$ is genuinely tripartite: it cannot be factored as an invariant on any pair of the three factors tensored with the trivial rep of the third.

When $n_i = n_j \geq 1$ and $n_k = 0$ then the coupling is genuinely bipartite. When $n_A = n_B = n_C = 0$ then there is a trivial irrep but no non-trivial coupling. In all other cases $(V_{n_A} \otimes V_{n_B} \otimes V_{n_C})^G = 0$.
\end{example}

\subsubsection{Two input two output processes}

\begin{proposition}[
restate=twoIntwoOut,
name={}
]\label{prop:twoIntwoOut}

  Given a two-input/two-output process space $P_{A_1A_2}^{B_1B_2}$, the corresponding invariant process space $\mathcal{A}(P_{A_1A_2}^{B_1B_2})$ fails to be locally tomographic if any one of the following conditions holds: 
    \begin{enumerate}
    \item
    \begin{enumerate}
        \item The inputs share a non-trivial irreducible representation, the outputs do not, and no input-output pair shares a non-trivial irreducible representation.

       \item The outputs share a non-trivial irreducible representation, the inputs do not, and no input-output pair shares a non-trivial irreducible representation.
      \item Only the input-output pair $(A_i,B_j)$ shares a non-trivial irreducible representation, while no other pair of wires does.
    \end{enumerate}
    \item
    \begin{enumerate}
        \item $V_A \cong (V_{A_1} \otimes V_{A_2})$ and $V_{B_i}$ share a non-trivial irreducible representation.
        \item $V_B \cong V_{B_1} \otimes V_{B_2}$ and $V_{A_i}$ share a non-trivial irreducible representation.
    \end{enumerate}
    \item
    \begin{enumerate}
        \item $V_{A_i}$ and $V_{A_{j}} \otimes V_{B_1} \otimes V_{B_2}$ share a non-trivial irreducible representation with $i \neq j$.
        \item $V_{B_i}$ and $V_{B_{j}} \otimes V_{A_1} \otimes V_{A_2}$ share a non-trivial irreducible representation with $i \neq j$.
    \end{enumerate}
    \end{enumerate}

Note that the situation where the bipartite input system $A_1A_2$ shares a common irrep with the bipartite system $B_1B_2$, while no pair of individual systems shares a common irrep, is already covered in 3(a) and 3(b), as we mention later in~\cref{rem:fourWireProcess}.

    In each of the cases failure of local tomography can be detected via
    \begin{enumerate}
    \item
    \begin{enumerate}
   \item Entangled invariant input states and separable effects suffice to distinguish every pair of distinct invariant processes, provided that the total input and output representations $V_A \cong V_{A_1} \otimes V_{A_2}$ and $V_B \cong V_{B_1} \otimes V_{B_2}$ share no non-trivial irreducible representation. If $V_A$ and $V_B$ share a non-trivial irreducible representation, then there exist distinct invariant processes whose distinction requires an ancillary system.

\item Separable input states and entangled invariant effects suffice to distinguish every pair of distinct invariant processes, provided that the total input and output representations $V_A \cong V_{A_1} \otimes V_{A_2}$ and $V_B \cong V_{B_1} \otimes V_{B_2}$ share no non-trivial irreducible representation. If $V_A$ and $V_B$ share a non-trivial irreducible representation, then there exist distinct invariant processes whose distinction requires an ancillary system.

        \item There exist two distinct invariant processes $\tilde f_1,\tilde f_2 \in \mathcal{A}(P^{B_1,B_2}_{A_1,A_2})$ which agree on all invariant input states and effects without ancillary systems, but can be distinguished using an ancillary system.
    \end{enumerate}
    \item

  \begin{enumerate}
        \item There exist distinct invariant processes $\tilde f_1,\tilde f_2 \in \mathcal{A}(P^{B_1,B_2}_{A_1,A_2})$ that agree on all entangled invariant input states and separable effects. Moreover, they remain indistinguishable even when entangled invariant effects are also allowed, and therefore require an ancillary system to be distinguished.

        \item There exist distinct invariant processes $\tilde f_1,\tilde f_2 \in \mathcal{A}(P^{B_1,B_2}_{A_1,A_2})$ that agree on all separable input states and entangled invariant effects. Moreover, they remain indistinguishable even when entangled invariant input states are also allowed, and therefore require an ancillary system to be distinguished.
    \end{enumerate}
    \item
    \begin{enumerate}
      \item There exist distinct invariant processes $\tilde f_1,\tilde f_2 \in \mathcal{A}(P^{B_1,B_2}_{A_1,A_2})$ whose distinction requires correlations across the partition $[A_i|A_jB_1B_2]$. In particular, 2(b) is recovered when the relevant non-trivial representation in the second block is supported entirely on $B_1B_2$.
       \item There exist distinct invariant processes $\tilde f_1,\tilde f_2 \in \mathcal{A}(P^{B_1,B_2}_{A_1,A_2})$  whose distinction requires correlations across the partition $[B_i|B_jA_1A_2]$. In particular, 2(a) is recovered when the relevant non-trivial representation in the second block is supported entirely on $A_1A_2$.
    \end{enumerate}

    \end{enumerate}

\end{proposition}
\blk
\proof 
Each of the six conditions implies that, for a suitable partition $[C_1|\cdots|C_K]$,
$\Im\mathcal R^{Prod}\subsetneq\Im\mathcal R^{[C_1|\cdots|C_K]}\subseteq\Im\mathcal R^{Cor}$. Indeed, a shared non-trivial irreducible representation gives an invariant in the corresponding joint block that is not generated by invariants of the separated blocks. By~\Cref{lem:trivial-irrep-system}, tensoring with non-zero invariant vectors of the remaining wires therefore gives $\Im\mathcal R^{Prod}\subsetneq\Im\mathcal R^{Cor}$. Hence, in every case, $\mathcal R^{Cor}\neq\mathcal R^{Prod}$.

For 1(a), if the total input and output representations share no non-trivial irreducible representation, then every $G$-equivariant map between them is supported only on their trivial sectors. Hence, $\Im\mathcal R^{Cor}=\Im\mathcal R^{Semi}$, and therefore $\mathcal R^{Cor}=\mathcal R^{Semi}$. By~\Cref{prop:entagledSuffice}, complete process tomography can therefore be achieved without ancillary systems. Moreover, since $B_1$ and $B_2$ share no non-trivial irreducible representation,~\Cref{lem:G_inv_GtimesG} gives $\mathcal R^{Semi}=\mathcal R^{[A_1A_2|B_1|B_2]}$, so entangled invariant input states together with separable effects suffice to distinguish every pair of distinct invariant processes. Since $A_1$ and $A_2$ share a non-trivial irreducible representation, $\Im\mathcal R^{Prod}\subsetneq\Im\mathcal R^{Semi}$, so separable input states do not suffice in general. Importantly, the condition that no input-output pair shares a non-trivial irreducible representation does not by itself imply that the total input and output representations share no non-trivial irreducible representation, since tensor products can contain irreducible representations that are absent from each individual factor. For example, for $G=\mathrm{SO}(3)$, let $V_{A_1}=V_{A_2}=V_0\oplus V_1$, $V_{B_1}=V_0\oplus V_2$, and $V_{B_2}=V_0$. Then $A_1$ and $A_2$ share the non-trivial irreducible representation $V_1$, while $B_1$ and $B_2$ share no non-trivial irreducible representation and no pair $(A_i,B_j)$ shares a non-trivial irreducible representation. Nevertheless, since $V_1\otimes V_1\cong V_0\oplus V_1\oplus V_2$, the total input representation $V_A\cong V_{A_1}\otimes V_{A_2}$ contains $V_2$, while the total output representation $V_B\cong V_{B_1}\otimes V_{B_2}$ also contains $V_2$. Thus, the original pairwise conditions in 1(a) alone do not imply $\mathcal R^{Cor}=\mathcal R^{Semi}$. If instead the total input and output representations share a non-trivial irreducible representation, there exist distinct invariant processes whose distinction requires an ancillary system. The same reasoning applies to 1(b), with inputs and outputs interchanged.

For 1(c), on $\mathcal{A}(P_{A_i}^{B_j})$ one has $\mathcal R^{Semi}=\mathcal R^{Prod}$. Thus, $\mathcal R^{Cor}\neq\mathcal R^{Prod}$ implies $\mathcal R^{Cor}\neq\mathcal R^{Semi}$, and by~\Cref{prop:ancillaNeeded} there exist distinct invariant processes that cannot be distinguished without an ancillary system but can be distinguished using one. 

For 2(a), since $V_A\cong V_{A_1}\otimes V_{A_2}$ and $V_{B_i}$ share a non-trivial irreducible representation, and by~\Cref{lem:trivial-irrep-system} $V_{B_j}^G\neq 0$, $V_A$ and $V_B\cong V_{B_1}\otimes V_{B_2}$ also share a non-trivial irreducible representation. Hence, $\mathcal R^{Cor}\neq\mathcal R^{Semi}$. By~\Cref{lem:InvariantSubprojection}, there therefore exist distinct invariant processes $\tilde f_1,\tilde f_2$ such that $\mathcal R^{Semi}(\tilde f_1)=\mathcal R^{Semi}(\tilde f_2)$. Since $\mathcal R^{[A_1A_2|B_1|B_2]}$ is a subprojection of $\mathcal R^{Semi}$, they also have the same image under $\mathcal R^{[A_1A_2|B_1|B_2]}$, and hence agree on entangled invariant input states and separable effects. Moreover, by~\Cref{prop:pairwiseProcessesAncilla}, they remain indistinguishable even when entangled invariant effects are allowed, and therefore require an ancillary system to be distinguished. The same reasoning applies to 2(b).

For 3(a), the shared non-trivial irreducible representation implies $\mathcal R^{[A_i|A_jB_1B_2]}\neq\mathcal R^{Cor}$. Hence, by~\Cref{lem:InvariantSubprojection}, there exist distinct invariant processes with the same image under $\mathcal R^{[A_i|A_jB_1B_2]}$, and thus their distinction requires correlations across the partition $[A_i|A_jB_1B_2]$. In particular, 2(b) is recovered as the special case in which the relevant correlations lie entirely within $B_1B_2$. The same reasoning applies to 3(b), with the partition $[B_i|B_jA_1A_2]$, where 2(a) is recovered as a special case.
\endproof

\subsubsection{General processes}
 We first establish general representation-theoretic conditions for the failure of local tomography in general multipartite processes, using partition projectors for this purpose. We then relate the status of process tomography in non-locally tomographic process spaces to these representation-theoretic conditions.
\begin{lemma}
Let $\mathcal{A}(P_A^B)$ be an invariant multipartite process space of a locally tomographic GPT, with $A=A_1,\ldots,A_n$ and $B=B_1,\ldots,B_m$. If a pair of wires $C_i$ and $C_j$ are such that $V_{C_i}$ and $V_{C_j}$ share a non-trivial irreducible representation, then $\mathcal{A}(P_A^B)$ fails to be locally tomographic.
\end{lemma}
\begin{proof}
   If $V_{C_i}$ and $V_{C_j}$ share a non-trivial irreducible representation, then $(V_{C_i} \otimes V_{C_j})^G \neq V_{C_i}^G \otimes V_{C_j}^G$. By~\Cref{lem:trivial-irrep-system}, $V_{C_k}^G\neq 0$ for every $k$. This implies $(V_{C_i} \otimes V_{C_j})^G \otimes \bigotimes_{k \neq i,j} V_{C_k}^G \neq \bigotimes_k V_{C_k}^G$. Therefore, $\bigotimes_k V_{C_k}^G \subsetneq (V_{C_i} \otimes V_{C_j})^G \otimes \bigotimes_{k \neq i,j} V_{C_k}^G \subseteq (\bigotimes_k V_{C_k})^G$, which implies that $\mathcal{A}(P_A^B)$ fails to be locally tomographic. Equivalently, defining $\mathcal R^{[C_iC_j|C_{k_1}|\cdots|C_{k_r}]}$ as the projector associated with the partition where $C_i$ and $C_j$ are grouped together and all other wires are separate, the above inclusions imply $\Im\mathcal R^{Prod}\subsetneq \Im\mathcal R^{[C_iC_j|C_{k_1}|\cdots|C_{k_r}]}\subseteq\Im\mathcal R^{Cor}$. Hence, $\mathcal R^{Prod}\neq\mathcal R^{Cor}$, and therefore local tomography fails.

\end{proof}
The above lemma gives a simple sufficient condition for the failure of local tomography, but it does not capture all multipartite possibilities. More general failures can arise from correlations involving larger collections of wires, which motivates introducing projectors associated with arbitrary partitions of the wires.

Given a multipartite process space $P_A^B$, with $A=A_1,\ldots,A_n$ and $B=B_1,\ldots,B_m$, we first collect all input and output wires into a single list by setting $C_i=A_i$ for $i=1,\ldots,n$ and $C_{n+j}=B_j$ for $j=1,\ldots,m$. Thus $C=\{C_1,\ldots,C_{n+m}\}$ denotes the full set of wires of the process.

We now group these wires into $K$ disjoint blocks, denoted by $[\cc_1|\cdots|\cc_K]$, where $\cc_i\subseteq C$, $\cc_i\cap\cc_j=\emptyset$ for $i\neq j$, and $\bigcup_i\cc_i=C$. Each block $\cc_i$ is acted on collectively by a single group element, while different blocks are acted on independently. This defines an action of $G^{\times K}$ associated with the partition.
\begin{align}
    &\scalebox{0.9}{$\Gamma_{[\cc_1|\cdots|\cc_K]}(g_1,\ldots,g_K)$}
    \nonumber\\&\vspace{0.3em}\nonumber\\
    :=&
    \scalebox{0.80}{$\bigotimes_{C_{i_1}\in\cc_1}\Gamma_{C_{i_1}}(g_1)
    \otimes
    \bigotimes_{C_{i_2}\in\cc_2}\Gamma_{C_{i_2}}(g_2)
    \otimes\cdots\otimes
    \bigotimes_{C_{i_K}\in\cc_K}\Gamma_{C_{i_K}}(g_K)$}\nonumber
    \\&\vspace{0.3em}\nonumber\\
    =&
    \scalebox{0.9}{$\bigotimes_{k=1}^K
    \bigotimes_{C_{i_k}\in\cc_k}
    \Gamma_{C_{i_k}}(g_k)$}.
\end{align}
For example, if $C_1,C_3\in\cc_1$ and $C_2\in\cc_2$, then $C_1$ and $C_3$ are acted on by the same group element $g_1$, while $C_2$ is acted on independently by $g_2$.

For each $K$-partition $[\cc_1|\cdots|\cc_K]$, we define the corresponding partition projector by averaging the associated $G^{\times K}$ action independently over each copy of the group. Writing $\mathbf g=(g_1,\ldots,g_K)$, this gives
\begin{equation}
   \mathcal R^{[\cc_1|\cdots|\cc_K]}(\cdot)
    :=
    \int_{G^{\times K}}
    \Gamma_{[\cc_1|\cdots|\cc_K]}(\mathbf g)(\cdot)\,
    \mathrm d\mu(g_1)\cdots\mathrm d\mu(g_K).
\end{equation}
As for the projectors introduced above, it is straightforward to verify that $\mathcal R^{[\cc_1|\cdots|\cc_K]}$ is idempotent and projects onto $(\bigotimes_{C_{i_1}\in\cc_1}V_{C_{i_1}})^G\otimes\cdots\otimes(\bigotimes_{C_{i_K}\in\cc_K}V_{C_{i_K}})^G$.
\begin{theorem}[
restate=LTfailureIrreps,
name={}
]\label{thm:LTfailureIrreps}
Let $\mathcal A(P_{A_1,\ldots,A_n}^{B_1,\ldots,B_m})$ be an invariant multipartite process space of a locally tomographic theory. Let $C=\{C_1,\ldots,C_{n+m}\}$ denote the set of all input and output wires, with $C_i=A_i$ for $i=1,\ldots,n$ and $C_{n+j}=B_j$ for $j=1,\ldots,m$, and for any $\cc\subseteq C$ define $V_{\cc}:=\bigotimes_{C_i\in\cc}V_{C_i}$. Then local tomography fails if and only if there exists a bipartition $[\cc_1|\cc_2]$ of $C$, with $\cc_1,\cc_2\neq\emptyset$, such that $V_{\cc_1}$ and $V_{\cc_2}$ share at least one non-trivial irreducible representation of $G$. Equivalently, local tomography fails exactly when the input and output wires, taken together, can be split into two non-empty groups whose joint representations share a non-trivial irreducible representation of $G$.
\end{theorem}
\proof
See~\cref{proof:LTfailureIrreps}.
\endproof

This result gives a practical way to certify the failure of local tomography in a given invariant process space. It suffices to find a bipartition of the wires such that the representations carried jointly by the two parts share a non-trivial irreducible representation. The simplest case occurs when two individual wires already share a non-trivial irreducible representation. Then any bipartition that places these two wires in different parts satisfies the condition above, since the remaining wires contain trivial representations. More generally, even when no pair of individual wires shares a non-trivial irreducible representation, tensor products of several wire representations may share one with another wire, or with another collection of wires. The theorem also covers these genuinely multipartite cases, including those encountered above for three-wire processes and for processes with two inputs and two outputs.

Having characterised the failure of local tomography in representation-theoretic terms, we now turn to its operational consequences. We begin by asking whether, given the representations carried by the systems, ancillary systems are required for process tomography. We have the following:

\begin{lemma}[
restate=WhenAncillaNeededIrreps,
name={}
]\label{prop:WhenAncillaNeededIrreps}

Let $\mathcal{A}(P_{A_1,\ldots,A_n}^{B_1,\ldots,B_m})$ be an invariant multipartite process space of a locally tomographic theory, with $V_A:=\bigotimes_{i=1}^n V_{A_i}$ and $V_B:=\bigotimes_{j=1}^m V_{B_j}$. Ancillary systems are required for complete process tomography if and only if $V_A$ and $V_B$ share at least one non-trivial irreducible representation of $G$. Otherwise, complete process tomography can be achieved without ancillary systems.
\end{lemma}
\proof
See~\cref{proof:WhenAncillaNeededIrreps}.
\endproof
\begin{remark}\label{rem:fourWireProcess}
At first sight, one might think that Lemma~\ref{prop:WhenAncillaNeededIrreps} misses shared-irrep requirements involving combinations of input and output systems. For example, suppose that in $\mathcal{A}(P^{B_1,B_2}_{A_1,A_2})$, $A_1A_2B_1$ and $B_2$ are genuine carriers (see~\cref{def:GeniuneCarrier}) of the same non-trivial irreducible representation $\lambda$. One might then expect ancillary systems to be required even if $A_1A_2$ and $B_1B_2$ appear to share no non-trivial irreducible representation.

There is, however, no such additional case. Since $A_1A_2B_1$ genuinely carries $\lambda$, there exist non-trivial irreducible representations $\mu$ carried by $V_{A_1A_2}$ and $\nu$ carried by $V_{B_1}$ such that $\lambda$ occurs in\footnote{Or, it could be the case that $U^{B_1}_{A_i}\cong V_{A_i}\otimes V_{B_1}$ contains $\mu$ and $V_{A_j}$ ($i,j = 1,2)$ contains $\nu$ such that $\lambda\in \mu\otimes \nu$.} $\mu\otimes\nu$,
hence
$\operatorname{Hom}_G(\lambda,\mu\otimes\nu)\neq 0$. By the canonical $G$-equivariant identification
$\operatorname{Hom}_G(\lambda,\mu\otimes\nu)\cong\operatorname{Hom}_G(\lambda\otimes\nu^*,\mu)$,
it follows that
$\operatorname{Hom}_G(\lambda\otimes\nu^*,\mu)\neq 0$. Since $\mu$ is irreducible and representations of compact groups are completely reducible, this implies that $\mu$ occurs in $\lambda\otimes\nu^*$. Moreover, for the real representations considered here, $\nu\cong\nu^*$, and hence $\mu$ occurs in $\lambda\otimes\nu$.

Since $B_1$ carries $\nu$ and $B_2$ carries $\lambda$, it follows that $V_{B_1B_2}$ carries $\mu$. But $\mu$ is also carried by $V_{A_1A_2}$. Thus $V_{A_1A_2}$ and $V_{B_1B_2}$ necessarily share the non-trivial irreducible representation $\mu$. Therefore, mixed shared-irrep requirements of this kind are already captured by the criterion of Lemma~\ref{prop:WhenAncillaNeededIrreps}, and there is no additional case requiring a refinement of that criterion.
\end{remark}

Equivalently, consider the bipartition that groups all input wires together and all output wires together. Ancillary systems are required exactly when the joint input and output representations share a non-trivial irreducible representation. This may arise either from a single input-output pair or only at the level of tensor products of several input and output representations.

We now consider the role of entangled invariant states and effects across partitions of the input and output systems, respectively, whether or not ancillary systems are required:
\begin{proposition}[
restate=DisjointIrrep,
name={}
]\label{prop:DisjointIrrep}
Let $\mathcal{A}(P_{A_1,\ldots,A_n}^{B_1,\ldots,B_m})$ be an invariant multipartite process space. Consider a partition of the input wires into blocks, and let $\cc_1$ and $\cc_2$ be two disjoint non-empty unions of these blocks. Then invariant input states that remain separable across $[\cc_1|\cc_2]$ are insufficient for complete process tomography if and only if $V_{\cc_1}$ and $V_{\cc_2}$ share a non-trivial irreducible representation of $G$. In this case, entangled invariant input states across $[\cc_1|\cc_2]$ are required. The analogous statement holds for output wires and invariant effects.
\end{proposition}
\proof
See~\cref{proof:DisjointIrrep}.
\endproof

When ancillary systems are not required, the entanglement requirements identified above can be combined to obtain a sufficient scheme for complete process tomography, as we make precise below. When ancillary systems are required, these entanglement requirements remain necessary whenever the corresponding shared-irrep conditions hold, but are no longer sufficient by themselves, since ancillary systems carrying the relevant shared irreducible representations are additionally required to probe the non-trivial irreducible sectors shared by the input and output representations.

When a non-trivial irreducible representation $\lambda$ appears in the joint representation of several systems, it need not arise from those systems collectively. It may already be present in a smaller collection, with the remaining systems contributing only invariant sectors. We therefore distinguish the $\lambda$-content that is genuinely associated with the full collection from that inherited from proper subcollections.

\begin{definition}[Genuine carrier of an irreducible representation]\label{def:GeniuneCarrier}
Let $\cc$ be a non-empty collection of systems and let $\lambda$ be a non-trivial irreducible representation of $G$. Denote by $V_{\cc}^{(\lambda)}$ the $\lambda$-isotypic component of $V_{\cc}$, and define
\begin{equation}\label{eq:GeniuneCarrier}
    I_{\cc}^{(\lambda)}
    :=
    \sum_{\emptyset\neq\cc'\subsetneq\cc}
    V_{\cc'}^{(\lambda)}
    \otimes
    V_{\cc\setminus\cc'}^G.
\end{equation}
We call $\cc$ a \emph{genuine carrier} of $\lambda$ if $ V_{\cc}^{(\lambda)}
    \neq    I_{\cc}^{(\lambda)}$.
\end{definition}

\begin{example}
Suppose that $A_1$ carries $\lambda$. Then $V_{A_1}\otimes V_{A_2}$ also contains copies of $\lambda$ obtained by tensoring the $\lambda$-sector of $A_1$ with $V_{A_2}^G$. These copies are inherited from $A_1$. If every occurrence of $\lambda$ in $V_{A_1}\otimes V_{A_2}$ is of this form, then $A_1A_2$ is not a genuine carrier of $\lambda$. If, however, $V_{A_1}\otimes V_{A_2}$ contains an additional $\lambda$-sector not contained in
$V_{A_1}^{(\lambda)}\otimes V_{A_2}^G
+
V_{A_1}^G\otimes V_{A_2}^{(\lambda)}$,
then $A_1A_2$ is a genuine carrier of $\lambda$.
\end{example}

\begin{example}
Suppose that $A_1A_2$ carries $\lambda$. Then $A_1A_2A_3A_4$ contains inherited copies of $\lambda$ whenever $A_3A_4$ contributes an invariant sector. Such copies do not make $A_1A_2A_3A_4$ a genuine carrier. If, on the other hand, $V_{A_1}\otimes V_{A_2}\otimes V_{A_3}\otimes V_{A_4}$ contains a $\lambda$-sector that is not inherited from any proper subcollection, then $A_1A_2A_3A_4$ is a genuine carrier of $\lambda$.
\end{example}

\begin{definition}[Shared-irrep tomography blocks for inputs and outputs]
Let $\mathcal{A}=\{A_1,\ldots,A_n\}$ be the input wires and $\mathcal{B}=\{B_1,\ldots,B_m\}$ the output wires.

We construct the input blocks using only shared-irrep requirements among input wires. For each such requirement $[\mathcal{C}_1|\mathcal{C}_2]$, group together all wires in $\mathcal{C}_1\cup\mathcal{C}_2$. Whenever two groups have an input wire in common, merge them. Continue until no two groups overlap. The resulting groups are the input shared-irrep tomography blocks; any input wire that appears in no requirement forms a singleton block.

Equivalently, two input wires belong to the same block if and only if they can be joined by a finite chain of shared-irrep requirements, with each consecutive pair of requirements sharing at least one input wire. The
resulting partition is denoted by
\begin{equation}
\Pi_{\mathrm{in}}^{\mathrm{SI}}
:=
[\mathcal{A}_1|\cdots|\mathcal{A}_r].
\end{equation}

We define the output blocks in the same way, using only shared-irrep requirements among output wires. Any output wire that appears in no requirement forms a singleton block. We denote the resulting partition by
\begin{equation}
\Pi_{\mathrm{out}}^{\mathrm{SI}}
:=
[\mathcal{B}_1|\cdots|\mathcal{B}_s].
\end{equation}

For each input block $\mathcal{A}_i$, the representation $V_{\mathcal{A}_i}$ shares no non-trivial irreducible representation with the joint representation of the remaining input wires. The analogous property holds for each output block.
\end{definition}

\begin{example}
Suppose instead that $A_1A_2$ and $A_3A_4$ are disjoint genuine carriers of the same non-trivial irreducible representation $\lambda$. They then define the shared-irrep requirement $[A_1A_2|A_3A_4]$.

\end{example}

\begin{definition}[Shared-irrep tomography blocks for inputs and outputs]
We construct these blocks separately for the input and output systems. For the input systems, two systems are connected if they occur in the same shared-irrep requirement $[\cc_1|\cc_2]$, or are linked by a chain of shared-irrep requirements such that every two consecutive requirements overlap in at least one input system. A shared-irrep tomography block is a maximal connected collection of input systems; input systems appearing in no requirement form singleton blocks.

Equivalently, for each block $\mathcal{A}_i$, the representation $V_{\mathcal{A}_i}$ shares no non-trivial irreducible representation with the joint representation of the remaining input systems.

This gives the partition
\begin{equation}
\Pi_{\mathrm{in}}^{\mathrm{SI}} := [\mathcal{A}_1|\cdots|\mathcal{A}_r].
\end{equation}
Applying the same construction to the output systems gives
\begin{equation}
\Pi_{\mathrm{out}}^{\mathrm{SI}} := [\mathcal{B}_1|\cdots|\mathcal{B}_s].
\end{equation}
\end{definition}

\begin{example}
Suppose that $A_1$ and $A_2$ define a shared-irrep requirement through $\lambda_1$, while $A_2$ and $A_3$ define another shared-irrep requirement through $\lambda_2$. The corresponding cuts are $  [A_1|A_2]$ and $    [A_2|A_3]$.

Since the two requirements overlap at $A_2$, the three systems form the single shared-irrep tomography block $A_1A_2A_3$.
\end{example}

\begin{example}
Suppose that $A_1A_2$ and $A_3A_4$ define the shared-irrep requirement $[A_1A_2|A_3A_4]$. Then $A_1,A_2,A_3,A_4$ form a single shared-irrep tomography block.
\end{example}
\begin{proposition}[
restate=SharedIrrepTomography,
name={Shared-irrep tomography blocks and process tomography}
]\label{prop:SharedIrrepTomography}
Suppose that ancillary systems are not required for complete process tomography. Then complete process tomography can be performed using invariant input states satisfying the following:
\begin{enumerate}
    \item the input states are separable between distinct blocks of the shared-irrep tomography partition $\Pi_{\mathrm{in}}^{\mathrm{SI}}$.
\vspace{0.4em}
    \item within each block, for every shared-irrep requirement $[\cc_1|\cc_2]$ defining that block, the tomography scheme includes entangled invariant input states across $[\cc_1|\cc_2]$. The states associated with different shared-irrep requirements need not be the same: each requirement may be satisfied by a different set of input states.
\end{enumerate}
The analogous statement holds for invariant effects with respect to the output partition $\Pi_{\mathrm{out}}^{\mathrm{SI}}$.
\end{proposition}
\proof
See~\cref{proof:SharedIrrepTomography}.
\endproof

\begin{example}
Consider three input systems $A_1,A_2,A_3$. Suppose that $A_1$ and $A_2$ define a shared-irrep requirement through a non-trivial irreducible representation $\lambda_1$, while $A_2$ and $A_3$ define another through a non-trivial irreducible representation $\lambda_2$, and suppose that there are no further shared-irrep requirements. The two requirements are therefore $[A_1|A_2]$ and $[A_2|A_3]$.

Since the two requirements overlap at $A_2$, the corresponding shared-irrep tomography block is $A_1A_2A_3$. Moreover, the absence of any further shared-irrep requirements implies that every invariant sector is generated by one of these two couplings, so that
\begin{equation}
\scalebox{0.85}{$
(V_{A_1}\otimes V_{A_2}\otimes V_{A_3})^G
=
(V_{A_1}\otimes V_{A_2})^G\otimes V_{A_3}^G
+
V_{A_1}^G\otimes (V_{A_2}\otimes V_{A_3})^G
$}.
\end{equation}
Consequently, a tomographically complete family of invariant input states can be chosen from states entangled on $A_1A_2$ and separable from $A_3$, together with states entangled on $A_2A_3$ and separable from $A_1$. Hence, genuinely tripartite-entangled input states are not required for complete process tomography.
\end{example}
\begin{example}
Consider four input systems $A_1,A_2,A_3,A_4$. Suppose that $A_1A_2$ and $A_3A_4$ are disjoint genuine carriers of the same non-trivial irreducible representation $\lambda$, giving the shared-irrep requirement $ [A_1A_2|A_3A_4]$.

Then $A_1,A_2,A_3,A_4$ belong to the same shared-irrep tomography block. Complete process tomography requires entangled invariant input states across $[A_1A_2|A_3A_4]$. This does not imply that the states must be genuinely four-partite entangled across every bipartition of $A_1A_2A_3A_4$; the required entanglement is specifically that associated with the shared-irrep cut $[A_1A_2|A_3A_4]$.
\end{example}

\begin{example}
As a contrasting four-partite case, suppose that $A_1A_2$ is a genuine carrier of a non-trivial irreducible representation $\lambda$, but there is no disjoint genuine carrier of $\lambda$, either an individual system or a multipartite collection. Then $\lambda$ defines no shared-irrep requirement. Consequently, it imposes no entanglement requirement and does not, by itself, place $A_1$ and $A_2$ in the same shared-irrep tomography block.
\end{example}
The construction above gives a direct prescription for determining the required entanglement structure when ancillary systems are not needed. For the input systems, identify all non-trivial irreducible representations and their genuine carriers. Whenever two disjoint genuine carriers carry the same non-trivial irreducible representation, record the corresponding shared-irrep requirement, and merge overlapping requirements to obtain the blocks of $\Pi_{\mathrm{in}}^{\mathrm{SI}}$. Complete process tomography can then be performed using invariant input states that are separable between distinct blocks, while including states entangled across each shared-irrep requirement. A genuine carrier with no disjoint genuine carrier of the same irreducible representation imposes no such requirement. Applying the same procedure to the output systems yields $\Pi_{\mathrm{out}}^{\mathrm{SI}}$ and the corresponding requirements on invariant effects.

Finally, we turn our attention to the case when ancillary systems are needed. For this purpose, we first define:
\begin{definition}[Shared-irrep tomography blocks]
Let $\mathcal{C}=\{C_1,\ldots,C_{n+m}\}$ denote the collection of all input and output wires, with $C_i=A_i$ for $i=1,\ldots,n$ and $C_{n+j}=B_j$ for $j=1,\ldots,m$. Two systems are said to be connected if they occur in the same shared-irrep requirement $[C_1|C_2]$, where $C_1,C_2\subseteq\mathcal{C}$ are disjoint genuine carriers of the same non-trivial irreducible representation, or if they can be joined by a chain of overlapping shared-irrep requirements. A shared-irrep tomography block is a maximal collection of mutually connected systems. A system that occurs in no shared-irrep requirement forms a singleton block. We denote the resulting partition of $\mathcal{C}$ by
\begin{equation}
\Pi_{\mathcal{C}}^{\mathrm{SI}}:=[\mathcal{C}_1|\cdots|\mathcal{C}_r].
\end{equation}
\end{definition}
We expect that, whenever ancillary systems are required, their role is to connect non-trivial irreducible sectors shared between the input and output parts of the mixed blocks of $\Pi_{\mathcal C}^{\mathrm{SI}}$. Thus, each such shared irrep $\lambda$ should be probed by invariant states and effects coupled through an ancillary system carrying $\lambda$. We now make this statement precise.

The necessity of ancillary systems has previously appeared in the special case of fermionic process tomography under parity superselection~\cite{zhang2021fermionic}. The representation-theoretic results here characterise the corresponding ancillary requirements for general invariant subtheories obtained from locally tomographic GPTs.

\begin{proposition}[
restate=AncillaAssisitedTesters,
name={Ancilla-assisted testers from $\Pi_{\mathcal C}^{\mathrm{SI}}$}
]\label{prop:AncillaAssisitedTesters}
Let
\begin{equation}
\Pi_{\mathcal C}^{\mathrm{SI}}
=
[\mathcal C_1|\cdots|\mathcal C_r].
\end{equation}
For each block $\mathcal C_k$, let $\mathcal A_k$ and $\mathcal B_k$ denote,
respectively, the input and output systems contained in $\mathcal C_k$, and define
\begin{equation}
\Lambda_k
:=
\left\{
\lambda:
\lambda\text{ occurs in both }
V_{\mathcal A_k}
\text{ and }
V_{\mathcal B_k},
\quad \lambda\neq\mathbf 1
\right\}.
\end{equation}
If $\mathcal A_k=\varnothing$ or $\mathcal B_k=\varnothing$, set
$\Lambda_k=\varnothing$.

The block $\mathcal C_k$ requires no ancillary system if
$\Lambda_k=\varnothing$. If $\lambda\in\Lambda_k$, the corresponding
$\lambda$-sector requires invariant states on $\mathcal A_kE_{k,\lambda}$
and invariant effects on $\mathcal B_kE_{k,\lambda}$, where
$E_{k,\lambda}$ carries $\lambda$.

Let
\begin{equation}
K_+:=\{k\mid\Lambda_k\neq\varnothing\},
\qquad
K_0:=\{k\mid\Lambda_k=\varnothing\}.
\end{equation}
For every $k\in K_+$ and every $\lambda_k\in\Lambda_k$, choose an ancillary
system $E_{k,\lambda_k}$ carrying $\lambda_k$. Such a choice is always possible. Since $\lambda_k\in\Lambda_k$, the irrep $\lambda_k$ occurs in both
$V_{\mathcal A_k}$ and $V_{\mathcal B_k}$, so one may, for example, take
$E_{k,\lambda_k}$ to be a copy of the composite system $\mathcal A_k$. For every choice
$(\lambda_k)_{k\in K_+}$ with $\lambda_k\in\Lambda_k$, process tomography
of an invariant process $\widetilde f$ can then be implemented using testers
of the form
\begin{equation}
\begin{aligned}
&\left(
\bigotimes_{k\in K_+}\widetilde e_{k,\lambda_k}
\otimes
\bigotimes_{k\in K_0}\widetilde e_k
\right)
\circ
\left(
\widetilde f
\otimes
\bigotimes_{k\in K_+}\mathcal I_{E_{k,\lambda_k}}
\right)
\\
&\qquad\circ
\left(
\bigotimes_{k\in K_+}\widetilde s_{k,\lambda_k}
\otimes
\bigotimes_{k\in K_0}\widetilde s_k
\right).
\end{aligned}
\end{equation}
where, for every $k\in K_+$,
$\widetilde s_{k,\lambda_k}$ and $\widetilde e_{k,\lambda_k}$ are invariant
states on $\mathcal A_kE_{k,\lambda_k}$ and invariant effects on
$\mathcal B_kE_{k,\lambda_k}$, respectively, while for every $k\in K_0$,
$\widetilde s_k$ and $\widetilde e_k$ are invariant states on
$\mathcal A_k$ and invariant effects on $\mathcal B_k$, respectively.
A tomographically complete family is obtained by taking the union of these
tester families over every choice $(\lambda_k)_{k\in K_+}$ with
$\lambda_k\in\Lambda_k$, that is, by including all testers obtained from
every such choice of shared irreducible representations.
\end{proposition} 
\proof
See~\cref{proof:AncillaAssisitedTesters}.
\endproof

\begin{example}
Consider $\mathcal{A}(P^{B_1,B_2}_{A_1,A_2})$ with
$\Pi_{\mathcal C}^{\mathrm{SI}}=[A_1|B_1|A_2B_2]$, and suppose that $\lambda$ is the only non-trivial irrep shared by $V_{A_2}$ and $V_{B_2}$. Then, for any two distinct invariant processes $\tilde f_1\neq\tilde f_2$, there exist invariant states $\tilde s,\tilde s_\lambda$ and invariant effects $\tilde e,\tilde e_\lambda$ such that
\begin{equation}
\input{Diagrams/7_LT142.tikz}\neq\input{Diagrams/7_LT143.tikz}\;,
\end{equation}
where $E_\lambda$ carries the irrep $\lambda$.

If, in addition, $A_1$ and $B_1$ share a non-trivial irrep $\mu$, then the block decomposition is no longer $[A_1|B_1|A_2B_2]$. In that case, the tester must also resolve the $\mu$-sector. Thus, instead of the ancilla-free state $\tilde s$ on $A_1$ and effect $\tilde e$ on $B_1$, one requires an invariant state on $A_1E_\mu$ and an invariant effect on $B_1E_\mu$, where $E_\mu$ carries $\mu$. Equivalently, the tomography scheme must contain testers that couple both shared sectors, namely the $\lambda$-sector between $A_2$ and $B_2$ and the $\mu$-sector between $A_1$ and $B_1$.
\end{example}
\begin{remark}
By~\Cref{prop:process-tomography-from-local-tomography}, taking an ancillary system $E\cong A_1\cdots A_n$ always suffices for process tomography. This choice automatically contains every non-trivial irreducible representation shared by the total input and output representations, and hence satisfies the conditions of the preceding proposition. The proposition shows, however, that depending on which irreducible representations are shared by the total input and output representations, a smaller ancillary system may suffice.
\end{remark}

\addtocontents{toc}{\protect\RestoreSubsubsectionsToTOC}
\subsection{Quantifying the failure of local tomography}
\label{subsec:QuantifyingLTFailure}
Here we aim to quantify the failure of local tomography by measuring the additional invariant degrees of freedom that are inaccessible to local tomography. Such a quantitative measure provides a natural starting point for a resource-theoretic study of tomographic nonlocality, allowing different invariant theories to be compared according to the extent of their departure from local tomography.

\begin{definition}[Tomographic excess]
For an invariant multipartite process space, we define its \emph{tomographic excess} as
\begin{equation}
\Delta_{\mathrm{LT}}
:=
\dim\operatorname{Im}\mathcal R^{\mathrm{Cor}}
-
\dim\operatorname{Im}\mathcal R^{\mathrm{Prod}}.
\end{equation}
\end{definition}

Here, $\dim\operatorname{Im}\mathcal R$ denotes the dimension of the subspace onto which $\mathcal R$ projects. Since
$\operatorname{Im}\mathcal R^{\mathrm{Prod}}
\subseteq
\operatorname{Im}\mathcal R^{\mathrm{Cor}}$,
one has $\Delta_{\mathrm{LT}}\geq 0$, with
$\Delta_{\mathrm{LT}}=0$ if and only if local tomography holds. Thus,
$\Delta_{\mathrm{LT}}$ quantifies the number of invariant degrees of freedom that are inaccessible to local tomography.

Let
$V_A:=\bigotimes_{i=1}^n V_{A_i}$ and
$V_B:=\bigotimes_{j=1}^m V_{B_j}$. Since
$\operatorname{Im}\mathcal R^{\mathrm{Cor}}=\Hom_G(V_A,V_B)$, by the standard character formula for the dimension of an intertwiner space~\cite[Secs.~II.4 and II.6]{brocker_representations_1995}, we have
\begin{equation}
\dim\operatorname{Im}\mathcal R^{\mathrm{Cor}}
=
\int_G
\chi_{V_B}(g)\chi_{V_A}(g^{-1})\,\mathrm d\mu(g),
\end{equation}
where $\chi_V(g):=\operatorname{Tr}[\Gamma_V(g)]$ denotes the character of the representation on $V$, while
\begin{equation}
\dim\operatorname{Im}\mathcal R^{\mathrm{Prod}}
=
\prod_{i=1}^n\dim V_{A_i}^G
\prod_{j=1}^m\dim V_{B_j}^G.
\end{equation}
Hence,
\begin{equation}
\Delta_{\mathrm{LT}}
=
\int_G
\chi_{V_B}(g)\chi_{V_A}(g^{-1})\,\mathrm d\mu(g)
-
\prod_{i=1}^n\dim V_{A_i}^G
\prod_{j=1}^m\dim V_{B_j}^G,
\end{equation}
where $\dim V^G$ denotes the dimension of the $G$-invariant subspace of $V$.

\section{Examples of twirled and swirled theories}

In this section we show two key examples. We show how bilocal classical theory (BCT) arises from twirling standard classical theory, and how real quantum theory (RQT) arises from swirling standard quantum theory. These connections were previously discussed in~\cite{centeno2024twirled} and~\cite{ying2025whether}, respectively. Here, we study them within our fully compositional framework.

\subsection{Classical Theory}\label{sec:CT}

In this section we begin the exploration of the scope of possibilities for invariant classical worlds. To begin, we argue that every $T$-action on classical theory is physical, and hence that there are no genuinely swirled classical worlds. We then consider the special case of twirling with respect to the simplest non-trivial group, $\mathbb Z_2$, and see that this results in a theory known in the literature as bilocal classical theory (BCT) \cite{d2020classicality} -- a theory which is locally classical but which has a tomographically nonlocal composition rule. We then see how our $\mathsf{Twirl}$ construction allows us to unify BCT with standard classical theory, thereby equipping BCT with standard classical control systems which allow us to choose which BCT transformation to apply, and which encode the outcomes of measurements on BCT systems. 

There remain many interesting questions to explore in the context of twirled classical worlds, however, for example, working towards a classification and understanding the structure of arbitrary $G$-twirled classical worlds, and exploring twirled worlds arising from infinite dimensional classical theory.

For the purposes of this section, we give a succinct definition of classical theory:
\begin{definition}
We define the GPT, $\mathbf{CT}$, to be the symmetric monoidal category $\mathbf{FinSubStoch}$ in which the objects are natural numbers composed with multiplication and morphisms are $n\times m$ substochastic matrices.
\end{definition}

\subsubsection{All classical theory actions are physical}

For classical theory, we find that every $T$-action is physical, so every invariant subtheory obtained by averaging a $T$-action is twirled rather than swirled. This is perhaps surprising, since general $T$-actions need not be physical, yet classical theory rules out all such nonphysical possibilities. However, there are two key features of classical theory which lead to this result. 

Firstly, there is the well-known fact that classical theory is tomographically local, and hence the results of \cref{thm:structure-theorem} apply. Thus, every $T$-action on classical theory admits a system-level representation $\chi$.

Secondly, recalling that each $\chi_g$ is necessarily a symmetry of the state space of the GPT, we note that every symmetry of the classical state space is a physical transformation (specifically, a permutation of the classical pure states). Hence, every $\chi_g$ is physical, which means that the family $\chi$ defines an $S$-action. 

Putting these two together tells us that every $T$-action on classical theory is induced by $S$-actions, namely permutations, and is therefore physical. Consequently, every invariant subtheory obtained by averaging a classical $T$-action is a twirled subtheory, and there are no genuinely swirled classical worlds.

 \subsubsection{Bilocal classical theory}\label{sec:BCT}

In this subsection we show that the theory known as bilocal classical theory~\cite{d2020classicality}, $\mathbf{BCT}$, can be viewed as a $\mathbb Z_2$ twirled classical world. More specifically, the $\mathbb Z_2$ twirled classical world is actually the ontological model for bilocal classical theory which was defined in~\cite{soltani2026decoupling}. The connection between BCT and $\mathbb Z_2$-twirled classical theory was previously identified in~\cite{centeno2024twirled}, but here we establish it for the full compositional theory. This also answers, within the ontological-model formulation considered here, the question raised in~\cite{d2020classicality} of whether BCT can arise as a superselection of classical theory, thereby refuting the conjecture made there that this is not the case. There are two important consequences of this beyond the conceptual implications discussed in~\cite{centeno2024twirled}. On the one hand, it immediately gives us a way to generalise bilocal classical theory by using groups other than $\mathbb Z_2$. On the other hand, it immediately gives us a way to unify bilocal classical theory with standard classical theory via our unification of twirled subtheories.

Consider even-dimensional classical theory as our starting GPT. An arbitrary system in this GPT can be denoted by its dimension. Hence, we denote an arbitrary non-trivial system in even-dimensional classical theory as $2n$, where $n$ is an integer such that $n \geqslant 1$. We depict a system $2n$ as a doubled wire so that one has
\begin{equation}
    \input{Diagrams/9_BCT.tikz}=\input{Diagrams/9_BCT1.tikz}\;.
\end{equation}
In other words, we are dividing a $2n$-dimensional classical system into a classical $n$-dimensional subsystem and a classical binary $B$ subspace.

For each system we have an orthonormal basis of states and effects which we denote by:
\begin{equation}\label{eq:orthonormal-basis-bct}
       \input{Diagrams/9_BCT38.tikz}\;,
\end{equation}
where $0\leq i \leq n-1$ and $s\in \lbrace 0,1\rbrace$, and they satisfy
\begin{equation}
       \input{Diagrams/9_BCT39.tikz}=\delta_{i i'}\delta_{ss'},
\end{equation}

An arbitrary process $f\in P_{2n_1,\cdots,2n_p}^{2m_1,\cdots,2m_q}$ can then be written as a unique linear combination of these:
\begin{equation}\label{eq:generic-procees}
f = \sum_{\mathbf{i},\mathbf{s},\mathbf{j},\mathbf{t}}
f_{\mathbf{i},\mathbf{s}}^{\mathbf{j},\mathbf{t}}\scalebox{0.9}{$\input{Diagrams/9_BCT5.tikz}$}\;,
\end{equation}
where $\mathbf{i}, \mathbf{s}, \mathbf{j}$, and $\mathbf{t}$ denote collections of indices labelling the summation, with $f_{\mathbf{i},\mathbf{s}}^{\mathbf{j},\mathbf{t}} \in [0,1]$ and $\sum_{\mathbf{j},\mathbf{t}} f_{\mathbf{i},\mathbf{s}}^{\mathbf{j},\mathbf{t}} \leq 1$.

Next, let us define the relevant action of $\mathbb Z_2$ on these systems:
\begin{equation}\label{eq:BCT_Rep}
    \input{Diagrams/9_BCT9.tikz} = \input{Diagrams/9_BCT6.tikz}\;,
\end{equation}
where $g \in \lbrace 0,1 \rbrace$ denotes the elements of $\mathbb Z_2$. The NOT gate acts on the binary subspace as
\begin{equation}
   \scalebox{0.9}{$ \input{Diagrams/9_BCT7.tikz} = \input{Diagrams/9_BCT8.tikz}$}\;,
\end{equation}
with the summation being modulo two. Also, one can readily verify that
\begin{equation}\label{eq:inverse-binary}
    \input{Diagrams/9_BCT43.tikz}\;.
\end{equation}
Note that on a binary subspace, using XOR gate, we can rewrite
\begin{equation}
    \input{Diagrams/9_BCT40.tikz}= \input{Diagrams/9_BCT41.tikz}\;.
\end{equation}
Thus, one has
\begin{equation}\label{eq:xor-rep}
  \input{Diagrams/9_BCT9.tikz}= \input{Diagrams/9_BCT42.tikz}\;.
\end{equation}

We now generalise the copy on a binary subspace to spiders. We define a spider as
\begin{equation}\label{def:spiders}
  \scalebox{0.9}{$  \underbrace{\overbrace{ \input{Diagrams/9_BCT44.tikz}}^{m}}_{n}:= \sum_{s}\underbrace{\overbrace{ \input{Diagrams/9_BCT45.tikz}}^{m}}_{n}$}\;,
\end{equation}
where they obey the fusion rule
\begin{equation}\label{eq:fusion}
   \scalebox{0.8}{$ \underbrace{\overbrace{ \input{Diagrams/9_BCT46.tikz}}^{m}}_{n}=    \underbrace{\overbrace{ \input{Diagrams/9_BCT47.tikz}}^{m}}_{n}$}\;.
\end{equation}
We can use spiders to denote the completely mixed state of a binary space as
\begin{equation}\label{eq:mixed-state}
\input{Diagrams/9_BCT49.tikz} = \frac{1}{2}\;\input{Diagrams/9_BCT48.tikz}\;.
\end{equation}
We also denote
\begin{equation}
    \label{def:choi}\input{Diagrams/9_BCT50.tikz}:=\input{Diagrams/9_BCT51.tikz}\quad,\quad \input{Diagrams/9_BCT53.tikz}:=\input{Diagrams/9_BCT54.tikz}\;,
\end{equation}
where using this notation, we have that
\begin{equation}\label{eq:moving-legs}
    \scalebox{0.9}{$\input{Diagrams/9_BCT52.tikz}=\input{Diagrams/9_BCT55.tikz}$}\;,
\end{equation}
and
\begin{equation}\label{eq:xor-not}
\input{Diagrams/9_BCT56.tikz}=\input{Diagrams/9_BCT57.tikz}=\input{Diagrams/9_BCT58.tikz}\;.
\end{equation}
With these notations in hand, we can now prove that:
\begin{theorem}[
  restate=BCTasTwirl,
  name={}
]
\label{Thm:BCTasTwirl}
   $\mathbb Z_2$ twirling of even-dimensional classical theory with respect to the representation \cref{eq:BCT_Rep} gives (the ontological model of Ref.~\cite{soltani2026decoupling}) BCT.
\end{theorem}
\proof See \cref{Proof:BCTasTwirl}.
\endproof
Since every system in even-dimensional classical theory carries the same representation of $\mathbb Z_2$ in~\cref{eq:BCT_Rep}, $\mathbf{BCT}$ exhibits failure of local tomography for every process involving more than one even-dimensional classical system (See~\cref{thm:LTfailureIrreps}), either as input or output. For example, a valid $\mathbb Z_2$-invariant process in $\mathbf{BCT}$ (see \cref{Proof:BCTasTwirl}) is given by
\begin{equation}\label{eq:atomic-single}
\input{Diagrams/9_BCT82.tikz}\;.
\end{equation}
One cannot extract any information about $\sigma$ using only $\mathbb Z_2$-invariant local states on system $2n$ as in~\cref{eq:bct-state} and $\mathbb Z_2$-invariant local effects on system $2m$ as in~\cref{eq:bct-effect},  since one has
\begin{equation}
  \scalebox{0.9}{$ \input{Diagrams/9_BCT98.tikz}$}= \delta_{il}\delta_{ki'},
\end{equation}
but once one uses pure $\mathbb Z_2$-invariant bipartite states and effects, i.e., (see \cref{Proof:BCTasTwirl} for how these are obtained from~\cref{eq:atomic-bct})
\begin{equation}
\label{eq:bipartite-pure-state-bct}
\scalebox{0.9}{$\input{Diagrams/9_BCT19.tikz}\quad,\quad\input{Diagrams/9_BCT23.tikz}$}\;,
\end{equation}
one can extract information about $\sigma$, since one has
\begin{equation}
   \scalebox{0.9}{$\input{Diagrams/9_BCT99.tikz}$}=\delta_{il}\delta_{k i'}\delta_{j j'}\delta_{\sigma\tilde{\tau}},
\end{equation}
where $\tilde{\tau}=\tau\oplus\tau'$.

Having seen an instance of the failure of local tomography for processes, we now turn to tomography of states and effects in $\mathbf{BCT}$.
\begin{lemma}[
restate=BilocalStateTomography,
name={Bilocal state and effect tomography in $\mathbb Z_2$-invariant theories}
]\label{lem:BilocalStateTomography}
Let $\mathcal{A}(\mathcal{P})$ be an invariant subtheory of a locally tomographic GPT with $G=\mathbb Z_2$, and let $\lambda$ denote its unique non-trivial irreducible representation. For an invariant state space, if every non-trivial system carries $\lambda$, possibly with different multiplicities, then complete state tomography of such an invariant state space requires correlations across every pair of non-trivial systems, but can be performed using invariant effects that are parallel compositions of effects acting on at most two systems. Hence no genuinely tripartite or higher-partite effects are required. The analogous statement holds for effect tomography.
\end{lemma}
\proof
See~\cref{proof:BilocalStateTomography}
\endproof
This is the reason for the term ``bilocal'', as $\mathbf{BCT}$ admits state and effect tomography using bipartite states and effects as discussed in~\cite{d2020classicality}.

Note, however, that $\mathbf{BCT}$ satisfies process tomography, as it is obtained from classical theory, which is a locally tomographic GPT (see~\cref{prop:process-tomography-from-local-tomography}). To see this more explicitly, we can show that $\mathbf{BCT}$ admits a dynamically faithful state inherited from classical theory, namely the uniformly correlated state, which is a $\mathbb Z_2$-invariant dynamically faithful state. Using the spiders defined in~\eqref{def:spiders} and the notation defined in~\eqref{def:choi}, we can represent the uniformly correlated vector in even dimensional classical theory as
\begin{equation}
    \scalebox{0.9}{$\input{Diagrams/9_BCT92.tikz}:=\sum_{i,s}\input{Diagrams/9_BCT102.tikz}$}\;,
\end{equation}
where the action of $\mathbb Z_2$~\cref{eq:BCT_Rep} for any $g\in \mathbb Z_2$ gives
\begin{equation}
   \scalebox{0.9}{$ \input{Diagrams/9_BCT93.tikz}$}\;,
\end{equation}
and since on the binary subspaces one has
\begin{align}
     \scalebox{0.9}{$\input{Diagrams/9_BCT94.tikz}$}=& \scalebox{0.9}{$\sum_s   \input{Diagrams/9_BCT96.tikz}$}= \scalebox{0.9}{$\sum_s   \input{Diagrams/9_BCT95.tikz}$}= \scalebox{0.9}{$\input{Diagrams/9_BCT97.tikz}$}\;,
\end{align}
it follows that the uniformly correlated state vector remains invariant under the action of $\mathbb Z_2$
\begin{equation}
\scalebox{0.8}{$\input{Diagrams/9_BCT93.tikz}=\input{Diagrams/9_BCT92.tikz}$}\;.
\end{equation}
Hence, $\mathbf{BCT}$ admits the uniformly correlated state as a dynamically faithful state, and satisfies process tomography as well (see~\cref{cor:twirled-invariant-dynamically-faithful-states}). For example, for the process of the kind~\eqref{eq:atomic-single}, using the uniformly correlated state and a pure bipartite effect, one can extract all information needed to determine the process as one has
\begin{equation}
    \scalebox{0.9}{$\input{Diagrams/9_BCT100.tikz}$}=\delta_{lj}\delta_{ki}\delta_{\sigma\tau}.
\end{equation}
Since the relevant information can be extracted for the above process, it immediately follows that any arbitrary process in $\mathbf{BCT}$ within the process space of interest can be determined, as such processes are linear combinations of processes of this form. More generally, we have that:
\begin{lemma}[
restate=BilocalProcessTomography,
name={Bilocal process tomography in $\mathbb Z_2$-invariant theories}
]\label{lem:BilocalProcessTomography}
Let $\mathcal{A}(\mathcal{P})$ be an invariant subtheory of a locally tomographic GPT with $G=\mathbb Z_2$, and let $\lambda$ denote its unique non-trivial irreducible representation. For an invariant process space, if every non-trivial system carries $\lambda$, possibly with different multiplicities, then complete tomography of any process within such an invariant process space, with non-trivial input $A=A_1\cdots A_n$ and output $B=B_1\cdots B_m$, requires an ancillary system $E$ carrying $\lambda$, and any such $E$ suffices. Moreover, complete process tomography can be performed using invariant states on $AE$ and invariant effects on $BE$ that are parallel compositions of invariant states and effects acting on at most two systems. Hence no genuinely tripartite or higher-partite states or effects are required.
\end{lemma}
\proof
See~\cref{proof:BilocalProcessTomography}
\endproof

\subsubsection{Unification of BCT and CT}

Twirling classical theory (or at least even dimensional classical theory) with respect to the $\mathbb Z_2$ action of \cref{eq:BCT_Rep} led to BCT as a twirled world. However, for a classical system $n$ we could also pick the trivial action:
\begin{equation}
    \input{Diagrams/9_BCT72.tikz}\quad,\;\forall g\in \mathbb Z_2,
\end{equation}
and twirl with respect to this. Clearly, doing so leads to standard classical theory as a twirled world. We can, hence, unify both $\mathbf{CT}$ and $\mathbf{BCT}$ within $\mathsf{Twirl}_{Z_2}[\mathbf{CT}]$. Concretely, however, what we want to know is what the interactions between systems in $\mathbf{CT}$ and systems in $\mathbf{BCT}$ look like.

\begin{proposition}
    Interactions between a system in $\mathbf{CT}$ and a system in $\mathbf{BCT}$ can be written as
    \begin{equation}
        \scalebox{0.9}{$\input{Diagrams/9_BCT74.tikz}$}=\sum_{k,i,l,j,\sigma}\tilde{f}^{l,j,\sigma }_{k,i }\;\input{Diagrams/9_BCT77.tikz}\;,
    \end{equation}
    where $ \tilde{f}^{l,j,\sigma}_{k,i}\in [0,1]$, and $\sum_{l,j,\sigma} \tilde{f}^{l,j,\sigma}_{k,i }\leq 1$.
\end{proposition}
\proof
This follows exactly as the proof of \cref{Thm:BCTasTwirl}, where with the same logic one can easily generalise to the case when one has multipartite systems for both $\mathbf{BCT}$ and $\mathbf{CT}$.
\endproof

Special cases of this include (i) using standard classical systems to control which BCT process happens:
\begin{equation}
      \sum_{k}  \input{Diagrams/9_BCT73.tikz}\;,
\end{equation}
where each $f_k$ can be written as
\begin{equation}
     \scalebox{0.9}{$\input{Diagrams/9_BCT78.tikz}$}=\sum_{i,j,\sigma}\left(\tilde{f}_k\right)^{j,\sigma}_{i}\input{Diagrams/9_BCT79.tikz}\;,
\end{equation}
with $\left(\tilde{f}_k\right)^{j,\sigma}_{i}\in [0,1]$ and $\sum_{j,\sigma}\left(\tilde{f}_k\right)^{j,\sigma}_{i}\leq 1 $ for each $k$ and $i$,
and (ii) using standard classical systems to encode the outcome of a measurement on a BCT process:

\begin{equation}
\sum_k\input{Diagrams/9_BCT80.tikz}\;,
\end{equation}
where for each $e_k$ one has
\begin{equation}
  \input{Diagrams/9_BCT85.tikz}=\sum_{i}\tilde{f}_{k}^{i} \;\;\input{Diagrams/9_BCT84.tikz}\;,
  \end{equation}
  where $\tilde{f}_{k}^i\in [0,1]$ and $\sum_{i}\tilde{f}_k^i\leq n$.
Hence, we can view this unification as ``internalising'' the classical interface of BCT which, within the OPT framework in which it was originally formalised, is traditionally treated using ``external'' indices rather than as physical systems.

\subsection{Real quantum theory} \label{sec:RQT}

In Ref.~\cite{ying2025whether} it was argued that real quantum theory, $\mathbf{RQT}$, could be obtained by swirling standard complex quantum theory, $\mathbf{QT}$. Their approach was bottom up starting by defining complex conjugation of quantum pure states and building up to give a notion of complex conjugation of quantum operations. Here, we take a complementary top down perspective, starting with the observation that complex conjugation defines an involutive diagram preserving map $*:\mathbf{QT}\to \mathbf{QT}$.

To begin we can give a succinct definition of quantum theory as a GPT.
\begin{definition}
We define the GPT, $\mathbf{QT}$, to be the symmetric monoidal category in which the objects are labelled by finite-dimensional Hilbert spaces $\mathcal{H}$ composed with the tensor product and morphisms from $\mathcal{H}$ to $\mathcal{K}$ are completely positive trace-nonincreasing maps $\mathcal{E}:\mathcal{B}(\mathcal{H})\to \mathcal{B}(\mathcal{K})$, with the normalized morphisms given by CPTP maps.
\end{definition}

It follows immediately, for example from Selinger's CPM construction \cite{selinger2007dagger}, that complex conjugation defines an involutive diagram preserving map $*:\mathbf{QT}\to \mathbf{QT}$. We can use this in order to define a $T$-action of $\mathbb Z_2$ on $\mathbf{QT}$. Specifically, we take $\Xi_0:=I$ and $\Xi_1:=*$. The involutivity of $*$ is critical for this to be a valid representation, as it ensures that $\Xi_1\circ \Xi_1 = *\circ * = I = \Xi_0 = \Xi_{1\cdot 1}$. 

We can hence define a swirled world with respect to this representation of $\mathbb Z_2$ and it is clear that we obtain the following:
\begin{proposition}Swirling $\mathbf{QT}$ with respect to the representation of $\mathbb Z_2$ given by $\Xi_1:=*$ gives us $\mathcal{S}(\mathbf{QT})=\mathbf{RQT}$.
\end{proposition}
\proof
The processes in $\mathcal{S}(\mathbf{QT})$ are processes which are invariant under $*$, i.e., are those that are real. This is precisely the definition of $\mathbf{RQT}$.
\endproof

Since quantum theory is locally tomographic, if we denote the non-identity element of $\mathbb Z_2$ by $g = * = g^{-1}$, then using~\cref{eq:structure-theorem-swirling1} for an arbitrary quantum process on multipartite systems implies
\begin{equation}
  \scalebox{0.9}{$  \input{Diagrams/10_RQT11.tikz} = \input{Diagrams/10_RQT12.tikz}$}\;.
\end{equation}
For a state $\rho$ of a system $A$ associated with $\mathbb C^a$, this implies
\begin{equation}
\label{eq:partial-trasnpose}
     \input{Diagrams/10_RQT.tikz} = \input{Diagrams/10_RQT1.tikz}\;.
\end{equation}
Since we require this equation to yield $\bar{\rho}$ in a chosen basis, $\chi_*$ must be the linear invertible transformation that maps $\rho$ to its complex conjugate in the chosen basis; in a fixed basis this is simply matrix transposition. For a composite state $\rho$, this gives
\begin{equation}
    \input{Diagrams/10_RQT2.tikz} = \input{Diagrams/10_RQT3.tikz}\;,
\end{equation}
where each $\chi_*$ acting on a single subsystem is the partial transpose in the chosen basis of that subsystem.

For a POVM element $E$ of a system $A$, one similarly has
\begin{equation}
    \input{Diagrams/10_RQT4.tikz} = \input{Diagrams/10_RQT5.tikz}\;,
\end{equation}
where $\chi_*$ again denotes transposition in the chosen basis, as we require the above equation to output $\bar{E}$. For a POVM element on a composite system, one obtains
\begin{equation}
\input{Diagrams/10_RQT6.tikz} = \input{Diagrams/10_RQT7.tikz}\;.
\end{equation}

It is well known that $\mathbf{RQT}$ is a valid GPT, in particular, that it is tomographic, but it is also well known that it is not locally tomographic. For consistency, however, we will see how this matches with our abstract results.

To see that $\mathbf{RQT}$ admits the maximally entangled state, we simply need to show that the density matrix corresponding to the maximally entangled state $\frac{1}{\sqrt{a}}\sum_{i=0}^{a-1}\vert i \rangle_A\vert i \rangle_{A^*}$ is given by $\frac{1}{a}\sum_{i,j} \vert i \rangle_A\vert i \rangle_{A^*} \langle j \vert_A\langle j \vert_{A^*}$. This density matrix is invariant under conjugation, equivalently under transposition in this representation, and hence is a valid state in $\mathbf{RQT}$.

To see this diagrammatically, we use the vector representation of a density matrix. In quantum theory, for a quantum system $A$ associated to $\Cl^a$, an invariant dynamically faithful state is given by the maximally entangled state of system $AA^*$, where $A^*$ is the dual of $A$. One can always represent a density matrix as a vector $\rho \to \mathbf{vec}(\rho)$, where vectorization acts on each entry of the density matrix as $\vert i \rangle \langle j\vert\to\vert i \rangle \vert j \rangle $. Within this representation, the density matrix associated to the maximally entangled state of system $A A^*$ is $\frac{1}{a}\sum_{i,j=0}^{a-1} \vert i\rangle\vert j \rangle\vert i \rangle\vert j \rangle $, in which the maximally entangled vector can be represented diagrammatically using the double-wire notation of~\cite{CoeckeKissinger2017PicturingQuantumProcesses} as
\begin{equation}\scalebox{0.95}{$
\input{Diagrams/10_RQT13.tikz}:= \input{Diagrams/10_RQT14.tikz}:=\sum_{i,j}\input{Diagrams/10_RQT15.tikz}$},
\end{equation}
where, since the action of each $\chi_*$ is given by partial transpose, one has
\begin{align}
   \input{Diagrams/10_RQT16.tikz} =\input{Diagrams/10_RQT18.tikz}
   &=\input{Diagrams/10_RQT14.tikz}\\ &=\quad\input{Diagrams/10_RQT13.tikz}\;.
\end{align}
Thus, the maximally entangled state is a valid state of $\mathbf{RQT}$. Hence, $\mathbf{RQT}$ satisfies process tomography, as it contains a dynamically faithful state from quantum theory.

Note that, in the conventional matrix representation of density matrices, $\chi_*$ for every system $A$ is the partial transpose, and hence an unphysical transformation in quantum theory. Here the $\mathbb Z_2$-action should not be understood as a physical unitary representation on each system. Rather, it is the theory-level involution given by complex conjugation. Consequently, the usual rule that a unitary action $U_A$ on $A$ induces the conjugate action $\bar U_A$ on $A^*$ does not apply directly. In the chosen basis, the involution acts by transposition on density matrices, and on composite systems by the corresponding partial transposes. Thus the action on the dual copy is again given by the same conjugation involution, rather than by a new physical unitary action obtained by conjugating a unitary action on the original system.

It is well known that RQT is not locally tomographic ~\cite{Hardy2011Limited,barnum2020composites}. In our framework, this follows directly from the $\mathbb Z_2$ representation carried by its systems. The non-trivial element of $\mathbb Z_2$ acts by complex conjugation, equivalently by transposition in a fixed basis. For a qubit, this is particularly transparent from the Bloch representation $\rho=\frac12(I+r_x\sigma_x+r_y\sigma_y+r_z\sigma_z)$. Complex conjugation leaves $\sigma_x$ and $\sigma_z$ unchanged, but sends $\sigma_y\mapsto-\sigma_y$. Hence the $\sigma_y$-direction transforms according to the unique non-trivial irrep $\lambda$ of $\mathbb Z_2$, and invariance under complex conjugation forces $r_y=0$, giving precisely the real qubit state space. More generally, operators of the form $i(\vert i\rangle\langle j\vert-\vert j\rangle\langle i\vert)$ change sign under transposition, so every non-trivial RQT system carries $\lambda$. Here, by a system carrying $\lambda$, we mean that the corresponding ambient quantum representation contains $\lambda$. Consequently, any two non-trivial RQT systems share the same non-trivial irrep $\lambda$, and~\Cref{thm:LTfailureIrreps} recovers the failure of local tomography in RQT~\cite{Hardy2011Limited}.

It is also well known that RQT is bilocally tomographic~\cite{Hardy2011Limited}. Our representation-theoretic results recover this directly. Since RQT is obtained by swirling the locally tomographic theory QT with $G=\mathbb Z_2$, and every non-trivial RQT system carries the unique non-trivial irrep $\lambda$, the assumptions of \Cref{lem:BilocalStateTomography,lem:BilocalProcessTomography} are satisfied. Hence~\Cref{lem:BilocalStateTomography} recovers the known bilocal tomography of RQT for states, together with the analogous result for effects, while~\Cref{lem:BilocalProcessTomography} gives the
corresponding bilocal result for process tomography.

\section{Discussion}

\subsection{Summary}

Working throughout in a fully compositional setting for arbitrary finite-dimensional GPTs and compact groups, we developed a framework for describing group actions at the level of systems, process spaces, and the theory as a whole. At the system level, we introduced $S$-actions and distinguished whether they are physical or nonphysical. We further distinguished these from $P$-actions defined directly on the process spaces of a theory, and from $T$-actions on the theory as a whole. A suitable family of $S$-actions can induce $P$-actions and hence a $T$-action. However, a $P$-action need not itself be induced by $S$-actions; consequently, a $T$-action need not in general arise from $S$-actions. We use \emph{twirling} for averaging a physical $T$-action and \emph{swirling} for averaging a nonphysical $T$-action. The motivation for considering general $T$-actions arises from several considerations. First, such actions provide a natural and technically convenient way to formulate symmetries of non-locally tomographic theories. Although we propose a candidate notion of a physical $S$-action for such theories, working directly with system transformations is cumbersome, since the behaviour of a transformation on a system need not fully determine its behaviour when extended to arbitrary ancillary systems, so one must keep track of an entire family of ancillary extensions. A $T$-action avoids this difficulty by specifying the symmetry consistently on the whole theory from the outset. Second, some symmetries are more naturally viewed as symmetries of the theory as a whole rather than some non-physical transformation of individual systems. This is particularly the case when the symmetry has a simple and consistent action on all processes and their composition, while its action on individual systems is not itself physically realisable. Complex conjugation in quantum theory, for example, is naturally regarded in this way. Moreover, theory-level actions provide a route to formulating symmetries that need not arise from a specified collection of system-level actions, thereby equipping the GPT framework to take a broader class of symmetries into consideration.

A central question throughout this work was whether process tomography is preserved on passing to an invariant subtheory. We showed that this might not hold for arbitrary invariant subtheories, and identified two broad conditions that guarantee it. The first is that, for every system in an invariant subtheory, there exists an invariant state that is dynamically faithful with respect to that invariant subtheory, or, more generally, a vector in the span of the corresponding invariant state space that is dynamically faithful with respect to that invariant subtheory. In particular, this implies that every invariant subtheory obtained from a locally tomographic GPT satisfies process tomography. The second condition uses internal classical reference frames together with universal classical control, allowing the group information to be encoded within the physical theory itself. For system-induced twirled worlds, this construction also leads to an extension of incoherent simulation to arbitrary finite-dimensional GPTs and compact groups. We further developed a framework for unifying different system-induced twirled subtheories associated with different $S$-actions, together with an analogous, more general proposal formulated at the level of $P$-actions, which in particular unifies invariant subtheories associated with different $T$-actions.

We then specialise to invariant subtheories of locally tomographic GPTs. Using the structure results of Refs.~\cite{schmid2024structure,wagner2025structure}, we show that every $T$-action on a locally tomographic GPT can be represented by linear actions on the individual system vector spaces. These system-level transformations need not themselves be physical, even though the resulting $T$-action maps physical processes to physical processes. Thus, in the locally tomographic setting, twirling and swirling have the same system-level form, with the distinction between them being whether the corresponding $T$-action is physical, equivalently whether the representing $S$-actions are physical. For finite-dimensional classical theory, only the twirling case occurs, since every reversible symmetry of a classical state space is a physical permutation. 

We first characterise the failure of local tomography and its operational consequences in terms of relations between the relevant invariant projectors. We then use representation theory to determine when these projector conditions are satisfied, obtaining explicit criteria in terms of the representations carried by the systems. In particular, local tomography fails precisely when there exists a bipartition of the systems such that the representations carried by the two parts share a non-trivial irreducible representation. The representation-theoretic analysis also determines the resources required for complete process tomography when local tomography fails, including when separable invariant states and effects suffice, when entanglement within the invariant subtheory is necessary, and when ancillary systems are required. We further quantify the extent of this failure through the tomographic excess, which counts the invariant degrees of freedom that are inaccessible to local tomography. For $\mathbb Z_2$-invariant theories in which every non-trivial system carries the unique non-trivial irreducible representation, these results further imply bilocal state and effect tomography and give the corresponding bilocal form of process tomography.

Finally, we applied the framework to bilocal classical theory and real quantum theory. We showed that BCT arises as a $\mathbb Z_2$-twirled subtheory of classical theory and that our unification construction allows it to coexist and interact with standard classical theory. Real quantum theory instead provides a natural example of swirling. It is obtained from quantum theory by imposing invariance under complex conjugation, whose system-level representation on individual systems is transposition and need not correspond to a physical quantum transformation. These two examples illustrate how physical and nonphysical $T$-actions capture twirled and swirled theories within a common compositional framework, while the representation-theoretic results recover their characteristic failures of local tomography and the corresponding forms of tomography that are sufficient. 
\subsection{Future directions}

These results leave several natural directions for further investigation. Although we establish broad sufficient conditions for process tomography in invariant subtheories, a general characterisation of when such subtheories satisfy process tomography remains open. An important direction for future work is therefore to develop necessary and sufficient conditions for process tomography that apply beyond locally tomographic starting theories, and to determine whether the sufficient conditions established here can be weakened or removed. It would also be interesting to construct an explicit example of a well-defined, locally non-tomographic GPT whose invariant subtheory fails to satisfy process tomography. Our formulation of $\mathbf{BCT}$ can serve as a suitable toy theory for exploring this possibility. One can study the status of process tomography in invariant subtheories obtained from $\mathbf{BCT}$ and investigate its failure using our formulation of processes in $\mathbf{BCT}$.

The quantification of the failure of local tomography introduced here also provides a natural starting point for a resource-theoretic study of tomographic nonlocality. In particular, it would be interesting to identify suitable free operations and determine whether the tomographic excess, or suitable refinements of it, is monotonic under such operations.

The internal classical reference-frame construction also suggests several extensions. A natural direction is to replace the classical reference system with an internal quantum reference frame, where the group information can be carried coherently and correlated with the systems being described. This could provide a route towards extending coherent simulations from their present formulation for finite groups in quantum theory to compact groups, and potentially to arbitrary finite-dimensional GPTs. More broadly, this suggests developing a notion of internal GPT reference frames, in which an arbitrary physical system of a GPT can serve as a reference frame within an arbitrary GPT. This would allow one to investigate which aspects of reference-frame transformations depend specifically on classical or quantum structure, and which can be formulated more generally within the GPT framework.

A related direction is to enlarge the defining data of compositional GPTs by incorporating higher-order processes, such as testers or combs, as primitive objects. This could provide a more general formulation of process tomography in invariant subtheories, while also raising new questions concerning the compositional structure of such higher-order processes. Also, one may ask whether, within our compositional and process-theoretic approach, symmetry transformations can be applied locally to only part of a diagram rather than necessarily acting globally on the entire diagram. Developing such a formulation would require a general theory of locally applicable endofunctors. The framework of locally applicable transformations and higher-order processes developed in~\cite{wilson2026quantum} provides a natural starting point for this direction.

It would also be interesting to understand the categorical status of the unification construction $\mathrm{Twirl}_G$ more abstractly. At the level of systems and processes, this construction is closely related to the standard $G$-equivariantisation of a monoidal category for the trivial categorical $G$-action: systems are equipped with $G$-actions and processes are restricted to those equivariant with respect to the chosen actions \cite{EtingofEtAl2015}. This raises the question of whether $\mathrm{Twirl}_G$ admits a universal characterisation as some form of completion of a process theory. A suggestive analogy is provided by the Karoubi envelope: the forgetful functor from categories to semicategories admits a right adjoint, with the resulting monad on categories giving the Karoubi completion \cite{HoofmanMoerdijk1995}. One may ask whether $\mathrm{Twirl}_G$ similarly arises from an adjunction between suitable categories of process theories, with and without compatible $G$-structure, and hence whether it carries a natural monad or comonad structure. Since the process theories considered here are naturally enriched in their process spaces, the appropriate formulation is likely to be an enriched one, for example in terms of convex-enriched coloured PROPs. Such a formulation could provide a universal property for $\mathrm{Twirl}_G$ and clarify in what precise sense, if any, it constitutes a canonical completion of a process theory.

Another natural direction is to classify the invariant subtheories arising from familiar GPTs. For classical theory, a concrete starting point is to generalise the $\mathbb Z_2$ construction underlying $\mathbf{BCT}$ to $\mathbb Z_n$ and, more generally, to finite-group twirled classical worlds. Such a project could also make the relation between twirls for different groups explicit. For instance, whenever $m\mid n$, $\mathbb Z_m$ is a subgroup of $\mathbb Z_n$, so one can compare direct $\mathbb Z_n$ twirling with first imposing $\mathbb Z_m$ invariance and then the residual $\mathbb Z_{n/m}$ symmetry. Since in classical theory these actions are realised simply by permutations, this provides a particularly concrete setting in which to understand how invariant subtheories associated with a group, its subgroups, and its quotient groups are related. For Abelian groups, one could further study how the group structure determines which multipartite invariant correlations appear and, consequently, how many systems must be accessed jointly for tomography. A further direction is to apply our proposed unification of invariant subtheories to GPTs containing both classical and quantum systems, for example a theory in which quantum systems interact with classical wires. Since every $T$-action on classical theory is physical, imposing the relevant $\mathbb Z_2$ symmetry on the classical sector yields BCT as a twirled theory. On the quantum sector, however, the corresponding invariant theory could arise either from $\mathbb Z_2$ twirling or from $\mathbb Z_2$ swirling. This would give hybrid theories in which $\mathbf{BCT}$ interacts with either a twirled or a swirled quantum world, making the resulting structure of entanglement and tomography particularly interesting to investigate.

Another direction is to refine the distinction between physical and nonphysical symmetry actions by asking whether a nonphysical action could become physical in a consistent extension of the theory. More precisely, after verifying that an action defines a symmetry of the theory, one may ask whether the theory can be consistently enlarged so that the action itself becomes physically implementable. This leads to a distinction between weak and strong nonphysicality: an action is weakly nonphysical if it is not physical in the original theory but can become physical in a consistent extension, and strongly nonphysical if no such extension is possible. Developing this distinction and a precise notion of such consistent extensions is left for future work.

A further direction for future work is to combine the symmetry framework developed here with the compositional approach to nonclassicality of~\cite{schmid2024structure}. In that framework, classical explainability is characterised through ontological models that respect the compositional structure of the theory. Incorporating $S$- and $P$-actions into this setting would allow one to ask how symmetry transformations, including changes of reference frame and gauge transformations, affect classical and nonclassical descriptions. In particular, one could study whether notions such as generalised noncontextuality are preserved under such transformations, whether a description that is nonclassical relative to one symmetry setting can become classically explainable in another, and what additional consistency conditions an ontological model must satisfy in order to respect both composition and symmetry.

The unification of invariant subtheories associated with general $T$-actions also suggests another direction for investigation. Incorporating $S$- and $T$-actions into this setting would allow one to ask how symmetry transformations, including changes of reference frame and gauge transformations, affect classical and nonclassical descriptions. A complete formulation may therefore require going beyond the process-theoretic framework used throughout this work, which we leave for future investigation.

Another obvious direction is to extend the framework beyond finite-dimensional GPTs and compact groups, including infinite-dimensional classical and quantum theories. Such an extension would require revisiting the role of Haar measure, the corresponding averaging procedures, and the notion of equivalence under group actions. Even within finite-dimensional GPTs, it would be useful to understand more systematically the conditions under which the integration procedures used throughout this work are well defined, and to generalise the integration results established here to broader notions of integration. We intend to address this question in forthcoming work~\cite{SalzgerSelbySoltaniIntegration}.

{\bf Acknowledgements---}%
We thank Yìlè Yīng and Daniel Centeno for their useful feedback on an early version of the manuscript. We also thank Robert Spekkens for insightful discussions and for suggesting refinements to the terminology of this paper.
SS,  ME, and JHS were supported by the National Science Centre, Poland (Opus project, Categorical Foundations of the Non-Classicality of Nature, project no. 2021/41/B/ST2/03149). ChatGPT was used to refine some of the proofs and arguments that had first been developed by the authors, including the proofs in~\cref{sec:OperationalRep} for the generalised multipartite case and the argument of~\Cref{lem:InvariantSubprojection}. It was also used to help refine the connection between the dimensional counting in~\Cref{subsec:QuantifyingLTFailure} and the corresponding representation-theoretic structure. AI was additionally used for proofreading and grammar checks.

\bibliography{reference.bib}

\addtocontents{toc}{\protect\HideSublevelsFromTOC}
\appendix
\crefalias{section}{appendix}
\crefalias{subsection}{appendix}
\crefalias{subsubsection}{appendix}
\raggedbottom
\setlength{\parskip}{1pt plus .3pt minus .3pt}

\setlength{\abovedisplayskip}{2.5pt plus .5pt minus .5pt}
\setlength{\belowdisplayskip}{2.5pt plus .5pt minus .5pt}
\setlength{\abovedisplayshortskip}{1pt plus .5pt minus .5pt}
\setlength{\belowdisplayshortskip}{1.5pt plus .5pt minus .5pt}
\setlength{\jot}{1pt}
\section{Proofs}
\subsection{Twirled worlds}
\subsubsection{\texorpdfstring{Proof of \cref{prop:twirledSubspaceBeingSubtheory}}{Proof of Proposition \ref*{prop:twirledSubspaceBeingSubtheory}}}\label{proof:twirled_is_subtheory}
\twirledSubspaceBeingSubtheory*
\begin{proof}
We verify the conditions of~\cref{def:subGPT} in order.

\noindent\textit{(i) Systems.}
The $S$-twirled subtheory has the same systems as the original theory $\mathcal{P}$, and hence its set of systems is a subset of $\textbf{Sys}[\mathcal{P}]$ and contains the trivial system $I$.

\noindent\textit{(ii) Subprocess spaces.}
By definition, for every pair of input and output systems, the corresponding process space of the $S$-twirled subtheory is the invariant subprocess space of the corresponding process space of the original theory, consisting precisely of the processes invariant under $S$-twirling. It therefore remains to verify conditions (a)--(d).

\noindent\textit{(a) Identities.}
For every system $A \in \textbf{Sys}[\mathcal{P}]$, the identity $\mathcal{I}_A$ is invariant under $S$-twirling, as we have
\begin{equation}
    \scalebox{0.85}{$\input{Diagrams/4_twirl75.tikz}=\input{Diagrams/4_twirl76.tikz}\quad\stackrel{\eqref{eq:invertibility-of-representation}}{=}\quad\input{Diagrams/4_twirl77.tikz}$}\;.
\end{equation}
Hence, for all $A \in \textbf{Sys}[\mathcal{T}(\mathcal{P})]$, one has $\mathcal{I}_A \in \mathcal{T}(P^A_A)$.

\noindent\textit{(b) Swaps.}
Swap can also be shown to be invariant under $S$-twirling, that is, for every $A, B \in \textbf{Sys}[\mathcal{P}]$, $\sigma_{A,B}$ remains invariant:
\begin{equation}
\scalebox{0.85}{$\input{Diagrams/4_twirl78.tikz}=\input{Diagrams/4_twirl79.tikz}=\input{Diagrams/4_twirl80.tikz}\stackrel{\eqref{eq:invertibility-of-representation}}{=}\input{Diagrams/4_twirl81.tikz}$}\;.
\end{equation}
Therefore, in an $S$-twirled subtheory $\mathcal{T}(\mathcal{P})$, for all $A, B \in \textbf{Sys}[\mathcal{T}(\mathcal{P})]$, one has $\sigma_{A,B} \in \mathcal{T}(P_{A,B}^{B,A})$.

\noindent\textit{(c) Scalars.}
Scalars under the action of $S$-twirling are invariant (\cref{cor:twirlingIsScalarPreserving}). Hence, the scalar space of an $S$-twirled subtheory $\mathcal{T}(\mathcal{P})$ is the same as that of the original theory $\mathcal{P}$, i.e., $\mathcal{T}(P^I_I) = P^I_I = \left(\{1\}, [0,1], \mathbb{R}\right)$.

\noindent\textit{(d) States and effects.}
First consider a trivial input system. Since the $S$-actions are physical, each $\pi_A^g$ is a reversible physical process, with physical inverse $\pi_A^{g^{-1}}$. Hence, it preserves normalisation of states: if $\pi_A^g(\omega)=p\omega'$ for normalised states $\omega,\omega'$ and $p\in[0,1]$, then applying the physical inverse gives $\omega=pq\omega''$ for some normalised $\omega''$ and $q\in[0,1]$, which implies $pq=1$ and hence $p=1$. Therefore, each $\pi_A^g$, and thus also $S$-twirling by convexity, maps normalised states to normalised states. Now let $s\in\mathcal{T}(P^A)$ be an invariant subnormalised state. Since $P^A$ is a state space, $s=p\omega$ for some normalised state $\omega$ and $p\in[0,1]$. For $p>0$, linearity and invariance give $p\omega=s=\mathcal{T}(s)=p\mathcal{T}(\omega)$, so that $\mathcal{T}(\omega)=\omega$. Conversely, every scalar multiple of an invariant normalised state is invariant by linearity. Hence, when restricted to a trivial input, the resulting subprocess space is a state space.

Now consider a trivial output system. We first note that the unit effect $u_A$ is invariant under each $\Pi_g$. Indeed, as shown above, the reversible physical $S$-actions map normalised states to normalised states, and hence the effect $\Pi_g(u_A)$ assigns probability one to every normalised state, just as $u_A$ does. By tomographic completeness of the original GPT, these effects are therefore equal, so $\Pi_g(u_A)=u_A$ for every $g\in G$, and consequently $\mathcal{T}(u_A)=u_A$. Finally, let $e\in\mathcal{T}(P_A)$ be an invariant effect. Since $P_A$ is an effect space, there exists a complementary effect $e^\perp\in P_A$ such that $e+e^\perp=u_A$. By linearity of $S$-twirling, $\mathcal{T}(e^\perp) =\mathcal{T}(u_A-e)=u_A-e=e^\perp$. Thus, the complement of every invariant effect is also invariant, and hence, when restricted to a trivial output, the resulting subprocess space is indeed an effect space.

\noindent\textit{(iii) Composition}
Parallel and sequential composition in the $S$-twirled subtheory are given by the corresponding composition maps of the original theory $\mathcal{P}$. Hence, they inherit bifunctoriality from $\mathcal{P}$.

\noindent\textit{(iv) Closure under composition.}
To show the closedness of composition, we need to show that for every process $f$ and $h$ in an $S$-twirled subtheory, the composition
\begin{equation}
\label{eq:twirled-composition}
 \scalebox{0.75}{$ \input{Diagrams/4_twirl18.tikz} $}
\end{equation}
is closed, i.e., remains invariant under $S$-twirling. This, in turn, ensures the closure of an arbitrary composition under $S$-twirling, as explained in the paragraph below~\cref{eq:composition-closure1}. This composition is indeed invariant under $S$-twirling, as one has
\begin{align}\label{eq:composition-closure}
&\scalebox{0.65}{$\input{Diagrams/4_twirl12.tikz}=\input{Diagrams/4_twirl13.tikz}=\input{Diagrams/4_twirl14.tikz}$}\nonumber\\
  \scalebox{0.65}{$=$}&\scalebox{0.65}{$ \input{Diagrams/4_twirl15.tikz}\;\;
     \stackrel{\eqref{eq:haar-measure-invariance}}{=}\input{Diagrams/4_twirl67.tikz}\;\;\stackrel{\eqref{eq:copy-discard}}{=}\;\;\input{Diagrams/4_twirl18.tikz}$}\;.
\end{align}
In particular, by trivialising suitable systems, this implies that sequential and parallel composition are indeed closed in an $S$-twirled subtheory.

Thus, all the conditions of~\cref{def:subGPT} are satisfied.
\end{proof}

\subsubsection{\texorpdfstring{Proof of \cref{prop:twirledcombtomography}}{Proof of Proposition \ref*{prop:twirledcombtomography}}}
\label{proof:twirledcombtomography}
\twirledCombTomography*
\begin{proof}

Process tomography completeness of the original GPT implies that for every pair of distinct processes
$\tilde{f} \in \mathcal{T}(P_A^B)$ and $\tilde{h} \in \mathcal{T}(P_A^B)$, there exist $s \in P^{A,E}$ and $e \in P_{B,E}$ in the original GPT $\mathcal{P}$ such that
\begin{equation}
\label{eq:process-tomography-twirl}
\scalebox{0.8}{$\input{Diagrams/2_theorysym14.tikz}$}\neq\scalebox{0.85}{$\input{Diagrams/4_twirl19.tikz}\Rightarrow\input{Diagrams/4_twirl20.tikz}\neq\input{Diagrams/4_twirl21.tikz}$}\;.
\end{equation}
Since $\mathcal{T}$ is scalar preserving (\cref{cor:twirlingIsScalarPreserving}), for  any $\tilde{f}\in\mathcal{T}(P^B_A)$, one has
\begin{equation*}
\scalebox{0.8}{$\input{Diagrams/4_twirl20.tikz}=\input{Diagrams/4_twirl24.tikz}=\input{Diagrams/4_twirl25.tikz}\;$}\;,
\end{equation*}
where using~\cref{eq:twirling-def} and diagram preservation of $\Pi_g$, one has
\begin{align}
\scalebox{0.67}{$\input{Diagrams/4_twirl25.tikz}$}=\scalebox{0.7}{$\input{Diagrams/4_twirl26.tikz}$}\nonumber\\
\stackrel{\eqref{eq:haar-measure-invariance}}{=}\scalebox{0.67}{$\input{Diagrams/4_twirl90.tikz}$}=\scalebox{0.7}{$\input{Diagrams/4_twirl91.tikz}$}\;.
\end{align}
Therefore, from~\eqref{eq:process-tomography-twirl}, we have that
\begin{equation}
\scalebox{0.67}{$
\input{Diagrams/4_twirl91.tikz} \neq
\input{Diagrams/4_twirl111.tikz}\quad .
$}
\end{equation}
We then identify
\begin{equation}\label{eq:comb-for-process-tomography}
\scalebox{0.67}{$\input{Diagrams/4_twirl112.tikz}$}\quad.
\end{equation} as being $S$-twirling applied to a comb.
\end{proof}

\subsubsection{\texorpdfstring{Proof of \cref{prop:pihatInvariance}}{Proof of Proposition \ref*{prop:pihatInvariance}}}
\label{app:pihat-invariance}
\piHatInvariance*
\proof
\begin{equation}
    \label{eq:pihat-invariance-proof}\scalebox{0.7}{$\input{Diagrams/15_internalRef81n.tikz}=\input{Diagrams/15_internalRef83.tikz}=
\input{Diagrams/15_internalRef84.tikz}\stackrel{\eqref{eq:rep-sequential-controlled}}{=}\input{Diagrams/15_internalRef85.tikz}$}\;.
\end{equation}
Now, considering the grey $G$-wires of the above diagram, one has
\begin{align}
    &\scalebox{0.6}{$\input{Diagrams/15_internalRef86.tikz}$}\stackrel{\eqref{eq:inverse-swap}}{=} \scalebox{0.6}{$\input{Diagrams/15_internalRef87.tikz}$}\stackrel{\eqref{eq:bialgebra}}{=}\scalebox{0.6}{$\input{Diagrams/15_internalRef88.tikz}$}
    \stackrel{\eqref{eq:associativity}}{=}\scalebox{0.6}{$\input{Diagrams/15_internalRef89.tikz}$}\nonumber\\=&\scalebox{0.6}{$\input{Diagrams/15_internalRef90.tikz}$}
    \stackrel{\eqref{copy-inverse}}{=}\scalebox{0.6}{$\input{Diagrams/15_internalRef91.tikz}$}\stackrel{\eqref{eq:identity-unit}}{=}\scalebox{0.6}{$\input{Diagrams/15_internalRef92.tikz}$}
    \stackrel{\eqref{eq:inverse-swap}}{=}\scalebox{0.6}{$\input{Diagrams/15_internalRef93.tikz}$} \nonumber\\
    \stackrel{\eqref{eq:inverse-swap}}{=}&\scalebox{0.6}{$\input{Diagrams/15_internalRef94.tikz}$}
    =\scalebox{0.6}{$\input{Diagrams/15_internalRef95.tikz}$}\stackrel{\eqref{eq:associativity}}{=}\scalebox{0.6}{$\input{Diagrams/15_internalRef96.tikz}$}
    \stackrel{\eqref{copy-inverse}}{=}\scalebox{0.6}{$\input{Diagrams/15_internalRef97.tikz}$}\stackrel{\eqref{eq:identity-unit}}{=}\scalebox{0.6}{$\input{Diagrams/15_internalRef98.tikz}$}\;.
\end{align}
Using this in~\cref{eq:pihat-invariance-proof} concludes the proof, as we have
\begin{equation}
   \scalebox{0.8}{$ \input{Diagrams/15_internalRef81.tikz}$}=    \scalebox{0.9}{$\input{Diagrams/15_internalRef99.tikz}$}=\scalebox{0.9}{$\input{Diagrams/15_internalRef82.tikz}$}\;.
\end{equation}
\endproof

\subsubsection{\texorpdfstring{Proof of \cref{thm:twirlInternalRef}}{Proof of Theorem \ref*{thm:twirlInternalRef}}}\label{app:twirlInternalRef}
\twirlInternalRef*
\proof
Using universal classical control, we can observe that in a twirled comb
\begin{equation}
\scalebox{0.7}{$\input{Diagrams/4_twirl112.tikz}$}\;,
\end{equation}
the state
\begin{equation}
\scalebox{0.7}{$\input{Diagrams/15_internalRef23.tikz}$}\;,
\end{equation}
can be thought of as a continuous function $G\to P^{A,E}$, and hence we have the associated classically controlled state in $P_G^{A,E}$, denoted by $\hat{\mathcal{S}}$, and depicted as
\begin{equation}
\label{eq:controlled-state}
\scalebox{0.7}{$\input{Diagrams/15_internalRef24.tikz}=\input{Diagrams/15_internalRef25.tikz}\;,\quad \forall g \in G$}.
\end{equation}
Similarly, the effect
\begin{equation}
\scalebox{0.7}{$\input{Diagrams/15_internalRef22.tikz}$}\;,
\end{equation}
can be thought of as a continuous function $G\to P_{B,E}$, so that one has the associated classically controlled effect in $P_{G,B,E}$, denoted by $\hat{\mathcal{E}}$, and depicted as
\begin{equation}
\scalebox{0.7}{$\input{Diagrams/15_internalRef20.tikz}=\input{Diagrams/15_internalRef21.tikz}\;,\quad \forall g \in G$}.
\end{equation}
We can now rewrite a twirled comb and then turn the classically controlled state into a state in $P^{G,A,E}$, the classically controlled effect into an effect in $P_{G,B,E}$ via
\begin{align}
\scalebox{0.7}{$\input{Diagrams/4_twirl112.tikz}
=\input{Diagrams/15_internalRef26.tikz}$}
\nonumber\\
=\scalebox{0.7}{$\input{Diagrams/15_internalRef27.tikz}
\quad\input{Diagrams/15_internalRef28.tikz}$}\;,
\end{align}
where one has
\begin{equation}\label{eq:state-s-prime}
\scalebox{0.7}{$\input{Diagrams/15_internalRef30.tikz}:=\input{Diagrams/15_internalRef29.tikz}$}\;,
\end{equation}
and
\begin{equation}
\scalebox{0.7}{$\input{Diagrams/15_internalRef32.tikz}:=\input{Diagrams/15_internalRef31.tikz}$}\;.
\end{equation}
Hence, a twirled comb is realisable by implementing $\mathcal{S}'$ and $\mathcal{E}'$. All that remains is to show that they are invariant under twirling.

First, note that we can write the state $\hat{\mathcal{S}}$ and the effect $\hat{\mathcal{E}}$ as
\begin{equation}
\label{eq:state-s}
  \scalebox{0.7}{$  \input{Diagrams/15_internalRef100.tikz}=     \input{Diagrams/15_internalRef101.tikz}$}\;,
\end{equation}
and
\begin{equation}
\label{eq:effect-e}
\scalebox{0.7}{$ \input{Diagrams/15_internalRef59.tikz}=\input{Diagrams/15_internalRef60.tikz}$}\;,
\end{equation}
so that for the state, we have
\begin{align}
\scalebox{0.65}{$\input{Diagrams/15_internalRef36.tikz}= \input{Diagrams/15_internalRef37.tikz} \stackrel{\eqref{eq:state-s}}{=}\input{Diagrams/15_internalRef74.tikz}$}\nonumber\\
    =\scalebox{0.65}{$\input{Diagrams/15_internalRef75.tikz}
     =\input{Diagrams/15_internalRef76.tikz}$}\nonumber\\
\stackrel{\eqref{eq:haar-invariance-twirl-controlled}}{=}\;\;\scalebox{0.65}{$\input{Diagrams/15_internalRef77.tikz}
\stackrel{\eqref{eq:state-s}}{=}\input{Diagrams/15_internalRef29.tikz}=\;\;\input{Diagrams/15_internalRef30.tikz}$}\;,
\end{align}
where in the third line we have used~\cref{cor:invariance-under-twirl-pihat}. For the effect, we have
\begin{align}
    \scalebox{0.7}{$\input{Diagrams/15_internalRef56.tikz}=\input{Diagrams/15_internalRef57.tikz}$}
=\scalebox{0.7}{$\input{Diagrams/15_internalRef102.tikz}$}\nonumber\\
\scalebox{0.7}{$
\input{Diagrams/15_internalRef103.tikz}
=\input{Diagrams/15_internalRef31.tikz}=\quad\input{Diagrams/15_internalRef32.tikz}$}\;,
\end{align}
where in the second line we have used~\cref{cor:invariance-under-twirl-pihat}.

Hence, a twirled comb is realisable using invariant states and effects within a GPT with internal classical system $G$ and universal classical control, which implies process tomography completeness (see~\cref{prop:twirledcombtomography}).
\endproof

\subsubsection{\texorpdfstring{Proof of \cref{prop:TwirlGTheory}}{Proof of Proposition \ref*{prop:TwirlGTheory}}}\label{proof:TwirlGTheory}
\twirlGTheory*
\proof
We verify the conditions of~\cref{def:GPT} in turn.

\noindent\textit{(i) Systems.}
By definition, the systems of $\mathsf{Twirl}_G[\mathcal{P}]$ are pairs $(A,\pi)$, where $A\in\mathbf{Sys}[\mathcal{P}]$ and $\pi$ is a system action on $A$, and include the trivial system equipped with the trivial action.

\noindent\textit{(ii) Process spaces.}
By definition, the process spaces of $\mathsf{Twirl}_G[\mathcal{P}]$ are the corresponding invariant subprocess spaces of the process spaces of $\mathcal{P}$.

\noindent\textit{(a) Identities.}
For every system $(A,\pi)\in \mathbf{Sys}\left[\mathsf{Twirl}_G\left[\mathcal{P}\right]\right]$ the identity is invariant as one has
\begin{equation}
    \scalebox{0.9}{$\input{Diagrams/4_twirl123.tikz}\stackrel{\eqref{eq:invertibility-of-representation}}{=}\input{Diagrams/4_twirl76.tikz}$}\;.
\end{equation}

\noindent\textit{(b) Swaps.}
Swap can also be shown to be invariant, that is, for any choice of $(A,\pi)$ and $(B,\chi)$, one has
\begin{align}
   \scalebox{0.90}{$ \input{Diagrams/4_twirl81.tikz}$}\stackrel{\eqref{eq:invertibility-of-representation}}{=}\scalebox{0.95}{$\input{Diagrams/4_twirl124.tikz}$}=\scalebox{0.95}{$\input{Diagrams/4_twirl125.tikz}$}\;.
\end{align}

\noindent\textit{(c) Scalars.}
Since scalars are invariant under any induced process action of the form~\cref{eq:appropriate-induced-process action} (proof similar to~\cref{cor:twirlingIsScalarPreserving}), $\mathsf{Twirl}_G\left[\mathcal{P} \right]$ contains scalar space $ P^I_I = \left(\{1\}, [0,1], \mathbb{R}\right)$ as well.

\noindent\textit{(d) States and effects.}
First consider a trivial input system in $\mathsf{Twirl}_G[\mathcal{P}]$. Since each $\pi_g$ is a reversible physical process, it preserves normalised states. Moreover, by the same argument as in~\cref{proof:twirled_is_subtheory}, a subnormalised state $s=p\omega$ in $\mathsf{Twirl}_G[\mathcal{P}]$ is invariant if and only if its normalised component $\omega$ is invariant. Hence, the corresponding subprocess space of $\mathsf{Twirl}_G[\mathcal{P}]$ is a state space. Now consider a trivial output system in $\mathsf{Twirl}_G[\mathcal{P}]$. The unit effect is invariant, and if $e$ is an invariant effect with complement $e^\perp$, then
\begin{equation}
\mathcal{T}(e^\perp)=\mathcal{T}(u_A-e)=u_A-e=e^\perp.
\end{equation}
Hence, the corresponding subprocess space of $\mathsf{Twirl}_G[\mathcal{P}]$ is an effect space.

\noindent\textit{(iii) Composition maps.}
Composition of processes in $\mathsf{Twirl}_G[\mathcal{P}]$ is given by composition of the underlying processes in $\mathcal{P}$, and therefore inherits associativity and normalisation preservation from $\mathcal{P}$. To show the closedness of composition, we need to show that for every process $f\in \mathsf{Twirl}_G[\mathcal{P}]^{(D,\chi_D),(C,\chi_C)}_{(A,\pi_A)}$
\begin{equation}
  \scalebox{0.8}{$ \input{Diagrams/4_twirl128.tikz}$} \;,
\end{equation}
and every process $h\in \mathsf{Twirl}_G[\mathcal{P}]^{(E,\chi_E)}_{(C,\chi_C),(B,\pi_{B})}$
\begin{equation}
   \scalebox{0.8}{$\input{Diagrams/4_twirl129.tikz}$} \;,
\end{equation}
the composition
\begin{equation}
\scalebox{0.9}{$\input{Diagrams/4_twirl18.tikz} $}
\end{equation}
is closed, i.e., is a process in $\mathsf{Twirl}_G[\mathcal{P}]^{(D,\chi_D),(E,\chi_{E})}_{(A,\pi_A),(B,\pi_B)}$. To show this, first, note that $f$ and $h$ are invariant under group action (proof similar to~\cref{prop:group-invariance}), i.e.,
\begin{equation}
 \label{eq:f-invariant}
    \scalebox{0.8}{$\input{Diagrams/4_twirl130.tikz}$}\;,
\end{equation}
and
\begin{equation}
 \label{eq:h-invariant}
  \scalebox{0.8}{$\input{Diagrams/4_twirl131.tikz}$}\;.
\end{equation}
Now, we can show that the closedness of composition indeed holds as one has
\begin{align}
\scalebox{0.80}{$\input{Diagrams/4_twirl126.tikz} \stackrel{\eqref{eq:invertibility-of-representation}}{=}\input{Diagrams/4_twirl132.tikz}$}\nonumber\\
=\scalebox{0.80}{$\input{Diagrams/4_twirl127.tikz}$}\nonumber\\
\stackrel{\eqref{eq:f-invariant},\eqref{eq:h-invariant}}{=}\scalebox{0.80}{$\input{Diagrams/4_twirl133.tikz}=\input{Diagrams/4_twirl18.tikz}$}\;.
\end{align}

\noindent\textit{(iv) Bifunctoriality.}
Since parallel and sequential composition in $\mathsf{Twirl}_G[\mathcal{P}]$ are inherited from $\mathcal{P}$, they obey bifunctoriality.

\noindent\textit{(v) Process tomography.}
To show why $\mathsf{Twirl}_G[\mathcal{P}]$ might not satisfy process tomography, consider two distinct invariant processes $\tilde f\neq \tilde h \in \mathsf{Twirl}_G[\mathcal{P}]^{(B,\chi)}_{(A,\pi)}$. Process tomography of $\mathcal{P}$ implies that there exist $e \in P_{B,E}$ and $s\in P^{A,E}$ such that
\begin{equation}
   \scalebox{0.7}{$ \input{Diagrams/4_twirl20.tikz}\neq\input{Diagrams/4_twirl21.tikz}$}\;,
\end{equation}
where using \cref{eq:invertibility-of-representation} for $\tilde f$ implies that for an arbitrary action $\xi$ of the group $G$ on system $E$, for any $g\in G$, one has
\begin{align}
\scalebox{0.70}{$\input{Diagrams/4_twirl20.tikz}$}=\scalebox{0.70}{$\input{Diagrams/4_twirl134.tikz}$}\nonumber\\
=\scalebox{0.70}{$\input{Diagrams/4_twirl136.tikz}$}=\scalebox{0.70}{$\input{Diagrams/4_twirl137.tikz}$}\;,
\end{align}
where in the last equality we have used invariance of $\tilde f$ under group action. This implies that
\begin{equation}
   \scalebox{0.80}{$\input{Diagrams/4_twirl137.tikz}\neq\input{Diagrams/4_twirl138.tikz}$} \;.
\end{equation}
Hence, combs of the form
\begin{equation}
 \scalebox{0.70}{$ \input{Diagrams/4_twirl139.tikz} $}\;,
\end{equation}
suffice for process tomography in $\mathsf{Twirl}_G[\mathcal{P}]$. Note, however, that it is not clear whether such combs can be realised using only states and effects within $\mathsf{Twirl}_G[\mathcal{P}]$, that is, there may not exist a state $s'$ and an effect $e'$
\begin{equation}
  \scalebox{0.70}{$\input{Diagrams/4_twirl139.tikz}  =  \input{Diagrams/4_twirl140.tikz}$}\;.
\end{equation}
Hence, $\mathsf{Twirl}_G[\mathcal{P}]$ may not satisfy process tomography.
\endproof
\subsubsection{\texorpdfstring{Proof of \cref{prop:IncoherenSimulation}}{Proof of Proposition \ref*{prop:IncoherenSimulation}}}\label{app:IncorherenSimulation}
\incoherentSimulation*
\proof
It is straightforward to verify that~\Cref{eq:incohorent-simulation} is linear, since for every collection of processes $f_i \in P^{B_1,\cdots,B_m}_{A_1,\cdots,A_n}$ one has $\$_\mathcal{T}(\sum_i r_i f_i)
=
\sum_i r_i \$_\mathcal{T}(f_i)$, where $r_i \in \Rl$ and $\sum_i r_i f_i \in P^{B_1,\cdots,B_m}_{A_1,\cdots,A_n}$.\\
Incoherent simulation maps every process to a $\mathcal{T}$-invariant process, as
\begin{align}
\label{eq:twirl-incohorent}
\scalebox{0.80}{$\input{Diagrams/4_twirl153.tikz}$}
\stackrel{\eqref{eq:twirl-controlled}}{=}
\scalebox{0.80}{$\input{Diagrams/4_twirl154.tikz}$}\nonumber\\
=
\scalebox{0.8}{$\input{Diagrams/4_twirl155.tikz}$}\nonumber\\
\stackrel{\eqref{eq:rep-sequential-controlled}}{=}
\scalebox{0.8}{$\input{Diagrams/4_twirl156.tikz}$}\;.
\end{align}
Now, since
\begin{equation}
\scalebox{0.9}{$\input{Diagrams/4_twirl157.tikz}
\stackrel{\eqref{eq:inverse-swap}}{=}
\input{Diagrams/4_twirl158.tikz}
\stackrel{\eqref{eq:associativity}}{=}
\input{Diagrams/4_twirl159.tikz}
\stackrel{\eqref{copy-inverse}}{=}
\input{Diagrams/4_twirl160.tikz}
\stackrel{\eqref{eq:identity-unit}}{=}
\input{Diagrams/4_twirl161.tikz}$}\;,
\end{equation}
then~\cref{eq:twirl-incohorent} gives
\begin{align}
\scalebox{0.80}{$\input{Diagrams/4_twirl153.tikz}$}
=
\scalebox{0.80}{$\input{Diagrams/4_twirl162.tikz}$}\nonumber\\
\stackrel{\eqref{eq:haar-invariance}}{=}
\quad\scalebox{0.80}{$\input{Diagrams/4_twirl163.tikz}$}
=
\quad\scalebox{0.80}{$\input{Diagrams/4_twirl151.tikz}$}\;.
\end{align}
One can also show that incoherent simulation is connectivity-preserving, as
\begin{align}
\scalebox{0.80}{$\input{Diagrams/4_twirl167.tikz}$}
=
\scalebox{0.80}{$\input{Diagrams/4_twirl168.tikz}$}\nonumber\\
\stackrel{\eqref{eq:invertibility-of-representation}}{=}
\scalebox{0.80}{$\input{Diagrams/4_twirl169.tikz}$}
=\scalebox{0.80}{$\input{Diagrams/4_twirl170.tikz}$}\;.
\end{align}
\endproof

\subsection{Invariant subtheories from \texorpdfstring{$T$}{T}-actions}
\subsubsection{\texorpdfstring{Proof of~\cref{prop:swirledSubspaceBeingSubtheory}}{Proof of Proposition \ref*{prop:swirledSubspaceBeingSubtheory}}}\label{proof:swirled_is_subtheory}
\swirledSubspaceBeingSubtheory*

\proof
We verify that $\mathcal{A}(\mathcal{P})$ satisfies the conditions in the definition of a subtheory,~\cref{def:subGPT}.

\noindent\textit{(i) Systems.}
By definition, $\mathcal{A}(\mathcal{P})$ has the same systems as $\mathcal{P}$, including the trivial system.

\noindent\textit{(ii) Subprocess spaces.}
For every pair of input and output systems, the corresponding process space of $\mathcal{A}(\mathcal{P})$ is the invariant subprocess space of the corresponding process space of $\mathcal{P}$.

\noindent\textit{(a) Identities.}
It follows from the diagram preservation of $\Xi$ as in~\cref{eq:swap-identity-preservation}, that for every system $A\in\textbf{Sys}[\mathcal{P}]$, the identity $\mathcal{I}_A$ is invariant
\begin{equation}
    \input{Diagrams/5_swirl11n.tikz}=\input{Diagrams/5_swirl12.tikz}=\input{Diagrams/4_twirl77.tikz}\;.
\end{equation}

\noindent\textit{(b) Swaps.}
The same holds for the swap as well, that is, for every pair of systems $A,B\in\textbf{Sys}[\mathcal{P}]$, for the swap $\sigma_{A,B}$, using diagram preservation of $\Xi$, one has
\begin{equation}
   \input{Diagrams/5_swirl13n.tikz} =\input{Diagrams/5_swirl14.tikz}=\input{Diagrams/4_twirl81.tikz}\;.
\end{equation}
Thus, for every $A\in \textbf{Sys}[\mathcal{P}]$, one has $\mathcal{I}_A\in \mathcal{A}(P^A_A)$, and for every $A,B\in \textbf{Sys}[\mathcal{P}]$, one has $\sigma_{A,B}\in\mathcal{A}(P_{A,B}^{B,A})$.

\noindent\textit{(c) Scalars.}
Similarly, scalars are invariant under averaging, so the scalar space of $\mathcal{A}(\mathcal{P})$ coincides with that of the original theory: $\mathcal{A}(P^I_I) = P^I_I = \left(\{1\}, [0,1], \mathbb{R}\right)$.

\noindent\textit{(d) States and effects.}
For a trivial input system, since each $\Xi_g$ is a process-space isomorphism, it follows as in the preceding arguments that it preserves normalisation of states. Hence, an invariant subnormalised state $s=p\omega$ has an invariant normalised component $\omega$, and conversely scalar multiples of invariant normalised states are invariant. Thus, the corresponding invariant subprocess space is a state space. For a trivial output system, $\Xi_g$ maps the unit effect $u_A$, which is the unique normalised effect, to a normalised effect. Since $u_A$ is the only normalised effect, it follows that $\Xi_g(u_A)=u_A$. The rest follows similarly: by linearity, the complement of every invariant effect is also invariant, so the corresponding invariant subprocess space is an effect space. 

\noindent\textit{(iii) Composition maps.}
Sequential and parallel composition in $\mathcal{A}(\mathcal{P})$ are given by the corresponding composition maps of the original theory $\mathcal{P}$, and hence inherit bifunctoriality from $\mathcal{P}$.

\noindent\textit{(iv) Closure under composition.}
Closure of composition also follows from~\cref{eq:haar-measure-invariance-swirl1}, with the same logic as for the twirled case in~\cref{eq:composition-closure}, except that one needs to replace $\Pi$ with $\Xi$. Thus, composing invariant processes yields a process that is again invariant under averaging. This, in turn, implies that both sequential and parallel composition in $\mathcal{A}(\mathcal{P})$ are well defined.
\endproof

\subsubsection{\texorpdfstring{Proof of~\cref{prop:xihatInvariance}}{Proof of Proposition \ref*{prop:xihatInvariance}}}\label{app:xihatInvariance}
\xiHatInvariance*
\proof
First, note that
 \begin{align}
   \scalebox{0.9}{$ \input{Diagrams/16_internalRefSwirl26.tikz}$}
    =\scalebox{0.9}{$ \input{Diagrams/16_internalRefSwirl37.tikz}$}
    \stackrel{\eqref{eq:seq-compoistion-xihat}}{=}
    \scalebox{0.9}{$\input{Diagrams/16_internalRefSwirl38.tikz}$}\;.
    \end{align}
    Since for a process space $P^B_A$ the action of $\hat{\Xi}_A^B$ is equal to $\hat{\Xi}$, we have that
    \begin{align}
\scalebox{0.9}{$\input{Diagrams/16_internalRefSwirl38.tikz}$}=\scalebox{0.9}{$\input{Diagrams/16_internalRefSwirl51.tikz}$}\stackrel{\eqref{eq:associativity}}{=}\scalebox{0.9}{$\input{Diagrams/16_internalRefSwirl39.tikz}$}\nonumber\\
    \stackrel{\eqref{copy-inverse}}{=}\scalebox{0.9}{$
\input{Diagrams/16_internalRefSwirl40.tikz}$}\stackrel{\eqref{eq:identity-unit}}{=}
\scalebox{0.9}{$\input{Diagrams/16_internalRefSwirl27.tikz}$}\;.
\end{align}
\endproof

\subsubsection{\texorpdfstring{Proof of~\cref{thm:swirlInternalRef}}{Proof of Theorem \ref*{thm:swirlInternalRef}}}\label{app:swirlInternalRef}
\swirlInternalRef*
\proof
The proof is analogous to that of~\cref{thm:twirlInternalRef}. In particular, an arbitrary averaged comb can be realised using a classically controlled state in $P^{G,A,E}$ and a classically controlled effect in $P_{G,B,E}$, such that
\begin{equation}
\scalebox{0.7}{$\input{Diagrams/16_internalRefSwirl5.tikz}
=
\input{Diagrams/15_internalRef28.tikz}$}\;.
\end{equation}
Here, one has
\begin{equation}
\label{eq:state-s-swirl}
\scalebox{0.7}{$\input{Diagrams/15_internalRef30.tikz}
:=
\input{Diagrams/16_internalRefSwirl6.tikz}$}\;,
\end{equation}
and
\begin{equation}
\scalebox{0.7}{$\input{Diagrams/15_internalRef32.tikz}
:=
\input{Diagrams/16_internalRefSwirl7.tikz}$}\;.
\end{equation}
Both can be shown to be invariant by following the argument of~\cref{thm:twirlInternalRef}, replacing $\mathcal{T}$ with $\mathcal{A}$, $\Pi$ with $\Xi$, and hence $\hat{\Pi}$ with $\hat{\Xi}$, and using~\cref{cor:invariance-under-swirl-xihat} in place of~\cref{cor:invariance-under-twirl-pihat}. Therefore, every averaged comb can be realised using invariant states and invariant effects.
\endproof

\subsection{Local tomography}
\subsubsection{\texorpdfstring{Proof of~\cref{lem:rcorIdempotence}}{Proof of Lemma \ref*{lem:rcorIdempotence}}}\label{proof:rcorIdempotence}
\rcorIdempotence*
\proof
Let us first denote the action of a group on $L$ as $\Pi_{\Gamma}(g) (L)$ and depict it as
\begin{equation}
  \scalebox{0.8}{$  \input{Diagrams/7_LT69.tikz}$}\;,
\end{equation}
where the action of $\Pi_\Gamma$ is defined similarly to ~\cref{def:theory-sym-induced-by-sys} with $\pi$ being replaced by $\Gamma$. Then, the action of the group on $\mathcal{R}^{Cor}(L)$ is given by
\begin{align}
   \scalebox{0.7}{$ \input{Diagrams/7_LT70.tikz}=    \input{Diagrams/7_LT71.tikz}$}\nonumber\\
    \stackrel{\eqref{eq:haar-invariance}}{=} \scalebox{0.7}{$\input{Diagrams/7_LT72.tikz}=\input{Diagrams/7_LT73.tikz}$}\;
\end{align}
which implies that every process is $G$-invariant when acted on by $\mathcal{R}^{Cor}$.

For the idempotence, we have
\begin{align}
    \scalebox{0.6}{$ \input{Diagrams/7_LT74.tikz}=    \input{Diagrams/7_LT75.tikz}$}=\scalebox{0.6}{$\input{Diagrams/7_LT76.tikz}$}\;,
\end{align}
where we have used the $G$-invariance of $\mathcal{R}^{Cor}(L)$.
\endproof

\subsubsection{\texorpdfstring{Proof of~\cref{lem:rprodIdempotence}}{Proof of Lemma \ref*{lem:rprodIdempotence}}}\label{proof:rprodIdempotence}
\rprodIdempotence*
\proof
For the first part, we have that
\begin{align}
   \scalebox{0.6}{$\input{Diagrams/7_LT77.tikz} =\;\; \input{Diagrams/7_LT78.tikz}$}\;\;
     \scalebox{0.6}{$\stackrel{\eqref{eq:rep-sequential}}{=}\;\;\input{Diagrams/7_LT79.tikz}$}\nonumber\\\scalebox{0.6}{$\stackrel{\eqref{eq:haar-invariance},\eqref{eq:haar-invariance-inverse}}{=}\quad\input{Diagrams/7_LT80.tikz}\quad=\quad
     \input{Diagrams/7_LT81.tikz}$}\;.
\end{align}
The idempotence follows from
\begin{align}
   \scalebox{0.6}{$ \input{Diagrams/7_LT82.tikz}= \;\input{Diagrams/7_LT83.tikz}\stackrel{\eqref{eq:rep-sequential}}{=}\;\input{Diagrams/7_LT84.tikz}$}\nonumber\\
   \scalebox{0.6}{$ \stackrel{\eqref{eq:haar-invariance},\eqref{eq:haar-invariance-inverse}}{=}\;\input{Diagrams/7_LT85.tikz}\;=\;\input{Diagrams/7_LT86.tikz}$}
\end{align}
\endproof

\subsubsection{\texorpdfstring{Proof of~\cref{lem:projection}}{Proof of Lemma \ref*{lem:projection}}}\label{proof:projection}
\projection*
\proof
We have
\begin{align}
    \scalebox{0.6}{$\input{Diagrams/7_LT87r.tikz}=\input{Diagrams/7_LT90.tikz}\stackrel{\eqref{eq:rep-sequential}}{=}\input{Diagrams/7_LT91.tikz}$}\nonumber\\
    \scalebox{0.6}{$\stackrel{\eqref{eq:haar-invariance},\eqref{eq:haar-invariance-inverse}}{=}\input{Diagrams/7_LT92.tikz}=\input{Diagrams/7_LT88.tikz}$}\;,
\end{align}
and
\begin{align}
    \scalebox{0.6}{$\input{Diagrams/7_LT89.tikz}=\input{Diagrams/7_LT93.tikz}\stackrel{\eqref{eq:rep-sequential}}{=}\input{Diagrams/7_LT94.tikz}$}\nonumber\\
\scalebox{0.6}{$\stackrel{\eqref{eq:haar-invariance},\eqref{eq:haar-invariance-inverse}}{=}\input{Diagrams/7_LT92.tikz}=\quad\input{Diagrams/7_LT88.tikz}$}\;.
\end{align}
\endproof

\subsubsection{\texorpdfstring{Proof of~\cref{lem:rsemiIdempotence}}{Proof of Lemma \ref*{lem:rsemiIdempotence}}}\label{proof:rsemiIdempotence}
\rsemiIdempotence*
\proof
For the first part, we have that
\begin{align}\label{eq:rsemiInvarinace}
    \scalebox{0.6}{$\input{Diagrams/7_LT98.tikz}=\input{Diagrams/7_LT99.tikz}\stackrel{\eqref{eq:rep-sequential}}{=}\input{Diagrams/7_LT100.tikz}$}\nonumber\\
    \scalebox{0.6}{$\stackrel{\eqref{eq:bialgebra}}{=}\input{Diagrams/7_LT101.tikz}\stackrel{\eqref{eq:haar-invariance},\eqref{eq:haar-invariance-inverse}}{=}$\input{Diagrams/7_LT102.tikz}=\input{Diagrams/7_LT103.tikz}}\;.
\end{align}
Note that here we are using the generalised version of~\cref{eq:bialgebra}, where instead of two black dots as outputs, one has $n$ black dots. For the second part, we find
\begin{align}
   \scalebox{0.6}{$ \input{Diagrams/7_LT104.tikz}=\input{Diagrams/7_LT105.tikz}\stackrel{\eqref{eq:rsemiInvarinace}}{=}\input{Diagrams/7_LT106.tikz}$}\nonumber\\
   \scalebox{0.6}{$=\input{Diagrams/7_LT107.tikz}$}\;,
\end{align}
\endproof

\subsubsection{\texorpdfstring{Proof of~\cref{lem:projectionSemi}}{Proof of Lemma \ref*{lem:projectionSemi}}}\label{proof:projectionSemi}
\projectionSemi*
\proof
\begin{align}
   \scalebox{0.6}{$\input{Diagrams/7_LT97.tikz}=\input{Diagrams/7_LT108.tikz}\stackrel{\eqref{eq:rep-sequential}}{=}\input{Diagrams/7_LT109.tikz}$} \nonumber\\
   \scalebox{0.6}{$\stackrel{\eqref{eq:bialgebra},\eqref{eq:haar-invariance-inverse}}{=}\input{Diagrams/7_LT110.tikz}\stackrel{\eqref{eq:haar-invariance}}{=}\input{Diagrams/7_LT111.tikz}=\input{Diagrams/7_LT96.tikz}$}\;.
\end{align}
\begin{align}
    \scalebox{0.6}{$\input{Diagrams/7_LT95.tikz}=\input{Diagrams/7_LT112.tikz}\stackrel{\eqref{eq:rep-sequential}}{=}\input{Diagrams/7_LT113.tikz}$}\nonumber\\
    \scalebox{0.6}{$\stackrel{\eqref{eq:bialgebra},\eqref{eq:haar-invariance-inverse}}{=}\input{Diagrams/7_LT114.tikz}\stackrel{\eqref{eq:haar-invariance}}{=}\input{Diagrams/7_LT111.tikz}=\input{Diagrams/7_LT96.tikz}$}\;.
\end{align}
\begin{align}
    \scalebox{0.6}{$\input{Diagrams/7_LT87.tikz}=\input{Diagrams/7_LT117.tikz}\stackrel{\eqref{eq:rep-sequential}}{=}\input{Diagrams/7_LT119.tikz}$}\nonumber\\
    \scalebox{0.6}{$\stackrel{\eqref{eq:haar-invariance},\eqref{eq:haar-invariance-inverse}}{=}\input{Diagrams/7_LT120.tikz}\stackrel{\eqref{eq:haar-invariance},\eqref{eq:haar-invariance-inverse}}{=}\input{Diagrams/7_LT88.tikz}$}.
\end{align}
\begin{align}
    \scalebox{0.6}{$\input{Diagrams/7_LT116.tikz}=\input{Diagrams/7_LT121.tikz}\stackrel{\eqref{eq:rep-sequential}}{=}\input{Diagrams/7_LT122.tikz}$}\nonumber\\
    \scalebox{0.6}{$\stackrel{\eqref{eq:haar-invariance},\eqref{eq:haar-invariance-inverse}}{=}\input{Diagrams/7_LT120.tikz}=\input{Diagrams/7_LT88.tikz}$}.
\end{align}
Note that here we are using the generalised version of~\cref{eq:bialgebra} as well.
\endproof

\subsubsection{\texorpdfstring{Proof of~\cref{lem:InvariantSubprojection}}{Proof of Lemma \ref*{lem:InvariantSubprojection}}}\label{proof:InvariantSubprojection}
\InvariantSubprojection*
    \begin{proof}
Since $Q(\cdot)\neq\mathcal{R}^{Cor}(\cdot)$, there exists an ambient
process vector $x$ such that
\begin{equation}
v:=\mathcal{R}^{Cor}(x)-Q(x)\neq0.
\end{equation}
Using the idempotence of $\mathcal{R}^{Cor}(\cdot)$ and $Q(\cdot)$,
together with the absorption identities~\eqref{eq:absorption}, one obtains
\begin{equation}
\scalebox{0.90}{$ \mathcal{R}^{Cor}(v)=\mathcal{R}^{Cor}\!\left(\mathcal{R}^{Cor}(x)-Q(x)\right)=\mathcal{R}^{Cor}(x)-Q(x)=v$},
\end{equation}
and
\begin{equation}
Q(v)=Q\!\left(\mathcal{R}^{Cor}(x)-Q(x)\right)=Q(x)-Q(x)=0.
\end{equation}
Hence, $v\in\operatorname{Im}\mathcal{R}^{Cor}$.

Since physical processes span the ambient process vector space and
$\mathcal{R}^{Cor}$ maps physical processes onto physical invariant
processes, one has $\operatorname{span}\left(\cs(P^{B_1,\cdots,B_m}_{A_1,\cdots,A_n})\right)=\operatorname{Im}\mathcal{R}^{Cor}$. Moreover, $\cs(P^{B_1,\cdots,B_m}_{A_1,\cdots,A_n})$ is a nonempty convex subset of a
finite-dimensional vector space, and every such convex set has nonempty
relative interior when considered within its affine hull. Since
$0\in \cs(P^{B_1,\cdots,B_m}_{A_1,\cdots,A_n})$, this affine hull coincides with
$\operatorname{span}\left(\cs(P^{B_1,\cdots,B_m}_{A_1,\cdots,A_n})\right)$.
Indeed, any affine subspace
containing $\cs(P^{B_1,\cdots,B_m}_{A_1,\cdots,A_n})$ also contains $0$, and is therefore a
linear subspace. Hence, one can choose a physical invariant process
$\tilde f\in\operatorname{relint}\left(\cs(P^{B_1,\cdots,B_m}_{A_1,\cdots,A_n})\right)$. By the defining property of the relative interior, one can move a sufficiently small amount from $\tilde f$ in either direction along any vector in
$\operatorname{span}\left(\cs(P^{B_1,\cdots,B_m}_{A_1,\cdots,A_n})\right)$ while remaining within
$\cs(P^{B_1,\cdots,B_m}_{A_1,\cdots,A_n})$. Since
$v\in\operatorname{span}\left(\cs(P^{B_1,\cdots,B_m}_{A_1,\cdots,A_n})\right)$ and $v\neq0$, there therefore exists $\varepsilon>0$ such that
\begin{align}
\tilde f_1:=\tilde f+\varepsilon v\in \cs(P^{B_1,\cdots,B_m}_{A_1,\cdots,A_n}),\\
\tilde f_2:=\tilde f-\varepsilon v\in \cs(P^{B_1,\cdots,B_m}_{A_1,\cdots,A_n}).
\end{align}
Since $v\neq0$, one has $\tilde f_1\neq\tilde f_2$. Finally, using
$Q(v)=0$, one obtains
\begin{align}
Q(\tilde f_1)
&=
Q(\tilde f)+\varepsilon Q(v)
=
Q(\tilde f),
\\
Q(\tilde f_2)
&=
Q(\tilde f)-\varepsilon Q(v)
=
Q(\tilde f).
\end{align}
Therefore, $Q(\tilde f_1)=Q(\tilde f_2)$.

This can be generalised to two projections $Q$ and $Q'$ satisfying $\Im Q \subsetneq \Im Q'$, for which there exist two distinct invariant vectors $\tilde v_1\neq \tilde v_2$ such that $Q'(\tilde v_1)\neq Q'(\tilde v_2)$, while $Q(\tilde v_1)=Q(\tilde v_2)$.

\end{proof}

\subsubsection{\texorpdfstring{Proof of~\cref{prop:fexistence}}{Proof of Proposition \ref*{prop:fexistence}}}\label{proof:fexistence}
\fexistence*
\proof
By~\cref{prop:LT_twirl_gen}, failure of local tomography implies that $\mathcal{R}^{Cor}(\cdot)\neq\mathcal{R}^{Prod}(\cdot)$. Since the physical processes span the corresponding process vector space, there must exist a process $f\in P_{A_1,\ldots,A_n}^{B_1,\ldots,B_m}$ such that
$\mathcal{R}^{Cor}(f)\neq\mathcal{R}^{Prod}(f)$. For the twirling case, we may therefore take $\tilde f_1=\mathcal{R}^{Cor}(f)$ and $\tilde f_2=\mathcal{R}^{Prod}(f)$. In the swirling case, $\mathcal{R}^{Prod}(f)$ need not be physical, so this choice cannot be used directly. However, we will show that the argument can nevertheless be generalised to the swirling case as well.

To see why these two invariant processes agree on all product (separable) invariant  states and effects, consider process space $P^{B_1,\cdots,B_m}_{A_1,\cdots,A_n}$, where using invariant local effect $\tilde e_j \in \mathcal{T}(P_{B_j})$ and invariant local states $\tilde s_i\in \mathcal{T}(P^{A_i})$, one has
\begin{align}\label{eq:localInvTomography}
     \scalebox{0.8}{$\input{Diagrams/7_LT41.tikz}$}
     =
     \scalebox{0.8}{$\input{Diagrams/7_LT42.tikz}$}
     \stackrel{\eqref{eq:rprod}}{=}
    \scalebox{0.8}{$ \input{Diagrams/7_LT43.tikz}$}\;.
\end{align}
Hence,
\begin{align}\label{eq:localInvTomography2}
   \scalebox{0.8}{$ \input{Diagrams/13_ubq36.tikz}$}
    =
   \scalebox{0.8}{$ \input{Diagrams/7_LT46.tikz}$}
   =
   \scalebox{0.75}{$ \input{Diagrams/7_LT44.tikz}$}
   \nonumber\\
   \stackrel{\eqref{eq:projection}}{=}
   \scalebox{0.8}{$ \input{Diagrams/7_LT47.tikz}$}
    =
    \scalebox{0.8}{$\input{Diagrams/13_ubq37.tikz}$}\;.
\end{align}
Note that for the process $\mathcal{R}^{Semi}(f)$, using the same logic, one can show that it agrees with both $\tilde f_1$ and $\tilde f_2$ on all product states.

For the swirling case, we use~\cref{lem:InvariantSubprojection} and set $Q:=\mathcal{R}^{Prod}$. Then, there exist $\tilde f_1,\tilde f_2\in\cs(P_{A_1,\ldots,A_n}^{B_1,\ldots,B_m})$ such that $\mathcal{R}^{Prod}(\tilde f_1)=\mathcal{R}^{Prod}(\tilde f_2)$. Applying the same reasoning as in~\cref{eq:localInvTomography} and~\cref{eq:localInvTomography2}, it follows that $\tilde{f}_1$ and $\tilde{f}_2$ agree on all product states and effects formed from local invariant states and effects.

The result here extends to convex combinations of product states and the corresponding sums of product effects by linearity, so that it is valid for separable states and effects as well.
\endproof

\subsubsection{\texorpdfstring{Proof of ~\cref{prop:entagledSuffice}}{Proof of Proposition \ref*{prop:entagledSuffice}}}\label{proof:entagledSuffice}
\proof
Consider the case $\mathcal{R}^{Cor}(\cdot)=\mathcal{R}^{Semi}(\cdot)$. For the twirling case, since $\tilde f\in\ct(P_{A_1,\ldots,A_n}^{B_1,\ldots,B_m})$ can be obtained by applying $\mathcal{R}^{Cor}$ to a process $f\in P_{A_1,\ldots,A_n}^{B_1,\ldots,B_m}$ from the original process space, this condition implies that any two distinct invariant processes can be written as $\tilde f_1:=\mathcal{R}^{Semi}(f_1)\neq\mathcal{R}^{Semi}(f_2)=:\tilde f_2$, where $f_1,f_2\in P_{A_1,\ldots,A_n}^{B_1,\ldots,B_m}$. Note that, in the swirling case, $\mathcal{R}^{Cor}$ yields a physical invariant process. Therefore, whenever the condition $\mathcal{R}^{Cor}(\cdot)=\mathcal{R}^{Semi}(\cdot)$ is satisfied, $\mathcal{R}^{Semi}$ also yields a physical process in the swirling case. Consequently, the proof of the ``if'' direction applies to the swirling case as well.

By local tomography of the original theory, there exist local states $s_i\in P^{A_i}$ and local effects $e_j\in P_{B_j}$ that distinguish these two processes, i.e.,
\begin{equation}\label{eq:fprod-fsemi}
    \scalebox{0.9}{$\input{Diagrams/7_LT48.tikz}
    \neq
    \input{Diagrams/7_LT54.tikz}$}\;.
\end{equation}
For the LHS, we have
\begin{align}
\scalebox{0.8}{$\input{Diagrams/7_LT48.tikz}$}
&=
\scalebox{0.8}{$\input{Diagrams/7_LT49.tikz}$}\nonumber\\
=
\scalebox{0.7}{$\input{Diagrams/7_LT50.tikz}$}
&=
\scalebox{0.7}{$\input{Diagrams/7_LT51.tikz}$}\;,
\end{align}
where we have used the idempotence of $\mathcal{R}^{Semi}$. By~\cref{def:TNLstateEffect}, we can rewrite the above equation as
\begin{equation}\label{eq:fprod-entangled}
   \scalebox{0.9}{$ \input{Diagrams/7_LT48.tikz}
    =
    \input{Diagrams/7_LT57.tikz}$}\;,
\end{equation}
so that from~\cref{eq:fprod-fsemi}, one has
\begin{equation}
   \scalebox{0.9}{$\input{Diagrams/7_LT57.tikz}
    \neq
    \input{Diagrams/13_ubq54.tikz}$}\;.
\end{equation}
Thus, these two processes can be distinguished using invariant states and effects of the form defined in~\cref{def:TNLstateEffect}.

For the other direction, suppose that, for the invariant process space of interest, process tomography can be achieved without ancillary systems. That is, any two distinct invariant processes can be distinguished by some invariant state $\tilde s$ and invariant effect $\tilde e$, without introducing ancillary systems. This does not require the distinguishing state and effect to be separable: as discussed above, there exist distinct invariant processes that cannot be distinguished using only separable invariant states and effects. Thus, in what follows, $\tilde s$ and $\tilde e$ are allowed to be arbitrary invariant states and effects.

Therefore, if two invariant processes yield identical statistics for every invariant state $\tilde s$ and invariant effect $\tilde e$, then one has the implication
\begin{align}
 \scalebox{0.9}{$ \input{Diagrams/7_LT123.tikz}
   =
    \input{Diagrams/7_LT124.tikz}  ,\forall \tilde s,
    \forall \tilde e $}\nonumber\\ \scalebox{0.9}{$\Rightarrow \input{Diagrams/7_LT125.tikz}=\input{Diagrams/7_LT126.tikz}$}\;.
\end{align}
Now consider $\tilde f_1:=\mathcal{R}^{Cor}(f)$ and $\tilde f_2:=\mathcal{R}^{Semi}(f)$ for a process $f\in P_{A_1,\ldots,A_n}^{B_1,\ldots,B_m}$. Then one has
\begin{align}\label{eq:fcor-fsemi-equivalence}
      \scalebox{0.8}{$ \input{Diagrams/7_LT127.tikz}$}= \scalebox{0.8}{$\input{Diagrams/7_LT59.tikz}$}
    =
     \scalebox{0.75}{$\input{Diagrams/7_LT61.tikz}$}\nonumber\\
\stackrel{\eqref{eq:projectionSemi}}{=}\scalebox{0.8}{$\input{Diagrams/7_LT62.tikz}$}=\scalebox{0.8}{$\input{Diagrams/7_LT63.tikz}$}\;,
\end{align}
and hence $\mathcal{R}^{Cor}(f)=\mathcal{R}^{Semi}(f)$ for every $f\in P_{A_1,\ldots,A_n}^{B_1,\ldots,B_m}$. Since physical processes span the corresponding ambient vector space, it follows that $\mathcal{R}^{Cor}(\cdot)=\mathcal{R}^{Semi}(\cdot)$.

Note that the proof of the ``only if'' direction applies directly only to the twirling case, since the equation above requires $\mathcal{R}^{Semi}(f)$ to be a physical process. However, we can generalise the argument to the swirling case as well. To this end, consider a non-zero vector $v\in\spann \left(\cs(P_{A_1,\ldots,A_n}^{B_1,\ldots,B_m})\right)$. Since $\cs(P_{A_1,\ldots,A_n}^{B_1,\ldots,B_m})$ is a nonempty convex subset of a finite-dimensional vector space containing $0$, by the same argument as in~\cref{lem:InvariantSubprojection}, for any non-zero $v\in\spann\left(\cs(P_{A_1,\ldots,A_n}^{B_1,\ldots,B_m})\right)$,
there exist a physical invariant process $\tilde f_0$ and a sufficiently small
$\epsilon>0$ such that $\tilde f_0+\epsilon v$ and
$\tilde f_0-\epsilon v$ are distinct physical invariant processes. Since there exist an invariant state $\tilde s$ and an invariant effect $\tilde e$ that can distinguish these two processes, we have
\begin{align}
\tilde e\circ(\tilde f_0+\epsilon v)\circ \tilde s
\neq
\tilde e\circ(\tilde f_0-\epsilon v)\circ \tilde s
\Rightarrow
\tilde e\circ v\circ\tilde s\neq 0,
\end{align}
so that testers of the above form separate every non-zero invariant vector $v\in\spann \left(\cs(P_{A_1,\ldots,A_n}^{B_1,\ldots,B_m})\right)$.

Now, consider $f^{Cor}=\mathcal{R}^{Cor}(f)$ for some $f\in P_{A_1,\ldots,A_n}^{B_1,\ldots,B_m}$. Since $\mathcal{R}^{Semi}$ is a subprojection of $\mathcal{R}^{Cor}$, one has $\mathcal{R}^{Semi}(f)\in \mathsf{Im}\mathcal{R}^{Cor}=\spann \left(\cs(P_{A_1,\ldots,A_n}^{B_1,\ldots,B_m})\right)$, which, using~\cref{eq:projectionSemi} and following the same logic as in~\cref{eq:fcor-fsemi-equivalence}, gives
\begin{align}
\tilde e\circ\mathcal{R}^{Cor}(f)\circ \tilde s
&=
\tilde e\circ\mathcal{R}^{Semi}\left(\mathcal{R}^{Cor}(f)\right)\circ \tilde s
\nonumber\\
&\stackrel{\eqref{eq:projectionSemi}}{=}
\tilde e\circ\mathcal{R}^{Semi}(f)\circ \tilde s.
\end{align}
Both $\mathcal{R}^{Cor}(f)$ and $\mathcal{R}^{Semi}(f)$ lie in $\spann \left(\cs(P_{A_1,\ldots,A_n}^{B_1,\ldots,B_m})\right)$ and yield identical values under any such tester without ancilla. Since, as shown above, these testers separate this invariant vector space, it follows that
$\mathcal{R}^{Cor}(f)=\mathcal{R}^{Semi}(f)$ for every physical process $f\in P_{A_1,\ldots,A_n}^{B_1,\ldots,B_m}$. Finally, since physical processes span their corresponding ambient real vector space, we obtain $\mathcal{R}^{Cor}(\cdot)=\mathcal{R}^{Semi}(\cdot)$.
\endproof

\subsubsection{\texorpdfstring{Proof of~\cref{prop:pairwiseProcessesEntangled}}{Proof of Proposition \ref*{prop:pairwiseProcessesEntangled}}}\label{proof:pairwiseProcessesEntangled}
\pairwiseProcessesEntangled*

\proof
If $\mathcal{R}^{Semi}(\tilde f_1)\neq \mathcal{R}^{Semi}(\tilde f_2)$, then, by local tomography of the original theory, one has
\begin{equation}
      \scalebox{0.8}{$ \input{Diagrams/7_LT128.tikz}\neq\input{Diagrams/7_LT129.tikz}$}\;
\end{equation}
where $s_i \in P^{A_i}$ and $e_j\in P_{B_j}$. Note that, even when $\mathcal{R}^{Semi}$ yields non-physical vectors, the above expression still holds. Indeed, local tomography of the original theory, together with the fact that physical processes span their corresponding ambient vector space, implies that local states and effects can also distinguish any two distinct vectors in this ambient vector space.

Applying the same reasoning as in the if direction of~\cref{prop:entagledSuffice}, this implies
\begin{equation}
   \scalebox{0.9}{$\input{Diagrams/7_LT57.tikz}
    \neq
    \input{Diagrams/13_ubq54.tikz}$}\;,
\end{equation}
where $\tilde s_{TNL}$ and $\tilde e_{TNL}$ are the invariant state and effect defined in~\cref{eq:entangled-state-effect}. Therefore, the two distinct processes can be distinguished using invariant states and effects of this form.

Conversely, suppose that the two processes can be distinguished using an invariant state $\tilde s$ and an invariant effect $\tilde e$, without ancillary systems. Then one has
\begin{equation}\label{eq:entangledDistinguish}
      \scalebox{0.9}{$ \input{Diagrams/7_LT130.tikz}\neq\input{Diagrams/7_LT131.tikz}$}\;.
\end{equation}
The left-hand side can be rewritten as
\begin{equation}
     \scalebox{0.8}{$ \input{Diagrams/7_LT130.tikz}=\input{Diagrams/7_LT132.tikz}=\input{Diagrams/7_LT133.tikz}$}\;.
\end{equation}
It therefore follows from~\cref{eq:entangledDistinguish} that
\begin{equation}
    \scalebox{0.8}{$ \input{Diagrams/7_LT133.tikz}\neq \input{Diagrams/7_LT134.tikz}$}\;,
\end{equation}
which implies that $\mathcal{R}^{Semi}(\tilde f_1)\neq \mathcal{R}^{Semi}(\tilde f_2)$, even when $\mathcal{R}^{Semi}$ yields a nonphysical vector.

\endproof

\subsubsection{\texorpdfstring{Proof of~\cref{prop:ancillaNeeded}}{Proof of Proposition \ref*{prop:ancillaNeeded}}}\label{proof:ancillaNeeded}
\ancillaNeeded*
\proof
If $\mathcal{R}^{Cor}(\cdot)\neq\mathcal{R}^{Semi}(\cdot)$, then, since physical process spaces span their corresponding ambient vector spaces, there exists a process $f\in P_{A_1,\ldots,A_n}^{B_1,\ldots,B_m}$ such that $\mathcal{R}^{Cor}(f)\neq\mathcal{R}^{Semi}(f)$. We therefore, for the twirling case, define $\tilde f_1:=\mathcal{R}^{Cor}(f)$ and $\tilde f_{2}:=\mathcal{R}^{Semi}(f)$. These two processes yield identical statistics for every invariant state $\tilde s$ and invariant effect $\tilde e$ when tested without ancillary systems, since
\begin{align}\label{eq:entangledTester}
      \scalebox{0.8}{$ \input{Diagrams/7_LT135.tikz}$}= \scalebox{0.8}{$\input{Diagrams/7_LT59.tikz}$}
  =
     \scalebox{0.75}{$\input{Diagrams/7_LT61.tikz}$}\nonumber\\
\stackrel{\eqref{eq:projectionSemi}}{=}\scalebox{0.8}{$\input{Diagrams/7_LT62.tikz}$}=\scalebox{0.8}{$\input{Diagrams/7_LT63.tikz}$}\;.
\end{align}
For the swirling case, we use~\cref{lem:InvariantSubprojection} and take $Q=\mathcal{R}^{Semi}$. Then, there exist distinct invariant processes $\tilde f_1\neq \tilde f_2\in \cs(P_{A_1,\ldots,A_n}^{B_1,\ldots,B_m})$ such that $\mathcal{R}^{Semi}(\tilde f_1)=\mathcal{R}^{Semi}(\tilde f_2)$. Using the same logic as in~\cref{eq:entangledTester}, this implies that no invariant state and invariant effect without ancillary systems can distinguish $\tilde f_1$ from $\tilde f_2$, since they yield the same statistic for both processes.

Nevertheless, since the original theory is locally tomographic,~\cref{prop:process-tomography-from-local-tomography} implies that these two processes can be distinguished using ancillary systems, i.e.,
\begin{equation}
\exists\,E,\tilde{s},\tilde{e}\ \text{such that}\
\scalebox{0.9}{$
\input{Diagrams/13_ubq57.tikz}
\neq
\input{Diagrams/13_ubq56.tikz}
$}\;,
\end{equation}
\endproof
\subsubsection{\texorpdfstring{Proof of~\cref{prop:pairwiseProcessesAncilla}}{Proof of Proposition \ref*{prop:pairwiseProcessesAncilla}}}\label{proof:pairwiseProcessesAncilla}
\pairwiseProcessesAncilla*
\proof
If $\mathcal{R}^{Semi}(\tilde f_1)= \mathcal{R}^{Semi}(\tilde f_2)$, then for any invariant state $\tilde s$ and invariant effect $\tilde e$ without ancillary systems one has
\begin{align}
     \scalebox{0.8}{$ \input{Diagrams/7_LT130.tikz}=\input{Diagrams/7_LT132.tikz}=\input{Diagrams/7_LT133.tikz}$}\nonumber\\
     =\scalebox{0.8}{$ \input{Diagrams/7_LT134.tikz}=\input{Diagrams/7_LT136.tikz}=\input{Diagrams/7_LT131.tikz}$}\;,
\end{align}
which remains true even when $\mathcal{R}^{Semi}$ yields a nonphysical vector. But since the original theory is locally tomographic, based on ~\cref{prop:process-tomography-from-local-tomography}, we have that
\begin{equation}
\exists\,E,\tilde{s},\tilde{e}\ \text{such that}\
\scalebox{0.9}{$
\input{Diagrams/7_LT137.tikz}
\neq
\input{Diagrams/7_LT138.tikz}
$}\;.
\end{equation}

The ``only if'' direction is the contrapositive of~\cref{prop:pairwiseProcessesEntangled}.
\endproof

\subsubsection{\texorpdfstring{Proof of~\cref{thm:LTfailureIrreps}}{Proof of Theorem \ref*{thm:LTfailureIrreps}}}\label{proof:LTfailureIrreps}
\LTfailureIrreps*
\begin{proof}
Throughout the proof, we use the isomorphism
$\Hom(\bigotimes_{i=1}^n V_{A_i},\bigotimes_{j=1}^m V_{B_j})
\cong \bigotimes_{i=1}^{n+m}V_{C_i}$ to identify the ambient process space with the tensor product of the wire spaces. In particular, this allows us to compare the images of $\mathcal R^{\mathrm{Cor}}$, $\mathcal R^{\mathrm{Prod}}$, and the partition projectors $\mathcal R^{[\cc_1|\cc_2]}$.

Suppose first that there exists a bipartition $[\cc_1|\cc_2]$ such that $V_{\cc_1}$ and $V_{\cc_2}$ share a non-trivial irreducible representation. By~\Cref{lem:G_inv_GtimesG}, one has
\begin{equation}
    (V_{\cc_1}\otimes V_{\cc_2})^G
    \neq
    V_{\cc_1}^G\otimes V_{\cc_2}^G.
\end{equation}
Since $\bigotimes_{C_i\in C}V_{C_i}^G\subseteq V_{\cc_1}^G\otimes V_{\cc_2}^G$, it follows that
\begin{equation}
    \left(\bigotimes_{C_i\in C}V_{C_i}\right)^G
    \neq
    \bigotimes_{C_i\in C}V_{C_i}^G.
\end{equation}
Thus $\mathcal R^{\mathrm{Cor}}\neq\mathcal R^{\mathrm{Prod}}$, and local tomography fails by \cref{prop:LT_twirl_gen}.

Conversely, suppose that local tomography fails. Then
\begin{equation}
    V_C^G
    \neq
    \bigotimes_{C_i\in C}V_{C_i}^G.
\end{equation}
Hence there exists a subset with this property, since in the worst case we may take the whole set $C$. As $C$ is finite, choose a subset $D\subseteq C$ of minimal size such that
\begin{equation}
    V_D^G
    \neq
    \bigotimes_{C_i\in D}V_{C_i}^G.
\end{equation}
Necessarily $|D|\geq2$. Write $D=D_1\sqcup D_2$, with $D_1,D_2\neq\emptyset$. Since $D_1$ and $D_2$ are proper subsets of $D$, minimality of $D$ gives
\begin{equation}
    V_{D_1}^G=\bigotimes_{C_i\in D_1}V_{C_i}^G,
    \qquad
    V_{D_2}^G=\bigotimes_{C_i\in D_2}V_{C_i}^G.
\end{equation}
Therefore
\begin{equation}
    (V_{D_1}\otimes V_{D_2})^G
    \neq
    V_{D_1}^G\otimes V_{D_2}^G,
\end{equation}
and~\Cref{lem:G_inv_GtimesG} implies that $V_{D_1}$ and $V_{D_2}$ share a non-trivial irreducible representation $\lambda$.

If $D=C$, then $[D_1|D_2]$ is already the required bipartition. Suppose instead that $D\subsetneq C$. By~\Cref{lem:trivial-irrep-system}, every wire $C_i\in C\setminus D$ contains a trivial representation, so their tensor product $V_{C\setminus D}$ also contains the trivial representation $\mathbf{1}$. Since $\lambda$ occurs in $V_{D_2}$, it follows that $\lambda\otimes\mathbf 1\cong\lambda$ occurs in $V_{D_2}\otimes V_{C\setminus D}=V_{C\setminus D_1}$. Hence $\lambda$ occurs in both $V_{D_1}$ and $V_{C\setminus D_1}$, so the bipartition $[D_1|C\setminus D_1]$ has the required shared non-trivial irreducible representation.
\end{proof}
\subsubsection{\texorpdfstring{Proof of~\cref{prop:WhenAncillaNeededIrreps}}{Proof of Lemma \ref*{prop:WhenAncillaNeededIrreps}}}\label{proof:WhenAncillaNeededIrreps}
 \WhenAncillaNeededIrreps*
\proof
Suppose that the total input representation $V_A$ and the total output representation $V_B$ share a single non-trivial irreducible representation $\lambda$. Then $V_A$ contains a copy of $\lambda$, and $V_B$ also contains a copy of $\lambda$. Since these two copies are equivalent irreducible representations, there exists a non-zero $G$-equivariant map $f_\lambda$ between them.

Because $f_\lambda$ is $G$-equivariant, it is invariant under the correlated action of the group on the input and output. Therefore, $f_\lambda\in\Im\mathcal R^{Cor}$, and hence $\mathcal R^{Cor}(f_\lambda)=f_\lambda\neq 0$.

On the other hand, $\mathcal R^{Semi}$ averages the total input and total output independently. In particular, the output averaging projects the image of $f_\lambda$ onto the $G$-invariant subspace of $V_B$. However, the image of $f_\lambda$ lies entirely in the non-trivial irreducible sector $\lambda$, and a non-trivial irreducible representation contains no non-zero $G$-invariant vectors. Hence, $\mathcal R^{Semi}(f_\lambda)=0$.

Therefore, there exists a map that is preserved by $\mathcal R^{Cor}$ but removed by $\mathcal R^{Semi}$, and hence $\mathcal R^{Cor}\neq\mathcal R^{Semi}$. By~\cref{prop:ancillaNeeded}, this implies that ancillary systems are required in order to fulfil process tomography.

The same argument extends directly to the case where $V_A$ and $V_B$ share more than one non-trivial irreducible representation. For each shared irreducible representation $\lambda_k$, there exists a non-zero equivariant map $f_{\lambda_k}$ supported on that sector. Each such map is preserved by $\mathcal R^{Cor}$ and removed by $\mathcal R^{Semi}$. Hence, the existence of one or more common non-trivial irreducible representations implies $\mathcal R^{Cor}\neq\mathcal R^{Semi}$.

Conversely, if $V_A$ and $V_B$ do not share a non-trivial representation, then by~\Cref{lem:G_inv_GtimesG}, under the canonical identification, one has $\Hom_G(V_A,V_B)=\Hom(V_A^G,V_B^G)$. Therefore both $\calr^\Semi$ and $\calr^\Cor$ have the same image. Since they are both orthogonal projectors, $\calr^\Semi=\calr^\Cor$, and hence by~\Cref{prop:entagledSuffice} ancillary systems are not needed for process tomography.
\endproof
\subsubsection{\texorpdfstring{Proof of~\cref{prop:DisjointIrrep}}{Proof of Proposition \ref*{prop:DisjointIrrep}}}\label{proof:DisjointIrrep}
\DisjointIrrep*
\begin{proof}
We prove the statement for input states. The proof for output effects is identical.

Let $\Pi_{\mathrm{sep}}$ denote the partition in which $\cc_1$ and $\cc_2$ are separate blocks, and let $\Pi_{\mathrm{joint}}$ denote the same partition except that $\cc_1$ and $\cc_2$ are replaced by the single block $\cc_1\cup\cc_2$. All remaining input and output blocks are left unchanged. Writing $W$ for the tensor product of the invariant subspaces associated with these remaining blocks, we have
\begin{equation}
    \operatorname{Im}\mathcal R^{\Pi_{\mathrm{sep}}}
    =
V_{\cc_1}^G\otimes V_{\cc_2}^G\otimes W,
\end{equation}
whereas
\begin{equation}
    \operatorname{Im}\mathcal R^{\Pi_{\mathrm{joint}}}
    =
    (V_{\cc_1}\otimes V_{\cc_2})^G\otimes W.
\end{equation}
Since, by~\Cref{lem:trivial-irrep-system}, every wire contains a trivial representation, $W$ is non-zero.

Suppose first that $V_{\cc_1}$ and $V_{\cc_2}$ share a non-trivial irreducible representation. By~\Cref{lem:G_inv_GtimesG},
\begin{equation}
    (V_{\cc_1}\otimes V_{\cc_2})^G
    \neq
    V_{\cc_1}^G\otimes V_{\cc_2}^G.
\end{equation}
It follows that
\begin{equation}
    \operatorname{Im}\mathcal R^{\Pi_{\mathrm{sep}}}
    \subsetneq
    \operatorname{Im}\mathcal R^{\Pi_{\mathrm{joint}}}.
\end{equation}
By the same argument as in~\cref{lem:InvariantSubprojection}, there therefore exist distinct physical invariant processes $\tilde f_1$ and $\tilde f_2$ such that
\begin{equation}
    \mathcal R^{\Pi_{\mathrm{sep}}}(\tilde f_1)
    =
    \mathcal R^{\Pi_{\mathrm{sep}}}(\tilde f_2),
    \qquad
    \mathcal R^{\Pi_{\mathrm{joint}}}(\tilde f_1)
    \neq
    \mathcal R^{\Pi_{\mathrm{joint}}}(\tilde f_2).
\end{equation}
By the same logic as in~\cref{prop:pairwiseProcessesAncilla}, the first equality implies that $\tilde f_1$ and $\tilde f_2$ give the same statistics for every tester compatible with $\Pi_{\mathrm{sep}}$, and hence for every invariant input state that is separable across $[\cc_1|\cc_2]$. On the other hand, the second inequality implies that there exists a tester compatible with $\Pi_{\mathrm{joint}}$ that distinguishes $\tilde f_1$ and $\tilde f_2$. Since the only difference between $\Pi_{\mathrm{sep}}$ and $\Pi_{\mathrm{joint}}$ is that $\cc_1$ and $\cc_2$ are treated jointly, the input state of such a distinguishing tester cannot be separable across $[\cc_1|\cc_2]$. Hence some entangled invariant input state across $[\cc_1|\cc_2]$ is required for complete process tomography.

Conversely, suppose that $V_{\cc_1}$ and $V_{\cc_2}$ share no non-trivial irreducible representation. By~\Cref{lem:G_inv_GtimesG},
\begin{equation}
    (V_{\cc_1}\otimes V_{\cc_2})^G
    =
    V_{\cc_1}^G\otimes V_{\cc_2}^G.
\end{equation}
Therefore
\begin{equation}
    \operatorname{Im}\mathcal R^{\Pi_{\mathrm{sep}}}
    =
    \operatorname{Im}\mathcal R^{\Pi_{\mathrm{joint}}}.
\end{equation}
Hence treating $\cc_1$ and $\cc_2$ jointly provides no additional distinguishing power over input states that are separable across $[\cc_1|\cc_2]$. Thus entanglement across this bipartition is not required.

The same argument, with input states replaced by output effects, proves the corresponding statement for the output wires.
\end{proof}
\subsubsection{\texorpdfstring{Proof of~\cref{prop:SharedIrrepTomography}}{Proof of Proposition \ref*{prop:SharedIrrepTomography}}}\label{proof:SharedIrrepTomography}
\SharedIrrepTomography*
\begin{proof}
    Grouping all systems in each input block $\mathcal A_k$ into a single effective system gives
$V_{\mathrm{in}} \cong V_{\mathcal A_1}\otimes\cdots\otimes V_{\mathcal A_r}$.
Since each block shares no non-trivial irreducible representation with the tensor product of the remaining blocks, repeated application of~\Cref{lem:G_inv_GtimesG} gives, under this identification,
$(V_{\mathrm{in}})^G = (V_{\mathcal A_1})^G\otimes\cdots\otimes (V_{\mathcal A_r})^G$.
Hence the input part of $\mathcal R^{\mathrm{Semi}}$ coincides with the projector induced by $\Pi_{\mathrm{in}}^{\mathrm{SI}}$. The same argument shows that the output part of $\mathcal R^{\mathrm{Semi}}$ coincides with the projector induced by $\Pi_{\mathrm{out}}^{\mathrm{SI}}$.

Since ancillary systems are not required, $\mathcal R^{\mathrm{Cor}}=\mathcal R^{\mathrm{Semi}}$. Thus, $\mathcal R^{\mathrm{Cor}}$ is the projector obtained by applying the partition projector induced by $\Pi_{\mathrm{in}}^{\mathrm{SI}}$ on the inputs and that induced by $\Pi_{\mathrm{out}}^{\mathrm{SI}}$ on the outputs. Following the same reasoning as in~\cref{prop:entagledSuffice}, it follows that invariant input states and invariant output effects that are separable between distinct blocks suffice to distinguish any pair of distinct invariant processes.

For the second statement, consider a shared-irrep requirement $[C_1|C_2]$ defining one of the blocks. By definition, $C_1$ and $C_2$ are disjoint genuine carriers of the same non-trivial irreducible representation. Hence, by~\Cref{lem:G_inv_GtimesG},
$(V_{C_1}\otimes V_{C_2})^G \neq V_{C_1}^G\otimes V_{C_2}^G$.
Therefore, the projector in which $C_1$ and $C_2$ are treated jointly has strictly larger image than the projector in which they remain separated. Since ancillary systems are not required, $\mathcal R^{\mathrm{Cor}}=\mathcal R^{\mathrm{Semi}}$, and following the same reasoning as in~\cref{prop:entagledSuffice}, restricting the tomography scheme to invariant input states separable across $[C_1|C_2]$ would therefore be insufficient to distinguish all invariant processes. Hence the tomography scheme must include entangled invariant input states across $[C_1|C_2]$. Applying this argument independently to every shared-irrep requirement defining the block proves the second statement; the states associated with different requirements need not be the same.
\end{proof}
\subsubsection{\texorpdfstring{Proof of~\cref{prop:AncillaAssisitedTesters}}{Proof of Proposition \ref*{prop:AncillaAssisitedTesters}}}\label{proof:AncillaAssisitedTesters}
\AncillaAssisitedTesters*
\proof
For each block $\mathcal C_k$, group all input systems contained in $\mathcal C_k$ into the composite system $\mathcal A_k$, and all output systems into the composite system $\mathcal B_k$. Let
$\mathcal R_k^{Cor}$ and $\mathcal R_k^{Semi}$ denote, respectively, the correlated and semi-correlated projectors acting on $\Hom\!\left(V_{\mathcal A_k},V_{\mathcal B_k}\right)$.

By~\Cref{lem:vec_space_twirl_loc}, the invariant processes on this effective single-input/single-output process span $\Hom_G\!\left(V_{\mathcal A_k},V_{\mathcal B_k}\right)$ for both twirling and swirling. Using the $G$-equivariant identification
$\Hom(V_{\mathcal A_k},V_{\mathcal B_k})
\cong V_{\mathcal A_k}\otimes V_{\mathcal B_k}$, the images of
$\mathcal R_k^{Cor}$ and $\mathcal R_k^{Semi}$ are identified with $\left(V_{\mathcal A_k}\otimes V_{\mathcal B_k}\right)^G$ and
$V_{\mathcal A_k}^G\otimes V_{\mathcal B_k}^G$, respectively.

By~\Cref{lem:schur_red_rep}, one has
\begin{equation}
\scalebox{0.85}{$\Hom_G\!\left(V_{\mathcal A_k},V_{\mathcal B_k}\right)
\cong
\Hom\!\left(V_{\mathcal A_k}^G,V_{\mathcal B_k}^G\right)
\oplus
\bigoplus_{\lambda\in\Lambda_k}
\Hom_G\!\left(
V_{\mathcal A_k}^{(\lambda)},
V_{\mathcal B_k}^{(\lambda)}
\right)$}.
\end{equation}
Hence the non-trivial sectors common to the input and output of the block are precisely those labelled by $\lambda\in\Lambda_k$.

Suppose first that $\Lambda_k=\varnothing$. Then $V_{\mathcal A_k}$ and $V_{\mathcal B_k}$ share no non-trivial irreducible representation, and therefore~\Cref{lem:G_inv_GtimesG} gives $\left(V_{\mathcal A_k}\otimes V_{\mathcal B_k}\right)^G
=
V_{\mathcal A_k}^G\otimes V_{\mathcal B_k}^G$. Thus $\mathcal R_k^{Cor}$ and $\mathcal R_k^{Semi}$ have the same image and, since they are the corresponding orthogonal projectors, $\mathcal R_k^{Cor}=\mathcal R_k^{Semi}$.

By~\Cref{prop:entagledSuffice}, invariant states on $\mathcal A_k$ and invariant effects on $\mathcal B_k$ therefore suffice without an ancillary system. This also covers $\mathcal A_k=\varnothing$ or $\mathcal B_k=\varnothing$, since the missing side is the trivial system.

Now let $\lambda\in\Lambda_k$. By~\Cref{lem:schur_red_rep}, there exists
\begin{equation}
0\neq
f_\lambda
\in
\Hom_G\!\left(
V_{\mathcal A_k}^{(\lambda)},
V_{\mathcal B_k}^{(\lambda)}
\right),
\end{equation}
which we regard as an element of
$\Hom(V_{\mathcal A_k},V_{\mathcal B_k})$ by taking it to vanish on the other isotypic components. Since $f_\lambda$ is $G$-equivariant, $\mathcal R_k^{Cor}(f_\lambda)=f_\lambda$. On the other hand, $\mathcal R_k^{Semi}$ averages the input and output independently. Since $\lambda$ is non-trivial,
\begin{equation}
\left(V_{\mathcal A_k}^{(\lambda)}\right)^G
=
\left(V_{\mathcal B_k}^{(\lambda)}\right)^G
=
0,
\end{equation}
and hence $\mathcal R_k^{Semi}(f_\lambda)=0$. Thus the correlated projector preserves the $\lambda$-sector whereas the semi-correlated projector removes it. In particular, one has $\mathcal R_k^{Cor}\neq\mathcal R_k^{Semi}$, and, by~\Cref{lem:InvariantSubprojection,prop:pairwiseProcessesAncilla}, this sector cannot in general be resolved using invariant states on $\mathcal A_k$ and invariant effects on $\mathcal B_k$ without an ancillary system.

Now, let $E$ be an ancillary system. Let $\widetilde s\in(V_{\mathcal A_k}\otimes V_E)^G$ be an invariant input state, and define
\begin{equation}
\widetilde s_\lambda
:=
(P_\lambda\otimes\mathcal I_E)\widetilde s,
\end{equation}
where $P_\lambda$ is the projector onto the $\lambda$-isotypic component $V_{\mathcal A_k}^{(\lambda)}$ of $V_{\mathcal A_k}$. Since $f_\lambda$ vanishes on every other isotypic component, it acts only on $\widetilde s_\lambda$. Hence
\begin{equation}
(f_\lambda\otimes\mathcal I_E)\widetilde s
=
(f_\lambda\otimes\mathcal I_E)\widetilde s_\lambda.
\end{equation}

Moreover, since $\widetilde s$ is $G$-invariant, applying the group action leaves it unchanged. As $P_\lambda$ commutes with the $G$-action, projecting onto the $\lambda$-isotypic component before or after applying the group gives the same result. Hence the projected component
$\widetilde s_\lambda=(P_\lambda\otimes\mathcal I_E)\widetilde s$
is also $G$-invariant. Since its $\mathcal A_k$-part lies in $V_{\mathcal A_k}^{(\lambda)}$, one has
\begin{equation}
\widetilde s_\lambda
\in
\left(
V_{\mathcal A_k}^{(\lambda)}
\otimes V_E
\right)^G.
\end{equation}
Hence, if this invariant subspace is zero, every invariant input state has $\widetilde s_\lambda=0$, and therefore $f_\lambda$ acts trivially on every invariant input state. Thus, probing the $\lambda$-sector requires
\begin{equation}\label{eq:anciallIrrep}
\left(
V_{\mathcal A_k}^{(\lambda)}
\otimes V_E
\right)^G
\neq0.
\end{equation}
Since $\lambda$ is non-trivial, $\left(V_{\mathcal A_k}^{(\lambda)}\right)^G=0$, and since $V_{\mathcal A_k}^{(\lambda)}$ contains only copies of $\lambda$,~\Cref{lem:G_inv_GtimesG} implies that~\eqref{eq:anciallIrrep} holds if and only if $V_E$ contains $\lambda$. The same argument applies to invariant effects on $\mathcal B_kE$.

Conversely, suppose that $E_{k,\lambda}$ carries $\lambda$. Consider two invariant processes whose difference has a non-zero component
\begin{equation}
0\neq
\Delta_\lambda
\in
\Hom_G\!\left(
V_{\mathcal A_k}^{(\lambda)},
V_{\mathcal B_k}^{(\lambda)}
\right).
\end{equation}
Since $V_{\mathcal A_k}^{(\lambda)}$ is a direct sum of copies of $V_\lambda$, there is a copy of $V_\lambda$ on which $\Delta_\lambda$ is non-zero. Since $E_{k,\lambda}$ also carries $\lambda$, consider the invariant Choi vector between this copy and a copy of $V_\lambda$ in $E_{k,\lambda}$, that is $\Omega_\lambda
\in
\left(V_\lambda\otimes V_\lambda\right)^G$.

By~\Cref{eq:choi-isomorphism,cor:Inv_Choi}, the Choi vector is invariant and dynamically faithful on this sector. Therefore $\left(
\Delta_\lambda\otimes
\mathcal I_{E_{k,\lambda}}
\right)
\Omega_\lambda
\neq0$. Hence, if the two invariant processes are $\widetilde f$ and $\widetilde h$,
\begin{equation}
\left(
\widetilde f\otimes
\mathcal I_{E_{k,\lambda}}
\right)
\Omega_\lambda
\neq
\left(
\widetilde h\otimes
\mathcal I_{E_{k,\lambda}}
\right)
\Omega_\lambda.
\end{equation}
Since $\Omega_\lambda$ belongs to the span of the invariant state space on $\mathcal A_kE_{k,\lambda}$, the same spanning argument as in~\Cref{prop:dynamically-faithful-vector} implies that there exists a physical invariant state $\widetilde s_{k,\lambda}$ for which the two resulting invariant output states are distinct. By state tomography, there is therefore an invariant effect $\widetilde e_{k,\lambda}$ on $\mathcal B_kE_{k,\lambda}$ that distinguishes them. Hence an ancillary system carrying $\lambda$ suffices to resolve the $\lambda$-sector.

It remains to combine the blocks. Let $W_k:=V_{\mathcal C_k}$. By the defining separation property of $\Pi_{\mathcal C}^{\mathrm{SI}}$, $W_k$ shares no non-trivial irreducible representation with the joint representation of the remaining blocks. Repeated application of~\Cref{lem:G_inv_GtimesG} therefore gives
\begin{equation}
\left(
\bigotimes_{k=1}^r W_k
\right)^G
=
\bigotimes_{k=1}^r W_k^G.
\end{equation}
Using again the corresponding $G$-equivariant isomorphism, i.e., $\Hom(V,W)\cong V\otimes W$, this means that the image of the correlated projector on the full process space coincides with the image of the partition projector associated with
\begin{equation}
\Pi_{\mathcal C}^{\mathrm{SI}}
=
[\mathcal C_1|\cdots|\mathcal C_r].
\end{equation}
It follows that tomography can be performed independently on each block, as in the preceding shared-irrep block argument.

Using the notation of the proposition, for every $k\in K_0$ use invariant
states on $\mathcal A_k$ and invariant effects on $\mathcal B_k$. For every
$k\in K_+$ and every $\lambda_k\in\Lambda_k$, choose an ancillary system
$E_{k,\lambda_k}$ carrying $\lambda_k$, and use invariant states on
$\mathcal A_kE_{k,\lambda_k}$ and invariant effects on
$\mathcal B_kE_{k,\lambda_k}$. For every choice
$(\lambda_k)_{k\in K_+}$ with $\lambda_k\in\Lambda_k$, their parallel
composition gives testers of the form
\begin{equation}
\begin{aligned}
&\left(
\bigotimes_{k\in K_+}\widetilde e_{k,\lambda_k}
\otimes
\bigotimes_{k\in K_0}\widetilde e_k
\right)
\circ
\left(
\widetilde f
\otimes
\bigotimes_{k\in K_+}\mathcal I_{E_{k,\lambda_k}}
\right)
\\
&\qquad\circ
\left(
\bigotimes_{k\in K_+}\widetilde s_{k,\lambda_k}
\otimes
\bigotimes_{k\in K_0}\widetilde s_k
\right).
\end{aligned}
\end{equation}
Taking the union of these tester families over every choice
$(\lambda_k)_{k\in K_+}$ with $\lambda_k\in\Lambda_k$, that is, including
all testers obtained from every such choice of shared irreducible
representations, resolves every $\lambda_k\in\Lambda_k$ in every block
$k\in K_+$, while the blocks $k\in K_0$ are resolved without ancillary
systems. Hence this family gives complete process tomography.

\endproof

\subsection{Bilocal classical theory}

\subsubsection{\texorpdfstring{Proof of~\cref{Thm:BCTasTwirl}}{Proof of Theorem \ref*{Thm:BCTasTwirl}}}\label{Proof:BCTasTwirl}
\BCTasTwirl*
\proof
To begin, noting that twirling $\mathcal{T}$ acts linearly and that every process in the even-dimensional classical theory can be written uniquely as a linear combination of the form~\cref{eq:generic-procees}, we simply need to know the action on processes of the form
\begin{equation}
    \scalebox{0.8}{$\input{Diagrams/9_BCT60.tikz}:=\input{Diagrams/9_BCT5.tikz}$}\;.
\end{equation}

Next, we will see that twirling such processes leads to all of the processes in Def.~2.1 of Ref.~\cite{soltani2026decoupling}, and hence, that this $\mathbb Z_2$ twirled classical world is indeed equivalent to BCT as it is exactly the ontological model for BCT defined in Ref.~\cite{soltani2026decoupling}.

To begin, let us consider twirling an arbitrary such process:
\begin{align}
    \scalebox{0.8}{$\input{Diagrams/9_BCT59.tikz}$}=\scalebox{0.9}{$\sum_{g}\frac{1}{|\mathbb Z_2|}$}\scalebox{0.8}{$\input{Diagrams/9_BCT61.tikz}$}\nonumber\\
    \stackrel{\eqref{eq:xor-rep},\eqref{eq:inverse-binary}}{=}\scalebox{0.8}{$\input{Diagrams/9_BCT62.tikz}$}\;.
\end{align}
This implies that one only needs to consider twirling on binary subspaces, so that we have
\begin{align}
    \scalebox{0.80}{$\input{Diagrams/9_BCT63.tikz}$}
    \stackrel{\eqref{eq:mixed-state},\eqref{eq:fusion}}{=}
    \quad\scalebox{0.80}{$\frac{1}{2}\input{Diagrams/9_BCT65.tikz}$}\\
    \stackrel{\eqref{eq:moving-legs}}{=}
    \;\;\scalebox{0.8}{$\frac{1}{2}\;\;\input{Diagrams/9_BCT66.tikz}$}
    \stackrel{\eqref{eq:xor-not}}{=}\;\;\;\;\scalebox{0.8}{$\frac{1}{2}\input{Diagrams/9_BCT67.tikz}$}\\
    \stackrel{\eqref{eq:xor-not}}{=}
    \;\;\scalebox{0.8}{$\frac{1}{2}\input{Diagrams/9_BCT68.tikz}$}
    \stackrel{\eqref{eq:fusion}}{=}\;\;\scalebox{0.8}{$\frac{1}{2}\input{Diagrams/9_BCT69.tikz}$}\nonumber\\
    =\;\;\scalebox{0.8}{$\frac{1}{2}\input{Diagrams/9_BCT70.tikz}$}\;.
\end{align}
This gives us
\begin{equation}
\label{eq:atomic-bct}
   \scalebox{0.8}{$
   \input{Diagrams/9_BCT59.tikz}= \frac{1}{2}\input{Diagrams/9_BCT71.tikz}$}\;.
\end{equation}
 \blk
This is precisely Eq.~82 of Ref.~\cite{soltani2026decoupling}. Specialising this to the case where we have no inputs then gives Eq.~78 of Ref.~\cite{soltani2026decoupling} and to the case where we have no outputs gives Eq.~80 of Ref.~\cite{soltani2026decoupling}.  Note that for the specific case of single input and single output,~\cref{eq:atomic-bct} reduces to
\begin{equation}
\scalebox{0.8}{$  \frac{1}{2}\input{Diagrams/9_BCT75.tikz}$}\;,
\end{equation}
which implies that an arbitrary process of $\mathbf{BCT}$ can be written as
\begin{equation}
 \scalebox{0.8}{$\input{Diagrams/9_BCT81.tikz}  =\sum_{i,s,j,t}\frac{f^{j,t}_{i,s}}{2}\input{Diagrams/9_BCT75.tikz}$}\;,
\end{equation}
where $f^{j,t}_{i,s}\in[0,1]$, and $\sum_{j,t}f^{j,t}_{i,s}\leq 1$. Note that one can reparametrize as $\sigma:=s\oplus t$, and
\begin{equation}
\tilde{f}^{j,\sigma}_{i}:=\frac{f^{j,t}_{i,s}+f^{j,t\oplus1}_{i,s\oplus 1}}{2}=\frac{\sum_{u}f^{j,u}_{i,u\oplus \sigma}}{2},
\end{equation}
where $\tilde{f}^{j,\sigma}_{i}\in [0,1]$, and $\sum_{j,\sigma} \tilde{f}^{j,\sigma}_{i}\leq 1$ as we have
\begin{equation}
\scalebox{0.9}{$\sum_{j,\sigma}\tilde f_i^{j,\sigma}
=\frac12\sum_{j,\sigma,u}f_{i,u\oplus\sigma}^{j,u}
=\frac12\sum_{s,j,t}f_{i,s}^{j,t}
\leq \frac12\sum_{s}1
=1$},
\end{equation}
which gives
\begin{equation}
\scalebox{0.8}{$ \input{Diagrams/9_BCT81.tikz}  =\sum_{i,j,\sigma}\tilde{f}_{i}^{j,\sigma}\input{Diagrams/9_BCT82i.tikz}$}\;.
\end{equation}
For a pure state of a single system,~\cref{eq:atomic-bct} gives
\begin{equation}
    \scalebox{0.8}{$\input{Diagrams/9_BCT86.tikz}$}\;,
\end{equation}
so that an arbitrary state $s$ in $\mathbf{BCT}$ is given as
\begin{equation}
\label{eq:bct-state}
  \scalebox{0.8}{$  \input{Diagrams/9_BCT87.tikz}=\sum_{i}\tilde{f}^i \input{Diagrams/9_BCT86.tikz}$}\;,
\end{equation}
where $\tilde{f}^i\in [0,1]$ and $\sum_{i}\tilde f^i\leq 1$.

For a pure effect of a single system,~\cref{eq:atomic-bct} gives
\begin{equation}
 \scalebox{0.8}{$  \frac{1}{2}  \input{Diagrams/9_BCT16.tikz}$}\;,
\end{equation}
 where for every arbitrary effect $e$, one has
\begin{equation}\label{eq:bct-effect}
 \scalebox{0.8}{$ \input{Diagrams/9_BCT83.tikz}=\sum_{i}\tilde{f}_{i} \;\input{Diagrams/9_BCT84.tikz}$}\;,
  \end{equation}
 with $\tilde{f}_i\in [0,1]$ and $\sum_{i}\tilde{f}_i\leq n$.
\endproof
\subsubsection{\texorpdfstring{Proof of~\cref{lem:BilocalStateTomography}}{Proof of Lemma \ref*{lem:BilocalStateTomography}}}\label{proof:BilocalStateTomography}
\BilocalStateTomography*
\proof

For the proof we begin by determining the invariant and
$\lambda$-isotypic sectors from the $\mathbb Z_2$ action. We then identify
the genuine carriers of $\lambda$, determine the pairwise correlations
required for tomography, show that the full invariant space is spanned by
one-system and two-system invariant sectors, and finally pass to physical
invariant effects to conclude tomography.

Write $\mathbb Z_2=\{e,g\}$, where $g$ is the non-trivial element. Over the reals, $\mathbb Z_2$ has only two irreducible representations: the trivial representation
$\mathbf 1$, on which $g$ acts as $+1$, and the non-trivial
one-dimensional representation $\lambda$, on which $g$ acts as $-1$.
Hence, for every system $C_i$ in a multipartite system $C_1\cdots C_n$, one has
\begin{equation}
V_{C_i}
=
V_{C_i}^G
\oplus
V_{C_i}^{(\lambda)},
\end{equation}
where $g$ acts as $+1$ on $V_{C_i}^G$ and as $-1$ on
$V_{C_i}^{(\lambda)}$. In particular,
$\lambda\otimes\lambda\cong\mathbf 1$.

We now expand the joint representation as
\begin{equation}
V_{C_1\cdots C_n}
=
\bigotimes_{i=1}^n
\left(
V_{C_i}^G\oplus V_{C_i}^{(\lambda)}
\right).
\end{equation}
Let $S$ be the set
of labels $i$ for which $V_{C_i}^{(\lambda)}$ is chosen. Thus $|S|$ is
exactly the number of $\lambda$-factors in that term. We denote this term by
\begin{equation}
T_S
:=
\left(
\bigotimes_{i\in S}V_{C_i}^{(\lambda)}
\right)
\otimes
\left(
\bigotimes_{j\notin S}V_{C_j}^G
\right).
\end{equation}
Hence $V_{C_1\cdots C_n}
=
\bigoplus_{S\subseteq\{1,\ldots,n\}}T_S$.

For example, for two systems,
\begin{equation}
\begin{aligned}
S=\varnothing
&:\quad
T_S=V_{C_1}^G\otimes V_{C_2}^G,
&& |S|=0,
\\
S=\{1\}
&:\quad
T_S=V_{C_1}^{(\lambda)}\otimes V_{C_2}^G,
&& |S|=1,
\\
S=\{2\}
&:\quad
T_S=V_{C_1}^G\otimes V_{C_2}^{(\lambda)},
&& |S|=1,
\\
S=\{1,2\}
&:\quad
T_S=V_{C_1}^{(\lambda)}\otimes V_{C_2}^{(\lambda)},
&& |S|=2.
\end{aligned}
\end{equation}
Thus $S$ runs over all subsets of $\{1,\ldots,n\}$, with each subset
specifying exactly which systems contribute their $\lambda$-sector.
Accordingly, $|S|$ is the number of $\lambda$-factors in $T_S$. For two systems, the non-trivial element $g$ acts on the four terms as
\begin{equation}
\begin{aligned}
V_{C_1}^G\otimes V_{C_2}^G
&\longmapsto (+1)(+1)=+1,\\
V_{C_1}^{(\lambda)}\otimes V_{C_2}^G
&\longmapsto (-1)(+1)=-1,\\
V_{C_1}^G\otimes V_{C_2}^{(\lambda)}
&\longmapsto (+1)(-1)=-1,\\
V_{C_1}^{(\lambda)}\otimes V_{C_2}^{(\lambda)}
&\longmapsto (-1)(-1)=+1.
\end{aligned}
\end{equation}
Thus each $\lambda$-factor contributes one factor of $-1$. In the general
term $T_S$, there are exactly $|S|$ such factors, so $g$ acts on $T_S$ as
$(-1)^{|S|}$. Hence
\begin{equation}
T_S\subseteq V_{C_1\cdots C_n}^G
\quad\Longleftrightarrow\quad
|S|\ \text{is even},
\end{equation}
while
\begin{equation}
T_S\subseteq V_{C_1\cdots C_n}^{(\lambda)}
\quad\Longleftrightarrow\quad
|S|\ \text{is odd}.
\end{equation}

We first determine the genuine carriers of $\lambda$. The bipartite case
illustrates the argument. For two systems,
\begin{equation}
V_{C_1C_2}^{(\lambda)}
=
\left(
V_{C_1}^{(\lambda)}\otimes V_{C_2}^G
\right)
\oplus
\left(
V_{C_1}^G\otimes V_{C_2}^{(\lambda)}
\right).
\end{equation}
These are precisely the two terms containing one $\lambda$-factor. The
first is inherited from $C_1$, since $C_2$ contributes only its invariant
sector, while the second is inherited from $C_2$. Indeed,
\begin{equation}
I_{C_1C_2}^{(\lambda)}
=
V_{C_1}^{(\lambda)}\otimes V_{C_2}^G
+
V_{C_1}^G\otimes V_{C_2}^{(\lambda)}.
\end{equation}
Thus
$V_{C_1C_2}^{(\lambda)}=I_{C_1C_2}^{(\lambda)}$, so $C_1C_2$ is not a
genuine carrier of $\lambda$ (See~\cref{def:GeniuneCarrier} for $I_{C_1C_2}^{(\lambda)}$).

The same reasoning applies to an arbitrary collection $\mathcal C$
containing more than one system. Since $V_{\mathcal C}^{(\lambda)}$ is
the direct sum of the terms $T_S$ with $|S|$ odd, consider one such
$T_S$ and choose any $i\in S$. The label $i$ singles out one system in
$S$, the index $j\in S\setminus\{i\}$ runs over the remaining systems in
$S$, and the index $k\notin S$ runs over the systems outside $S$. Thus
$C_i$ and every $C_j$ contribute their $\lambda$-sectors, while every
$C_k$ contributes its invariant sector. Hence
\begin{equation}
T_S
=
V_{C_i}^{(\lambda)}
\otimes
\left[
\left(
\bigotimes_{j\in S\setminus\{i\}}V_{C_j}^{(\lambda)}
\right)
\otimes
\left(
\bigotimes_{k\notin S}V_{C_k}^{G}
\right)
\right].
\end{equation}
The factor in brackets contains $|S|-1$ $\lambda$-factors. Since $|S|$
is odd, $|S|-1$ is even, so this factor is invariant. Therefore $T_S
\subseteq
V_{C_i}^{(\lambda)}
\otimes
V_{\mathcal C\setminus\{C_i\}}^G$.

Since $\{C_i\}$ is a non-empty proper subcollection of $\mathcal C$, this
subspace is included in the sum defining
$I_{\mathcal C}^{(\lambda)}$ (see~\Cref{def:GeniuneCarrier,eq:GeniuneCarrier}),
\begin{equation}
I_{\mathcal C}^{(\lambda)}
=
\sum_{\emptyset\neq\mathcal C'\subsetneq\mathcal C}
V_{\mathcal C'}^{(\lambda)}
\otimes
V_{\mathcal C\setminus\mathcal C'}^G.
\end{equation}
Thus every $T_S$ contributing to $V_{\mathcal C}^{(\lambda)}$ is contained
in $I_{\mathcal C}^{(\lambda)}$. Since
$V_{\mathcal C}^{(\lambda)}$ is the direct sum of all such $T_S$, it
follows that \begin{equation}
    V_{\mathcal C}^{(\lambda)}
\subseteq
I_{\mathcal C}^{(\lambda)}.
\end{equation}

For the converse, consider any subspace $V_{\mathcal C'}^{(\lambda)}
\otimes
V_{\mathcal C\setminus\mathcal C'}^G$ appearing in $I_{\mathcal C}^{(\lambda)}$. The non-trivial element $g$ acts as $-1$ on
$V_{\mathcal C'}^{(\lambda)}$ and as $+1$ on
$V_{\mathcal C\setminus\mathcal C'}^G$. Hence, for any
$v\in V_{\mathcal C'}^{(\lambda)}$ and
$w\in V_{\mathcal C\setminus\mathcal C'}^G$,
\begin{equation}
\Gamma_{\mathcal C}(g)(v\otimes w)
=
\left(\Gamma_{\mathcal C'}(g)v\right)
\otimes
\left(\Gamma_{\mathcal C\setminus\mathcal C'}(g)w\right)
=
-(v\otimes w).
\end{equation}
Thus every vector in
$V_{\mathcal C'}^{(\lambda)}
\otimes
V_{\mathcal C\setminus\mathcal C'}^G$
transforms according to $\lambda$. Therefore $V_{\mathcal C'}^{(\lambda)}
\otimes
V_{\mathcal C\setminus\mathcal C'}^G
\subseteq
V_{\mathcal C}^{(\lambda)}$.

Since every subspace appearing in the sum defining
$I_{\mathcal C}^{(\lambda)}$ is contained in
$V_{\mathcal C}^{(\lambda)}$, any vector in
$I_{\mathcal C}^{(\lambda)}$ is a finite sum of vectors that each belong
to $V_{\mathcal C}^{(\lambda)}$. Because
$V_{\mathcal C}^{(\lambda)}$ is a vector subspace and is therefore closed
under addition, that sum also belongs to
$V_{\mathcal C}^{(\lambda)}$. Hence
\begin{equation}
I_{\mathcal C}^{(\lambda)}
\subseteq
V_{\mathcal C}^{(\lambda)}.
\end{equation}
Therefore $V_{\mathcal C}^{(\lambda)}
=
I_{\mathcal C}^{(\lambda)}$.

Thus, for every collection $\mathcal C$ containing more than one system,
the entire $\lambda$-isotypic component is inherited from proper
subcollections. By the definition of genuine carrier, such a collection
is therefore not a genuine carrier of $\lambda$. Hence the only possible
genuine carriers are the individual systems.

For a single system $C_i$, $I_{C_i}^{(\lambda)}$
(see~\Cref{def:GeniuneCarrier}) contains no terms, because $\{C_i\}$ has
no non-empty proper subcollection. Hence
$I_{C_i}^{(\lambda)}=0$. In other words, for a single system there is no
smaller collection of systems from which a $\lambda$-sector could be
inherited. Since $C_i$ carries $\lambda$ by assumption,
$V_{C_i}^{(\lambda)}\neq0$. Therefore its $\lambda$-sector is necessarily
genuine, and $C_i$ is a genuine carrier of $\lambda$.

Thus the genuine carriers of $\lambda$ are precisely the individual
non-trivial systems. Since every such system carries the same
non-trivial irrep $\lambda$, every pair $C_i,C_j$ of non-trivial systems
defines the shared-irrep requirement $[C_i|C_j]$. Since there are no
other genuine carriers, there are no other shared-irrep requirements.

Having identified the genuine carriers, we now determine the pairwise
correlations required for tomography.

For any such pair, one has $V_{C_i}^{(\lambda)}\neq0$ and
$V_{C_j}^{(\lambda)}\neq0$. Hence
$V_{C_i}^{(\lambda)}\otimes V_{C_j}^{(\lambda)}$ is non-zero. Moreover,
$g$ acts as $-1$ on each factor, and therefore as
$1$ on their tensor product. Thus
\begin{equation}
V_{C_i}^{(\lambda)}
\otimes
V_{C_j}^{(\lambda)}
\subseteq
\left(
V_{C_i}\otimes V_{C_j}
\right)^G.
\end{equation}
By~\Cref{prop:DisjointIrrep}, effects that are all separable across
$[C_i|C_j]$ are insufficient for complete state tomography. Therefore
the tomographically complete family must contain some invariant effect
that correlates $C_i$ and $C_j$. Since this holds for every pair of
non-trivial systems, correlations across every such pair are required.

We now show that correlations involving more than two systems are not
required. Since the $T_S$ form a direct-sum decomposition of
$V_{C_1\cdots C_n}$, each vector has a unique component in every $T_S$, i.e.,
\begin{equation}
v
=
\sum_{S\subseteq\{1,\ldots,n\}}v_S,
\qquad
v_S\in T_S.
\end{equation}
Since $g$ acts on $T_S$ as $(-1)^{|S|}$, one has $gv=v$ if and only if $v_S=0$ for every $S$ with $|S|$ odd. Thus
\begin{equation}
V_{C_1\cdots C_n}^G
=
\bigoplus_{\substack{S\subseteq\{1,\ldots,n\}\\ |S|\ {\rm even}}}
T_S.
\end{equation}

Consider one such invariant subspace $T_S$. Since $|S|$ is even, write $|S|=2m$. Thus exactly $2m$ systems contribute a $\lambda$-factor to $T_S$. We
divide their labels into $m$ disjoint pairs,
\begin{equation}
(i_1,j_1),\ldots,(i_m,j_m),
\end{equation}
so that
\begin{equation}
S=\{i_1,j_1,\ldots,i_m,j_m\}.
\end{equation}
Here $a=1,\ldots,m$ labels the pairs, with the $a$th pair consisting of
the systems $C_{i_a}$ and $C_{j_a}$.

To see how this leads to bilocal tomography, consider first the
four-system case with $S=\{1,2,3,4\}$. Then
\begin{equation}
T_{\{1,2,3,4\}}
=
V_{C_1}^{(\lambda)}
\otimes
V_{C_2}^{(\lambda)}
\otimes
V_{C_3}^{(\lambda)}
\otimes
V_{C_4}^{(\lambda)}.
\end{equation}
Grouping the systems into the pairs $(C_1,C_2)$ and $(C_3,C_4)$ gives
\begin{equation}
T_{\{1,2,3,4\}}
=
\left(
V_{C_1}^{(\lambda)}
\otimes
V_{C_2}^{(\lambda)}
\right)
\otimes
\left(
V_{C_3}^{(\lambda)}
\otimes
V_{C_4}^{(\lambda)}
\right).
\end{equation}
Each factor on the right is a bipartite invariant subspace, since $g$
acts on each pair as $(-1)(-1)=+1$.

Let
\begin{equation}
W_{12}
:=
V_{C_1}^{(\lambda)}\otimes V_{C_2}^{(\lambda)},
\qquad
W_{34}
:=
V_{C_3}^{(\lambda)}\otimes V_{C_4}^{(\lambda)}.
\end{equation}
Then $T_{\{1,2,3,4\}}
=
W_{12}\otimes W_{34}$.

Suppose that $\{w_{12}^r\}_r$ spans $W_{12}$ and
$\{w_{34}^s\}_s$ spans $W_{34}$. Any $x\in W_{12}$ and
$y\in W_{34}$ can therefore be written as $x=\sum_r\alpha_r w_{12}^r$ and $y=\sum_s\beta_s w_{34}^s$. Hence $x\otimes y
=
\sum_{r,s}\alpha_r\beta_s\,
w_{12}^r\otimes w_{34}^s$.

Each $w_{12}^r$ is an invariant vector on the pair $C_1C_2$, and each
$w_{34}^s$ is an invariant vector on the pair $C_3C_4$. Hence every
tensor $w_{12}^r\otimes w_{34}^s$ is formed by tensoring together two
bipartite invariant vectors, one on $C_1C_2$ and one on $C_3C_4$.
Therefore, for every $z\in T_{\{1,2,3,4\}}$, there exist coefficients
$\gamma_{rs}$ such that
\begin{equation}
z
=
\sum_{r,s}\gamma_{rs}\,
w_{12}^r\otimes w_{34}^s.
\end{equation}
Thus every vector in $T_{\{1,2,3,4\}}$ is a linear combination of
tensor products of invariant vectors, each involving only two systems.

Returning to the general even set
\begin{equation}
S=\{i_1,j_1,\ldots,i_m,j_m\},
\end{equation}
we may reorder the tensor factors in $T_S$ according to the chosen pairs,
giving
\begin{equation}
T_S
=
\left(
\bigotimes_{a=1}^{m}
\left(
V_{C_{i_a}}^{(\lambda)}
\otimes
V_{C_{j_a}}^{(\lambda)}
\right)
\right)
\otimes
\left(
\bigotimes_{k\notin S}V_{C_k}^G
\right).
\end{equation}
Here $k\notin S$ runs over the systems that do not contribute a
$\lambda$-sector to $T_S$, and therefore contribute their invariant
sector $V_{C_k}^G$. Thus the first tensor product runs over the $m$ chosen pairs of systems
inside $S$, while the second runs over all systems outside $S$.

For each $a=1,\ldots,m$,
\begin{equation}
V_{C_{i_a}}^{(\lambda)}
\otimes
V_{C_{j_a}}^{(\lambda)}
\subseteq
\left(
V_{C_{i_a}}\otimes V_{C_{j_a}}
\right)^G,
\end{equation}
because $g$ acts as $(-1)(-1)=+1$. For every $k\notin S$,
$V_{C_k}^G$ is already a one-system invariant subspace. Hence, as in the
four-partite case, $T_S$ is a tensor product of bipartite invariant
subspaces
$V_{C_{i_a}}^{(\lambda)}\otimes V_{C_{j_a}}^{(\lambda)}$
and one-system invariant subspaces $V_{C_k}^G$. Therefore every factor
in this decomposition involves at most two systems.

Choose a spanning family for each of these factors. By the same
tensor-product argument used in the four-system example, tensoring one
element from each spanning family gives a spanning family for $T_S$.
Thus every vector in $T_S$ is a linear combination of tensor products
whose individual factors belong to invariant spaces involving at most
two systems.

Repeating this for every even $S$, and using
\begin{equation}
V_{C_1\cdots C_n}^G
=
\bigoplus_{\substack{S\subseteq\{1,\ldots,n\}\\ |S|\ {\rm even}}}
T_S,
\end{equation}
gives a spanning family for the entire invariant space
$V_{C_1\cdots C_n}^G$.

It remains to pass from this spanning statement to physical invariant
effects and conclude tomography.

By~\Cref{lem:vec_space_twirl_loc}, the invariant effect space on each
one-system or two-system block is spanned by physical invariant effects.
Therefore every vector appearing in the factors above can be written as
a linear combination of physical invariant effects on the corresponding
one-system or two-system block. Expanding the tensor products then shows
that the full multipartite invariant effect space is spanned by tensor
products of such physical invariant effects. Since the tensor product
of effects acting on disjoint systems is their parallel composition, the
full invariant effect space is spanned by parallel compositions of
physical invariant effects, each acting on at most two systems.

Finally, by~\Cref{prop:state-tomography,prop:state-tomography-swirl},
invariant effects distinguish invariant states. Therefore, if two
invariant states agree on every parallel composition of invariant effects
acting on at most two systems, they agree on a spanning set of the full
invariant effect space, and hence on every invariant effect. They are
therefore equal.

The analogous statement holds for effect tomography as well.
\endproof
\subsubsection{\texorpdfstring{Proof of~\cref{lem:BilocalProcessTomography}}{Proof of Lemma \ref*{lem:BilocalProcessTomography}}}\label{proof:BilocalProcessTomography}
\BilocalProcessTomography*
\proof
Since the total input $A=A_1\cdots A_n$ and total output
$B=B_1\cdots B_m$ are non-trivial systems, both carry $\lambda$ by
assumption. Hence $V_A$ and $V_B$ share the non-trivial irrep $\lambda$,
so~\Cref{prop:WhenAncillaNeededIrreps} implies that an ancillary system
is required for complete process tomography.

By~\Cref{lem:BilocalStateTomography}, the special representation
structure of $\mathbb Z_2$ implies that the genuine carriers of
$\lambda$ are precisely the individual non-trivial systems, and every
pair of such systems defines a shared-irrep requirement. Hence all
non-trivial input and output systems belong to a single shared-irrep
tomography block. Since $\lambda$ is the only non-trivial irrep of
$\mathbb Z_2$,~\Cref{prop:AncillaAssisitedTesters} implies that the
ancillary system must carry $\lambda$, and that any ancillary system
$E$ carrying $\lambda$ suffices.

Moreover, by~\Cref{lem:BilocalStateTomography}, again using the special
$\mathbb Z_2$ structure, the invariant states on $AE$ and invariant
effects on $BE$ can be spanned by parallel compositions of invariant
states and effects acting on at most two systems.
Since~\Cref{prop:AncillaAssisitedTesters} gives complete process
tomography in terms of invariant states on $AE$ and invariant effects
on $BE$, linearity allows these states and effects to be replaced by
their bilocal spanning components. Thus complete process tomography
can be performed using only such bilocal states and effects. Hence no
genuinely tripartite or higher-partite states or effects are required.
\endproof

\section{Group representation theory for process theories}\label{app:group-rep-theory}
In this appendix, we collect the representation-theoretic results for
finite-dimensional real representations that are used throughout the
paper. For further details on finite-dimensional real representations of compact groups, see~\cite[Chapter~II.6]{brocker_representations_1995}.
\begin{lemma}[Schur's lemma for real representations]\label{lem:schur}
Let $\Gamma_V: G \to \GL(V)$ and $\Gamma_W: G \to \GL(W)$ be irreducible real representations. 
The space of $G$-intertwiners, defined as

\begin{align}
    &\mathrm{Hom}_G(V, W) \nonumber
    \\&:= \{\, T \in \mathrm{Hom}(V, W) 
    \mid T \circ \Gamma_V(g) = \Gamma_W(g) \circ T, \;\forall g \in G \,\},
\end{align}

satisfies
\begin{equation}
    \mathrm{Hom}_G(V, W) \cong
    \begin{cases}
        \{0\}, & \text{if } \Gamma_V \not\cong \Gamma_W,\\[1ex]
        K = \Rl, \Cl, \Hl , & \text{if } \Gamma_V \cong \Gamma_W ,   
    \end{cases}
\end{equation}
where $K$ is the type of the common real irreducible representation. $\mathrm{Hom}_G(V, W) \cong_\Rl  \Rl, \Cl, \Hl$  implies that the vector space of intertwiners
has dimension 1,2 or 4 as a real vector space.

\end{lemma}

\begin{proof}

Recall first that, for irreducible complex representations, Schur's
lemma states that the space of intertwiners between inequivalent
representations is zero, while every intertwiner from an irreducible
representation to itself is a scalar multiple of the identity. The
kernel--image argument underlying the first of these statements is
independent of the underlying field and therefore applies equally to
real representations.

Indeed, let $0\neq T\in\mathrm{Hom}_G(V,W)$. Since $T$ is a
$G$-intertwiner, both $\ker T\subseteq V$ and
$\operatorname{im}T\subseteq W$ are $G$-invariant subspaces. Since
$\Gamma_V$ and $\Gamma_W$ are irreducible, one has
$\ker T=\{0\}$ and $\operatorname{im}T=W$. Hence $T$ is an invertible
$G$-intertwiner. It follows that if
$\Gamma_V\not\cong\Gamma_W$, then
$\mathrm{Hom}_G(V,W)=\{0\}$.

Suppose now that $\Gamma_V\cong\Gamma_W$, and let $S:V\to W$ be an
invertible $G$-intertwiner. Composition with $S^{-1}$ gives a
real-linear isomorphism
$\mathrm{Hom}_G(V,W)\cong_\Rl\mathrm{Hom}_G(V,V)$.
By the same irreducibility argument, every non-zero element of
$\mathrm{Hom}_G(V,V)$ is invertible. Moreover, the inverse of a
$G$-intertwiner is again a $G$-intertwiner. Hence
$\mathrm{Hom}_G(V,V)$ is a finite-dimensional division algebra over
$\Rl$. For an irreducible real representation, this division algebra
is isomorphic to $\Rl$, $\Cl$, or $\Hl$, according to whether the
representation is of real, complex, or quaternionic type,
respectively~\cite[Theorem~6.7]{brocker_representations_1995}.
Therefore
$\mathrm{Hom}_G(V,W)\cong_\Rl\Rl,\Cl$, or $\Hl$, and hence the space of
intertwiners has dimension $1$, $2$, or $4$ as a real vector space.

\end{proof}

A (reducible) representation $\Gamma_V: G \to \GL(V)$ has an isotypic decomposition:
\begin{align}
    V &\cong \bigoplus_\lambda V_{(\lambda)}
 \cong \bigoplus_\lambda V_\lambda^{\oplus d_\lambda} \cong \bigoplus_\lambda \Rl^{d_\lambda} \otimes V_\lambda,\\
 \Gamma_V &\cong \bigoplus_\lambda \Gamma_{(\lambda)}
 \cong \bigoplus_\lambda \Gamma_\lambda^{\oplus d_\lambda} \cong \bigoplus_\lambda \I_{d_\lambda} \otimes \Gamma_\lambda,
 \end{align}
 where $V_\lambda$ are mutually non-isomorphic irreducible representations. The isomorphism $V_{(\lambda)} \cong V_\lambda^{\oplus d_\lambda}$ is non-unique. The non-uniqueness of $V_{(\lambda)} \cong V_\lambda^{\oplus d_\lambda}$ is made apparent using $  V_{(\lambda)} \cong \Rl^{d_\lambda} \otimes V_\lambda$: a choice of basis $\{e_{i_\lambda}\}_{i_\lambda = 1}^{d_\lambda}$ induces a decomposition $V_{(\lambda)} = \sum_{i_\lambda} e_{i_\lambda} \otimes V_\lambda = \bigoplus_{i_\lambda} V_{i_\lambda}$. Different choices of bases induce different direct sum decompositions of $V_{(\lambda)}$ into isomorphic irreducible representations.

\begin{lemma}[Schur's lemma for real reducible representations]\label{lem:schur_red_rep}
     Let
    $\Gamma_V: G \to \GL(V)$ and $\Gamma_W: G \to \GL(W)$
    be real representations of $G$, with
    isotypic decompositions
    \begin{align}
        V &\cong \bigoplus_{\lambda \in \Lambda} \Rl^{d_\lambda} \otimes V_\lambda,
        \\
        W &\cong \bigoplus_{\mu \in \Lambda} \Rl^{e_\mu} \otimes V_\mu,
    \end{align}
    then 
    \begin{align}
        \Hom_G(V, W) &\cong
        \bigoplus_{\lambda \in \Lambda}
        M_{e_\lambda \times d_\lambda}(K_\lambda) \\
        & \cong \bigoplus_{\lambda \in \Lambda}  M_{e_\lambda \times d_\lambda} \otimes_\Rl K_\lambda
    \end{align}
    where $K_\lambda :=\End_G(V_\lambda)$ is the Schur division
    algebra of $V_\lambda$ (isomorphic to $\Rl$, $\Cl$, or $\Hl$). In particular, $\Hom_G(V, W) \neq 0$ if and
    only if $\Gamma_V$ and $\Gamma_W$ share at least one irreducible representation
    (i.e. there exists $\lambda$ with $d_\lambda > 0$ and $e_\lambda > 0$).
\end{lemma}

\begin{proof}
Using the isotypic decompositions of $V$ and $W$, every
$G$-intertwiner decomposes into maps between their isotypic components.
Hence
\begin{equation}
\mathrm{Hom}_G(V,W)
\cong
\bigoplus_{\lambda,\mu\in\Lambda}
\mathrm{Hom}_G
\left(
\Rl^{d_\lambda}\otimes V_\lambda,
\Rl^{e_\mu}\otimes V_\mu
\right).
\end{equation}

Fix $\lambda,\mu\in\Lambda$, and choose bases
$\{e_{i_\lambda}\}_{i_\lambda=1}^{d_\lambda}$ of
$\Rl^{d_\lambda}$ and
$\{f_{j_\mu}\}_{j_\mu=1}^{e_\mu}$ of
$\Rl^{e_\mu}$. Every real-linear map
$T:\Rl^{d_\lambda}\otimes V_\lambda
\to\Rl^{e_\mu}\otimes V_\mu$
is uniquely determined by real-linear maps
$T_{j_\mu i_\lambda}:V_\lambda\to V_\mu$ satisfying
\begin{equation}
T(e_{i_\lambda}\otimes v)
=
\sum_{j_\mu=1}^{e_\mu}
f_{j_\mu}\otimes T_{j_\mu i_\lambda}(v).
\end{equation}
Since $G$ acts trivially on the multiplicity spaces
$\Rl^{d_\lambda}$ and $\Rl^{e_\mu}$, the map $T$ is
$G$-equivariant if and only if
\begin{equation}
T_{j_\mu i_\lambda}\circ\Gamma_\lambda(g)
=
\Gamma_\mu(g)\circ T_{j_\mu i_\lambda}
\end{equation}
for every $g\in G$ and every $i_\lambda,j_\mu$. Thus
\begin{equation}
\scalebox{0.9}{$\mathrm{Hom}_G
\left(
\Rl^{d_\lambda}\otimes V_\lambda,
\Rl^{e_\mu}\otimes V_\mu
\right)
\cong
M_{e_\mu\times d_\lambda}(\Rl)
\otimes_\Rl
\mathrm{Hom}_G(V_\lambda,V_\mu)$}.
\end{equation}

By~\Cref{lem:schur},
\begin{equation}
\mathrm{Hom}_G(V_\lambda,V_\mu)
\cong
\begin{cases}
\{0\}, & \lambda\neq\mu,\\[1ex]
K_\lambda, & \lambda=\mu,
\end{cases}
\end{equation}
where $K_\lambda=\mathrm{End}_G(V_\lambda)$. Therefore all terms with
$\lambda\neq\mu$ vanish, and
\begin{equation}
\mathrm{Hom}_G(V,W)
\cong
\bigoplus_{\lambda\in\Lambda}
M_{e_\lambda\times d_\lambda}(\Rl)
\otimes_\Rl K_\lambda.
\end{equation}

For each $\lambda$, there is a real-linear isomorphism
\begin{equation}
M_{e_\lambda\times d_\lambda}(\Rl)
\otimes_\Rl K_\lambda
\cong
M_{e_\lambda\times d_\lambda}(K_\lambda).
\end{equation}
Here $M_{e_\lambda\times d_\lambda}(K_\lambda)$ denotes the real
vector space of $e_\lambda\times d_\lambda$ matrices whose entries lie
in $K_\lambda$. Since $K_\lambda=\mathrm{End}_G(V_\lambda)$ consists
of real-linear $G$-equivariant endomorphisms of $V_\lambda$, these
matrices represent real-linear $G$-intertwiners between the
corresponding isotypic components. In particular, when
$K_\lambda\cong\Cl$ or $\Hl$, its elements are still being regarded
here as real-linear maps, and the tensor product is taken over $\Rl$.

Consequently,
\begin{equation}
\mathrm{Hom}_G(V,W)
\cong
\bigoplus_{\lambda\in\Lambda}
M_{e_\lambda\times d_\lambda}(K_\lambda).
\end{equation}
Since $K_\lambda$ contains the identity endomorphism of $V_\lambda$,
the summand
$M_{e_\lambda\times d_\lambda}(K_\lambda)$ is non-zero if and only if
$d_\lambda>0$ and $e_\lambda>0$. Hence
$\mathrm{Hom}_G(V,W)\neq0$ if and only if $\Gamma_V$ and $\Gamma_W$
share at least one irreducible representation.
\end{proof}

\begin{lemma}\label{lem:G_inv_GtimesG}
    $(V \otimes W)^G \neq V^G \otimes W^G$ if and only if $V$ and $W$ share at least one non-trivial representation.
\end{lemma}

\begin{proof}
    Using $V\otimes W \cong \Hom(V,W)$ we have:
    \begin{align}
        (V\otimes W )^G \cong&  \bigoplus_{\lambda \in \Lambda}  M_{e_\lambda \times d_\lambda} \otimes_\Rl K_\lambda \\
        \cong&  M_{e_0 \times d_0} \oplus \bigoplus_{\lambda \in \Lambda / 0}  M_{e_\lambda \times d_\lambda} \otimes_\Rl K_\lambda
    \end{align}
    Moreover,
    \begin{align}
        V^G \otimes W^G \cong \Rl^{d_0} \otimes V_0 \otimes  \Rl^{e_0} \otimes V_0 \cong M_{e_0 \times d_0}
    \end{align}
    Therefore $(V\otimes W )^G \neq V^G\otimes W^G$ if and only if there exists a $\lambda \neq 0$ such that $d_\lambda, e_\lambda \neq 0$.
\end{proof}

\subsection{Examples}

\begin{example}[Real irreducible representation of real type] Consider the standard representation
\begin{equation}
\Gamma:SO(3)\to GL(\Rl^3),\qquad \Gamma(R)=R,
\end{equation}
where
\begin{equation}
SO(3)=\{R\in M_3(\Rl)\mid R^{\mathsf T}R=I,\ \det R=1\}.
\end{equation}
This is real irreducible (i.e., there are no invariant subspaces in
$\Rl^3$ apart from $0$ and $\Rl^3$). Moreover, the only real-linear
maps commuting with every $R\in SO(3)$ are scalar multiples of the
identity. Hence
\begin{equation}
\mathrm{End}_G(\Rl^3)=\{aI\mid a\in\Rl\}\cong\Rl.
\end{equation}
Therefore, the standard representation of $SO(3)$ is of real type.
\end{example}

\begin{example}[Real irreducible representation of complex type]
Consider the fundamental representation of $\SO(2)$:
\begin{align}
    \theta \mapsto  R(\theta) = \begin{pmatrix}
        \cos \theta & -\sin \theta \\
        \sin \theta & \cos \theta
    \end{pmatrix}
\end{align}
This is real irreducible (i.e., there are no invariant subspaces in $\mathbb R^2$ apart from $0$ and $\mathbb R^2$); however, both $I$ and $J$ commute with the representation where,
\begin{align}
    J = \begin{pmatrix}
        0 & -1 \\
        1 & 0 
    \end{pmatrix}.
\end{align}
The intertwiner space is therefore isomorphic to $\Cl \cong \Rl^2$. 

The above follows because the complexification of $R(\theta)$ is complex reducible:
\begin{align}
    L R(\theta) L^{-1} = \begin{pmatrix}
        e^{i \theta} & 0 \\
        0 & e^{-i \theta}
    \end{pmatrix}
\end{align}
\end{example}

\begin{example}[Real irreducible representation of quaternionic type]
    Consider first the fundamental representation of $\SU(2)$ on $\Cl^2$ and the real isomorphisms $\Cl^2 \to \Rl^4$, $(z_1, z_2) \mapsto (x_1, y_1,x_2,y_2)$ where $z_j = x_j + i y_j$. Under this map the fundamental representation of $\SU(2)$ maps to a real irreducible representation of $\SU(2)$ on $\Rl^4$.

    Let us define the following operators on $\Cl^2 \cong \Rl^4$:
    \begin{align}
        \I&: (z_1,z_2) \mapsto (z_1,z_2) ,\\
        I&:  (z_1,z_2) \mapsto (i z_1,i z_2) ,\\
        J&: (z_1,z_2) \mapsto (-\bar z_2 , \bar z_1), \\
        K&: (z_1, z_2) \mapsto (- i \bar z_2, i \bar z_1).
    \end{align}
    We observe that both $\I$ and $I$ are complex linear whilst $J$ and $K$ are conjugate linear. All four of these are real linear:
    \begin{align}
        \I&: (x_1, y_1,x_2,y_2)\mapsto (x_1, y_1,x_2,y_2),\\
        I&:  (x_1, y_1,x_2,y_2) \mapsto (-y_1, x_1,-y_2,x_2) ,\\
        J&: (x_1, y_1,x_2,y_2)\mapsto (-x_2, y_2,x_1,-y_1), \\
        K&: (x_1, y_1,x_2,y_2)\mapsto (-y_2, -x_2,-y_1,x_1).
    \end{align}

    Both $\I$ and $I$ are proportional (as operators on a complex space) to the identity and commute with the fundamental representation, which is consistent with Schur's lemma for complex representations. $J$ and $K$ are not proportional to the identity but commute with the representation, which is not in contradiction with Schur's lemma since they are not complex-linear operators. 

    Hence the complex space of complex linear intertwiners is one dimensional, and the complex space of conjugate complex intertwiners is one dimensional, as can be seen from $I = i \I$ and $K = i J$. As a real vector space of real intertwiners this is a four dimensional vector space. 

    The above can be derived from the fact that under the complexification $\Rl^4 \to \Cl^4$ the representation is complex reducible and of the form $U \oplus U$, which by Schur's lemma has a complex intertwiner space of dimension $4$.
\end{example}

\begin{example}[the Reynolds (i.e., averaging) operators for real irreps of complex type]
Consider the fundamental representation of $\SO(2)$:
\begin{align}
    \theta \mapsto R(\theta)
    =
    \begin{pmatrix}
        \cos\theta & -\sin\theta \\
        \sin\theta & \cos\theta
    \end{pmatrix}.
\end{align}
Then, for
\[
L=\begin{pmatrix}a&b\\c&d\end{pmatrix},
\]
the Reynolds (averaging) operator on $\End(V)$ is
\begin{align}
    \mathcal{R}_{\End(V)}(L)
    &= \int R(\theta)LR(-\theta)\,d\theta \\
    &= \frac12
    \begin{pmatrix}
        a+d & b-c\\
        c-b & a+d
    \end{pmatrix}.
\end{align}
On the other hand,
\begin{align}
    \mathcal{R}\circ L\circ\mathcal{R} 
    &= \iint R(\theta')LR(-\theta)\,d\theta\,d\theta' \\
    &= 0,
\end{align}
where $\calr:V \to V^G$ is given by $\calr:= \int R(\theta)\,d\theta=0$.
\end{example}

\section{Canonical representation of locally tomographic theories}\label{app:RepofTLGPT}

\FVectRep*
\begin{proof}
Throughout the proof, we take $U_A^B=\spann(P_A^B)$ for every process space. We first note that local tomography then separates not only physical processes, but also their linear span. Since $P_A^B$ is convex, contains $0$, and spans $U_A^B$, every $x\in U_A^B$ can be written as $x=\lambda(f-h)$ for some $f,h\in P_A^B$ and $\lambda\geq0$, by separating the positive and negative parts of a linear combination of physical processes and rescaling them into $P_A^B$. If $\lambda=0$, then $x=0$. Otherwise, if $e_B\circ x\circ s^A=0$ for all physical $s^A$ and $e_B$, then $f$ and $h$ give identical local statistics. Local tomography therefore gives $f=h$, and hence $x=0$. The same argument applies to process spaces with arbitrary multipartite inputs and outputs.

$(\Rightarrow)$ Suppose that $\cP$ is locally tomographic. We first show that states and effects are dual to one another. Composition of a state $s^A\in U^A$ with an effect $e_A\in U_A$ gives a scalar $e_A\circ s^A\in\Rl$. Since composition is bilinear, every effect $e_A$ defines a linear functional on $U^A$. We therefore obtain a linear map $\iota_A:U_A\to(U^A)^*$ given by $\iota_A(e_A)(s^A)=e_A\circ s^A$.

The map $\iota_A$ is injective. Indeed, if $\iota_A(e_A)=\iota_A(e_A')$ for two physical effects, then $e_A\circ s^A=e_A'\circ s^A$ for every physical state $s^A$. Local tomography applied to the two effects therefore gives $e_A=e_A'$. By the observation above, if $\iota_A(x)=0$ for some $x\in U_A$, then $x=0$. Hence $\iota_A$ is injective.

The same argument with states and effects interchanged gives an injective map $U^A\to(U_A)^*$. Hence $\dim U_A\leq\dim U^A$ and $\dim U^A\leq\dim U_A$. Since the spaces are finite dimensional, they have the same dimension, and therefore $ U_A\cong(U^A)^*$.

We next determine the state space of a composite system. Since parallel composition defines a bilinear map $\boxtimes:U^A\times U^B\to U^{A,B}$, the universal property of the tensor product gives a unique linear map $\alpha_{A,B}:U^A\otimes U^B\to U^{A,B}$ satisfying $\alpha_{A,B}(s^A\otimes s^B)=s^A\boxtimes s^B$.

We first show that $\alpha_{A,B}$ is injective. Suppose that $\alpha_{A,B}(x)=0$. Take arbitrary effects $e_A\in U_A$ and $e_B\in U_B$. Applying the product effect $e_A\boxtimes e_B$ to $\alpha_{A,B}(x)$ gives zero. By bifunctoriality, applying $e_A\boxtimes e_B$ after parallel composition of states is the same as applying $e_A$ and $e_B$ separately. Thus
\begin{equation}
    \bigl(\iota_A(e_A)\otimes\iota_B(e_B)\bigr)(x)=0
    \qquad
    \forall e_A\in U_A,\ e_B\in U_B.
\end{equation}

Choose bases $\{e_i\}_i$ of $U_A$ and $\{f_j\}_j$ of $U_B$. Since $\iota_A$ and $\iota_B$ are isomorphisms, $\{\iota_A(e_i)\}_i$ and $\{\iota_B(f_j)\}_j$ are bases of $(U^A)^*$ and $(U^B)^*$, respectively. Hence $\{\iota_A(e_i)\otimes\iota_B(f_j)\}_{i,j}$ is a basis of $(U^A)^*\otimes(U^B)^*\cong(U^A\otimes U^B)^*$. We have already shown that every such product functional vanishes on $x$, so every functional in $(U^A\otimes U^B)^*$ vanishes on $x$. Therefore $x=0$. Thus $\alpha_{A,B}$ is injective. Its domain is $U^A\otimes U^B$ and its codomain is $U^{A,B}$, so finite dimensionality implies that the dimension of the domain cannot exceed that of the codomain. Since $\dim(U^A\otimes U^B)=\dim U^A\dim U^B$, we obtain
\begin{equation}
    \dim U^{A,B}
    \geq
    \dim (U_A\otimes U_B)^*
    =
    \dim U_A\dim U_B.
\end{equation}
We now obtain the opposite inequality from local tomography. A bipartite state is completely determined by the probabilities obtained from local effects. The map $U^{A,B}\to(U_A\otimes U_B)^*$ sends each bipartite state $s^{A,B}$ to the functional which records its value on every product of local effects,
\begin{equation}
    e_A\otimes e_B
    \longmapsto
    (e_A\boxtimes e_B)\circ s^{A,B}.
\end{equation}
Thus, if two physical states $s^{A,B}$ and $t^{A,B}$ have the same image under this map, then
\begin{equation}
    (e_A\boxtimes e_B)\circ s^{A,B}
    =
    (e_A\boxtimes e_B)\circ t^{A,B}
\end{equation}
for every pair of physical local effects $e_A$ and $e_B$. These scalars are exactly the probabilities obtained by applying the two local effects to the bipartite states. Thus the two states give the same local statistics for every choice of local effects, and local tomography implies $s^{A,B}=t^{A,B}$. By the observation at the beginning of the proof, the same conclusion holds for arbitrary $x,y\in U^{A,B}$. Indeed, if $x$ and $y$ have the same image, then $(e_A\boxtimes e_B)\circ(x-y)=0$ for every pair of physical local effects $e_A$ and $e_B$. The observation above therefore gives $x-y=0$, so $x=y$. Hence the map is injective. Since an injective linear map cannot have a domain of larger dimension than its codomain, and $\dim (U_A\otimes U_B)^*=\dim U_A\dim U_B$, we obtain
\begin{equation}
    \dim U^{A,B}\leq\dim U_A\dim U_B.
\end{equation}
Since $\dim U_A=\dim U^A$ and $\dim U_B=\dim U^B$, the two bounds coincide, and hence $\dim U^{A,B}=\dim U^A\dim U^B$.

The map $\alpha_{A,B}$ is injective and its domain and codomain have the same dimension, so it is an isomorphism. Therefore $U^{A,B}\cong U^A\otimes U^B$. The same argument, with states and effects exchanged, applies to composite effect spaces. Repeating these arguments gives
\begin{equation}
    U^{A_1,\ldots,A_n}
    \cong
    \bigotimes_{i=1}^n U^{A_i}\;,
       \qquad   U_{A_1,\ldots,A_n}
    \cong
    \bigotimes_{i=1}^n U_{A_i}\;.
\end{equation}
We use these isomorphisms from now on to identify composite state and effect spaces with the corresponding tensor products. Recall that every $U_A^B$ is a finite-dimensional real vector space, so in particular each $U^A$ is an object of $\mathbf{FVect}_{\mathbb{R}}$.

We now define the representation in $\mathbf{FVect}_{\mathbb{R}}$. On systems, set $\cm(A):=U^A$. For each pair of systems $A$ and $B$, define the linear map $\cm_A^B:U_A^B\to\Hom(U^A,U^B)$ by $\cm_A^B(x)(s^A):=x\circ s^A$ for every $x\in U_A^B$ and $s^A\in U^A$. This map is linear because sequential composition is bilinear.

When the input and output systems are clear from context, we suppress the labels $A$ and $B$ and simply write $\cm$ instead of $\cm_A^B$. Thus, for $x\in U_A^B$, $\cm(x)$ means $\cm_A^B(x)$. In particular, for a physical process $f_A^B\in P_A^B$, $\cm(f_A^B)$ is the linear map $U^A\to U^B$ given by $\cm(f_A^B)(s^A)=f_A^B\circ s^A$.

The map $\cm_A^B$ is injective. Suppose that $\cm_A^B(x)=0$. Then $x\circ s^A=0$ for every physical state $s^A$. Composing the resulting output with any physical effect $e_B$ gives $e_B\circ x\circ s^A=0$ for every $s^A$ and $e_B$. These are precisely the local testers appearing in local tomography for a process from $A$ to $B$. Hence, by the observation at the beginning of the proof, $x=0$.

It remains to show that the image of $\cm_A^B$ is the whole of $\Hom(U^A,U^B)$. Take a physical effect $e_A\in P_A$ and a physical state $s^B\in P^B$. Their parallel composition $e_A\boxtimes s^B$ is a physical process from $A$ to $B$. Its action on an input vector $v\in U^A$ is $v\longmapsto \iota_A(e_A)(v)s^B$. Thus, for every input $v\in U^A$, the output is a scalar multiple of the same fixed vector $s^B$. Hence the image of this map is contained in the one-dimensional subspace $\spann\{s^B\}$. Since the rank of a linear map is the dimension of its image, any nonzero map of this form has rank one. Thus measure-and-prepare processes are represented by rank-one linear maps.

The physical effects span $U_A$, which we have identified with $(U^A)^*$, and the physical states span $U^B$. Thus, for any $\varphi\in(U^A)^*$ and $w\in U^B$, we can write $\varphi$ as a linear combination of functionals coming from physical effects and $w$ as a linear combination of physical states. Expanding the product $\varphi(v)w$ therefore expresses the rank-one map $v\mapsto\varphi(v)w$ as a linear combination of images of physical measure-and-prepare processes. Thus an arbitrary rank-one map need not itself be the image of a single physical measure-and-prepare process, but it is always a linear combination of such images.

Now every linear map $T:U^A\to U^B$ is itself a sum of rank-one maps. Indeed, if $\{v_i\}_i$ is a basis of $U^A$ with dual basis $\{\varphi_i\}_i$, then $T(v)=\sum_i\varphi_i(v)T(v_i)$, where each summand $v\mapsto\varphi_i(v)T(v_i)$ is rank one. Hence $T$ is a linear combination of images of physical measure-and-prepare processes. Taking the same linear combination of the corresponding physical measure-and-prepare processes gives an element $x\in U_A^B$, and linearity of $\cm_A^B$ gives $\cm_A^B(x)=T$. Thus every $T\in\Hom(U^A,U^B)$ has a preimage under $\cm_A^B$, so $\cm_A^B$ is surjective. Together with injectivity, this gives $U_A^B\cong\Hom(U^A,U^B)$. Applying this to composite inputs and outputs, and using the tensor-product identifications above, gives
\begin{equation}
    U_{A_1,\ldots,A_n}^{B_1,\ldots,B_m}
    \cong
    \Hom\left(
    \bigotimes_{i=1}^n U^{A_i},
    \bigotimes_{j=1}^m U^{B_j}
    \right).
\end{equation}

We finally check that this representation respects the compositional structure. Here $\cm$ denotes the representation of the whole theory, assigning $\cm(A)=U^A$ to each system $A$ and $\cm(f_A^B):U^A\to U^B$ to each physical process $f_A^B$. By contrast, $\cm_A^B:U_A^B\to\Hom(U^A,U^B)$ denotes its linear extension to the whole ambient process space, so $\cm(f_A^B)=\cm_A^B(f_A^B)$ for every physical process.

First consider sequential composition. Let $h\in P_A^B$ and $f\in P_B^C$. For every state $s^A\in U^A$, the definition of $\cm$ and associativity of sequential composition give
\begin{align}
 \cm(f\circ h)(s^A)
    &=
    (f\circ h)\circ s^A
    =
    f\circ\bigl(h\circ s^A\bigr)
    \nonumber\\
    &=
    \cm(f)\bigl(\cm(h)(s^A)\bigr).
\end{align}
Thus $\cm(f\circ h)$ and $\cm(f)\circ\cm(h)$ have the same action on every $s^A\in U^A$, and hence $\cm(f\circ h)=\cm(f)\circ\cm(h)$.

Now consider parallel composition, with $f\in P_A^B$ and $g\in P_C^D$. For product states $s^A\boxtimes t^C$, bifunctoriality gives
\begin{equation}
    (f\boxtimes g)\circ(s^A\boxtimes t^C)
    =
    (f\circ s^A)\boxtimes(g\circ t^C).
\end{equation}
Recall that the identification $U^{A,C}\cong U^A\otimes U^C$ is given by the isomorphism $\alpha_{A,C}$ satisfying $\alpha_{A,C}(s^A\otimes t^C)=s^A\boxtimes t^C$, and similarly for $U^{B,D}$. Therefore, the linear map representing $f\boxtimes g$, expressed on the tensor-product spaces, is
\begin{equation}
    \alpha_{B,D}^{-1}
    \circ
    \cm_{A,C}^{B,D}(f\boxtimes g)
    \circ
    \alpha_{A,C}.
\end{equation}
Its action on a product state is
\begin{align}
    &\alpha_{B,D}^{-1}
    \left(
        \cm_{A,C}^{B,D}(f\boxtimes g)
        \bigl(\alpha_{A,C}(s^A\otimes t^C)\bigr)
    \right)
    \nonumber\\
    &=
    \alpha_{B,D}^{-1}
    \left(
        (f\boxtimes g)\circ(s^A\boxtimes t^C)
    \right)
    \nonumber\\
    &=
    \alpha_{B,D}^{-1}
    \left(
        (f\circ s^A)\boxtimes(g\circ t^C)
    \right)
    \nonumber\\
    &=
    \cm(f)(s^A)\otimes\cm(g)(t^C),
\end{align}
where the second equality follows from bifunctoriality. Using the tensor-product identifications fixed above, this shows that $\cm(f\boxtimes g)$ and $\cm(f)\otimes\cm(g)$ agree on every product state $s^A\otimes t^C$. Since such product states span $U^A\otimes U^C$, the two linear maps agree on the whole space. Therefore
\begin{equation}
    \cm(f\boxtimes g)
    =
    \cm(f)\otimes\cm(g).
\end{equation}

The identity process is also preserved. Indeed, for every $s^A\in U^A$, $\cm(\mathcal{I}_A)(s^A)=\mathcal{I}_A\circ s^A=s^A$, so $\cm(\mathcal{I}_A)$ is the identity linear map on $U^A$. Similarly, for the swap process $\sigma_{A,B}$, its defining property gives $\sigma_{A,B}\circ(s^A\boxtimes t^B)=t^B\boxtimes s^A$ for all states $s^A$ and $t^B$. Under the tensor-product identifications fixed above, this means $\cm(\sigma_{A,B})(s^A\otimes t^B)=t^B\otimes s^A$. This is exactly the action of the usual swap map $U^A\otimes U^B\to U^B\otimes U^A$ on product states. Since product states span $U^A\otimes U^B$, the two maps are equal.

Finally, consider the scalar process space $U_I^I=\mathbb{R}$. Write $\cm(1)=c\cdot\operatorname{id}_{\mathbb{R}}$ for some $c\in\mathbb{R}$. Since $1\circ1=1$ and $\cm$ preserves sequential composition, one has $c=c^2$. Injectivity of $\cm$ implies $c\neq0$, and hence $c=1$. Therefore $\cm(1)=\operatorname{id}_{\mathbb{R}}$. By linearity, $\cm(r)=r\cdot\operatorname{id}_{\mathbb{R}}$ for every $r\in\mathbb{R}$, so $\cm$ preserves scalars.

Therefore $\cm$ preserves sequential and parallel composition, identities, swaps, and scalars. Since each $\cm_A^B$ is linear, $\cm$ is linear on every process space, and hence $\cm$ is a linear, scalar-preserving, diagram-preserving representation of $\mathcal{P}$ in $\mathbf{FVect}_{\mathbb{R}}$.

Since each $\cm_A^B$ is injective, $\cm$ gives an isomorphism between $\cP$ and its image in $\mathbf{FVect}_{\Rl}$. Moreover, since $U_A^B=\spann(P_A^B)$ and $\cm_A^B$ is linear, we have $  \cm_A^B(U_A^B)= \spann\bigl(\cm(P_A^B)\bigr)$.
But we have already shown that $\cm_A^B(U_A^B)=\Hom(U^A,U^B)$. Hence
\begin{equation}
    \spann\bigl(\cm(P_A^B)\bigr)
    =
    \Hom(U^A,U^B).
\end{equation}
Thus the physical processes in the image span the full corresponding spaces of linear maps. Hence the image of $\cm$ is a spanning subtheory of $\mathbf{FVect}_{\Rl}$, and $\cP$ is isomorphic to it.

$(\Leftarrow)$ Suppose that $\cP$ is isomorphic to a spanning subtheory $\mathcal{Q}$ of $\mathbf{FVect}_{\Rl}$. Let $f,h:\bigotimes_iV_{A_i}\to\bigotimes_jV_{B_j}$ be two processes in $\mathcal{Q}$ with the same local statistics. For physical local states $s_i$ and effects $e_j$, write $s:=\bigotimes_i s_i$ and $e:=\bigotimes_j e_j$. Having the same local statistics means that $e\circ f\circ s=e\circ h\circ s$ for every such $s$ and $e$.

Since the physical product states span $\bigotimes_iV_{A_i}$, any $x$ can be written as $x=\sum_k c_k s_k$ with physical product states $s_k$. For any physical product effect $e$, the assumed equality of local statistics gives $e\circ f\circ s_k=e\circ h\circ s_k$ for every $k$. Hence, by linearity,
\begin{equation}
    e\circ f\circ x
    =
    \sum_k c_k(e\circ f\circ s_k)
    =
    \sum_k c_k(e\circ h\circ s_k)
    =
    e\circ h\circ x.
\end{equation}

The physical product effects span $(\bigotimes_jV_{B_j})^*$, so by linearity every functional on the output space takes the same value on $f(x)$ and $h(x)$. Hence $f(x)=h(x)$ for every $x$, and therefore $f=h$. Thus $\mathcal{Q}$ is locally tomographic, and hence so is $\cP$.
\end{proof}
\end{document}